\documentclass[a4paper,11pt]{article}
\usepackage{jheppub} 

\usepackage{ifthen}
\usepackage[utf8]{inputenc}
\usepackage{amsmath,amssymb,mathtools,amsthm,stmaryrd,quiver}
\usepackage{tikz}
\usepackage{tikz-cd}
\usetikzlibrary{arrows.meta,calc,decorations.pathmorphing,decorations.markings,patterns.meta}
\usepackage{caption}
\usepackage{subcaption}
\usepackage{hyperref}
\tikzcdset{scale cd/.style={every label/.append style={scale=#1},
    cells={nodes={scale=#1}}}}
\usepackage{enumitem}
\usepackage{subcaption} 

\tikzset{
  smallbox/.style={
    box,
    inner sep=0.3ex,
    minimum size=0pt,
    outer sep=0pt,
    text height=1.5ex,
    text depth=.25ex
  }  
}
\tikzset{
    every picture/.style={draw=black,thick}  
}

\input{macros.sty}
\newcommand \cau {\mathrm{Cau}}
\newcommand \mayseven[1]{}
\newcommand \wtT {{\widetilde T}}
\newcommand \conv {\circledcirc}
\newcommand \coevC {\Id_\cC^\spadesuit}
\newcommand \evC {\Id_\cC^\heartsuit}
\newcommand \oplax {\mathrm{oplax}}
\newcommand \co {\mathrm{co}}
\newcommand \bLTamb {\overline{\mathrm{LTamb}}}
\newcommand \LTamb {\mathrm{LTamb}}
\newcommand {\natiso}{\overset{\sim}{\Rightarrow}}

\newcommand {\six}[1]{\ignorespaces}
\newcommand \bcV {\overline{\cV}}

\newcommand \isom			{\overset{\sim}{\to}}
\newcommand \rep {\mathrm{Rep}}

\newcommand\vect{\mathrm{Vect_{f.d.}}}
\newcommand\Vect{\mathrm{Vect}}

\newcommand\Mor{\mathrm{Mor}}
\newcommand \nat {\mathrm{Nat}}

\newcommand{\enf}[3][]{{}_{#1} [#2,#3]}
\renewcommand{\id}{1}
\renewcommand{\Id}{1}

\usetikzlibrary{shadows.blur} 
\tikzset{
  wire/.style    ={line width=1pt},
  box/.style     ={draw,minimum height=4mm,minimum width=10mm,very thick,fill=white},
  bullet/.style  ={circle,inner sep=.7pt,fill},
  curly/.style   ={decorate, decoration={snake,amplitude=.4mm,segment length=2.2mm}},
  dot/.style     ={circle,inner sep=1.2pt,fill},
  ->-/.style     ={decoration={markings,mark=at position .55 with {\arrow{Stealth}}},
                   postaction={decorate}}
}

\usepackage{chngcntr}  

\usepackage[color,all,2cell,arrow,matrix]{xy} \UseAllTwocells \SilentMatrices
\newcommand \ladj           {\dashv}
\newcommand \radj           {\vdash}
\newcommand \copr           {\mathrm{copr}}
\newcommand \diagram		{\xymatrix}      
\newcommand \os             {\otimes}
\newcommand \defdtobe       {\coloneq}
\renewcommand{\:}           {\colon}
\newcommand \arprof         {\nrightarrow}

\newcommand {\eqnn}[1]       {\begin{equation}#1\end{equation}}

\newcommand {\ctikz}[1]{\begin{array}{c}\begin{tikzpicture}#1\end{tikzpicture}\end{array}}

\newcommand{\mynode}[5]{
    \draw[fill=white] (#1+#3,#2+#4)rectangle(#1-#3,#2-#4) node[midway] {$#5$};
}
\newcommand {\onenodenolabeledge}[1]{
    \ctikz{ 
        \draw (1.5,0)--(-0.5,0);
        \mynode{0.5}{0}{0.3}{0.3}{#1}
    }
    }
\newcommand {\twonodenolabeledge}[2]{
    \ctikz{[scale=1]
        \draw (1.5,0)--(-1.5,0);
        \mynode{0.5}{0}{0.3}{0.3}{#2}
        \mynode{-0.5}{0}{0.3}{0.3}{#1}
    }
}
\newcommand {\threenodenolabeledge}[3]{
\ctikz{[scale=1]
    \draw (2.5,0)--(-1.5,0);
    \mynode{1.5}{0}{0.3}{0.3}{#3}
    \mynode{0.5}{0}{0.3}{0.3}{#2}
    \mynode{-0.5}{0}{0.3}{0.3}{#1}
}
}
\newcommand {\fournodenolabeledge}[4]{
\ctikz{[scale=1]
    \draw (3.5,0)--(-1.5,0);
    \mynode{2.5}{0}{0.3}{0.3}{#4}
    \mynode{1.5}{0}{0.3}{0.3}{#3}
    \mynode{0.5}{0}{0.3}{0.3}{#2}
    \mynode{-0.5}{0}{0.3}{0.3}{#1}
}
}
\newcommand{\basefiveturn}{
    \draw[color3] (0,0)--(-1,0);
    \draw[color3] (-1,0) arc (270:90:0.5);
    \draw[color3] (-1,1)--(1.5,1);
    \draw[color3] (1.5,1) arc (-90:90:0.5);
    \draw[color3] (1.5,2)--(0.5,2);
    \draw[color3] (0.5,2) arc (270:90:0.5);
    \draw[color3] (0.5,3) -- (3,3);
    \draw[color3] (3,3) arc (-90:90:0.5);
    \draw[color3] (3,4)--(2,4);
    \draw[color3] (2,4) arc (270:90:0.5);
    \draw[color3] (2,5)--(4.5,5);
}
\newcommand{\basethreeturn}{
    \draw[color3] (0,0)--(-1,0);
    \draw[color3] (-1,0) arc (270:90:0.5);
    \draw[color3] (-1,1)--(1.5,1);
    \draw[color3] (1.5,1) arc (-90:90:0.5);
    \draw[color3] (1.5,2)--(0.5,2);
    \draw[color3] (0.5,2) arc (270:90:0.5);
    \draw[color3] (0.5,3) -- (3,3);
}
\newcommand{\baseoneturn}{
    \draw[color3] (0,0)--(-1,0);
    \draw[color3] (-1,0) arc (270:90:0.5);
    \draw[color3] (-1,1)--(1.5,1);
}
\newcommand{\basetwoturn}{
    \draw[color3] (0,0)--(-2.5,0);
    \draw[color3] (-2.5,0) arc (270:90:0.5);
    \draw[color3] (-2.5,1)--(-1.5,1);
    \draw[color3] (-1.5,1) arc (-90:90:0.5);
    \draw[color3] (-1.5,2)--(-4,2);
}

\newcommand{\myparabola}[5]{ 
	\begin{scope}[xshift = #2 cm, yshift = #3 cm]
		\draw (0,0) parabola bend ({((#4-#2)+(#5-#3)/#1)/2},{((#4-#2)+(#5-#3)/#1)*((#4-#2)+(#5-#3)/#1)/(4*(#4-#2)/#1)}) ({#4-#2},{#5-#3});
	\end{scope}
}
\newcommand{\mygrid}[2][off]{
  \ifthenelse{\equal{#1}{on}}{%
    #2%
  }{}%
}
\tikzset{
    nattrans/.style={
        double equal sign distance,
        -{Implies[length=6pt,open]},
        line width=0.6pt,
        shorten >=3pt, shorten <=3pt
      },
    nat/.style={
        double equal sign distance,
        -{Implies[length=6pt,open]},
        line width=0.6pt,
        shorten >=3pt, shorten <=3pt,
        execute at begin node=$\scriptstyle,
        execute at end node=$
      }
}

\definecolor{Mybrown}{RGB}{223,145,16} 
\definecolor{Mygray}{RGB}{155,155,155}

\colorlet{color3}{Mybrown}
\colorlet{color4}{red}
\colorlet{color1}{cyan!50!gray!40}
\colorlet{color2}{blue!40!gray!40}

\makeatletter
\gdef\@fpheader{}
\makeatother

\title{Pro-Tensor Network}

\author[a,1]{Gen Yue,\note{These authors contributed equally.}}
\author[a,1]{Ansi Bai,}
\author[a]{Linqian Wu,}
\author[a,b,2]{and Tian Lan\note{Corresponding author.}}
\affiliation[a]{Department of Physics, The Chinese University of Hong Kong, Shatin, New Territories, Hong Kong, China}
\affiliation[b]{The State Key Laboratory of Quantum Information Technologies and Materials, The Chinese University of Hong Kong, Shatin, New Territories, Hong Kong, China}
\emailAdd{tlan@cuhk.edu.hk}

\abstract{
We introduce the \emph{pro-tensor network}, a categorification of the tensor network, as a fully rigorous yet graphically transparent framework for studying the collection of many many-body theories, which we dub \emph{many-many-body theory}. We provide a comprehensive toolbox for the graphical calculations using pro-tensor networks. As applications, we recover the Levin-Wen model as a  ``uniform" pro-tensor network and generalize a result of Kitaev and Kong by characterizing particles as modules over promonads. One can also interpret the string-net pro-tensor network as the space of symmetric tensor networks, thus our framework also applies to the study of generalized symmetry and topological holography. Notably, our generalization dispenses with the assumptions of semisimplicity, finiteness, and rigidity, potentially facilitating the exploration of many-body physics beyond these constraints.
}

\begin{document}
\maketitle
\flushbottom

\section{Introduction}

In many-body theory, it is a central problem to understand phases of matter and phase transitions. It is worth noting that a \emph{phase} is not a single theory or model, but a \emph{collection} of many-body theories; the properties usually associated to a phase or a phase transition, such as symmetry, superconductivity, topological order, critical exponents, etc., are \emph{universal} across a collection of many-body theories. It is becoming more and more important to study these universal properties, and a suitable framework or even a new paradigm is in urgent need. In a series of talks by Kong~\cite{KongTalk}, such idea has been emphasized as the ``global many-body theory". Here ``global'' refers to the entire landscape of many-body theories; however, in physics literature, it might be confused with ``global'' (vs local) in a single system (such as global symmetry). To avoid confusion, we will use a much more straightforward phrase, \emph{many-many-body theory}, for the perspective of studying the universal properties of many many-body theories all together.

Historically, the idea of ``collecting many systems and study them all together" has been proved very successful. When there are many independent systems, instead of studying a single system among them (which might involve occasionality), it is common to the study the probability distribution of their properties. This means that one is secretly considering an ensemble of these systems. Such idea lies underneath every statistical science. In physics, particularly, statistical mechanics and quantum mechanics are within such paradigm.

For a generic quantum many-body system, possibly with interactions between parts, the second-quantization formalism was successful. In the second-quantization perspective, one focuses on the collective motion modes (such as phonons) of the entire many-body system. When the interactions between these collective modes are weak, solving the many-body system is feasible. 

We see a magical coincidence: the (first) quantization deals with the probabilistic description of (secretly many independent) quantum particles; the second-quantization deals with the properties of many-body systems. Now we want to further upgrade our perspective to many-many-body theory, and we might call our efforts as exploring the third-quantization.
The mathematics we rely on is higher category theory, and specifically, categorification. The most naive example of categorification is
\begin{enumerate}
    \item The collection of many elements form a set (0-category);
    \item The collection of sets (0-categories) form a 1-category;
    \item The collection of (1-)categories form a 2-category;
    \item So on and so forth. In general the collection of $n$-categories form an $(n+1)$-category.
\end{enumerate}
In this naive example, categorification means studying the collection of structures in the previous level. Conceptually, quantization and categorification are along the same direction.

In this work we approach the study of many-many-body theory via \emph{tensor network}, which might be the most successful tool for presenting and calculating many-body systems, and also have broad applications beyond physics, for example, in machine learning and AI science. The framework we need is the categorification of tensor network. Naively speaking, we would study the space of a collection of tensor networks. The crucial mathematical object in our work is the profunctor, and we thus call the categorified tensor network as the \emph{pro-tensor network}. 
The pro-tensor network is a graphical, intuitive, but also fully rigorous framework, which we believe well-suited for the many-many-body theory. 

The pro-tensor network naturally incorporates \emph{generalized symmetries}~\cite{Kong_Zheng_2018,Ji_2020, Kong_2020Gapless,Kong_2020,Chen_2020,Kong_2020Classi, Lichtman_2021,Kong_2022,KongZheng_2022,Chatterjee_2023,Moradi_2023, freed2024,KongZheng_2024,Lan_2024,XuZhang_2024,WenYe2024, Bhardwaj2024,LanYueWang_2024,Jia_2024,cordova2024,cordova2024representationtheorysolitons,choi2026generalizedtubealgebrassymmetryresolved,Jones_2024,Lan_2025,Huang2025,evans2026}. Its advantage is that it provides a workable playground for non-finite or non-semisimple symmetries, a subject of intense current research interest. We introduce the necessary mathematical backgrounds and the graphical conventions for building and computing a pro-tensor network. We then restrict to a special class of pro-tensor networks that are uniform in the sense that every pro-tensor in the network is generated by the tensor product of a given monoidal category $\cC$. If one views $\cC$ as the charge category of a (generalized) symmetry, the corresponding pro-tensor network is literally the space of all symmetric tensor networks.   Such pro-tensor networks are manifestly re-triangulation invariant. When $\cC$ is a unitary fusion category, we use $\cC$-symmetric pro-tensor networks to recover the Levin-Wen model with input $\cC$~\cite{Levin_2005}. For this reason, we also call our construction the \emph{$\cC$ string-net pro-tensor network}.
The fact that the same pro-tensor network admits two physical interpretations, either as symmetry or as topological order, is in natural agreement with topological holography, which states more generally the duality between a $d$-dimensional symmetry and its associated $(d+1)$-dimensional topological order~\cite{Kong_Zheng_2018,Ji_2020, Kong_2020Gapless,Kong_2020,Chen_2020,Kong_2020Classi, Lichtman_2021,Kong_2022,KongZheng_2022,Chatterjee_2023,Moradi_2023, freed2024,KongZheng_2024,XuZhang_2024,WenYe2024, Bhardwaj2024,LanYueWang_2024,Jones_2024,Lan_2025,Huang2025,Lan_2024}. Therefore, this work may enhance our understanding of $(1+1)$D systems with potentially non-finite or non-semisimple symmetries.

Mathematically, pro-tensor networks are built upon the classical theory of \emph{profunctors} (also known as distributors or bimodules) and \emph{enriched categories}~\cite{Benabou_1973, Lawvere_1973, Kelly_1982, Street_1983, Benabou_2000}. Profunctors generalize the notion of bimodules and provide a natural setting for a category-theoretic linear algebra where vector spaces are replaced by $\cV$-categories and linear maps by $\cV$-profunctors. The contraction of pro-tensors corresponds to the coend construction, a categorical analogue of tensor contraction. By working in the setting of $\cV$-enriched categories for a suitable cosmos $\cV$, we obtain a framework that encompasses not only finite-dimensional vector spaces but also, for instance, super vector spaces and infinite-dimensional vector spaces, thereby transcending the usual finiteness and semisimplicity assumptions.

On the other hand, profunctors have recently emerged as a powerful tool in computer science, particularly in the study of ``profunctor optics''~\cite{boisseau2020,Clarke_2024}. The promonad $\cM\bbH_\cC\cN$ studied in this work is precisely the mathematical structure underlying the ``profunctor optics'' construction, which elegantly captures the notion of the ``double'', ``center'' or ``centralizer'' of a monoidal or module category~\cite{pastro2007doublesmonoidalcategories, Tambara2006DistributorsOA,Lopez_Franco_2007}. This connection provides an alternative, computationally-minded viewpoint on our algebraic constructions and hints at deeper structural analogies between renormalization in physics and data transformation patterns in computer science.

Finally, we use several analogies to place our framework of pro-tensor network on appropriate ground, and to specify what questions it \textit{can} and \textit{cannot} answer. The first analogy is between a traditional theory of a (many-body) system and a pro-tensor network (a many-many-body theory). In a traditional theory, there are \textit{kinematic} ingredients (lattice, Hilbert spaces, observables, quantum field operators, etc.), and also \textit{dynamics}, usually governed by a Hamiltonian, Lagrangian, etc.~ On the kinematic part, a pro-tensor network still requires a network, which can be thought of as a discrete spacetime background, some variant of lattice, but the degrees of freedom living on the spacetime background get \emph{categorified}. The \textit{dynamics} part is much more different: indeed, we should better say \emph{renormalization}, instead of dynamics, in a pro-tensor network. Technically, renormalization is computed by the (partial) contraction of pro-tensors, and is the defining data of a pro-tensor network, just like that a Hamiltonian is a defining data of a quantum lattice model. Specific dynamics (Hamiltonian, Lagrangian, etc.) is no longer a first citizen; instead, renormalization, which can be considered as evolution or dynamics in the ``theory space'', becomes the first citizen in the pro-tensor network.
These being said, we emphasize again that the pro-tensor network aims to answer questions about universal properties of phases, not about individual systems or dynamics. It works at a different level of ``more is different''. A good analogy here is quantum mechanics, which only predicts the probability distribution of measurements. For a single experiment, quantum mechanics can say almost nothing, except that whether the experiment result is ``valid" or ``impossible'' (the valid result has to lie within the spectrum of observables, and the probability can not be zero). Similarly, a many-many-body theory can say almost nothing about details of a single many-body system, except that whether a certain single many-body system is ``valid'' or ``impossible''. This is, however, still useful, as one can rule out ``impossible'' many-body systems and speed up calculations or searching.

We now state a non-trivial application of pro-tensor network which we study in great detail in this paper: we reproduce and generalize a key result from \cite{Kitaev_Kong_2012}, used by Kitaev and Kong in classifying topological excitations in Levin-Wen models as module functors. The classification by Kitaev and Kong \cite{Kitaev_Kong_2012, Kong_2012} is fundamental, as it provides the first example of 2+1D topological orders with fully-specified defects in each codimension -- formally, the first condensation complete 2-category of topological defects \cite{Kong_Wen_2014, Johnson_Freyd_2022}. To state our contribution in a more concrete way, we begin by recalling the result we improve upon:
\begin{theorem}[\cite{Kitaev_Kong_2012, Bai_Zhang_2025}]\label{thm.old_Kitaev_Kong}
    Given a fusion category $\cC$ and finite semisimple left $\cC$-modules $\cM$ and $\cN$, there is an algebra $A^\cC_{\cM,\cN}$ such that there is a $\C$-linear equivalence $\rep(A^\cC_{\cM,\cN})\cong\enf[\cC]{\cM}{\cN}$, where $\enf[\cC]{\cM}{\cN}$ denotes the $\C$-linear category of $\cC$-module functors from $\cM$ to $\cN$. Moreover, $A^\cC_{\cM,\cM}$ is a weak Hopf algebra and the equivalence $\rep(A^\cC_{\cM,\cM})\cong\enf[\cC]{\cM}{\cM}$ is monoidal.
\end{theorem}
A fusion category is a finite semisimple $\C$-linear rigid monoidal category. In this work, we reformulate Theorem \ref{thm.old_Kitaev_Kong} using the language of pro-tensors and promonads and generalize Theorem \ref{thm.old_Kitaev_Kong} to cover the cases of non-finite, non-semisimple and non-rigid\footnote{And if one wishes, non-$\C$-linear} categories:
\begin{theorem}[Theorem \ref{thm.gen_Kitaev_Kong} and Theorem~\ref{thm.mon_Kitaev_Kong}]\label{mainthm.gen.Kitaev_Kong}
    Let $\cV$ be the category of $\C$-vector spaces or more generally a Bénabou cosmos. Let $\cC$ be a $\cV$-enriched monoidal category, and $\cM,\cN$ be left $\cC$-modules. Then there is a promonad $\MCN$ such that there exists an equivalence $\bLmd_{\MCN}(*)\cong  {}_\cC\bvprof^\lax(\cM,\cN)$, where $\bLmd_{\MCN}(*)$ is the $\cV$-category of modules over $\MCN$ from $*$ and ${}_\cC\bvprof^\lax(\cM,\cN)$ is the $\cV$-category of lax left $\cC$-module profunctors $\cM\arprof\cN$. Moreover, $\MCM$ is a probimonad and $\bLmd_{\MCM}(*)\cong  {}_\cC\bvprof^\lax(\cM,\cM)$ is a monoidal equivalence.
\end{theorem}
The terminologies in Theorem \ref{mainthm.gen.Kitaev_Kong} is explained in the main text. We believe that the reason why a generalization as Theorem \ref{mainthm.gen.Kitaev_Kong} is possible is that a correct language is found, namely pro-tensors. Indeed, the original proof in \cite{Kitaev_Kong_2012} employs an ``extended diagrammatic calculus" that is geometrically intuitive yet the diagrammatic operations lack a firm ground. 
On the other hand, the rigorous proof in \cite{Bai_Zhang_2025} is lengthy and lacks the geometric insight. In our treatment of the generalized Kitaev-Kong theorem, we find that pro-tensors and promonads provide a framework that can accommodate the extended diagrammatic calculus in \cite{Kitaev_Kong_2012}, thereby stating and proving the result in a way both diagrammatic and rigorous. For example, we give a precise interpretation of the diagrammatic equation \cite[Eq.(26)]{Kitaev_Kong_2012} in Definition \ref{dfn.promonad_module}.

The rest of this paper is organized as follows. 
In Section~\ref{sec.idea}, we intuitively introduce the idea of pro-tensor and pro-tensor network. In Section~\ref{sec.network}, we finish the formal definition of pro-tensor network, and provide the toolbox for graphical calculations. In Section \ref{sec.LW}, we apply pro-tensor network to obtain the ground-state Hilbert space and the Hamiltonian of the Levin-Wen model. In Section \ref{sec.main_thm}, we prove Theorem \ref{mainthm.gen.Kitaev_Kong}, our main theorem, as an application. In Section~\ref{sec.TubeHolo}, we discuss the topological holography from the pro-tensor network perspective.

\section{Pro-tensor network: A first encounter}\label{sec.idea}
In this section, we introduce pro-tensor networks without resorting to technical details. In Section \ref{sec.network}, we make the discussion in this section rigorous by filling in the omitted details.

We begin by recalling the concept of tensor networks. A tensor network is composed of nodes and edges. For simplicity, we restrict our consideration to oriented tensor networks, in which all edges are directed from right to left, and vertical edges are excluded. However, to maintain diagrammatic cleanliness, arrowheads indicating direction are systematically omitted. 

\begin{figure}[htbp]
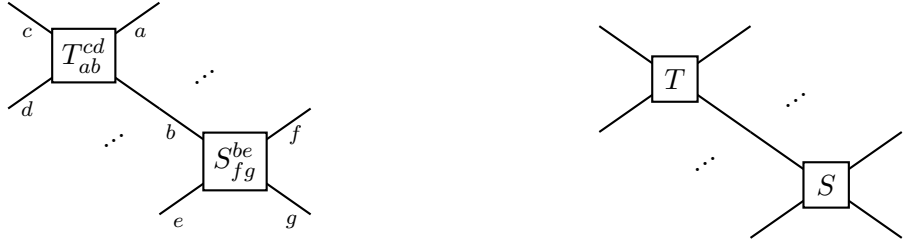

  \centering
  \begin{subfigure}{0.45\textwidth}
    \centering
    \[
        \ctikz{
            \draw (-2,1.4)--(0,0);
            \draw (-2,0)--(0,1.4);
            \draw (0,0)--(2,-1.4);
            \draw (0,-1.4)--(2,0);
            \draw (0.25,-1.5)node{$\scriptstyle e$};
            \draw (-0.25,1)node{$\scriptstyle a$};
            \draw (-1.75,-0)node{$\scriptstyle d$};
            \draw (1.8,-0.35)node{$\scriptstyle f$};
            \draw (-1.75,1)node{$\scriptstyle c$};
            \draw (0.15,-0.3)node{$\scriptstyle b$};
            \draw (1.75,-1.5)node{$\scriptstyle g$};
            \node[draw, rectangle,fill=white] at (-1,0.7) {$T_{ab}^{cd}$};
            \node[draw, rectangle,fill=white] at (1,-0.7) {$S_{fg}^{be}$};
            \draw (.5,0.35)node{.};
            \draw (.6,0.42)node{.};
            \draw (.7,0.49)node{.};
            \draw (-.5,-0.35)node{.};
            \draw (-.6,-0.42)node{.};
            \draw (-.7,-0.49)node{.};
        }
    \]
    \caption{$T_{ab}^{cd}$ and $S_{fg}^{be}$ are complex numbers.}
    \label{fig.tn.1}
  \end{subfigure}
  \hspace{0.05\textwidth}
  \begin{subfigure}{0.45\textwidth}
    \centering
    \[
        \ctikz{
            \draw (-2,1.4)--(0,0);
            \draw (-2,0)--(0,1.4);
            \draw (0,0)--(2,-1.4);
            \draw (0,-1.4)--(2,0);
            \mynode{-1}{0.7}{0.3}{0.3}{T}
            \mynode{1}{-0.7}{0.3}{0.3}{S}
            \draw (.5,0.35)node{.};
            \draw (.6,0.42)node{.};
            \draw (.7,0.49)node{.};
            \draw (-.5,-0.35)node{.};
            \draw (-.6,-0.42)node{.};
            \draw (-.7,-0.49)node{.};
        }
    \]
    \caption{$T$ and $S$ are linear maps $V\os V\to V\os V$.}
    \label{fig.tn.2}
  \end{subfigure}
  \caption{Two interpretations of the data assigned to the same tensor network. Here, both nodes in our displayed tensor network have 2 incoming edges and 2 outgoing edges.}
  \label{fig.tn}
\end{figure}

There are certain data assigned to a tensor network. We present two formulations of these assignments, both crucial for our exposition of pro-tensor networks. \textbf{Interpretation 1 (Labeling perspective):} Each edge $L$ is labeled by elements from a finite set $E_L=\{a,b,c,\cdots\}$. For simplicity, we assume $E_L=\colon E$ is identical for all edges $L$. Furthermore, for each node and for every possible labeling of its incident edges, a complex number is assigned. An illustration is provided in Figure~\ref{fig.tn.1}. \textbf{Interpretation 2 (Linear algebra perspective):} Each edge $L$ is associated with a finite-dimensional vector space $V_L$. For simplicity, we assume $V_L=\colon V$ is the same for all edges $L$. In this view, each node with $n$ incoming edges and $m$ outgoing edges is assigned a linear map $T\:V^{\os n}\to V^{\os m}$. See Figure~\ref{fig.tn.2} for an illustration.

The two interpretations are equivalent. Indeed, in view from Interpretation 1, the vector space $V$ in Interpretation 2 is precisely the vector space with basis $E$, and the assignment $\{T_{ab}^{cd}\}_{a,b,c,d\in E}$ uniquely determines a linear map $V^{\os 2}\to V^{\os 2}$ whose matrix elements in this basis are given by
\[
    \langle cd |T|ab\rangle = T_{ab}^{cd},\quad\forall a,b,c,d\in E.
\]
The converse direction, constructing the node assignments from a given linear map, follows straightforwardly by reading its matrix elements in a fixed basis.

Let us note additionally that $T$ can be canonically identified with an element in the tensor product space $V^\ast\os V^\ast\os V\os V$,  i.e.,  $T$ is a ``tensor'', hence the name ``tensor network''.

\begin{figure}[htp]
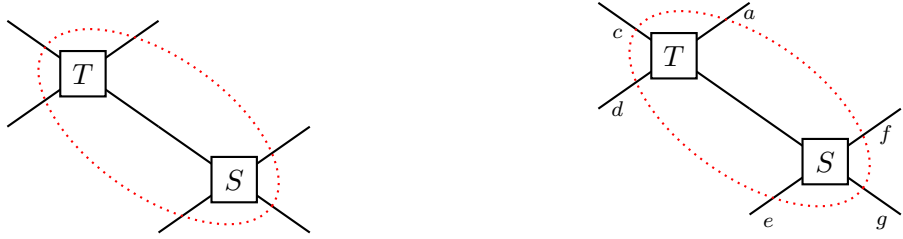

    \begin{subfigure}{0.45\textwidth}
    \centering
    \[
        \ctikz{
            \draw (-2,1.4)--(0,0);
            \draw (-2,0)--(0,1.4);
            \draw (0,0)--(2,-1.4);
            \draw (0,-1.4)--(2,0);
            \mynode{-1}{0.7}{0.3}{0.3}{T}
            \mynode{1}{-0.7}{0.3}{0.3}{S}
            \pgfmathsetmacro{\distance}{veclen(2,1.4)*1.5} 
            \pgfmathsetmacro{\angle}{atan2(-0.7-0.7,1-(-1))} 
            \draw[rotate around={\angle:(0,0)}, color4, dotted] 
            (0,0) ellipse ({\distance/2} and {\distance*0.25});
        }
    \]
    \caption{Dotted area represents a contracted tensor}
    \label{fig.contract_tn.1}
  \end{subfigure}
  \hspace{0.05\textwidth}
  \begin{subfigure}{0.45\textwidth}
    \centering
    \[
        \ctikz{
            \draw (-2,1.4)--(0,0);
            \draw (-2,0)--(0,1.4);
            \draw (0,0)--(2,-1.4);
            \draw (0,-1.4)--(2,0);
            \mynode{-1}{0.7}{0.3}{0.3}{T}
            \mynode{1}{-0.7}{0.3}{0.3}{S}
            \pgfmathsetmacro{\distance}{veclen(2,1.4)*1.5} 
            \pgfmathsetmacro{\angle}{atan2(-0.7-0.7,1-(-1))} 
            \draw[rotate around={\angle:(0,0)}, color4, dotted] 
            (0,0) ellipse ({\distance/2} and {\distance*0.25});
            \draw (0.25,-1.5)node{$\scriptstyle e$};
            \draw (-0,1.25)node{$\scriptstyle a$};
            \draw (-1.75,-0)node{$\scriptstyle d$};
            \draw (1.8,-0.35)node{$\scriptstyle f$};
            \draw (-1.75,1)node{$\scriptstyle c$};
            \draw (1.75,-1.5)node{$\scriptstyle g$};
        }
    \]
    \caption{Illustrating the tensor contraction in Interpretation 1, where $a,f,g,c,d,e\in E$.}
    \label{fig.contract_tn.2}
  \end{subfigure}
    \caption{Tensor contraction}
    \label{fig.contract_tn}
\end{figure}

A fundamental operation in tensor networks is the contraction of two tensors that share common edge(s). In Figure~\ref{fig.contract_tn.1}, the dotted area represents a tensor formed by contracting two constituent tensors, resulting in a tensor with 3 incoming and 3 outgoing edges. Under Interpretation 1, its associated coefficients are computed as
\begin{equation}\label{eq.contract_tn}
    \sum_{b\in E}T^{cd}_{ab} S^{be}_{fg}
\end{equation}
under the label assignment in Figure~\ref{fig.contract_tn.2}.

On the other hand, under Interpretation 2, the same contracted tensor corresponds to the composition of linear maps:
\[
    \diagram{
    V\os V\os V \ar[r]^-{1\os S} & V\os V\os V \ar[r]^-{T\os 1} & V\os V\os V
}.
\]
It can be verified that these two constructions yield identical results under the equivalence between the two interpretations. Finally, we note that there is a third way to expressing the contraction of $T$ and $S$. Namely, since $T,S\in V^\ast\os V^\ast\os V\os V$, the pairing of one copy of $V^\ast$ with one copy of $V$ yields from $T\os S$ a tensor living in $V^\ast\os V^\ast\os V^\ast\os V\os V\os V$. We leave it to the reader to work out the details of this perspective.

We conclude our short introduction to tensor networks, and we're now prepared to categorify the above ideas. We extensively use the idea that a categorification of the notion of complex number is that of $\C$-vector space, while a categorification of the notion of $\C$-vector space is that of a $\C$-linear category \cite{Kapranov_Voevodsky_1994, Baez_Dolan_1995, Gaiotto_Johnson_Freyd_2025}. 

As a categorified version of tensor networks, pro-tensor networks are likewise built upon a directed graph comprising nodes and edges, with the same convention that edges are oriented from right to left. However, this time each edge is assigned a $\C$-linear category rather than a $\C$-vector space. We still need to be more precise. To this end, we present two equivalent formulations of the data defining a pro-tensor network, paralleling the two interpretations of classical tensor networks. 

\begin{figure}[htbp]
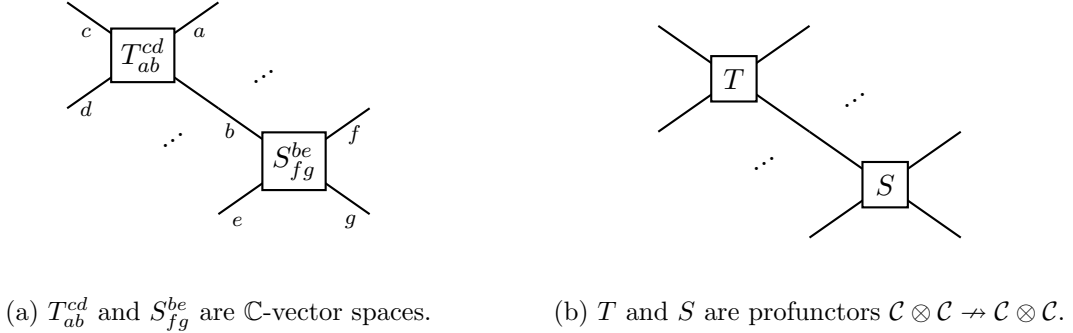

  \centering
  \begin{subfigure}{0.45\textwidth}
    \centering
    \[
        \ctikz{
            \draw (-2,1.4)--(0,0);
            \draw (-2,0)--(0,1.4);
            \draw (0,0)--(2,-1.4);
            \draw (0,-1.4)--(2,0);
            \draw (0.25,-1.5)node{$\scriptstyle e$};
            \draw (-0.25,1)node{$\scriptstyle a$};
            \draw (-1.75,-0)node{$\scriptstyle d$};
            \draw (1.8,-0.35)node{$\scriptstyle f$};
            \draw (-1.75,1)node{$\scriptstyle c$};
            \draw (0.15,-0.3)node{$\scriptstyle b$};
            \draw (1.75,-1.5)node{$\scriptstyle g$};
            \node[draw, rectangle,fill=white] at (-1,0.7) {$T_{ab}^{cd}$};
            \node[draw, rectangle,fill=white] at (1,-0.7) {$S_{fg}^{be}$};
            \draw (.5,0.35)node{.};
            \draw (.6,0.42)node{.};
            \draw (.7,0.49)node{.};
            \draw (-.5,-0.35)node{.};
            \draw (-.6,-0.42)node{.};
            \draw (-.7,-0.49)node{.};
        }
    \]
    \caption{$T_{ab}^{cd}$ and $S_{fg}^{be}$ are $\C$-vector spaces.}
    \label{fig.ptn.1}
  \end{subfigure}
  \hspace{0.05\textwidth}
  \begin{subfigure}{0.45\textwidth}
    \centering
    \[
        \ctikz{
            \draw (-2,1.4)--(0,0);
            \draw (-2,0)--(0,1.4);
            \draw (0,0)--(2,-1.4);
            \draw (0,-1.4)--(2,0);
            \mynode{-1}{0.7}{0.3}{0.3}{T}
            \mynode{1}{-0.7}{0.3}{0.3}{S}
            \draw (.5,0.35)node{.};
            \draw (.6,0.42)node{.};
            \draw (.7,0.49)node{.};
            \draw (-.5,-0.35)node{.};
            \draw (-.6,-0.42)node{.};
            \draw (-.7,-0.49)node{.};
        }
    \]
    \caption{$T$ and $S$ are profunctors $\cC\os\cC\arprof\cC\os\cC$.}
    \label{fig.ptn.2}
  \end{subfigure}
  \caption{Two interpretations of some data assigned to a pro-tensor network.}
  \label{fig.ptn}
\end{figure}

\textbf{Interpretation 1' (Labeling perspective):} Each edge $L$ is labeled by objects from a $\C$-linear category $\cC_L$. We assume $\cC_L=\colon \cC$ is identical for all edges $L$.\footnote{\label{fn.uniformity_assumption}This assumption is made here for simplicity and is not inherent to the general definition of a pro-tensor network.} In addition, for each node and for every possible labeling of its incident edges, a vector space (as opposed to a complex number) is assigned. See Figure~\ref{fig.ptn.1} for an illustration for this part of the data. Crucially, a node carries additional data in the form of linear maps between the vector spaces associated to this node. For example, for the vector spaces $\{T_{ab}^{cd}\}_{a,b,c,d\in\cC}$ in Figure~\ref{fig.ptn.1}, there is a linear map $T(f): T_{ab}^{cd} \to T_{a'b}^{cd}$ associated to a morphism $f\:a\to a'$ in $\cC$ and any three objects $b,c,d\in\cC$, which is required to be functorial in $a$ in an obvious sense; a precise definition appears in Section \ref{sub.ptn}. An analogous structure applies to the variables $b$, $c$, and $d$, with a technical note that the dependence on $c$ and $d$ (or in general, any outgoing-edge variables) is \emph{contravariant}. That is, for a morphism $g: c \to c'$, the induced linear map is $T(g): T_{ab}^{c'd} \to T_{ab}^{cd}$. Together with these linear maps, the vector spaces $\{T_{ab}^{cd}\}_{a,b,c,d\in\cC}$ constitute the complete data assigned to a node.

\textbf{Interpretation 2' (Categorified linear algebra perspective):} Each edge $L$ is associated with a $\C$-linear category $\cC_L$; again, we assume $\cC_L=\colon \cC$ is the same for all edges $L$. In this view, a node with $n$ incoming edges and $m$ outgoing edges is assigned a \emph{($\C$-)profunctor} $T\:\cC^{\os n}\arprof \cC^{\os m}$. See Figure~\ref{fig.ptn.2} for an illustration. 

Profunctors are familiar mathematical objects \cite{Benabou_1973, Benabou_2000}, whose precise meaning will be recalled in Section \ref{sub.ptn}. Here, we only remark that a profunctor $T\:\cC^{\os 2}\arprof\cC^{\os 2}$, by definition, consists of the collection $\{T_{ab}^{cd}\}_{a,b,c,d\in\cC}$ of vector spaces along with the linear maps $T(f)$, $T(g)$, etc. described in Interpretation 1'. Hence in particular the above two interpretations are equivalent. While a reader's inclination towards concreteness or abstractness may render one interpretation more intuitive than the other---especially when connecting to classical tensor networks---we emphasize that both perspectives are complementary and encourage maintaining this binary viewpoint.

From now on, we use the word \emph{pro-tensors} interchangeably with profunctors to emphasize its role as the object assigned to a node in the categorified tensor network. A natural question arises: if a $\C$-linear category is the categorification of a vector space, then since tensors are linear maps, shouldn't a categorified tensor be a \emph{$\C$-linear functor}, rather than a \emph{$\C$-profunctor}? To address this, we first note that $\C$-linear functors constitute a special case of $\C$-profunctors, as will be illustrated in Example \ref{ex.companion_and_conjoin}. However, this observation alone does not justify the necessity of profuncors. 

The justification we offer, while somewhat ad hoc, is rooted in a specific perspective on classical tensors. We aim to categorify the viewpoint that treats tensors not as arbitrary linear maps, but as specific ones of the form $\C\to (V^\ast)^{\os n}\os V^{\os m}$. This perspective crucially relies on the duality of finite-dimensional vector spaces, which can be used for defining contraction of tensors by our analysis of contraction in tensor networks. It is well-known, however, that such duality does not hold for generic $\C$-linear categories. Fortunately, within the framework of profunctors, any $\C$-linear category $\cC$ possesses a dual given by its opposite category $\cC^\op$. Consequently, a general pro-tensor $T\:\cC^{\os n}\arprof\cC^{\os m}$ can indeed be treated as a pro-tensor $*\arprof (\cC^\op)^{\os n}\os\cC^{\os m}$, where $*$ denotes the trivial $\C$-linear category; see Section \ref{sub.duality} for a formal discussion. This allows us to define the contraction of pro-tensors in a manner completely analogous to that of tensors.

We now proceed to define this contraction. Suppose Figure~\ref{fig.contract_tn} depicts a segment of a pro-tensor network. Then the dotted area automatically obtains a structure of a pro-tensor, called the contracted pro-tensor, which has the associated vector spaces given by the coend construction
\[
    \int^{b\in \cC}T_{ab}^{cd}\os S_{fg}^{be},\quad\forall a,f,g,c,d,e\in \cC.
\]
A more detailed discussion of coend appears in Section \ref{sub.contraction}. For now, it suffices to understand that it is a generalized form of ``summation'' over the object $b$, which precisely reflects the usage of the pairing between $\cC^\op$ and $\cC$ in the contraction; notice the analogy with \eqref{eq.contract_tn}.

By contracting pro-tensors, a pro-tensor network represents a single pro-tensor uniquely. This is analogous to the fact that, by contraction, a tensor network represents a single linear map uniquely. Nevertheless, as in the case of tensor networks, in pro-tensor networks the contracted pro-tensor is not the only thing that matters; the local pro-tensors distributed across the network also play a role.

Last but not least, due to its categorical nature, in pro-tensor networks, we're not only concerned with a single pro-tensor network, but also a \emph{map} between two such networks. A map between two pro-tensor networks is a homomorphism between the contracted pro-tensors of the respect networks; a homomorphism between pro-tensors is defined in Definition \ref{dfn.profunctor}. Such maps can always be expressed quantitatively and carry useful information. Physically, these maps can be used to encode the ``evolution'' in the many-many-body theory space; the examples considered in this paper are mostly renormalization maps. We will return to such aspect in Sections \ref{sec.LW} and \ref{sec.main_thm}.



\begin{remark}
    The pro-tensor networks introduced thus far can be more precisely termed $\C$-pro-tensor networks. We now outline a natural generalization: for any symmetric monoidal category $\cV$ satisfying suitable conditions (such as being a Bénabou cosmos, see Section \ref{sub.enrichedbg}), one can define $\cV$-pro-tensor networks. Here, a $\C$-linear category is replaced by a $\cV$-enriched category, while a $\C$-profunctor is replaced by a $\cV$-profunctor. Specifically, in this latter context, the symbols $S^{be}_{fg}$ and $T^{cd}_{ab}$ in Figure~\ref{fig.ptn.1} represent objects of $\cV$, while $T(f)\:T^{cd}_{ab}\to T^{cd}_{a'b}$ etc. become morphisms in $\cV$. The original $\C$-pro-tensor network corresponds to the case $\cV=\Vect$, where $\Vect$ denotes the category of (not necessarily finite) $\C$-vector spaces. Since we wish to study the case where $\cV$ is the category of super vector spaces in future, and since the passage from $\C$-pro-tensor network to $\cV$-pro-tensor network is direct, this is the generality we adopt. Nevertheless, as we will emphasize again in Section \ref{sec.network}, readers less fluent with category theory may safely keep the concrete case $\cV=\Vect$ in mind.
\end{remark}


\begin{remark}
There exists a further generalization of $\cV$-pro-tensor network to a categorified tensor network in any monoidal 2-category \cite{Day_Street_1997}, a direction we do not explore in details. Although this generalization may appear abstract, it parallels the fact that tensor networks can be defined not only in $\vect$, but in any monoidal category \cite{Joyal_Street_1991}, despite that only the former is recalled in this section. Table \ref{tab.tensor_and_pro-tensor} is a concise overview of this hierarchy, where various tensor networks are listed; the present work focuses exclusively on the entry marked with ($\star$).
\begin{table}[htp]
    \centering
    \begin{tabular}{|p{3.5cm}|p{5cm}|p{5cm}|}
        \hline
        Mathematical object assigned to a node& ordinary & categorified\\
        \hline
        ordinary & tensor (=linear  map) & $\cV$-pro-tensor (=$\cV$-profunctor) ($\star$) \\
        \hline
        generalized & morphism in a monoidal category & 1-morphism in a monoidal 2-category \\
        \hline
    \end{tabular}
    \caption{Mathematical objects assigned to a node in various tensor/pro-tensor network frameworks. }
    \label{tab.tensor_and_pro-tensor}
\end{table}
\end{remark}

\section{Pro-tensor network: Basic ingredients and tools}
\label{sec.network}
We base the mathematical foundation of pro-tensor networks on the theory of enriched profunctors (also known as distributors or bimodules), which was well-established in the 1970s \cite{Justesen_1968, Day_1970_b, Benabou_1973, Lawvere_1973, Kelly_1982, Street_1981, Street_1983}. In this section, we recall this theory, and subsequently introduce the basic definitions and key features of pro-tensor networks. 

In Section \ref{sub.enrichedbg}, we recall basic enriched category theory and fix notations. In Section \ref{sub.ptn}, we introduce profunctors and the definition of pro-tensor network; to be short, a pro-tensor network is an oriented directed graph with enriched categories for a fixed base assigned to edges and profunctors assigned to nodes (Definition \ref{dfn.ptn_net}). In Sections \ref{sub.contraction} and \ref{sub.map} respectively, we set up the two most important notions used in pro-tensor network theory:
\begin{itemize}
\item Contraction of pro-tensors.
\item Maps between pro-tensor networks.
\end{itemize}
In Sections \ref{sub.duality}-\ref{sub.pmd}, we touch upon some more advanced topics of profunctor theory. Although these topics do not appear in the definition of a generic pro-tensor network, they're 
useful for the study of this paper. To be specific, in Section \ref{sub.duality}, we discuss dualities. In Section \ref{sub.mon_VcatVModule}, we review the pro-tensor formalism of monoidal $\cV$-categories and module $\cV$-categories. In Section \ref{sub.pmd}, we introduce the notion of promonads and probimonads. This section do not contain new mathematical materials. 


\six{In this section, we employ the standard framework of enriched categories and enriched profunctors \cite{Benabou_1973, Kelly_1982} to formalize the concepts introduced in Section \ref{sec.idea}. We proceed as follows: in Section \ref{sub.enrichedbg}, we set up the enriched background; in Section \ref{sub.ptn}, we provide a precise definition of pro-tensors; in Section \ref{sub.contraction}, we review the definition and key properties of the contraction of pro-tensors; and finally, in Section \ref{sub.duality}, we gather these pieces together and introduce the 2-category of pro-tensors, denoted as $\vprof$.}

\subsection{Enriched categories}
\label{sub.enrichedbg}
\six{Let $\cV=(\cV,\os,\one)$ be a symmetric monoidal category. We will need the notion of $\cV$-enriched categories, $\cV$-functors and $\cV$-natural transformations 
recalled in Appendix \ref{sub.enriched.sec.app}.
}

The theory of enriched categories is well-known \cite{Kelly_1982} (we also recommend \cite{Borceux_Stubbe_2000}, \cite[Section 2]{Street_2004} or \cite[Appendix A]{Riehl_Verity_2022} for short expositions).\mayseven{concise overviews).} 
In this subsection, we review some basic aspects of enriched category theory and set up the notation along the way.

Let $\cV=(\cV,\os,\one)$ be a symmetric monoidal category, which we call the \emph{base} of our enrichment. For $X,Y\in\cV$, we denote the symmetric braiding by $\tau_{X,Y}\:X\os Y\isom Y\os X$. As is customary, we assume that $\cV$ is strict without loss of generality. A \emph{$\cV$-enriched category}, or simply a \emph{$\cV$-category} $\cC$ consists of a collection $\ob(\cC)$ of \emph{objects} in $\cC$, an object $\cC(a,b)\in\cV$ for every $a,b\in\ob(\cC)$ called the \emph{hom object from $a$ to $b$}, and two families of morphisms $\{\circ_\cC^{abc}\:\cC(b,c)\os\cC(a,b)\to\cC(a,c)\}_{a,b,c\in\ob(\cC)}$ and $\{i_\cC^a\:\one\to\cC(a,a)\}_{a\in\ob(\cC)}$ in $\cV$, the members of which are called the \emph{compositions} and \emph{units} respectively. The compositions and units are required to fulfill the usual associativity and unitality axioms. For a $\cV$-category $\cC$, we use $a\in\cC$ as a shorthand for the expression $a\in\ob(\cC)$. For $a,b,c\in\cC$, we often write $\circ_\cC^{abc}$ as $\circ^{abc}$ and $i_\cC^a$ as $i^a$ if the subscript is clear from the context. For $\cV$-categories $\cC$ and $\cD$, a \emph{$\cV$-functor} $F\:\cC\to\cD$ consists of a map $F\:\ob(\cC)\to\ob(\cD)$ together with a family $\{F_{ab}\:\cC(a,b)\to\cD(Fa,Fb)\}_{a,b\in\cC}$ of morphisms in $\cV$ such that $\circ_\cD^{FaFbFc}\circ (F_{bc}\os F_{ab})=F_{ac}\circ\circ_\cC^{abc}$ and $i_\cD^{Fa}=F_{aa}\circ i^a_\cC$ for every $a,b,c\in\cC$. For $\cV$-functors $F,G\:\cC\to\cD$, a \emph{$\cV$-natural transformation} is a family $\alpha=\{\alpha_a\:\one\to D(Fa,Ga)\}_{a\in\cC}$ of morphisms in $\cV$ such that $\circ_\cD^{FaFbGb}\circ (\alpha_b\os F_{ab})=\circ_\cD^{FaGaGb}\circ(G_{ab}\os\alpha_a)$ for every $a,b\in\cC$. There exists an ordinary category $\Vcat(\cC,\cD)$ of $\cV$-functors $\cC\to\cD$ and $\cV$-natural transformations between them. 

\begin{remark}
    If $\cV=\set$, $\cV$-enriched categories (/functors/natural transformations) are ordinary categories (/functors/natural transformations). If $\cV=\Vect$, the category of $\C$-vector spaces, then a $\cV$-category is a $\C$-linear category, and a $\cV$-functor is a $\C$-linear functor. 
    Reader less fluent with enriched category theory may \textbf{adopt the concrete choice $\cV=\Vect$ whenever deemed helpful.} Nevertheless, we maintain generality in the exposition, not only because we wish to include potential applications of pro-tensor networks enriched in other bases such as $\sVect$ (the category of super vector spaces) or certain representation categories of groups, but also because we believe that, as will become clear in later subsections, the natural setting for the theory of pro-tensor networks is the $\cV$-enriched category theory for a reasonably general $\cV$.
\end{remark}

\six{It makes no harm for readers who are not familiar to enriched categories to take $\cV = \Vect$, the category of vector spaces
(over complex numbers $\C$) and linear maps; the $\Vect$-enriched categories are simply $\C$-linear categories and all the prefixes $\cV-$ can be replaced for $\C$-linear. Nevertheless, we maintain generality in the exposition, as future applications may invovle pro-tensor networks enriched in other base categories such as }

Given any $\cV$-category $\cC$, there is an ordinary category $\underline{\cC}$ associated with it, called the \emph{underlying category of $\cC$}, defined by $\ob(\underline{\cC})\defdtobe\ob(\cC)$ and 
\[\underline{\cC}(a,b)\coloneq\cV(\one,\cC(a,b)),\quad\forall a,b\in\underline{\cC}.
\]
The compositions and units of $\underline{\cC}$ are all induced from those in $\cC$.

We emphasize several ways of constructing new $\cV$-categories from old ones. If $\cC$ is a $\cV$-category, there is a $\cV$-category $\cC^\op$ with $\ob(\cC)^\op\coloneq\ob(\cC)$ and $\cC^\op(a,b)\coloneq\cC(b,a)$ for $a,b\in\cC^\op$, called the \emph{opposite $\cV$-category of $\cC$}. If $\cC$ and $\cD$ are $\cV$-categories, then there exists a $\cV$-category $\cC\os \cD$, called the \emph{tensor product of $\cC$ and $\cD$}, defined by $\ob(\cC\os\cD)\coloneq\ob(\cC)\times\ob(\cD)$ and $(\cC\os\cD)((c,d),(c',d'))\coloneq \cC(c,c')\os\cD(d,d')$ for $(c,d),(c',d')\in\cC\os\cD$.
Finally, when $\cV$ has enough limits, the ordinary category $\Vcat(\cC,\cD)$ can be enriched in $\cV$ by defining the hom object to be certain limit in $\cV$ \cite[Proposition 2.3.30]{Borceux_Stubbe_2000} (See also Apppendix~\ref{app.enriched_structure} for an analogous treatment). We denote the resulting $\cV$-category by $[\cC,\cD]$. We have $\underline{[\cC,\cD]}=\Vcat(\cC,\cD)$.

We end this subsection by discussing some additional requirements on $\cV$ for the theory of profunctors to work nicely. First, we require $\cV$ is \emph{closed}, meaning that for any pair of objects $a,b\in\cV$, there is an object $[a,b]\in\cV$ and a family of bijections
\eqnn{\label{eq.closedness_iso}
    \cV(c\os a,b)\isom \cV(c,[a,b])
}
natural in $c\in\cV$.\footnote{Equivalently, the functor $-\os a\:\cV\to\cV$ admits a right adjoint for every $a\in\cV$.} The object $[a,b]$ is called the \emph{internal hom} from $a$ to $b$. 
Note that $\Vect$ is a closed category, with $[V,W]$ given by the vector space $\Hom(V,W)$ of linear maps from $V$ to $W$ for $V,W\in\Vect$. Similarly, the category $\vect\subset\Vect$ of finite-dimensional vector spaces is also closed. When the base $\cV$ is closed,
there is a $\cV$-enriched category $\bcV$ with $\ob(\bcV)\defdtobe\ob(\cV)$ and $\bcV(a,b)\defdtobe[a,b]$ \cite[Proposition 2.3.29]{Borceux_Stubbe_2000}. $\bcV$ is commonly known as ``$\cV$ enriched in itself''. It can be verified that $\underline{\bcV}\cong\cV$.
\six{We say $\cV$ is \emph{closed} if for any object $a\in \cV$, $-\otimes a:\cV\rightarrow \cV$ admits a right adjoint $[a,-]$, referred to as the internal hom. 
We take the assumption that $\cV$ is closed, so that $\cV$ can be promoted to a $\cV$-category by defining the hom-object as the internal hom $[a,b]$.  We denote such $\cV$-category by $\bcV$ (Example \ref{ex.self_enriched}).}
In the case of $\cV=\Vect$, the existence of $\overline{\Vect}$ corresponds to the tautological fact that $\Vect$ can be viewed as a $\Vect$-enriched category.

Secondly, we require $\cV$ to be sufficiently well-behaved with respect to limits and colimits, meaning that all relevant limits and colimits must exist in $\cV$, and that colimits behave well with respect to the tensor product of $\cV$. The latter condition is already ensured by the closedness assumption on $\cV$. The existence of the ``relevant'' limits serves to guarantee that the aforementioned $\cV$-category $[\cC,\cD]$ of $\cV$-functors and some analogous constructions are well-defined. The necessity of the existence of certain colimits, on the other hand, will only be clear in Section \ref{sub.contraction} where we introduce compositions of profunctors. A category $\cV$ is called complete if it has all small limits and cocomplete if it has all small colimits. Therefore, a layman choice is to require $\cV$ to be both complete and cocomplete; a symmetric monoidal closed category with these properties is called a \emph{(Bénabou) cosmos}. Typical examples of cosmos include $\Vect$ and $\sVect$. We will indeed adopt this setting while establishing the foundations of $\cV$-pro-tensor network and proving general theorems, thus $\Vect$-pro-tensor network and $\sVect$-pro-tensor network are automatically established. However, we should emphasize that the general framework is also applicable for non-cosmos $\cV$ such as $\vect$ as long as these ``relevant'' limits and colimits exist. For example, if the $\vect$-categories in consideration are all finite semisimple, then all ends and coends labelled by these $\vect$-categories exist, affording a well-defined $\vect$-pro-tensor network theory (a fact which essentially follows from \cite[Lemma 2.5]{huang20242charactertheoryfinite2groups}). Nevertheless, the general framework for $\cV$ being a cosmos does allow the possibility of considerations of $\Vect$-categories that are neither finite nor semisimple. The above reasoning justifies our assumption in the exposition that $\cV$ is a cosmos.

\subsection{Profunctors and the definition of pro-tensor network\six{Profunctors, or what is assigned to a node}}\label{sub.ptn}
In this subsection, we introduce the notion of profunctors, provide examples, and present the definition of pro-tensor network. 

Let $\cV$ be a cosmos. As we emphasize in the previous subsection, the reader may assume $\cV=\Vect$ or even $\cV=\vect$ whenever helpful. 
\begin{definition}\label{dfn.profunctor}
    Let $\cC,\cD$ be $\cV$-categories. A \emph{($\cV$-)profunctor} $F\:\cC\arprof\cD$ is a $\cV$-functor $\cD^\op\os\cC\to\bcV$. Let $F,F'\:\cC\arprof\cD$ be $\cV$-profunctors. A \emph{($\cV$-)profunctor homomorphism} is a $\cV$-natural transformation $F\Rightarrow F'$. Profunctors $\cC\arprof\cD$ and profunctor homomorphisms form a category which we denote by $\vprof(\cC,\cD)$. 
\end{definition}

Given profunctors $F,F'\:\cC\arprof\cD$, we use $\vprof(\cC,\cD)(F,F')$, or simply $\Hom(F,F')$, to denote the set of profunctor homomorphisms from $F$ to $F'$.
\mayseven{When $\cV=\Vect$ or $\vect$, we sometimes also use the term $\C$-profunctor for $\cV$-profunctor.}

\mayseven{By unpacking data, one finds that Definition \ref{dfn.profunctor} is equivalent to the following definition, where the ``bimodules'' point of view is emphasized. A profunctor $\cC\arprof\cD$ is equivalent consists of a family of objects in $\cV$ that carries \emph{``left actions''} from $\cC$ and \emph{``right actions'' } from $\cD$. To be precise, it consists the following data:}
By unpacking data, one finds that a profunctor $\cC\arprof\cD$ is precisely a family of objects in $\cV$ carrying compatible left actions from $\cC$ and right actions from $\cD$, i.e., is a ``bimodule'' between $\cC$ and $\cD$. Definitions \ref{dfn.prof_unpack} and \ref{dfn.homo_unpack} below spell out this perspective, giving equivalent formulations of a profunctor and a profunctor homomorphism, respectively.
\begin{definition}\label{dfn.prof_unpack}
A \emph{($\cV$-)profunctor} $F\:\cC\arprof\cD$ consists of the following data:
\begin{itemize}
    \item For every $c\in\cC,d\in\cD$, an object $F(d,c)\in\cV$. The objects $\{F(d,c)\}_{c\in\cC,d\in\cD}$ are called the \emph{effects on objects} of the profunctor $F$.
    \item For every $c,c'\in\cC,d\in\cD$, a morphism
    \[
        F_l^{dcc'}\:\cC(c,c')\os F(d,c)\to F(d,c')
    \]
    in $\cV$. The morphisms $\{F_l^{dcc'}\}_{c,c'\in\cC,d\in\cD}$ are called the \emph{(left) $\cC$-action} on $F$.
    \item For every $c\in\cC,d,d'\in\cD$, a morphism
    \[
        F_r^{dd'c}\:F(d',c)\os \cD(d,d')\to F(d,c)
    \]
    in $\cV$. The morphisms $\{F_r^{dd'c}\}_{c\in\cC,d,d'\in\cD}$ are called the \emph{(right) $\cD$-action} on $F$.
\end{itemize}
The following properties should be satisfied.
\begin{enumerate}
    \item The left actions are associative and unital: for every $c,c',c''\in\cC$, the diagrams 
    \[
    \diagram@C=3pc{
        \cC(c',c'')\os\cC(c,c')\os F(d,c) \ar[d]_{\circ^{cc'c''}_\cC\os 1} \ar[r]^-{1\os F_l^{dcc'}} & \cC(c',c'')\os F(d,c') \ar[d]^{F_l^{dc'c''}} \\
        \cC(c,c'')\os F(d,c) \ar[r]_-{F_l^{dcc''}} & F(d,c'')
    }
    \quad
    \diagram{
        F(d,c) \ar[d]_{i^c_\cC\os 1} \ar[rd]^-1 \\
        \cC(c,c)\os F(d,c) \ar[r]_-{F_l^{dcc}} & F(d,c)
    }
    \]
    in $\cV$ commute.
    \item The right actions are associative and unital: for every $d,d',d''\in\cD$, the diagrams
    \[
    \diagram@C=3pc{
        F(d'',c)\os \cD(d',d'')\os \cD(d,d') \ar[d]_{1\os\circ_\cD^{dd'd''}} \ar[r]^-{F_r^{d'd''c}\os 1} & F(d',c)\os \cD(d,d') \ar[d]^{F_r^{dd'c}}\\
        F(d'',c)\os \cD(d,d'') \ar[r]_-{F_r^{dd''c}} & F(d,c)
    }
    \quad
    \diagram{
        F(d,c) \ar[d]_{1\os i^d_\cD} \ar[rd]^-1 \\
        F(d,c)\os \cD(d,d) \ar[r]_-{F_r^{ddc}} & F(d,c)
    }
    \]
    in $\cV$ commute.
    \item The left and right actions commute with each other: for every $c,c'\in\cC,d,d'\in\cD$, the diagram
    \[
    \diagram@C=3pc{
    \cC(c,c')\os F(d',c)\os \cD(d,d') \ar[d]_{1\os F_r^{dd'c}} \ar[r]^-{F_l^{d'cc'}\os 1} & F(d',c')\os \cD(d,d') \ar[d]^{F_r^{dd'c'}}\\
    \cC(c,c')\os F(d,c) \ar[r]_-{F_l^{dcc'}} & F(d,c')
    }.
    \]
    in $\cV$ commute.
\end{enumerate}
\end{definition}
\begin{definition}\label{dfn.homo_unpack}
    Let $F,G\:\cC\arprof\cD$ be $\cV$-profunctors. A \emph{profunctor homomorphism} $\phi\:F\Rightarrow G$ is a family of morphisms $\{\phi_{d,c}\:F(d,c)\to G(d,c)\in\Mor(\cV)\}_{d\in\cD,c\in\cC}$ between the effects of $F$ and $G$ on objects commuting with the left and right actions, i.e., render the following diagrams commutative for any $c,c'\in\cC,d,d'\in\cD$:
    \[
        \diagram{
            \cC(c,c')\os F(d,c) \ar[d]_{\id\os \phi_{d,c}} \ar[r]^-{F^{dcc'}_l} & F(d,c') \ar[d]^{\phi_{d,c'}}  \\
            \cC(c,c')\os G(d,c) \ar[r]_-{G^{dcc'}_l} & G(d,c')
        }
        \qquad
        \diagram{
            F(d',c)\os \cD(d,d') \ar[d]_{\phi_{d',c}\os\id} \ar[r]^-{F^{dd'c}_r} & F(d,c) \ar[d]^{\phi_{d,c}} \\
            G(d',c)\os \cD(d,d') \ar[r]_-{G^{dd'c}_r} & G(d,c)
        }.
    \]
\end{definition}

Profunctors indeed generalize the notion of bimodules. To see this, we assume $\cV=\Vect$ for the time being. Let $\cC,\cD$ be one-object $\cV$-categories. Let $*$ denote the unique object in $\cC$ and also $\cD$. Then one can see that a profunctor $\cC\arprof\cD$ is nothing but a $\cC(*,*)$-$\cD(*,*)$-bimodule, where $\cC(*,*)$ and $\cD(*,*)$ are the $\C$-algebras determining the $\cV$-categories. More examples of profunctors are provided as follows; we usually define a profunctor by only specifying its effects on objects if the left and right actions are evident to define.

\begin{example}[Identity profunctor]
\label{ex.idprof}
    Given any $\cV$-category $\cC$, there is an \emph{identity profunctor} $\Id_\cC\:\cC\arprof\cC$ defined by $\Id_\cC(a,b)\defdtobe \cC(a,b)$ with $(\Id_\cC)_l^{abc}\defdtobe\circ_\cC^{abc}$ and $(\Id_\cC)_r^{abc}\defdtobe \circ_\cC^{abc}$ for $a,b,c\in\cC$.
    \six{This pro-tensor can be drawn as  
    \[
        \ctikz{[scale = 0.5]
            \draw (-2.5,0)--(0,0);
            \draw (2.5,0)--(0,0);
            \mynode{0}{0}{0.5}{0.5}{\Id_\cC};
        },
    \]
    while in most cases, we simply refrain ourselves from drawing it explicitly as node:
    \[
        \ctikz{[scale = 0.5]
            \draw (-2.5,0)--(0,0);
            \draw (2.5,0)--(0,0);
        },
    \]
    a convention which will be justified in Section \ref{sub.contraction}.}
\end{example}
\begin{example}[Companions and conjoins]\label{ex.companion_and_conjoin}
    Given a $\cV$-functor $H\:\cC\to\cD$, there are two profunctors,  $H_\ast\:\cC\arprof\cD$ given by $(H_\ast)(d,c)\defdtobe \cD(d,Hc)$, and $H^\ast\:\cD\arprof\cC$ given by $(H^\ast)(c,d)\defdtobe \cD(Hc,d)$. They're called the \emph{companion} and \emph{conjoin} of $H$, respectively. A useful relation between the companion and the conjoin is given in Section \ref{sub.duality}.
\end{example}
\begin{example}[Constructing new profunctor from old ones, I]\label{ex.transpose_of_profunctor}
    Definition \ref{dfn.profunctor} lends itself to a tautological construction of new profunctors from old ones. Namely, a profunctor $F\:\cC\arprof \cD$ is by definition also a profunctor $F^\op:\cD^\op\arprof\cC^\op$, as well as a profunctor  $F^\heartsuit\:*\arprof\cC^\op\os\cD$ and a profunctor $F^\spadesuit\:\cC\os\cD^\op\arprof*$. Here, $*$ is the trivial $\cV$-category, i.e., the unique $\cV$-category with a single object (by a slight abuse of notation, also denoted by $*$) such that $*(*,*)=\one$, and satisfying $*\os\cC\cong\cC\cong\cC\os*$ for any $\cV$-category $\cC$. Especially, the profunctor $\Id_\cC$ in Example \ref{ex.idprof} gives rise to two profunctors below, which will be needed in the future:
    \[
        \evC\:\ast\arprof\cC^\op\os\cC,\quad (a,b)\mapsto \cC(b,a);\quad\qquad \coevC\:\cC\os\cC^\op\arprof\ast,\quad (b,a)\mapsto \cC(a,b).
    \]
\end{example}
\begin{example}[Constructing new profunctors from old ones, II]\label{ex.tensor_of_profun}
    Let $F\:\cC\arprof \cD$ and $G\:\cC'\arprof\cD'$ be profunctors. Then there is a profunctor $F\os G\:\cC\os\cC'\arprof\cD\os\cD'$ with 
    \[
        (F\os G)((d,d'),(c,c'))\defdtobe F(d,c)\os G(d',c'),\quad \forall c\in\cC,c'\in\cC',d\in\cD,d'\in\cD'.
    \]
    $F\os G$ is the \emph{tensor product} of $F$ and $G$.
\end{example}

Now we define the notion of pro-tensor network.
\begin{definition}\label{dfn.dir_gph}
    An \emph{(open) directed graph} $(V,E,s,t)$ consists of a set $V$ of \emph{vertices}, a set 
    \(E=E_{\mathrm{int}}\coprod E_\partial^-\coprod E_\partial^+\)
    of \emph{edges}, and two maps 
    \(s\:E_{\mathrm{int}}\coprod E_\partial^+\to V\) and \( t\:E_{\mathrm{int}}\coprod E_\partial^-\to V\). For an edge $e$, $s(e)$ is called its \emph{source} and $t(e)$ its \emph{target}. The edges in $E_{\mathrm{int}}$ are referred to as \emph{internal edges}, those in $E_\partial^-$ as \emph{input external edges} and those in $E_\partial^+$ as \emph{output external edges}. We always assume that $V$ is embedded in a plane, and moreover, there is a global orientation in the sense that every edge progresses from right to left \footnote{This can be made slightly more formal by requiring the existence of a level function $\ell : V \to \mathbb{Z}$ such that for every edge $e:u \to v$, we have $\ell(u) < \ell(v)$.}.
\end{definition}
One can see from above definition that each internal edge $e$ has both a source and a target, whereas input external edges possess only targets and output external edges only sources.

\begin{definition}[Pro-tensor Network]\label{dfn.ptn_net}
A \textbf{($\cV$-)pro-tensor network} consists of the following data:
\begin{enumerate}
\item A directed graph $(V,E,s,t)$.
\item An assignment of a $\cV$-category $\cC_e$ to each edge $e\in E$.
\item An assignment of a $\cV$-profunctor 
\[
F_v : \bigotimes_{e \in t^{-1}(v)} \cC_e \arprof\bigotimes_{e'\in s^{-1}(v)} \cC_{e'}
\]
to each vertex $v \in V$, where the tensor products are taken according to the top-to-bottom order on both $t^{-1}(v)$ and $s^{-1}(v)$, and $\bigotimes_{e\in\emptyset}\cC_e\defdtobe*$.
\end{enumerate}
\end{definition}
Let us provide some examples of pro-tensor networks, through the process of which we fix our conventions of drawing a pro-tensor network.
\begin{example}[Standard pro-tensor network presentation of a profunctor]\label{ex.tautological_ptn}
    A very simple type of pro-tensor networks have the underlying directed graph determined by 
    \[
        V=\{v\},\quad E_{\mathrm{int}}=\emptyset,\quad E_\partial^-=\{e^-\},\quad E_\partial^+=\{e^+\}.
    \]
    One can check that the data of this pro-tensor network is uniquely determined by the profunctor $F\:\cC\arprof\cD$ assigned to $v$, in which case the edges $e^-,e^+$ must be assigned $\cC$ and $\cD$ respectively. We will draw this pro-tensor network as
    \eqnn{\label{eq.tautological_ptn}
        \ctikz{[scale = 0.5]
            \draw (-2.5,0)--(0,0);
            \node at (-1.5,-0.5) {$\cD$};
            \draw (2.5,0)--(0,0);
            \node at (1.5,-0.5) {$\cC$};
            \mynode{0}{0}{0.6}{0.6}{F}
        }.
    }
  When the assignments $\cC$ and $\cD$ are clear from context, to reduce notational clutter, we will either (1) simply suppress these labels from the graphs or (2) suppress the labels and introduce an additional coloring to indicate the distinctions between different assignments of the edges (see, e.g., Definition \ref{def.moduleVcat} or Definition \ref{dfn.promonad_module}). Note that \eqref{eq.tautological_ptn} can be viewed as a ``standard'' pro-tensor network presentation of the profunctor $F$. Every profunctor admits such a standard pro-tensor network presentation; this holds, in particular, for the profunctors considered in Examples \ref{ex.idprof}, \ref{ex.companion_and_conjoin}, \ref{ex.transpose_of_profunctor} and \ref{ex.tensor_of_profun}.
\end{example}
We henceforth also refer to a profunctor as a \emph{pro-tensor}, having in mind its standard pro-tensor network presentation \eqref{eq.tautological_ptn}. This terminology is analogous to the use of ``tensor'' for linear maps in the study of tensor networks.
\begin{example}\label{ex.evC_and_coevC_graph_notation}
    The profunctor $\evC$ in Example \ref{ex.transpose_of_profunctor} can also be drawn as
    \[
        \ctikz{[scale = 1, baseline=(current bounding box.center)]
            \draw (1,1.5)--node[anchor=south]{$\cC^\op$}(2.5,1.5);
            \draw (1,0.5)--node[anchor=north]{$\cC$}(2.5,0.5);   
            \draw (2.5,0.5) arc (-90:90:0.5cm);
            \mynode{3}{1}{0.3}{0.3}{\evC};
        },
        \quad\text{or simply}\quad
        \ctikz{[scale = 1, baseline=(current bounding box.center)]
            \draw (1,1.5)--node[anchor=south]{$\cC^\op$}(2.5,1.5);
            \draw (1,0.5)--node[anchor=north]{$\cC$}(2.5,0.5);   
            \draw (2.5,0.5) arc (-90:90:0.5cm);
        },
    \]
    which is a pro-tensor network with 0 incoming edges and 2 outgoing edges. Similarly, the profunctor $\coevC$ in Example \ref{ex.transpose_of_profunctor} can be drawn as 
    \[
        \ctikz{[scale = 1, xscale=-1, baseline=(current bounding box.center)]
            \draw (1,1.5)--node[anchor=south]{$\cC$}(2.5,1.5);
            \draw (1,0.5)--node[anchor=north]{$\cC^\op$}(2.5,0.5);   
            \draw (2.5,0.5) arc (-90:90:0.5cm);
            \mynode{3}{1}{0.3}{0.3}{\coevC};
        },
        \quad\text{or simply}\quad
        \ctikz{[scale = 1, xscale=-1, baseline=(current bounding box.center)]
            \draw (1,1.5)--node[anchor=south]{$\cC$}(2.5,1.5);
            \draw (1,0.5)--node[anchor=north]{$\cC^\op$}(2.5,0.5);   
            \draw (2.5,0.5) arc (-90:90:0.5cm);
        }.
    \]
    The justification for this graphical convention is offered by Proposition \ref{prp.zigzagger}.
\end{example}

\subsection{Contraction of pro-tensors}\label{sub.contraction}
In this subsection, we use composition of profunctors to define the contraction of pro-tensors, and discuss the basic properties of this operation. 

As the arrow notation indicates, given profunctors $F\:\cC\arprof\cD$ and $G\:\cD\arprof\cE$, it is possible to obtain a profunctor $G\bullet F\:\cC\arprof\cE$ as their \emph{composition}.

The effect of $G\bullet F$ on objects is defined as
\eqnn{\label{eq.compose_prof}
(G\bullet F)(e,c)\defdtobe \int^{d\in \cD}F(d,c)\os G(e,d),\quad c\in\cC,e\in\cE,
}
where $\int^{d\in\cD}F(d,c)\os G(e,d)$ denotes the coend of the profunctor $F(-,c)\os G(e,-)\:\cD\arprof\cD$. For definition of coend, we refer the reader to Definition \ref{dfn.coend}; see Lemma \ref{lem.coend_Vect_prot} for the computation of any coend in the case $\cV=\Vect$. For now, it suffices to understand that $\int^{d\in\cD}F(d,c)\os G(e,d)$ is the universal object in $\cV$ with respect to $\{F(d,c)\os G(e,d)\}_{d\in\cD}$ when the action of $\cD$ is ``integrated out''\footnote{However, the actual relation between coend and the integral in mathematical analysis seems to be obscure \cite{Loregian_2021}.}. More precisely, $\int^{d\in\cD}F(d,c)\os G(e,d)$ is equipped with a family of morphisms 
\[
    \{\copr_d\:F(d,c)\os G(e,d)\to \int^{d\in\cD}F(d,c)\os G(e,d)\}_{d\in\cD}
\]
in $\cV$ called the \emph{coprojections}, such that the pair $(\int^{d\in\cD}F(d,c)\os G(e,d),\{\copr_d\}_{d\in\cD})$ satisfies the following universal property: given any object $W\in \cV$ equipped with a family of morphisms  
\begin{equation*}
   \{\theta_d\:F(d,c)\ot G(e,d)\rightarrow W\}_{d\in \cD}
\end{equation*}
that is compatible with $\cD$-actions\footnote{
Here, an object $w$ equipped with a family of morphisms $\{\theta_d\:F(d,c)\os G(e,d)\to w\}_{d\in\cD}$ is said to be compatible with $\cD$-actions if the diagram
 \[
        \diagram{
            F(d',c)\os \cD(d,d')\os G(e,d) \ar[r]^-{\id\os G_l^{edd'}} \ar[d]_{F_r^{dd'c}\os\id} & F(d',c)\os G(e,d') \ar[d]^{\theta_{d'}} \\
            F(d,c)\os G(e,d) \ar[r]_-{\theta_d} & w
        }
\]
is commutative for any $d,d'\in\cD$.
}, $\theta_d$ factors uniquely through the coend via $\copr_d$:
\[ \theta_d = (F(d,c)\ot G(e,d)\xrightarrow{\copr_d}(G\bullet F)(e,c)\xlongrightarrow{\exists ! \underline{\theta}} W);\]
moreover the pair $(\int^{d\in\cD}F(d,c)\os G(e,d),\{\copr_d\}_{d\in\cD})$ itself is compatible with $\cD$-actions.

Therefore, although a coend can be an abstract object that is in general difficult to manipulate directly, any morphism out of the coend into an object $w$ can be understood concretely in terms of the corresponding family of morphisms \(\{\theta_d\}_{d\in \cD}\).
\begin{remark}
    It is helpful to draw analogy with quotient map. Given a vector space $V$ and a subspace $W\subset V$, the quotient map is a vector space $V/W$ together with a linear map $\pi\:V\to V/W$ satisfying the following universal property: $\pi(W)=0$, and moreover, for any pair $(Q,q\:V\to Q)$ of a vector space and a linear map satisfying $q(W)=0$, there exists a unique linear map $\underline{q}\:V/W\to Q$ rendering $\underline{q}\circ \pi=q$. Thus, as part of the structure of the quotient map, $\pi$ has \textbf{codomain} $V/W$. On the other hand, to induce a linear map with \textbf{domain} $V/W$, one precisely needs a linear map $q\:V\to Q$ satisfying an additional property (i.e. $q(W)=0$). This is analogous to our case, where the structure morphisms $\copr_d\:F(d,c) \os G(e,d)\to \int^{d\in\cD}F(d,c)\os G(e,d)$ has codomain the coend $\int^{d\in\cD} F(d,c)\os G(e,d)$, while to induce a morphism with domain $\int^{d\in\cD}F(d,c)\os G(e,d)$, one precisely needs a family of morphisms $\{F(d,c)\os G(e,d)\to w\}_{d\in\cD}$ satisfying an additional property (i.e., being compatible with $\cD$-action). Thus the morphisms $\copr_d$'s can be viewed as generalized quotient map.
\end{remark}

Note that there are well-defined left $\cC$-action and right $\cE$-action on the objects \eqref{eq.compose_prof}, so that $G\bullet F$ becomes a well-defined profunctor $\cC\arprof\cE$. In addition, given profunctor homomorphisms $\alpha\:F\Rightarrow F'$ and $\beta\:G\Rightarrow G'$, there is a well-defined profunctor homomorphism $\beta\bullet\alpha\:G\bullet F\Rightarrow G'\bullet F'$ enjoying certain nice properties. These constructions are all given in Section \ref{sub.compose_of_prof.app}.

Before we move on, we turn to several abstract yet useful aspects of the coends:
\begin{enumerate}
    \item \eqref{eq.compose_prof} requires the existence of coends. If $\cV$ is a cosmos, we can see that coends always exist as long as $\cD$ is small (Proposition \ref{prp.existence_of_coend_cosmos}). If $\cV$ is not a cosmos yet admits all finite limits and colimits, and $\cD$ is a $\cV$-category equivalent to the Cauchy completion of a $\cV$-category $\cD'$ where $\cD'$ has finitely many objects, then all $\cD$-indexed coends in $\cV$ also exists (Proposition \ref{prp.finite_coend}).
    \item There is a Fubini's theorem which holds for coends, stating that two coends can ``exchange'', resulting in the associativity of profunctor compositions:
    \[
    \begin{multlined}
         (H\bullet G)\bullet F
         \cong         
         H\bullet (G\bullet F),\quad\forall F\:\cB\arprof\cC,G\:\cC\arprof\cD,H\:\cD\arprof\cE.
    \end{multlined}
    \]
    \item The representable functors $\cC(-,a)\:\cC^\op\to\bcV$ and $\cC(a,-)\:\cC\to\bcV$ behave like ``Dirac delta function'' in the ``integration'' of coends:
    \eqnn{\label{eq.coYoneda}
        \int^{c\in\cC}\cC(c,a)\os F(b,c)=F(b,a),\qquad \int^{c\in\cC}G(c,d)\os \cC(a,c)=G(a,d),
    }
    where $F\:\cC\arprof\cB$ and $G\:\cD\arprof\cC$ are profunctors and $b\in\cB,d\in\cD$, and the coprojections are given by the left and right actions of the categories.
\end{enumerate}
\begin{remark}\label{rmk.generalizetensor}
    We emphasize again the analogy of the coend formula with the expression
    \[
        \sum_{b}M_{c,b}N_{b,a}
    \]
    with $M,N$ being matrices. Besides the formal similarity, we also want to point out that the matrix multiplication data is already contained in the coend. Note that matrix multiplication corresponds to the composition of linear maps, which is exactly the universal map associated to the composition of the identity profunctor $\Id_\Vect$ of $\Vect$ with itself,
    \[ \copr_Y: \Vect(Y,Z)\otimes\Vect(X,Y)\to \int^{V\in\Vect} \Vect(V,Z)\otimes\Vect(X,V)=\Vect(X,Z),\]
    i.e. for any $f:X\to Y$ and $g:Y\to Z$, \[\copr_Y(g\otimes f)=g\circ f.\]
    For a general profunctor $F:\cC\nrightarrow \cD, G:\cD\nrightarrow \cE$ with objects $c,d,e\in \cC,\cD,\cE$, respectively, $F(d,c)$ generalize the concept of \emph{space} of linear maps from $c$ to $d$, and similar for $G(e,d)$. The morphism $\copr_d:F(d,c) \ot G(e,d) \rightarrow \int^{d\in\cD} F(d,c)\ot G(e,d)=(G\bullet F)(e,c)$ then generalizes the \emph{composition} of linear maps.
\end{remark}

Now we're ready to define contraction of a pro-tensor network. The contraction of a pro-tensor network refers to a process of extracting a single pro-tensor, which we call the \emph{contracted pro-tensor}, out of a given pro-tensor network; this process is the categorification of that of extracting a tensor out of a tensor network.

\begin{figure}[htbp]
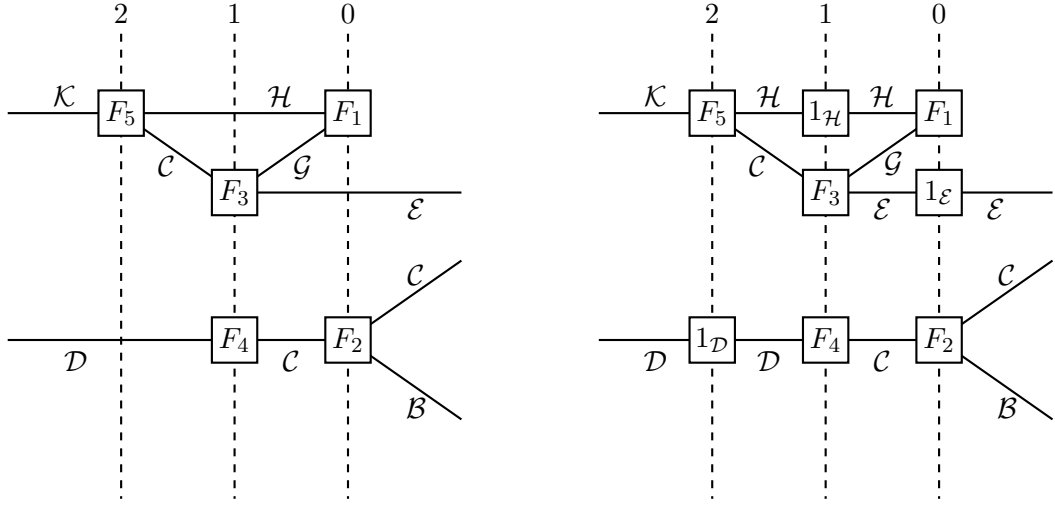

    \begin{subfigure}{0.45\textwidth}
    \[
    \ctikz{[scale=1.5]
        \draw[dashed] (1,2.7)node[anchor=south]{0}--(1,-1.4);
        \draw[dashed] (0,2.7)node[anchor=south]{1}--(0,-1.4);
        \draw[dashed] (-1,2.7)node[anchor=south]{2}--(-1,-1.4);
        \draw (1,2) -- (-1,2) node[pos=0.3,above,yshift=-0.05cm]{$\cH$} --(-2,2) node[pos=0.5,above,yshift=-0.05cm]{$\cK$};
        \draw (1,2)--(0,1.3) node[pos=0.4,below]{$\cG$} --(-1,2) node[pos=0.6,below]{$\cC$};
        \draw (2,1.3)--(0,1.3) node[pos=0.2,below,yshift=0.05cm]{$\cE$};
        \draw (2,0.7)--(1,0) node[pos=0.4,above,yshift=-0.05cm]{$\cC$}-- (0,0) node[pos=0.5,below]{$\cC$} -- (-2,0) node[pos=0.7,below]{$\cD$} ;
        \draw (2,-0.7)--(1,0) node[pos=0.4,below]{$\cB$};
        \mynode{1}{2}{0.2}{0.2}{F_1}
        \mynode{1}{0}{0.2}{0.2}{F_2}
        \mynode{0}{1.3}{0.2}{0.2}{F_3}
        \mynode{0}{0}{0.2}{0.2}{F_4}
        \mynode{-1}{2}{0.2}{0.2}{F_5}
    }
    \]
    \caption{A pro-tensor network to be contracted}
    \label{subfig1.fig.contract_precise}
    \end{subfigure}
    \hspace{0.05\textwidth}
    \begin{subfigure}{0.45\textwidth}
    \[
    \ctikz{[scale=1.5]
        \draw[dashed] (1,2.7)node[anchor=south]{0}--(1,-1.4);
        \draw[dashed] (0,2.7)node[anchor=south]{1}--(0,-1.4);
        \draw[dashed] (-1,2.7)node[anchor=south]{2}--(-1,-1.4);
        \draw (1,2) -- (0,2)node[pos=0.5,above,yshift=-0.05cm]{$\cH$} -- (-1,2) node[pos=0.5,above,yshift=-0.05cm]{$\cH$} --(-2,2) node[pos=0.5,above,yshift=-0.05cm]{$\cK$};
        \draw (1,2)--(0,1.3) node[pos=0.4,below,yshift=0.05cm]{$\cG$} --(-1,2) node[pos=0.6,below]{$\cC$};
        \draw (2,1.3)--(1,1.3)node[pos=0.5,below,yshift=0.05cm]{$\cE$}--(0,1.3) node[pos=0.5,below,yshift=0.05cm]{$\cE$};
        \draw (2,0.7)--(1,0) node[pos=0.4,above,yshift=-0.05cm]{$\cC$}-- (0,0) node[pos=0.5,below]{$\cC$} -- (-1,0) node[pos=0.5,below]{$\cD$}-- (-2,0) node[pos=0.5,below]{$\cD$} ;
        \draw (2,-0.7)--(1,0) node[pos=0.4,below]{$\cB$};
        \mynode{1}{2}{0.2}{0.2}{F_1}
        \mynode{1}{0}{0.2}{0.2}{F_2}
        \mynode{0}{1.3}{0.2}{0.2}{F_3}
        \mynode{0}{0}{0.2}{0.2}{F_4}
        \mynode{-1}{2}{0.2}{0.2}{F_5}
        \mynode{-1}{0}{0.2}{0.2}{\Id_\cD}
        \mynode{1}{1.3}{0.2}{0.2}{\Id_\cE}
        \mynode{0}{2}{0.2}{0.2}{\Id_\cH}
    }
    \]
    \caption{First step of contraction}
    \label{subfig2.fig.contract_precise}
    \end{subfigure}
    \caption{Illustration of contraction of a pro-tensor network}
    \label{fig.contract_precise}
\end{figure}

We believe that it is enough to illustrate the contraction process through examples. Figure \ref{subfig1.fig.contract_precise} displays a pro-tensor network with five vertices, three incoming edges, two outcoming edges and four internal edges (all coming from right to left). The vertices are assigned profunctors $F_1\:*\arprof\cH\os\cG,F_2\:\cC\os\cB\arprof\cC,F_3\:\cG\os\cE\arprof\cC,F_4\:\cC\arprof\cD,F_5\:\cH\os\cC\arprof\cK$ respectively. The pro-tensor network is equipped with a ``level map'', such that each vertex $v$ has a level $\ell(v)\in\{0,1,2\}$, indicated by the number above the dashed line passing through the vertex. 

To obtain the contracted pro-tensor, we conduct the following three steps:
\begin{enumerate}
    \item Create a vertex in each crossing of a dashed line and an edge and assign it the identity profunctor (Example \ref{ex.idprof}) for the $\cV$-category on the edge, see Figure \ref{subfig2.fig.contract_precise} for an illustration. 
    \item For each level $n\in\{0,1,2\}$, obtain the tensor product $F^{(n)}$ of profunctors assigned to vertices on this level from top to down, i.e., $F^{(n)}\defdtobe \bigotimes_{v\in\ell^{-1}(n)}F_v$, where $F_v$ is the profunctor assigned to the vertex $v$. In our example, these tensor products read
    \[
        F^{(0)}=F_1\os \Id_\cE\os F_2,\qquad F^{(1)}=\Id_\cH\os F_3\os F_4,\qquad F^{(2)}=F_5\os\Id_\cD.
    \]
    \item The contracted pro-tensor $F^{(ct)}$ is defined as 
    \[
        F^{(ct)}\defdtobe F^{(2)}\bullet F^{(1)}\bullet F^{(0)}\:\cE\os\cC\os\cB\arprof\cK\os\cD.
    \]    
\end{enumerate}
\begin{remark}
    The dashed lines in Figure \ref{subfig1.fig.contract_precise} and \ref{subfig2.fig.contract_precise} are included only for expository purpose, and will often be omitted in practice. 
\end{remark}

\begin{remark}
    There is a ``Feynman rule''-like way of defining the contracted pro-tensor, which is equivalent to the one above thanks to the Fubini's theorem (Proposition \ref{prp.Fubini}) and \eqref{eq.coYoneda}. In this approach, the computation of the profunctor $F^{(ct)}$ is even more direct, given by the following two steps:
    \begin{enumerate}
        \item Write down a symbol besides each edge, which represents an object from the $\cV$-category assigned to this edge. For example, from Figure \ref{subfig1.fig.contract_precise} we obtain
        \[
            \ctikz{[scale=1]
                \draw (1,2) -- (-1,2) node[pos=0.3,above,yshift=-0.05cm]{$h$} --(-2,2) node[pos=0.5,above,yshift=-0.05cm]{$k$};
                \draw (1,2)--(0,1.3) node[pos=0.4,below]{$g$} --(-1,2) node[pos=0.6,below]{$c''$};
                \draw (2,1.3)--(0,1.3) node[pos=0.2,below,yshift=0.05cm]{$e$};
                \draw (2,0.7)--(1,0) node[pos=0.4,above,yshift=-0.05cm]{$c$}-- (0,0) node[pos=0.5,below]{$c'$} -- (-2,0) node[pos=0.7,below]{$d$} ;
                \draw (2,-0.7)--(1,0) node[pos=0.4,below]{$b$};
                \mynode{1}{2}{0.3}{0.3}{F_1}
                \mynode{1}{0}{0.3}{0.3}{F_2}
                \mynode{0}{1.3}{0.3}{0.3}{F_3}
                \mynode{0}{0}{0.3}{0.3}{F_4}
                \mynode{-1}{2}{0.3}{0.3}{F_5}
            }.
        \]
        Note that a symbol assigned to an internal edge cannot be assigned to any other edge (internal or external), even if both edges are assigned the same $\cV$-category. 
        \item We write down the object in $\cV$ for each vertex given by the evaluation of the assigned profunctor at the assigned edge labels. For example, for the vertex in the upper-left corner of our example, the object reads $F_5(k,h,c'')\in\cV$. The other vertices correspond to objects $F_1(h,g,\ast),F_2(c',c,b),F_3(c'',g,e)$ and $F_4(d,c')$ respectively. Next, tensor these objects together and write down a coend formula which ``integrate out'' all ``internal degree of freedoms'', i.e., the objects assigned to all internal edges:
        \eqnn{\label{eq.Feynman}
            \int^{c',c''\in\cC}\int^{h\in\cH,\;g\in\cG}F_1(h,g,*)\os F_2(c',c,b)\os F_3(c'',g,e)\os F_4(d,c')\os F_5(k,h,c'').
        }
        Note that \eqref{eq.Feynman} is a concrete object in $\cV$ that can be computed. Then, we define \eqref{eq.Feynman} as the effect on objects of profunctor $F^{(ct)}$ at the external labels we wrote down, i.e., $F^{(ct)}(k,d,e,c,b)$. The action part of $F^{(ct)}$ is induced from the universal property of coends. This completes our alternative definition of $F^{(ct)}$.
    \end{enumerate}
\end{remark}

\begin{remark}
    Graphically, we need both a top-to-bottom orientation to define the tensor product between $\cV$-categories and a right-to-left orientation to define the composition of profunctors. One can always assign these orientations locally to a node (in a small region), but the entire network may not come with an overall orientation, which causes potential ambiguity to pro-tensor contraction. For example, on an open disk with holes, one may find ambiguities on how to arrange the order of tensor product between the labeling categories in the holes. Fortunately, due the nice properties of profunctors, such as symmetric braiding and duality, the ambiguities can be easily overcome. A typical example we will encounter is the pro-tensor network $\cM\Omega_\cC\cN$ and we will explain more subtlety then. See also the discussion in Appendix~\ref{sub.is_probimonad}.
\end{remark}

\begin{remark}
\label{rmk.ptn_as_qft}

A $(\Vect)$-pro-tensor network may be viewed as a discrete quantum field theory, or equivalently as a lattice model. The edges and vertices encode the discretized space, while the pro-tensors supported at the vertices represent local quantum degrees of freedom. After fixing objects on all external edges, the contracted pro-tensor $F^{(ct)}$ gives the total state space. The actions of the edge categories on the pro-tensors supported at the vertices serve to \emph{entangle} these local degrees of freedom. In the special case where every edge is assigned the trivial category $\ast$, each vertex pro-tensor $F_v$ is simply a vector space, and the contracted pro-tensor reduces to the tensor product space
$F^{(ct)}=\bigotimes_v F_v$.
In the general case, where the edge-category actions are nontrivial, \(F^{(ct)}\) is not necessarily a tensor product space. However, the theory still admits a well-defined notion of locality, as will be explained in Section~\ref{sub.map}.

\end{remark}

\begin{remark}
\label{rmk.RealspaceRG}

Contraction in a pro-tensor network merges the supporting vertices of the pro-tensors into a single vertex and simultaneously assembles the corresponding pro-tensors into a new pro-tensor. This operation is analogous to real-space block-spin renormalization: as space is coarse-grained to a larger length scale, the local degrees of freedom are transformed accordingly. Although the contracted pro-tensor, and hence the total state space $F^{(ct)}$, is independent of the particular contraction process, some of the local data encoded by the original vertex pro-tensors is discarded during contraction.

\end{remark}


\subsection{Maps between pro-tensor networks and the locality of pro-tensor network}\label{sub.map}
In this subsection, we define maps between pro-tensor networks and discuss their properties. Let us first recall that between two profunctors, we can define the notion of homomorphism (Definition \ref{dfn.profunctor} or Definition \ref{dfn.homo_unpack}).
\begin{definition}
    Let $\Sigma_1,\Sigma_2$ be pro-tensor networks whose external data coincide, that is, their sets $E_\partial^-$ of incoming edges coincide and their sets $E_\partial^+$ of outgoing edges coincide, and moreover, the assigned $\cV$-categories to those edges are the same. Let $F^{(ct)}_i\:\bigotimes_{e\in E_\partial^-}\cC_e\arprof\bigotimes_{e'\in E_\partial^+}\cC_{e'}$ be the contracted pro-tensor of $\Sigma_i$ for $i=1,2$. Then a \emph{map between pro-tensor networks} from $\Sigma_1$ to $\Sigma_2$ is a profunctor homomorphism $F^{(ct)}_1\Rightarrow F^{(ct)}_2$.
\end{definition}
In what follows, we provide basic examples of maps between pro-tensor networks, discuss their properties, and describe the operations that can be performed on them. 
\begin{example}[Standard pro-tensor network presentation of a profunctor homomorphism]\label{ex.tautological_ptn_map}
    This example shares the same spirit as Example \ref{ex.tautological_ptn}. Since the contracted pro-tensor of the pro-tensor network \eqref{eq.tautological_ptn} is obviously $F$ itself, any profunctor homomorphism $\gamma\:F\Rightarrow G\:\cC\arprof\cD$ can be viewed as a map between pro-tensor networks
    \[
        \ctikz{[scale = 0.5,baseline=-1ex]
            \draw (-2.5,0)--(0,0);
            \node at (-1.5,-0.5) {$\cD$};
            \draw (2.5,0)--(0,0);
            \node at (1.5,-0.5) {$\cC$};
            \mynode{0}{0}{0.6}{0.6}{F}
        }
        \;\stackrel{\gamma}{\Rightarrow}\;
        \ctikz{[scale = 0.5,baseline=-1ex]
            \draw (-2.5,0)--(0,0);
            \node at (-1.5,-0.5) {$\cD$};
            \draw (2.5,0)--(0,0);
            \node at (1.5,-0.5) {$\cC$};
            \mynode{0}{0}{0.6}{0.6}{F}
        }.
    \]
\end{example}
\begin{example}\label{ex.unitor_in_Vprof}
    Let $F\:\cC\arprof\cD$ be a profunctor. By \eqref{eq.coYoneda}, there are canonical invertible maps of pro-tensor networks
    \[
        \twonodenolabeledge{F}{\Id_\cC}
        \;\stackrel{\rho_F}{\natiso}\;
        \onenodenolabeledge{F}
        \;\stackrel{\lambda_F^{-1}}{\natiso}\;
        \twonodenolabeledge{\Id_\cD}{F}.
    \]
    This justifies the terminology of ``identity profunctor''.
\end{example}
\begin{example}\label{ex.swapper}
    Let $F\:\cC\arprof\cD$ and $G\:\cE\arprof\cH$ be profunctors. Then there are canonical invertible maps of pro-tensor networks
    \[
        \ctikz{[scale=1]
            \draw (1,0.5)--(-1,0.5);
            \draw (1,-0.5)--(-1,-0.5);
            \mynode{-0.33}{0.5}{0.3}{0.3}{F}
            \mynode{0.33}{-0.5}{0.3}{0.3}{G}
        }
        \;\stackrel{\sigma_{F,G}}{\natiso}\;
        \ctikz{[scale=1]
            \draw (1,0.5)--(-1,0.5);
            \draw (1,-0.5)--(-1,-0.5);
            \mynode{0}{0.5}{0.3}{0.3}{F}
            \mynode{0}{-0.5}{0.3}{0.3}{G}
        }
        \;\stackrel{{\sigma'}_{F,G}^{-1}}{\natiso}\;
        \ctikz{[scale=1]
            \draw (1,0.5)--(-1,0.5);
            \draw (1,-0.5)--(-1,-0.5);
            \mynode{0.33}{0.5}{0.3}{0.3}{F}
            \mynode{-0.33}{-0.5}{0.3}{0.3}{G}
        }.
    \]
\end{example}
The most important message we want to convey is that ``local pro-tensor network map'' induces a ``global pro-tensor network map'' in a functorial way.

\begin{theorem}\label{thm.single_locally_induce}
    Let $\gamma\:F\Rightarrow F'\:\cC\arprof\cD$ be a profunctor homomorphism. Then $\gamma$ induces a map
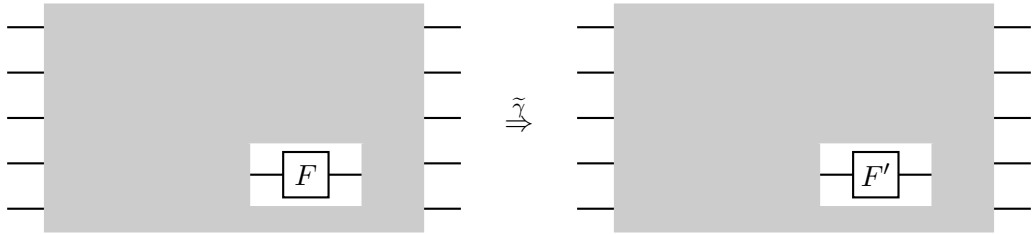
\[
    \ctikz{
        \foreach \y in {-1.2, -0.6, 0.0, 0.6, 1.2} {
            \draw (-3.0, \y) -- (3.0, \y);
        }
        \draw[gray!40,fill=gray!40] (-2.5,-1.5)rectangle(2.5,1.5);
        \def\xmiddle{0.95cm}
        \def\ymiddle{-0.75cm}
        \draw[white,fill=white](\xmiddle-0.75cm,\ymiddle+0.4cm)rectangle(\xmiddle+0.75cm,\ymiddle-0.4cm);
        \draw (\xmiddle-0.75cm,\ymiddle)--(\xmiddle+0.75cm,\ymiddle);
        \mynode{\xmiddle}{\ymiddle}{0.3cm}{0.3cm}{F}
        \draw[gray!40,fill=gray!40] (\xmiddle-0.75cm,\ymiddle+0.4cm) rectangle (\xmiddle-0.85cm,\ymiddle-0.4cm);  
        \draw[gray!40,fill=gray!40] (\xmiddle+0.75cm,\ymiddle+0.4cm) rectangle (\xmiddle+0.85cm,\ymiddle-0.4cm); 
    }
    \quad
    \stackrel{\widetilde{\gamma}}{\Rightarrow}
    \quad
    \ctikz{
        \foreach \y in {-1.2, -0.6, 0.0, 0.6, 1.2} {
            \draw (-3.0, \y) -- (3.0, \y);
        }
        \draw[gray!40,fill=gray!40] (-2.5,-1.5)rectangle(2.5,1.5);
        \def\xmiddle{0.95cm}
        \def\ymiddle{-0.75cm}
        \draw[white,fill=white](\xmiddle-0.75cm,\ymiddle+0.4cm)rectangle(\xmiddle+0.75cm,\ymiddle-0.4cm);
        \draw (\xmiddle-0.75cm,\ymiddle)--(\xmiddle+0.75cm,\ymiddle);
        \mynode{\xmiddle}{\ymiddle}{0.3cm}{0.3cm}{F'}
        \draw[gray!40,fill=gray!40] (\xmiddle-0.75cm,\ymiddle+0.4cm) rectangle (\xmiddle-0.85cm,\ymiddle-0.4cm);  
        \draw[gray!40,fill=gray!40] (\xmiddle+0.75cm,\ymiddle+0.4cm) rectangle (\xmiddle+0.85cm,\ymiddle-0.4cm); 
    }
\]
between two pro-tensor networks that differ precisely at one vertex, which is assigned $F$ and $F'$ respectively (the gray part represents the other part of the pro-tensor network). Moreover, the assignment $\gamma\mapsto\widetilde{\gamma}$ is functorial, i.e., there is $\widetilde{\id_F}=\id$, and $\widetilde{\gamma'\cdot\gamma}=\widetilde{\gamma'}\cdot\widetilde{\gamma}$ for any profunctors $F,F',F'':\cC\nrightarrow \cD$ and any homomorphisms $\gamma\:F\Rightarrow F',\gamma'\:F'\Rightarrow F''$.
\begin{proof}
    By definition of contraction of pro-tensor, it is enough to consider two special cases: the two pro-tensor networks read
    \[      
        \ctikz{[scale=1]
            \draw (1,0.5)--(-1,0.5);
            \draw (1,-0.5)--(-1,-0.5);
            \mynode{0}{0.5}{0.3}{0.3}{G}
            \mynode{0}{-0.5}{0.3}{0.3}{F}
        }
        \quad\text{and}\quad
        \ctikz{[scale=1]
            \draw (1,0.5)--(-1,0.5);
            \draw (1,-0.5)--(-1,-0.5);
            \mynode{0}{0.5}{0.3}{0.3}{G}
            \mynode{0}{-0.5}{0.3}{0.3}{F'}
        },
    \]
    respectively, for some profunctor $G\:\cB\to\cE$ (or the dual case where $G$ is put below $F$ and $F'$),  
    or the two pro-tensor networks read
    \[        
        \twonodenolabeledge{H}{F}
        \quad\text{and}\quad
        \twonodenolabeledge{H}{F'},
    \]
    respectively, for some profunctor $H\:\cD\to\cK$ (or the dual case where $F$ is to the left of some $H'\:\cB\arprof\cC$). 
    In the first case, it is direct to see that $\gamma\:F\Rightarrow F'$ induces a profunctor homomorphism $\id_G\os\gamma\:G\os F\Rightarrow G\os F'$. In the second case, note that by universal property of coends, $\gamma\:F\Rightarrow F'$ induces a profunctor homomorphism $\id_H\bullet \gamma\:H\bullet F\Rightarrow H\bullet F'$ in a functorial way,
    \begin{equation*}
        \widetilde{ \gamma'\cdot \gamma} = (\id_H\cdot \id_H)\bullet (\gamma'\cdot \gamma)  = (\id_H\bullet\gamma')\cdot (\id_H\bullet \gamma) = \widetilde{\gamma'}\cdot\widetilde\gamma
    \end{equation*}
    where the second equality is due to the interchanging law of pro-tensor homomorphism, which is proved in Proposition \ref{Prop.interchangeing}.
\end{proof}
\end{theorem}

Theorem \ref{thm.single_locally_induce} can be generalized as follows.
\begin{theorem}\label{thm.locally_induce}
    Let $\gamma$ be a map between pro-tensor networks as follows.
\[
\ctikz{
    \def\amp{0.1cm}
    \def\waveLen{0.6cm}
    \foreach \y in {0.6,0,-0.6} {
            \draw (-1.2, \y) -- (1.2, \y);
    }
    \fill[color2]         
        decorate[decoration={snake, amplitude=\amp, segment length=\waveLen}]{(0.75,1) -- (0.75,-1)} -- (-0.05,-1) -- (-0.05,1) -- cycle;
    \begin{scope}[xscale=-1]
        \fill[color2]         
        decorate[decoration={snake, amplitude=\amp, segment length=\waveLen}]{(0.75,1) -- (0.75,-1)} -- (-0.05,-1) -- (-0.05,1) -- cycle;
    \end{scope}
}
\;\stackrel{\gamma}{\Rightarrow}\;
\ctikz{
    \foreach \y in {0.6,0,-0.6} {
        \draw (-1.2, \y) -- (1.2, \y);
    }
    \draw[color2, fill=color2] (0.75,1)rectangle(-0.75,-1);
}.
\]
Then $\gamma$ induces a map
\[
\ctikz{
    \foreach \y in {-1.2, -0.6, 0.0, 0.6, 1.2} {
        \draw (2.5, \y) -- (3.0, \y);
        \draw (-2.5, \y) -- (-3.0, \y);
    }
    \def\xmiddle{0.9cm}
    \def\ymiddle{-0.6cm}
    \draw[gray!40,fill=gray!40] (-2.5,-1.5)rectangle(2.5,1.5);
    \draw[white,fill=white] (\xmiddle-0.6cm,\ymiddle-0.6cm)rectangle(\xmiddle+0.6cm,\ymiddle+0.6cm);
    \begin{scope}[xshift=\xmiddle,yshift=\ymiddle,scale=0.5]
        \def\amp{0.05cm}
        \def\waveLen{0.3cm}
        \foreach \y in {0.6,0,-0.6} {
            \draw (-1.2, \y) -- (1.2, \y);
        }
        \fill[color2]         
            decorate[decoration={snake, amplitude=\amp, segment length=\waveLen}]{(0.75,1) -- (0.75,-1)} -- (-0.05,-1) -- (-0.05,1) -- cycle;
        \begin{scope}[xscale=-1]
            \fill[color2]         
            decorate[decoration={snake, amplitude=\amp, segment length=\waveLen}]{(0.75,1) -- (0.75,-1)} -- (-0.05,-1) -- (-0.05,1) -- cycle;
        \end{scope}
    \end{scope}
    \draw[gray!40,fill=gray!40] (\xmiddle-0.6cm-0.05cm,\ymiddle-0.6cm-0.05cm)rectangle(\xmiddle-0.6cm+0.05cm,\ymiddle+0.6cm+0.05cm);
    \draw[gray!40,fill=gray!40] (\xmiddle+0.6cm-0.05cm,\ymiddle-0.6cm-0.05cm)rectangle(\xmiddle+0.6cm+0.05cm,\ymiddle+0.6cm+0.05cm);
}
\;\stackrel{\widetilde{\gamma}}{\Rightarrow}\;
\ctikz{
    \foreach \y in {-1.2, -0.6, 0.0, 0.6, 1.2} {
        \draw (2.5, \y) -- (3.0, \y);
        \draw (-2.5, \y) -- (-3.0, \y);
    }
    \def\xmiddle{0.9cm}
    \def\ymiddle{-0.6cm}
    \draw[gray!40,fill=gray!40] (-2.5,-1.5)rectangle(2.5,1.5);
    \draw[white,fill=white] (\xmiddle-0.6cm,\ymiddle-0.6cm)rectangle(\xmiddle+0.6cm,\ymiddle+0.6cm);
    \begin{scope}[xshift=\xmiddle,yshift=\ymiddle,scale=0.5]
        \foreach \y in {0.6,0,-0.6} {
            \draw (-1.2, \y) -- (1.2, \y);
        }
        \draw[color2, fill=color2] (0.75,1)rectangle(-0.75,-1);
    \end{scope}
    \draw[gray!40,fill=gray!40] (\xmiddle-0.6cm-0.05cm,\ymiddle-0.6cm-0.05cm)rectangle(\xmiddle-0.6cm+0.05cm,\ymiddle+0.6cm+0.05cm);
    \draw[gray!40,fill=gray!40] (\xmiddle+0.6cm-0.05cm,\ymiddle-0.6cm-0.05cm)rectangle(\xmiddle+0.6cm+0.05cm,\ymiddle+0.6cm+0.05cm);
}
\]
between pro-tensor networks that differ exactly at the unshaded area. Moreover, the assignment $\gamma\mapsto\widetilde{\gamma}$ is functorial.
\end{theorem}

These maps between pro-tensor networks induced by maps between local pro-tensor networks are ``local'' in a precise way we discuss now.
\begin{lemma}\label{lem.local}
    Let $\gamma\:F\Rightarrow F'\:\cC\arprof\cD$ and $\theta\:G\Rightarrow G'\:\cB\arprof\cE$ be profunctor homomorphisms. Then the following diagram
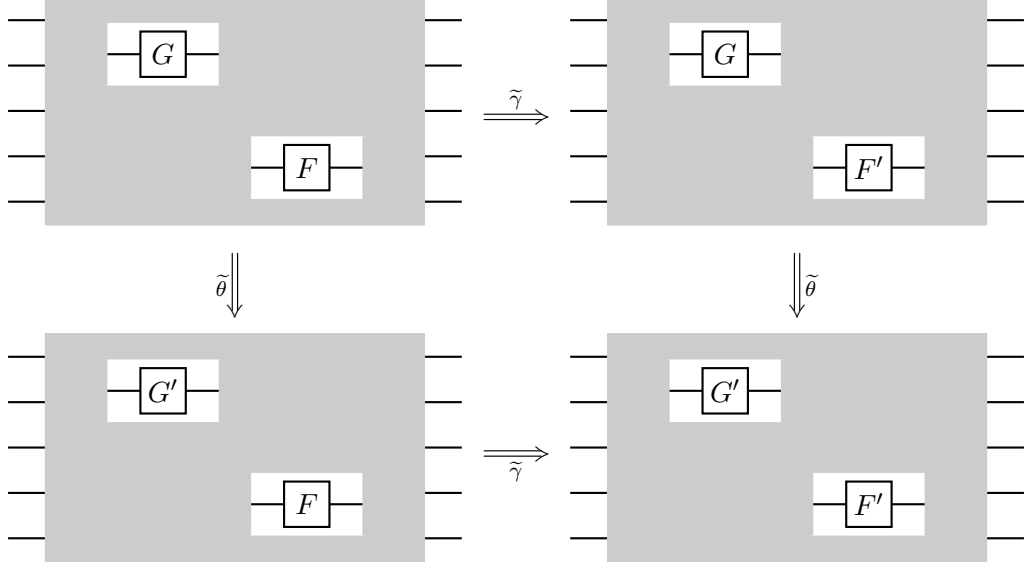
\[
    \diagram@M=0.5pc{
    \ctikz{
        \foreach \y in {-1.2, -0.6, 0.0, 0.6, 1.2} {
            \draw (-3.0, \y) -- (3.0, \y);
        }
        \draw[gray!40,fill=gray!40] (-2.5,-1.5)rectangle(2.5,1.5);
        \def\xmiddle{0.95cm}
        \def\ymiddle{-0.75cm}
        \def\xmiddletwo{-0.95cm}
        \def\ymiddletwo{0.75cm}
        \draw[white,fill=white](\xmiddle-0.75cm,\ymiddle+0.4cm)rectangle(\xmiddle+0.75cm,\ymiddle-0.4cm);
        \draw[white,fill=white](\xmiddletwo-0.75cm,\ymiddletwo+0.4cm)rectangle(\xmiddletwo+0.75cm,\ymiddletwo-0.4cm);
        \draw (\xmiddle-0.75cm,\ymiddle)--(\xmiddle+0.75cm,\ymiddle);
        \draw (\xmiddletwo-0.75cm,\ymiddletwo)--(\xmiddletwo+0.75cm,\ymiddletwo);
        \mynode{\xmiddle}{\ymiddle}{0.3cm}{0.3cm}{F};
        \mynode{\xmiddletwo}{\ymiddletwo}{0.3cm}{0.3cm}{G};
        \draw[gray!40,fill=gray!40] (\xmiddle-0.75cm,\ymiddle+0.4cm) rectangle (\xmiddle-0.85cm,\ymiddle-0.4cm);  
        \draw[gray!40,fill=gray!40] (\xmiddle+0.75cm,\ymiddle+0.4cm) rectangle (\xmiddle+0.85cm,\ymiddle-0.4cm);
        \draw[gray!40,fill=gray!40] (\xmiddletwo-0.75cm,\ymiddletwo+0.4cm) rectangle (\xmiddletwo-0.85cm,\ymiddletwo-0.4cm);  
        \draw[gray!40,fill=gray!40] (\xmiddletwo+0.75cm,\ymiddletwo+0.4cm) rectangle (\xmiddletwo+0.85cm,\ymiddletwo-0.4cm); 
    }
    \ar@{=>}[r]^-{\widetilde{\gamma}} \ar@{=>}[d]_-{\widetilde{\theta}} & 
    \ctikz{
        \foreach \y in {-1.2, -0.6, 0.0, 0.6, 1.2} {
            \draw (-3.0, \y) -- (3.0, \y);
        }
        \draw[gray!40,fill=gray!40] (-2.5,-1.5)rectangle(2.5,1.5);
        \def\xmiddle{0.95cm}
        \def\ymiddle{-0.75cm}
        \def\xmiddletwo{-0.95cm}
        \def\ymiddletwo{0.75cm}
        \draw[white,fill=white](\xmiddle-0.75cm,\ymiddle+0.4cm)rectangle(\xmiddle+0.75cm,\ymiddle-0.4cm);
        \draw[white,fill=white](\xmiddletwo-0.75cm,\ymiddletwo+0.4cm)rectangle(\xmiddletwo+0.75cm,\ymiddletwo-0.4cm);
        \draw (\xmiddle-0.75cm,\ymiddle)--(\xmiddle+0.75cm,\ymiddle);
        \draw (\xmiddletwo-0.75cm,\ymiddletwo)--(\xmiddletwo+0.75cm,\ymiddletwo);
        \mynode{\xmiddle}{\ymiddle}{0.3cm}{0.3cm}{F'};
        \mynode{\xmiddletwo}{\ymiddletwo}{0.3cm}{0.3cm}{G};
        \draw[gray!40,fill=gray!40] (\xmiddle-0.75cm,\ymiddle+0.4cm) rectangle (\xmiddle-0.85cm,\ymiddle-0.4cm);  
        \draw[gray!40,fill=gray!40] (\xmiddle+0.75cm,\ymiddle+0.4cm) rectangle (\xmiddle+0.85cm,\ymiddle-0.4cm);
        \draw[gray!40,fill=gray!40] (\xmiddletwo-0.75cm,\ymiddletwo+0.4cm) rectangle (\xmiddletwo-0.85cm,\ymiddletwo-0.4cm);  
        \draw[gray!40,fill=gray!40] (\xmiddletwo+0.75cm,\ymiddletwo+0.4cm) rectangle (\xmiddletwo+0.85cm,\ymiddletwo-0.4cm); 
    }
    \ar@{=>}[d]^-{\widetilde{\theta}} \\
    \ctikz{
        \foreach \y in {-1.2, -0.6, 0.0, 0.6, 1.2} {
            \draw (-3.0, \y) -- (3.0, \y);
        }
        \draw[gray!40,fill=gray!40] (-2.5,-1.5)rectangle(2.5,1.5);
        \def\xmiddle{0.95cm}
        \def\ymiddle{-0.75cm}
        \def\xmiddletwo{-0.95cm}
        \def\ymiddletwo{0.75cm}
        \draw[white,fill=white](\xmiddle-0.75cm,\ymiddle+0.4cm)rectangle(\xmiddle+0.75cm,\ymiddle-0.4cm);
        \draw[white,fill=white](\xmiddletwo-0.75cm,\ymiddletwo+0.4cm)rectangle(\xmiddletwo+0.75cm,\ymiddletwo-0.4cm);
        \draw (\xmiddle-0.75cm,\ymiddle)--(\xmiddle+0.75cm,\ymiddle);
        \draw (\xmiddletwo-0.75cm,\ymiddletwo)--(\xmiddletwo+0.75cm,\ymiddletwo);
        \mynode{\xmiddle}{\ymiddle}{0.3cm}{0.3cm}{F};
        \mynode{\xmiddletwo}{\ymiddletwo}{0.3cm}{0.3cm}{G'};
        \draw[gray!40,fill=gray!40] (\xmiddle-0.75cm,\ymiddle+0.4cm) rectangle (\xmiddle-0.85cm,\ymiddle-0.4cm);  
        \draw[gray!40,fill=gray!40] (\xmiddle+0.75cm,\ymiddle+0.4cm) rectangle (\xmiddle+0.85cm,\ymiddle-0.4cm);
        \draw[gray!40,fill=gray!40] (\xmiddletwo-0.75cm,\ymiddletwo+0.4cm) rectangle (\xmiddletwo-0.85cm,\ymiddletwo-0.4cm);  
        \draw[gray!40,fill=gray!40] (\xmiddletwo+0.75cm,\ymiddletwo+0.4cm) rectangle (\xmiddletwo+0.85cm,\ymiddletwo-0.4cm); 
    }
    \ar@{=>}[r]_-{\widetilde{\gamma}}  &
    \ctikz{
        \foreach \y in {-1.2, -0.6, 0.0, 0.6, 1.2} {
            \draw (-3.0, \y) -- (3.0, \y);
        }
        \draw[gray!40,fill=gray!40] (-2.5,-1.5)rectangle(2.5,1.5);
        \def\xmiddle{0.95cm}
        \def\ymiddle{-0.75cm}
        \def\xmiddletwo{-0.95cm}
        \def\ymiddletwo{0.75cm}
        \draw[white,fill=white](\xmiddle-0.75cm,\ymiddle+0.4cm)rectangle(\xmiddle+0.75cm,\ymiddle-0.4cm);
        \draw[white,fill=white](\xmiddletwo-0.75cm,\ymiddletwo+0.4cm)rectangle(\xmiddletwo+0.75cm,\ymiddletwo-0.4cm);
        \draw (\xmiddle-0.75cm,\ymiddle)--(\xmiddle+0.75cm,\ymiddle);
        \draw (\xmiddletwo-0.75cm,\ymiddletwo)--(\xmiddletwo+0.75cm,\ymiddletwo);
        \mynode{\xmiddle}{\ymiddle}{0.3cm}{0.3cm}{F'};
        \mynode{\xmiddletwo}{\ymiddletwo}{0.3cm}{0.3cm}{G'};
        \draw[gray!40,fill=gray!40] (\xmiddle-0.75cm,\ymiddle+0.4cm) rectangle (\xmiddle-0.85cm,\ymiddle-0.4cm);  
        \draw[gray!40,fill=gray!40] (\xmiddle+0.75cm,\ymiddle+0.4cm) rectangle (\xmiddle+0.85cm,\ymiddle-0.4cm);
        \draw[gray!40,fill=gray!40] (\xmiddletwo-0.75cm,\ymiddletwo+0.4cm) rectangle (\xmiddletwo-0.85cm,\ymiddletwo-0.4cm);  
        \draw[gray!40,fill=gray!40] (\xmiddletwo+0.75cm,\ymiddletwo+0.4cm) rectangle (\xmiddletwo+0.85cm,\ymiddletwo-0.4cm); 
    }
    }
\]
of maps between pro-tensor networks is commutative.
\begin{proof}
    Similar to the proof of Theorem \ref{thm.single_locally_induce}, it suffices to consider the case where $G$ and $F$ lies in the same vertical line or horizontal line. 
    The commutativity of the diagram in the two cases follows from the functoriality of the tensor product $\ot$ and the Proposition \ref{Prop.interchangeing} again, respectively.

\end{proof}
\end{lemma}

Generalizing Lemma \ref{lem.local}, we obtain the following theorem.
\begin{theorem}[Locality of pro-tensor networks]\label{thm.local}
Let $\theta$ be another map between pro-tensor network as follows
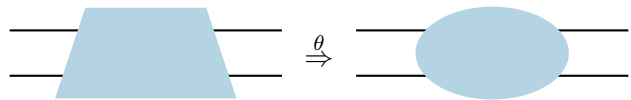
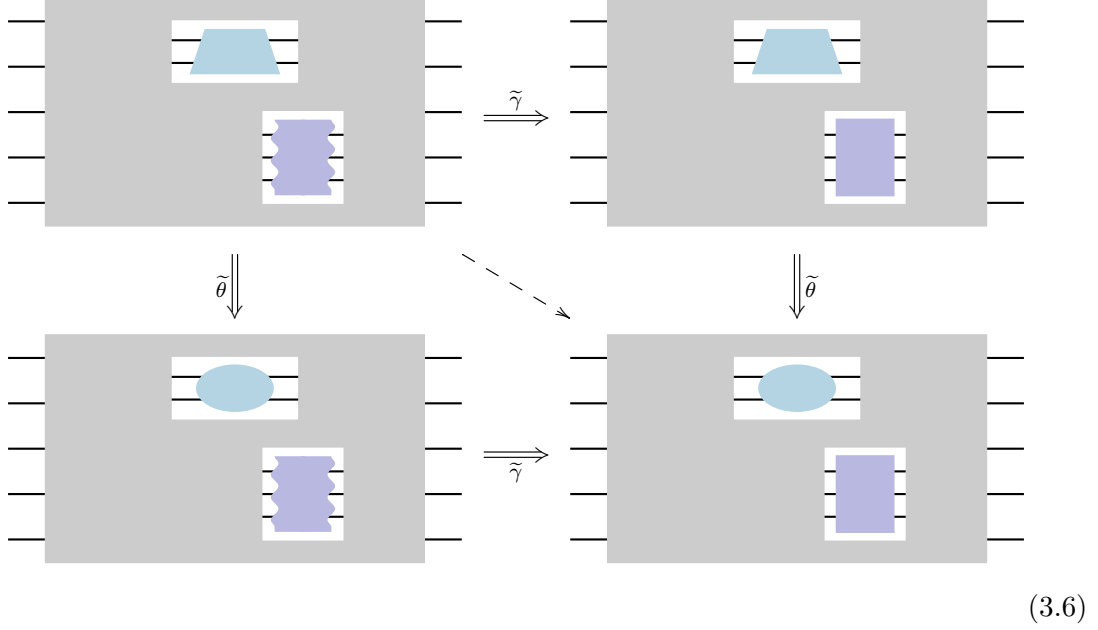
\[
    \ctikz{
        \foreach \y in {0.3,-0.3} {
            \draw (-1.8, \y) -- (1.8, \y);
        }
        \fill[color1] (-0.8,0.6) -- (0.8,0.6) -- (1.2,-0.6) -- (-1.2,-0.6) -- cycle;
    }
    \;\stackrel{\theta}{\Rightarrow}\;
    \ctikz{
        \foreach \y in {0.3,-0.3} {
            \draw (-1.8, \y) -- (1.8, \y);
        }
        \draw [color1,fill=color1] (0,0) circle[x radius = 1cm, y radius = 0.6cm];
    }
\]
Then the diagram 
\eqnn{\label{eq.thm.local}
\diagram@M=0.5pc{
    \ctikz{
        \foreach \y in {-1.2, -0.6, 0.0, 0.6, 1.2} {
            \draw (2.5, \y) -- (3.0, \y);
            \draw (-2.5, \y) -- (-3.0, \y);
        }
        \def\xmiddle{0.9cm}
        \def\ymiddle{-0.6cm}
        \def\xmiddletwo{0.0cm}
        \def\ymiddletwo{0.8cm}
        \draw[gray!40,fill=gray!40] (-2.5,-1.5)rectangle(2.5,1.5);
        \draw[white,fill=white] (\xmiddle-0.6cm,\ymiddle-0.6cm)rectangle(\xmiddle+0.6cm,\ymiddle+0.6cm);
        \draw[white,fill=white] (\xmiddletwo-0.9cm,\ymiddletwo-0.4cm) rectangle (\xmiddletwo+0.9cm,\ymiddletwo+0.4cm);
        \begin{scope}[xshift=\xmiddle,yshift=\ymiddle,scale=0.5]
            \def\amp{0.05cm}
            \def\waveLen{0.3cm}
            \foreach \y in {0.6,0,-0.6} {
                \draw (-1.2, \y) -- (1.2, \y);
            }
            \fill[color2]         
                decorate[decoration={snake, amplitude=\amp, segment length=\waveLen}]{(0.75,1) -- (0.75,-1)} -- (-0.05,-1) -- (-0.05,1) -- cycle;
            \begin{scope}[xscale=-1]
                \fill[color2]         
                decorate[decoration={snake, amplitude=\amp, segment length=\waveLen}]{(0.75,1) -- (0.75,-1)} -- (-0.05,-1) -- (-0.05,1) -- cycle;
            \end{scope}
        \end{scope}
        \begin{scope}[xshift=\xmiddletwo,yshift=\ymiddletwo,scale=0.5]
            \foreach \y in {0.3,-0.3} {
            \draw (-1.8, \y) -- (1.8, \y);
            }
            \fill[color1] (-0.8,0.6) -- (0.8,0.6) -- (1.2,-0.6) -- (-1.2,-0.6) -- cycle;
        \end{scope}
        \draw[gray!40,fill=gray!40] (\xmiddle-0.6cm-0.05cm,\ymiddle-0.6cm-0.05cm)rectangle(\xmiddle-0.6cm+0.05cm,\ymiddle+0.6cm+0.05cm);
        \draw[gray!40,fill=gray!40] (\xmiddle+0.6cm-0.05cm,\ymiddle-0.6cm-0.05cm)rectangle(\xmiddle+0.6cm+0.05cm,\ymiddle+0.6cm+0.05cm);
        \draw[gray!40,fill=gray!40] (\xmiddletwo-0.9cm-0.05cm,\ymiddletwo-0.4cm-0.05cm)rectangle(\xmiddletwo-0.9cm+0.05cm,\ymiddletwo+0.4cm+0.05cm);
        \draw[gray!40,fill=gray!40] (\xmiddletwo+0.9cm-0.05cm,\ymiddletwo-0.4cm-0.05cm)rectangle(\xmiddletwo+0.9cm+0.05cm,\ymiddletwo+0.4cm+0.05cm);
    }
    \ar@{-->}[rd]\ar@{=>}[d]_{\widetilde{\theta}} \ar@{=>}[r]^-{\widetilde{\gamma}} & 
    \ctikz{
        \foreach \y in {-1.2, -0.6, 0.0, 0.6, 1.2} {
            \draw (2.5, \y) -- (3.0, \y);
            \draw (-2.5, \y) -- (-3.0, \y);
        }
        \def\xmiddle{0.9cm}
        \def\ymiddle{-0.6cm}
        \def\xmiddletwo{0.0cm}
        \def\ymiddletwo{0.8cm}
        \draw[gray!40,fill=gray!40] (-2.5,-1.5)rectangle(2.5,1.5);
        \draw[white,fill=white] (\xmiddle-0.6cm,\ymiddle-0.6cm)rectangle(\xmiddle+0.6cm,\ymiddle+0.6cm);
        \draw[white,fill=white] (\xmiddletwo-0.9cm,\ymiddletwo-0.4cm) rectangle (\xmiddletwo+0.9cm,\ymiddletwo+0.4cm);
        \begin{scope}[xshift=\xmiddle,yshift=\ymiddle,scale=0.5]
            \foreach \y in {0.6,0,-0.6} {
                \draw (-1.2, \y) -- (1.2, \y);
            }
            \draw[color2, fill=color2] (0.75,1)rectangle(-0.75,-1);
        \end{scope}
        \begin{scope}[xshift=\xmiddletwo,yshift=\ymiddletwo,scale=0.5]
            \foreach \y in {0.3,-0.3} {
            \draw (-1.8, \y) -- (1.8, \y);
            }
            \fill[color1] (-0.8,0.6) -- (0.8,0.6) -- (1.2,-0.6) -- (-1.2,-0.6) -- cycle;
        \end{scope}
        \draw[gray!40,fill=gray!40] (\xmiddle-0.6cm-0.05cm,\ymiddle-0.6cm-0.05cm)rectangle(\xmiddle-0.6cm+0.05cm,\ymiddle+0.6cm+0.05cm);
        \draw[gray!40,fill=gray!40] (\xmiddle+0.6cm-0.05cm,\ymiddle-0.6cm-0.05cm)rectangle(\xmiddle+0.6cm+0.05cm,\ymiddle+0.6cm+0.05cm);
        \draw[gray!40,fill=gray!40] (\xmiddletwo-0.9cm-0.05cm,\ymiddletwo-0.4cm-0.05cm)rectangle(\xmiddletwo-0.9cm+0.05cm,\ymiddletwo+0.4cm+0.05cm);
        \draw[gray!40,fill=gray!40] (\xmiddletwo+0.9cm-0.05cm,\ymiddletwo-0.4cm-0.05cm)rectangle(\xmiddletwo+0.9cm+0.05cm,\ymiddletwo+0.4cm+0.05cm);
    }
    \ar@{=>}[d]^{\widetilde{\theta}} \\
    \ctikz{
        \foreach \y in {-1.2, -0.6, 0.0, 0.6, 1.2} {
            \draw (2.5, \y) -- (3.0, \y);
            \draw (-2.5, \y) -- (-3.0, \y);
        }
        \def\xmiddle{0.9cm}
        \def\ymiddle{-0.6cm}
        \def\xmiddletwo{0.0cm}
        \def\ymiddletwo{0.8cm}
        \draw[gray!40,fill=gray!40] (-2.5,-1.5)rectangle(2.5,1.5);
        \draw[white,fill=white] (\xmiddle-0.6cm,\ymiddle-0.6cm)rectangle(\xmiddle+0.6cm,\ymiddle+0.6cm);
        \draw[white,fill=white] (\xmiddletwo-0.9cm,\ymiddletwo-0.4cm) rectangle (\xmiddletwo+0.9cm,\ymiddletwo+0.4cm);
        \begin{scope}[xshift=\xmiddle,yshift=\ymiddle,scale=0.5]
            \def\amp{0.05cm}
            \def\waveLen{0.3cm}
            \foreach \y in {0.6,0,-0.6} {
                \draw (-1.2, \y) -- (1.2, \y);
            }
            \fill[color2]         
                decorate[decoration={snake, amplitude=\amp, segment length=\waveLen}]{(0.75,1) -- (0.75,-1)} -- (-0.05,-1) -- (-0.05,1) -- cycle;
            \begin{scope}[xscale=-1]
                \fill[color2]         
                decorate[decoration={snake, amplitude=\amp, segment length=\waveLen}]{(0.75,1) -- (0.75,-1)} -- (-0.05,-1) -- (-0.05,1) -- cycle;
            \end{scope}
        \end{scope}
        \begin{scope}[xshift=\xmiddletwo,yshift=\ymiddletwo,scale=0.5]
            \foreach \y in {0.3,-0.3} {
                \draw (-1.8, \y) -- (1.8, \y);
            }   
            \draw [color1,fill=color1] (0,0) circle[x radius = 1cm, y radius = 0.6cm];
        \end{scope}
        \draw[gray!40,fill=gray!40] (\xmiddle-0.6cm-0.05cm,\ymiddle-0.6cm-0.05cm)rectangle(\xmiddle-0.6cm+0.05cm,\ymiddle+0.6cm+0.05cm);
        \draw[gray!40,fill=gray!40] (\xmiddle+0.6cm-0.05cm,\ymiddle-0.6cm-0.05cm)rectangle(\xmiddle+0.6cm+0.05cm,\ymiddle+0.6cm+0.05cm);
        \draw[gray!40,fill=gray!40] (\xmiddletwo-0.9cm-0.05cm,\ymiddletwo-0.4cm-0.05cm)rectangle(\xmiddletwo-0.9cm+0.05cm,\ymiddletwo+0.4cm+0.05cm);
        \draw[gray!40,fill=gray!40] (\xmiddletwo+0.9cm-0.05cm,\ymiddletwo-0.4cm-0.05cm)rectangle(\xmiddletwo+0.9cm+0.05cm,\ymiddletwo+0.4cm+0.05cm);
    }
    \ar@{=>}[r]_-{\widetilde{\gamma}} & 
    \ctikz{
        \foreach \y in {-1.2, -0.6, 0.0, 0.6, 1.2} {
            \draw (2.5, \y) -- (3.0, \y);
            \draw (-2.5, \y) -- (-3.0, \y);
        }
        \def\xmiddle{0.9cm}
        \def\ymiddle{-0.6cm}
        \def\xmiddletwo{0.0cm}
        \def\ymiddletwo{0.8cm}
        \draw[gray!40,fill=gray!40] (-2.5,-1.5)rectangle(2.5,1.5);
        \draw[white,fill=white] (\xmiddle-0.6cm,\ymiddle-0.6cm)rectangle(\xmiddle+0.6cm,\ymiddle+0.6cm);
        \draw[white,fill=white] (\xmiddletwo-0.9cm,\ymiddletwo-0.4cm) rectangle (\xmiddletwo+0.9cm,\ymiddletwo+0.4cm);
        \begin{scope}[xshift=\xmiddle,yshift=\ymiddle,scale=0.5]
            \foreach \y in {0.6,0,-0.6} {
                    \draw (-1.2, \y) -- (1.2, \y);
                }
            \draw[color2, fill=color2] (0.75,1)rectangle(-0.75,-1);    
        \end{scope}
        \begin{scope}[xshift=\xmiddletwo,yshift=\ymiddletwo,scale=0.5]
            \foreach \y in {0.3,-0.3} {
                \draw (-1.8, \y) -- (1.8, \y);
            }   
            \draw [color1,fill=color1] (0,0) circle[x radius = 1cm, y radius = 0.6cm];
        \end{scope}
        \draw[gray!40,fill=gray!40] (\xmiddle-0.6cm-0.05cm,\ymiddle-0.6cm-0.05cm)rectangle(\xmiddle-0.6cm+0.05cm,\ymiddle+0.6cm+0.05cm);
        \draw[gray!40,fill=gray!40] (\xmiddle+0.6cm-0.05cm,\ymiddle-0.6cm-0.05cm)rectangle(\xmiddle+0.6cm+0.05cm,\ymiddle+0.6cm+0.05cm);
        \draw[gray!40,fill=gray!40] (\xmiddletwo-0.9cm-0.05cm,\ymiddletwo-0.4cm-0.05cm)rectangle(\xmiddletwo-0.9cm+0.05cm,\ymiddletwo+0.4cm+0.05cm);
        \draw[gray!40,fill=gray!40] (\xmiddletwo+0.9cm-0.05cm,\ymiddletwo-0.4cm-0.05cm)rectangle(\xmiddletwo+0.9cm+0.05cm,\ymiddletwo+0.4cm+0.05cm);
    }
}
}
of maps between pro-tensor networks is commutative.
\end{theorem}
Theorem \ref{thm.local} shows that maps between pro-tensor networks induced by local pro-tensor networks supported on disjoint regions commute with each other, which we refer to as the locality property of pro-tensor networks.

Subsequently, to emphasize locality, for two local pro-tensor network maps $\theta$ and $\gamma$ as in \eqref{eq.thm.local}, we denote the induced global map (the dashed line in \eqref{eq.thm.local}) as $\theta\star \gamma$, which admits various equivalent expressions $\theta\star\gamma=\widetilde{\theta}\cdot\widetilde{\gamma}=\widetilde{\gamma}\cdot\widetilde{\theta}=\id\os(\id\bullet\theta\bullet \id)\os\id\os(\id\bullet\gamma\bullet\id)\os\id=\cdots$. Moreover, when it does not cause confusion, we often label the map $\widetilde{\gamma}$ between pro-tensor networks by $\gamma$ and $\widetilde{\theta}$ by $\theta$.

We end this subsection by stating the naturality of the canonical maps of pro-tensor networks in Examples \ref{ex.unitor_in_Vprof} and \ref{ex.swapper}.
\begin{theorem}\label{thm.nat_unitor_swapper}
    \begin{enumerate}
        \item Let $\gamma\:F\Rightarrow F'\:\cC\arprof\cD$ be profunctor homomorphisms. Then the following diagram of pro-tensor networks is commutative:
        \eqnn
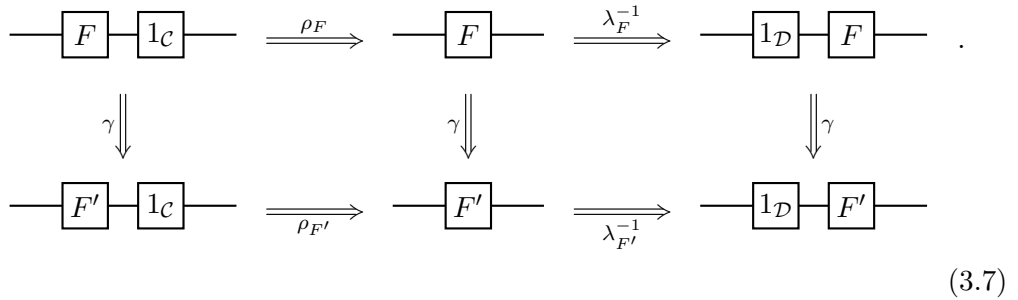
{\label{eq1.thm.nat_unitor_swapper}
            \diagram@M=0.75pc@R=2pc@C=3pc{
                \twonodenolabeledge{F}{\Id_\cC}
                \ar@{=>}[d]_{\gamma} \ar@{=>}[r]^-{\rho_F}  &
                \onenodenolabeledge{F}
                \ar@{=>}[d]_{\gamma} \ar@{=>}[r]^-{\lambda_F^{-1}} &
                \twonodenolabeledge{\Id_\cD}{F}
                \ar@{=>}[d]^{\gamma} \\
                \twonodenolabeledge{F'}{\Id_\cC}
                \ar@{=>}[r]_-{\rho_{F'}} & 
                \onenodenolabeledge{F'}
                \ar@{=>}[r]_-{\lambda_{F'}^{-1}} &
                \twonodenolabeledge{\Id_\cD}{F'}
            }.
        }
        \item Let $\gamma\:F\Rightarrow F'$ and $\theta\:G\Rightarrow G'$ be profunctor homomorphisms. Then the following diagram of pro-tensor networks is commutative:
        \eqnn{\label{eq2.thm.nat_unitor_swapper}
            \diagram@M=0.75pc@R=1.5pc@C=5pc{
                \ctikz{[scale=1]
                    \draw (1,0.5)--(-1,0.5);
                    \draw (1,-0.5)--(-1,-0.5);
                    \mynode{-0.33}{0.5}{0.3}{0.3}{F}
                    \mynode{0.33}{-0.5}{0.3}{0.3}{G}
                }
                \ar@{=>}[d]_{\gamma\star\theta} \ar@{=>}[r]^-{\sigma_{F,G}}  &
                \ctikz{[scale=1]
                    \draw (1,0.5)--(-1,0.5);
                    \draw (1,-0.5)--(-1,-0.5);
                    \mynode{0}{0.5}{0.3}{0.3}{F}
                    \mynode{0}{-0.5}{0.3}{0.3}{G}
                }
                \ar@{=>}[d]_{\gamma\star\theta} \ar@{=>}[r]^-{{\sigma'}_{F,G}^{-1}} &
                \ctikz{[scale=1]
                    \draw (1,0.5)--(-1,0.5);
                    \draw (1,-0.5)--(-1,-0.5);
                    \mynode{0.33}{0.5}{0.3}{0.3}{F}
                    \mynode{-0.33}{-0.5}{0.3}{0.3}{G}
                }
                \ar@{=>}[d]^{\theta\star\gamma} \\
                \ctikz{[scale=1]
                    \draw (1,0.5)--(-1,0.5);
                    \draw (1,-0.5)--(-1,-0.5);
                    \mynode{-0.33}{0.5}{0.3}{0.3}{F'}
                    \mynode{0.33}{-0.5}{0.3}{0.3}{G'}
                }
                \ar@{=>}[r]_-{\sigma_{F',G'}} & 
                \ctikz{[scale=1]
                    \draw (1,0.5)--(-1,0.5);
                    \draw (1,-0.5)--(-1,-0.5);
                    \mynode{0}{0.5}{0.3}{0.3}{F'}
                    \mynode{0}{-0.5}{0.3}{0.3}{G'}
                }
                \ar@{=>}[r]_-{{\sigma'}_{F',G'}^{-1}} &
                \ctikz{[scale=1]
                    \draw (1,0.5)--(-1,0.5);
                    \draw (1,-0.5)--(-1,-0.5);
                    \mynode{0.33}{0.5}{0.3}{0.3}{F'}
                    \mynode{-0.33}{-0.5}{0.3}{0.3}{G'}
                }
            }.
        }
    \end{enumerate}
\end{theorem}
Theorem \ref{thm.nat_unitor_swapper} together with a set of coherence conditions satisfied by $\rho_F,\lambda_F,\sigma_{F,G},\sigma'_{F,G}$ dictate that the three pro-tensor networks in the first line of \eqref{eq1.thm.nat_unitor_swapper} (resp. \eqref{eq2.thm.nat_unitor_swapper}) are essentially the same. From now on, we will not distinguish among pro-tensor networks that can be connected using $\rho_F,\lambda_F$ and $\sigma_{F,G}$, and in most cases suppress these three kinds of maps.

\subsection{Duality aspects for pro-tensors}
\label{sub.duality}
In Sections \ref{sub.enrichedbg}-\ref{sub.map}, we have provided the most basic ingredients in pro-tensor network theory. In this and the next two subsections, we go on to use profunctor theory to introduce some more advanced tools.

In this subsection, we focus on two duality aspects of pro-tensors. Note that we do not distinguish between a profunctor $F$ and its standard pro-tensor network presentation defined in Example \ref{ex.tautological_ptn}; similarly, we do not distinguish between a profunctor homomorphism and its corresponding pro-tensor network map in Example \ref{ex.tautological_ptn_map}.

\begin{definition}
    The \emph{right adjoint} of a pro-tensor 
    \[
        \ctikz{[scale = 0.5]
            \draw (-2.5,0)--(0,0);
            \node at (-1.5,-0.5) {$\cD$};
            \draw (2.5,0)--(0,0);
            \node at (1.5,-0.5) {$\cC$};
            \mynode{0}{0}{0.6}{0.6}{F};
        }
    \]
    is a pro-tensor
    \[
        \ctikz{[scale = 0.5]
            \draw (-2.5,0)--(0,0);
            \node at (-1.5,-0.5) {$\cC$};
            \draw (2.5,0)--(0,0);
            \node at (1.5,-0.5) {$\cD$};
            \mynode{0}{0}{0.6}{0.6}{G};
        }
    \]
    equipped with two pro-tensor network maps 
    \[
        \epsilon\:            
        \twonodenolabeledge{F}{G}
        \Rightarrow
        \ctikz{[scale = 0.5, baseline = -0.5ex]
            \draw (0,0)--(3,0);
        }
        \quad\text{and}\quad
        \eta\:\ctikz{[scale = 0.5, baseline = -0.5ex]
            \draw (0,0)--(3,0);
        }
        \Rightarrow
        \twonodenolabeledge{G}{F}
    \]
    called the \emph{counit} and the \emph{unit} respectively, which renders the following two diagrams of pro-tensor networks commutative:
    \eqnn{\label{eq.dfn.adjoint}
        \diagram@C=1.25pc{
            \onenodenolabeledge{F}
            \ar@{=>}[rd]_-{\id_F}
            \ar@{=>}[r]^-{\id_F\star\eta}
            &
            \threenodenolabeledge{F}{G}{F}
            \ar@{=>}[d]^-{\epsilon\star\id_F}
            \\
            &
            \onenodenolabeledge{F}
        }
        \quad
        \diagram@C=1.25pc{
            \onenodenolabeledge{G}
            \ar@{=>}[rd]_-{\id_G}
            \ar@{=>}[r]^-{\eta\star\id_G}
            &
            \threenodenolabeledge{G}{F}{G}
            \ar@{=>}[d]^-{\id_G\star\epsilon}
            \\
            & 
            \onenodenolabeledge{G}
        }.
    }
    We also abbreviate the triple $(G,\epsilon,\eta)$ as $G$. If $G$ is the right adjoint of $F$, then we also write $F\ladj G$ or $G\radj F$, and we say $F$ is the \emph{left adjoint} of $G$; equivalently, $(F,G)$ form an \emph{adjunction}. The right and left adjoint of a pro-tensor $F$ is denoted by $F^R$ and $F^L$, respectively.
\end{definition}
\begin{remark}
    A pro-tensor need not possess a right (or left) adjoint. On the other hand, a right adjoint of a pro-tensor, if exists, is unique up to a canonical isomorphism of pro-tensors. Indeed, if $(G,\epsilon,\eta)$ and $(G',\epsilon',\eta')$ are both right adjoints of a pro-tensor $F\:\cC\arprof\cD$, then 
    \[
        \diagram{
            \onenodenolabeledge{G'} \ar@{=>}[r]^-{\eta} & \threenodenolabeledge{G}{F}{G'} \ar@{=>}[r]^-{\epsilon'} & \onenodenolabeledge{G}
        }
    \]
    provides an isomorphism of pro-tensors, with its inverse defined in an analogous way.
\end{remark}
\begin{example}\label{ex.conjoin_right_adj}
    Let $H\:\cC\to\cD$ be a $\cV$-functor. It is well-known that the companion $H_\ast\:\cC\arprof\cD$ of $H$ defined in Example \ref{ex.companion_and_conjoin} possesses a right adjoint, given by its conjoin $H^\ast\:\cD\arprof\cC$ \cite[Proposition 3.4.1]{Benabou_1973}. The unit $\eta\:\Id_\cC\Rightarrow H^\ast\bullet H_\ast$ is given by the $\cV$-functor structure of $H$, $H_{ab}:\cC(a,b)\to \cD(Ha,Hb)$. The counit $\epsilon\:H_\ast\bullet H^\ast\Rightarrow \Id_\cD$ is given by the composition in $\cD$
   \begin{equation}
\label{eq.counit}
\begin{tikzcd}
	{H_\ast\bullet H^\ast (d',d)} && {\cD(d',d)} \\
{\cD(d',H(x))\ot \cD(H(x),d)}
	\arrow["{\epsilon_{d',d}}", from=1-1, to=1-3]
	\arrow["{\copr_x}", from=2-1, to=1-1]
\arrow["\circ"', from=2-1, to=1-3] .
\end{tikzcd}
   \end{equation}   
   We also record the axioms the unit and the counit satisfy for future use: 
    \eqnn{\label{eq.ex.conjoin_right_adj}
        \diagram@C=1pc{
            \onenodenolabeledge{H_\ast}
            \ar@{=>}[rd]_-{\id}
            \ar@{=>}[r]^-{\id\star\eta}
            &
            \threenodenolabeledge{H_\ast}{H^\ast}{H_\ast}
            \ar@{=>}[d]^-{\epsilon\star\id}
            \\
            &
            \onenodenolabeledge{H_\ast}
        }
        \quad
        \diagram@C=1pc{
            \onenodenolabeledge{H^\ast}
            \ar@{=>}[rd]_-{\id}
            \ar@{=>}[r]^-{\eta\star\id}
            &
            \threenodenolabeledge{H^\ast}{H_\ast}{H^\ast}
            \ar@{=>}[d]^-{\id\star\epsilon}
            \\
            & 
            \onenodenolabeledge{H^\ast}
        }.
    }
\end{example}
In Section \ref{sub.mon_VcatVModule}, we will encounter plenty of examples of pro-tensor possessing a right adjoint, of the form described by Example \ref{ex.conjoin_right_adj}, where the right adjoints are simply given by the mirror reflection of the left adjoint, making them extremely convenient to handle in graphs.
\begin{proposition}[Transposing]\label{prp.transpose}
    Let $F\:\cC\arprof\cD$ be a pro-tensor with right adjoint $(F^R\:\cD\arprof\cC,\epsilon,\eta)$. 
    \begin{enumerate}
        \item \label{item1.prp.transpose} Let $G\:\cB\arprof\cC$ and $H\:\cB\arprof\cD$ be pro-tensors. There exists a canonical bijection of sets
        \[
            \Hom(F\bullet G,H)\cong\Hom(G,F^R\bullet H),
        \]
        where $\Hom(F,F')$ for two pro-tensors with same domain and codomain refers to the set of profunctor homomorphisms $F\Rightarrow F'$ as in Section \ref{sub.ptn}.
        \item \label{item2.prp.transpose} Let $L\:\cC\arprof\cE$ and $K\:\cD\arprof\cE$ be pro-tensors. There exists a canonical bijection of sets
        \[
            \Hom(L\bullet F^R,K) \cong \Hom(L,K\bullet F).
        \]
    \end{enumerate}
    \begin{proof}
        The construction of the bijections are provided by ``doing yoga'' with $\eta$ and $\epsilon$. We only take \ref{item1.prp.transpose} as an example. Let us construct a pair of maps 
        \[j\:\diagram{\Hom(F\bullet G,H)\ar@<0.6ex>[r] & \Hom(G,F^R\bullet H) \ar@<0.6ex>[l]\colon k}
        \]
        as follows. Given $\phi\in\Hom(F\bullet G,H)$,
        we set $j(\phi)$ as the pro-tensor network map
        \[
            \diagram{
            \onenodenolabeledge{G}
            \ar@{=>}[r]^-{\eta} & 
            \threenodenolabeledge{F^R}{F}{G}
            \ar@{=>}[r]^-{\phi} &
            \twonodenolabeledge{F^R}{H}
            }.
        \]
        Given $\psi\in\Hom(G,F^R\bullet H)$, we set $k(\psi)$ as the pro-tensor network map 
        \[
            \diagram{
                \twonodenolabeledge{F}{G}
                \ar@{=>}[r]^-{\psi} &
                \threenodenolabeledge{F}{F^R}{H}
                \ar@{=>}[r]^-\epsilon & 
                \onenodenolabeledge{H}
            }.
        \]
        It is easy to see that $j$ and $k$ are inverses to each other by the preceding axioms we established. More concretely, one can verify that the following diagram of pro-tensor network maps commute:
        \[
            \diagram{
                \twonodenolabeledge{F}{G}
                \ar@{=>}[r]^-{\eta} \ar@{=>}[rd]_-{1} \ar@{}[rrd]|(0.35)*!/u 4pt/{\labelstyle\eqref{eq.dfn.adjoint}}
                &
                \fournodenolabeledge{F}{F^R}{F}{G} 
                \ar@{}[rd]|{\text{(Theorem \ref{thm.local})}}
                \ar@{=>}[r]^-{\phi} \ar@{=>}[d]^{\epsilon}
                & 
                \threenodenolabeledge{F}{F^R}{H}
                \ar@{=>}[d]^{\epsilon}
                \\
                &
                \twonodenolabeledge{F}{G}
                \ar@{=>}[r]_-\phi
                &
                \onenodenolabeledge{H}
            }.  
        \]
    Thus $kj(\phi)=\phi$. Similarly, the equality $jk(\psi)=\psi$ can be seen from the following commutative diagram.
    \[
        \diagram{
            \onenodenolabeledge{G}
            \ar@{}[rd]|{\text{(Theorem \ref{thm.local})}}
            \ar@{}[rrd]|(0.65)*!/d 2pt/{\labelstyle\eqref{eq.dfn.adjoint}}
            \ar@{=>}[d]_-\eta \ar@{=>}[r]^-\psi
            &
            \twonodenolabeledge{F^R}{H}
            \ar@{=>}[d]_{\eta} \ar@{=>}[rd]^-1
            \\
            \threenodenolabeledge{F^R}{F}{G}
            \ar@{=>}[r]_-{\psi} & 
            \fournodenolabeledge{F^R}{F}{F^R}{H}
            \ar@{=>}[r]_-\epsilon
            & \twonodenolabeledge{F^R}{H}
        }
    \]
    \end{proof}
\end{proposition}

The bijections in Proposition \ref{prp.transpose} are called \emph{transposing}, and the image of a pro-tensor network map under these bijections can be called a \emph{transpose} of the original map. 

As a consequence of Proposition \ref{prp.transpose}, if $F,G\:\cC\arprof\cD$ both admits a right adjoint, we can obtain a ``double transposing'' map
\eqnn{\label{eq.double_transpose}
    (-)^R\:\Hom(F,G)\isom \Hom(\Id_\cC,F^R\bullet G)\isom\Hom(G^R,F^R),\quad\phi\mapsto\phi^R.
}
The pro-tensor network map $\phi^R$ is also called the \emph{transpose} of $\phi$ by abuse of terminology. The transpose satisfies the following two properties which are not hard to prove:
\begin{itemize}
    \item The following diagrams commute:
    \eqnn{\label{eq.right_transpose}
        \diagram@C=1.25pc@R=0.9pc{
            \twonodenolabeledge{F}{G^R}
            \ar@{=>}[r]^-{\phi^R} \ar@{=>}[d]_{\phi}& 
            \twonodenolabeledge{F}{F^R}
            \ar@{=>}[d]^{\epsilon_F} \\
            \twonodenolabeledge{G}{G^R}
            \ar@{=>}[r]_-{\epsilon_G} &
            \ctikz{
                \draw (0,0)--(1.5,0);
            }
        }
        \quad
        \diagram@C=1.25pc@R=0.9pc{
        \ctikz{
                \draw (0,0)--(1.5,0);
            }
        \ar@{=>}[r]^-{\eta_F} \ar@{=>}[d]_{\eta_G} & 
        \twonodenolabeledge{F^R}{F} \ar@{=>}[d]^{\phi} \\
        \twonodenolabeledge{G^R}{G} \ar@{=>}[r]_-{\phi^R} &
        \twonodenolabeledge{F^R}{G}
        }
    }
    where $(i^R,\epsilon_i,\eta_i)$ denotes the right adjoint of $i$ for $i=F,G$.
    \item If $F,G,H\:\cC\arprof\cD$ all admit right adjoints, and $\phi\:F\Rightarrow G,\psi\:G\Rightarrow H$ are pro-tensor network maps, then 
    \eqnn{\label{eq2.right_transpose}
    \phi^R\circ\psi^R=(\psi\circ\phi)^R.
    }
\end{itemize}

\six{
$\cV$-categories, $\cV$-functors, $\cV$-natural transformations form a 2-category $\Vcat$. In summary of Section~\ref{sub.ptn} and Section~\ref{sub.contraction},
$\cV$-categories, $\cV$-profunctors and profunctor homomorphisms form a 2-category $\vprof$. The identity 1-morphism on a $\cV$-category $\cC$ is $\Id_\cC$ defined in Example~\ref{ex.idprof}, composition of 1-morphisms is contraction of pro-tensors discussed in Section \ref{sub.contraction}, composition of 2-morphisms is composition of pro-tensor homomorphisms defined in Definition~\ref{def.comp_profun_homo}. It is possible to canonically enrich hom-categories $\vprof(\cM,\cN)$ over $\cV$ and the result will be denoted by $\bvprof(\cM,\cN)$ (see Appendix \ref{app.enriched_structure} for details).
}

We now explore a different aspect of duality in profunctor theory that will lead eventually to an analogue of the two isomorphisms in Proposition \ref{prp.transpose}. Recall $\evC$ and $\coevC$ from Example \ref{ex.evC_and_coevC_graph_notation}.

\begin{proposition}\label{prp.zigzagger}
    There exist invertible pro-tensor network maps 
    \[
    \chi_\cC\:
    \ctikz{[scale=0.5,yscale=-1,baseline = (current bounding box.center)]
        \draw  (-1,0)--(2,0);
        \draw  node at (0.5,0.4) {$\cC$};
        \draw  (2,-2) arc (-90:90:1);
        \draw  (1,-2)--(2,-2);
        \draw  node at (1.5,-1.6) {$\cC^\op$};
        \draw  (1,-2) arc (90:270:1);
        \draw  (1,-4)--(4,-4);
        \draw  node at (2.5,-4.4) {$\cC$};
    }
    \stackrel{\sim}{\Rightarrow} 
    \ctikz{[baseline=2ex]
        \draw  (0,0)--(2,0)node[pos=0.5,below]{$\cC$};
    },
    \qquad
    \zeta_\cC\:\ctikz{[scale=0.5,baseline = (current bounding box.center)]
        \draw  (-1,0)--(2,0);
        \draw  node at (0.5,0.4) {$\cC^\op$};
        \draw  (2,-2) arc (-90:90:1);
        \draw  (1,-2)--(2,-2);
        \draw  node at (1.5,-1.6) {$\cC$};
        \draw  (1,-2) arc (90:270:1);
        \draw  (1,-4)--(4,-4);
        \draw  node at (2.5,-4.4) {$\cC^\op$};
    }
    \stackrel{\sim}{\Rightarrow} 
    \ctikz{[baseline=2ex]
        \draw  (0,0)--(2,0)node[pos=0.5,below]{$\cC^\op$};
    }
    \]
    \begin{proof}
    $\chi_\cC$ is given by $\int^{x\in\cC}\cC(a,x)\ot\cC(x,b)\cong \cC(a,b)$, and similarly for $\zeta_\cC$. Componentwise one can safely view both as the identity morphism.
    \end{proof}
\end{proposition}
Note that $\chi_\cC$ and $\zeta_\cC$ also renders the following two diagrams of pro-tensor network maps commutative:
\eqnn{\label{eq.zigzagger}
    \diagram{
    \ctikz{[scale=0.5,baseline = (current bounding box.center)]
        \draw  (-1,0)--(2,0);
        \draw  node at (0.5,0.4) {$\cC^\op$};
        \draw  (2,-2) arc (-90:90:1);
        \draw  (1,-2)--(2,-2);
        \draw  (1,-2) arc (90:270:1);
        \draw  (1,-4)--(2,-4);
        \draw (2,-4) arc (90:-90:1);
        \draw (2,-6)--(-1,-6);
    }
        \ar@/^2.5pc/@{=>}[r]^-{\id\star\chi_\cC} \ar@{=>}@/_2.5pc/[r]_-{\zeta_\cC\star\id} &
    \ctikz{[scale=0.5,baseline = (current bounding box.center)]
        \draw  (-1,0)--(2,0);
        \draw  node at (0.5,0.4) {$\cC^\op$};
        \draw  (2,-2) arc (-90:90:1);
        \draw  (-1,-2)--(2,-2);
    }
    }
    \qquad\qquad
    \diagram{
    \ctikz{[scale=0.5,xscale=-1,baseline = (current bounding box.center)]
        \draw  (-1,0)--(2,0);
        \draw  node at (0.5,0.4) {$\cC$};
        \draw  (2,-2) arc (-90:90:1);
        \draw  (1,-2)--(2,-2);
        \draw  (1,-2) arc (90:270:1);
        \draw  (1,-4)--(2,-4);
        \draw (2,-4) arc (90:-90:1);
        \draw (2,-6)--(-1,-6);
    }
        \ar@/^2.5pc/@{=>}[r]^-{\chi_\cC\star\id} \ar@{=>}@/_2.5pc/[r]_-{\id\star\zeta_\cC} &
    \ctikz{[scale=0.5,xscale=-1,baseline = (current bounding box.center)]
        \draw  (-1,0)--(2,0);
        \draw  node at (0.5,0.4) {$\cC$};
        \draw  (2,-2) arc (-90:90:1);
        \draw  (-1,-2)--(2,-2);
    }
    }.
}
In analogy with tensor network theory, Proposition \ref{prp.zigzagger} and \eqref{eq.zigzagger} precisely say that the $\cV$-category $\cC^\op$ plays the role of ``dual space'' of $\cC$.\footnote{Formally, $\cC^\op$ is the right dual of $\cC$ in the monoidal 2-category $(\vprof,\os,\ast)$ of $\cV$-categories, profunctors and profunctor homomorphisms.} It is well-known that for a dual vector space $V^\ast$ of a finite-dimensional vector space $V$, there exists a canonical bijection between the following two sets of tensors, where $U,V$ are vector spaces:
\eqnn{\label{eq.vector_space_adj}
    \Vect(V\os W,U)\cong \Vect(W,V^\ast\os U).
}
Analogously, we have the following equivalence of categories of pro-tensors.
\begin{theorem}\label{thm.higher_transpose}
    Let $\cC,\cD,\cE$ be $\cV$-categories. The assignments
    \[
        \ctikz{
            \draw (-1,0)--(0,0)node[pos=0.5,below]{$\cE$};
            \draw (0,0.15)--(1,0.15)node[pos=0.6,above]{$\cC$};
            \draw (0,-0.15)--(1,-0.15) node[pos=0.6,below]{$\cD$};
            \mynode{0}{0}{0.3}{0.3}{F}
        }
        \mapsto
        \ctikz{
            \draw (-1,0)--(0,0)node[pos=0.5,below]{$\cE$};
            \draw (0,0.15)--(0.5,0.15);
            \draw (0.5,0.15) arc (-90:90:0.25);
            \draw (0.5,0.65)--(-1,0.65);
            \draw (-0.4,0.9)node{$\cC^\op$};
            \draw (0,-0.15)--(1,-0.15) node[pos=0.6,below]{$\cD$};
            \mynode{0}{0}{0.3}{0.3}{F}
        }
        \qquad\text{and}\qquad
        \ctikz{[xscale=-1]
            \draw (-1,0)--(0,0)node[pos=0.5,below]{$\cD$};
            \draw (0,0.15)--(1,0.15)node[pos=0.6,above]{$\cC^\op$};
            \draw (0,-0.15)--(1,-0.15) node[pos=0.6,below]{$\cE$};
            \mynode{0}{0}{0.3}{0.3}{G}
        }
        \mapsto
        \ctikz{[xscale=-1]
            \draw (-1,0)--(0,0)node[pos=0.5,below]{$\cD$};
            \draw (0,0.15)--(0.5,0.15);
            \draw (0.5,0.15) arc (-90:90:0.25);
            \draw (0.5,0.65)--(-1,0.65);
            \draw (-0.5,0.9)node{$\cC$};
            \draw (0,-0.15)--(1,-0.15) node[pos=0.6,below]{$\cE$};
            \mynode{0}{0}{0.3}{0.3}{G}
        }
    \]
    extends to an equivalence between categories
    \eqnn{\label{eq.thm.higher_transpose}
        J\:\vprof(\cC\os\cD,\cE)\simeq \vprof(\cD,\cC^\op\os\cE)\colon K.
    }
    In the above, $\vprof(\cA,\cB)$ refers to the category of profunctors $\cA\arprof\cB$ and profunctor homomorphisms as defined in Definition \ref{dfn.profunctor}.
    \begin{proof}
        The proof is rather direct; both for completeness and to demonstrate the utility of the graphical calculus rules established till now, a detailed proof is included in Appendix \ref{sub.pf_of_higher_transpose}.
    \end{proof}
\end{theorem}
\begin{remark}
    The equivalence in \eqref{eq.thm.higher_transpose} is easy to establish using solely Definition \ref{dfn.profunctor}. After all, a profunctor $\cC\os\cD\arprof\cE$ is a $\cV$-funcor $\cE^\op\os\cC\os\cD\to\bcV$, which is the same as a profunctor $\cD\arprof\cC^\op\os\cE$ by definition. The purpose of Theorem \ref{thm.higher_transpose} is to realize this equivalence as a concrete manipulation of pro-tensor networks, which will be useful in our future graphical proofs (such as the proof of Theorem \ref{thm.unenriched_Kitaev_Kong}).
\end{remark}
\begin{remark}
    The composition $g\circ f$ of linear maps (tensors) $f\:U\to V$ and $g\:V\to W$ is also equal to
    \[
        (d_V\os\id_W)\circ(f\os g^\heartsuit)\:U\to W,
    \]
    where $g^\heartsuit\:\C\to V^\ast\os W$ is image of $g$ under \eqref{eq.vector_space_adj}, and $d_V\:V\os V^\ast\to \C$ is the canonical evaluation map $v\os\omega\mapsto \omega(v)$. Using the fact that $\chi_\cC$ is invertible, an analogous fact holds for pro-tensor network theory: the composition $G\bullet F$ of pro-tensors $F\:\cD\arprof\cC$ and $G\:\cC\arprof\cE$ is isomorphic to 
    \[
        (\coevC\os\Id_\cE)\bullet (F\os G^\heartsuit),
    \]
    where $G^\heartsuit$ is the image of $G$ under $J\:\vprof(\cC,\cE)\to\vprof(\ast,\cC^\op\os\cE)$ defined in Theorem \ref{thm.higher_transpose}.
\end{remark}


\subsection{Pro-tensor formalism of monoidal $\cV$-categories and module $\cV$-categories}\label{sub.mon_VcatVModule}

In this subsection, we review the pro-tensor perspective on a monoidal $\cV$-category $\cC$, the module $\cV$-categories over $\cC$, and comment on the rigidity of $\cC$. 


Monoidal $\cV$-categories are abundant in physical literature. For example, fusion categories and modular tensor categories in the theory of topological order and conformal field theory are all instances of monoidal $\Vect$-category. Now we introduce an equivalent definition of monoidal $\cV$-category using the language of profunctors.
\begin{definition}\label{dfn.mon_cat}
    A \emph{monoidal $\cV$-category} $(\cC,\boxtimes,I,\alpha,\lambda,\rho)$ consists of the following data
    \begin{itemize}
        \item A $\cV$-category $\cC$.
        \item Two $\cV$-functors $\boxtimes\:\cC\os\cC\to\cC$ and $I\:*\to\cC$ ($*\mapsto I(*):= \one$). The associated pro-tensors $\boxtimes_\ast\:\cC\os\cC\arprof\cC,\boxtimes^\ast\:\cC\arprof\cC\os\cC,I_\ast\:*\arprof\cC,I^\ast\:\cC\arprof*$ defined in Example \ref{ex.companion_and_conjoin} are denoted graphically by
        \[
    \ctikz{[scale=0.4,xscale=-1]
        \draw (1,0)--(0,0);
        \draw (0,0)--(-1,0.7);
        \draw (0,0)--(-1,-0.7);
    },\,
    \ctikz{[scale=0.4]
        \draw (1,0)--(0,0);
        \draw (0,0)--(-1,0.7);
        \draw (0,0)--(-1,-0.7);
    },\,
    \ctikz{[xscale=-1,scale=0.8]
        \draw (1,0)--(0,0);
        \draw[fill=white] (0,0)circle[radius=0.1];
    }
    \quad\text{and}\quad
    \ctikz{[scale=0.8]
        \draw (1,0)--(0,0);
        \draw[fill=white] (0,0)circle[radius=0.1];
    }
\]
respectively. The counit and unit of the adjunction $\boxtimes_\ast\ladj\boxtimes^\ast$ which exist by Example \ref{ex.conjoin_right_adj} are denoted by 
\[\epsilon_2\:
    \diagram{
        \ctikz{[scale=0.4]
            \draw (-2,0)--(-1,0);
            \draw (-1,0)--(0,0.7);
            \draw (-1,0)--(0,-0.7);
            \draw (2,0)--(1,0);
            \draw (1,0)--(0,0.7);
            \draw (1,0)--(0,-0.7);
        }
        \Rightarrow
        \ctikz{[scale=0.4]
            \draw (0,0)--(3,0);
            \draw [white] (0,-0.3)--(3,-0.3);
        }
    }
    \quad\text{and}\quad
    \eta_2\:
    \diagram{
        \ctikz{[scale=0.4]
            \draw (0,0.7)--(3,0.7);
            \draw (0,-0.7)--(3,-0.7);
        }
        \Rightarrow
        \ctikz{[scale=0.4]
            \draw (-1.5,0.7)--(-0.5,0);
            \draw (-1.5,-0.7)--(-0.5,0);
            \draw (-0.5,0)--(0.5,0);
            \draw (1.5,0.7)--(0.5,0);
            \draw (1.5,-0.7)--(0.5,0);
        }
    }
\]
respectively, and the counit and the unit of the adjunction $I_\ast\ladj I^\ast$ are denoted by
\[
    \epsilon_0\:
    \diagram{
        \ctikz{[scale=0.8]
            \draw (0,0)--(1,0);
            \draw (2,0)--(3,0);
            \draw[fill=white] (1,0)circle[radius=0.1];
            \draw[fill=white] (2,0)circle[radius=0.1];
        }
        \Rightarrow
        \ctikz{[scale=0.8]
            \draw (0,0)--(2,0);
            \draw [white] (0,-0.1)--(2,-0.1);
        }
    }
    \quad\text{and}\quad
    \eta_0\:
    \diagram{
        \ctikz{[scale=0.8]
            \draw[dashed] (0,0)--(1,0);
            \draw [white] (0,-0.08)--(1,-0.08);
        }
        \Rightarrow 
        \ctikz{[scale=0.8]
            \draw (0,0)--(1,0);
            \draw[fill=white] (0,0)circle[radius=0.1];
            \draw[fill=white] (1,0)circle[radius=0.1];
        }
    }
    \]
    respectively, where $\ctikz{[scale=0.8]
            \draw[dashed] (0,0)--(1,0);
            \draw [white] (0,-0.08)--(1,-0.08);
        }$ represents the unit pro-tensor on the $\cV$-category $*$. The pro-tensor network maps $\epsilon_2,\eta_2,\epsilon_0,\eta_0$ could be understood as part of the data of the monoidal $\cV$-category $\cC$.
        
        \item Three invertible maps of pro-tensor network 
        \[
        \alpha\:
        \diagram{
        \ctikz{[scale=0.4]
            \draw (-1,0)--(0,0);
            \draw (0,0)--(2,1.4);
            \draw (0,0)--(2,-1.4);
            \draw (1,0.7)--(2,0);
        }
        \Rightarrow 
        \ctikz{[scale=0.4,yscale=-1]
            \draw (-1,0)--(0,0);
            \draw (0,0)--(2,1.4);
            \draw (0,0)--(2,-1.4);
            \draw (1,0.7)--(2,0);
        }
        },\quad
        \lambda\:
        \diagram{
        \ctikz{[scale=0.4]
            \draw (-1,0)--(0,0);
            \draw (0,0)--(1,0.7);
            \draw[fill=white] (1,0.7)circle[radius=0.25];
            \draw (0,0)--(1,-0.7);
        }
        \Rightarrow 
        \ctikz{[scale=0.4]
            \draw[white] (1,0)circle[radius=0.1]; 
            \draw (-1,0)--(1,0);
        }
        }
        \quad\text{and}\quad
        \rho\:
        \diagram{
        \ctikz{[scale=0.4,yscale=-1]
            \draw (-1,0)--(0,0);
            \draw (0,0)--(1,0.7);
            \draw[fill=white] (1,0.7)circle[radius=0.25];
            \draw (0,0)--(1,-0.7);
        }
        \Rightarrow 
        \ctikz{[scale=0.4]
            \draw[white] (1,0)circle[radius=0.1]; 
            \draw (-1,0)--(1,0);
        }
        },
        \]
        rendering the following diagrams commutative:
        \eqnn{\label{eq.pentagon_and_triangle}
            \ctikz{[black,node distance=3cm, every node/.style={inner sep=0.5pt},scale=0.9]
            \node (A) at (162:2.25)  
                {$\ctikz{[scale=0.2]
                    \draw (-2,0)--(0,0);
                    \draw (0,0)--(3,2.1);
                    \draw (1,0.7)--(3,-0.7);
                    \draw (2,1.4)--(3,0.7);
                    \draw (0,0)--(3,-2.1);
                }$};
            \node (B) at (90:2.25)   
                {$\ctikz{[scale=0.2]
                    \draw (-2,0)--(0,0);
                    \draw (0,0)--(1,0.7)--(2,1.4)--(3,2.1);
                    \draw (2,-1.4)--(3,-0.7);
                    \draw (2,1.4)--(3,0.7);
                    \draw (0,0)--(3,-2.1);
                }$};
            \node (C) at (18:2.25)
                {$\ctikz{[scale=0.2,yscale=-1]
                    \draw (-2,0)--(0,0);
                    \draw (0,0)--(3,2.1);
                    \draw (1,0.7)--(3,-0.7);
                    \draw (2,1.4)--(3,0.7);
                    \draw (0,0)--(3,-2.1);
                }$};
            \node (D) at (234:2.25) 
                {$\ctikz{[scale=0.2]
                    \draw (-2,0)--(0,0);
                    \draw (0,0)--(3,2.1);
                    \draw (1,0.7)--(3,-0.7);
                    \draw (2,0)--(3,0.7);
                    \draw (0,0)--(3,-2.1);
                }$};
            \node (E) at (306:2.25)
                {$\ctikz{[scale=0.2,yscale=-1]
                    \draw (-2,0)--(0,0);
                    \draw (0,0)--(3,2.1);
                    \draw (1,0.7)--(3,-0.7);
                    \draw (2,0)--(3,0.7);
                    \draw (0,0)--(3,-2.1);
                }$};
            \draw[nattrans] (A) -- (B) node[midway, above=4pt, xshift=-3pt] {$\scriptstyle\alpha$};
            \draw[nattrans] (A) -- (D) node[midway, below=0pt, xshift=-10pt] {$\scriptstyle\alpha\star\id$};
            \draw[nattrans] (B) -- (C) node[midway, above=4pt, xshift=3pt] {$\scriptstyle\alpha$};
            \draw[nattrans] (D) -- (E) node[midway, below=5pt] {$\scriptstyle\alpha$};
            \draw[nattrans] (E) -- (C) node[midway, below=0pt, xshift=10pt] {$\scriptstyle\id\star\alpha$};
        }
            \qquad\qquad
            \begin{array}{c}
                \diagram@C=1.5pc@R=1.5pc{
                &
                \ctikz{[scale=0.25,yscale=-1]
                    \draw (-1,0)--(0,0);
                    \draw (0,0)--(2,1.4);
                    \draw (0,0)--(2,-1.4);
                }
                \\
                \ctikz{[scale=0.25]
                    \draw (-1,0)--(0,0);
                    \draw (0,0)--(2,1.4);
                    \draw (0,0)--(2,-1.4);
                    \draw (1,0.7)--(2,0);
                    \draw[fill=white] (2,0)circle[radius=0.2];
                }
                \ar@{=>}[ru]^-{\rho} \ar@{=>}[rr]_-{\alpha}  & & 
                \ctikz{[scale=0.25,yscale=-1]
                    \draw (-1,0)--(0,0);
                    \draw (0,0)--(2,1.4);
                    \draw (0,0)--(2,-1.4);
                    \draw (1,0.7)--(2,0);
                    \draw[fill=white] (2,0)circle[radius=0.2];
                }
                \ar@{=>}[lu]_-{\lambda}
            }
            \end{array}
            .
        }
    \end{itemize}
    \end{definition}

\begin{example}[$\alpha$ as a generalization of the $F$-symbols]\label{ex.F_symbol}
    In this example, we see how fusion categories are monoidal $\Vect$-categories (or $\vect$-categories) in our sense, along the way gaining a more concrete understanding of the pro-tensor network map $\alpha$. A fusion category $\cC$ clearly gives rise to a $\Vect$-category and two linear functors $\boxtimes\:\cC\os\cC\to\cC$ and $I\:*\to\cC$. We still need to provide the data $\alpha,\lambda$ and $\rho$. We take $\alpha$ as an example. The domain of $\alpha$ is the pro-tensor $\cC\os\cC\os\cC\arprof\cC$ sending $(d,a,b,c)\in\cC^\op\os\cC\os\cC\os\cC$ to 
\[
\cC(d,(a\boxtimes b)\boxtimes c),
\]
while the codomain of $\alpha$ sends $(d,a,b,c)$ to
\[
\cC(d,a\boxtimes(b\boxtimes c)).
\]
Hence $\alpha$ is by definition a family of isomorphisms 
\[
    \{\cC(d,(a\boxtimes b)\boxtimes c)\stackrel{\sim}{\to}\cC(d,a\boxtimes(b\boxtimes c))\}_{a,b,c,d\in\cC}
\]
satisfying certain property, which is provided by nothing but the ``$F$-symbols'' associated to a fusion category. Similar statements hold for $\lambda,\rho$, thus a fusion category, indeed any monoidal $\Vect$-category in the traditional sense, is a monoidal $\Vect$-category defined in Definition \ref{dfn.mon_cat}.
\end{example}
\begin{remark}
    A more traditional definition of a monoidal $\cV$-category is a pseudoalgebra object in the monoidal 2-category $(\Vcat,\os,\ast)$, that is, a sextuple $(\cC,\boxtimes,I,\hat{\alpha},\hat{\lambda},\hat{\rho})$, where $\cC,\boxtimes,I$ are the same as in Definition \ref{dfn.mon_cat}, and $\hat{\alpha}\:\boxtimes(\boxtimes\os\Id_\cC)\Rightarrow\boxtimes(\Id_\cC\os\boxtimes)$, $\hat{\lambda}\:\boxtimes(I\os\Id_\cC)\Rightarrow\Id_\cC$, $\hat{\rho}\:\boxtimes(\Id_\cC\os I)\Rightarrow\Id_\cC$ are $\cV$-natural transformations satisfying the pentagon and triangle identities. As alluded to in Example \ref{ex.F_symbol}, using enriched Yoneda lemma \cite[Corollary A.3.12]{Riehl_Verity_2022}, the data of $(\hat{\alpha},\hat{\lambda},\hat{\rho})$ and the data of $(\alpha,\lambda,\rho)$ can be canonically identified with each other. Thus Definition \ref{dfn.mon_cat} is equivalent to the more traditional definition. It is also using the traditional definition that it becomes apparent that the underlying category $\underline{\cC}$ of a monoidal $\cV$-category $\cC$ is naturally an ordinary monoidal category. Nevertheless, we retain Definition \ref{dfn.mon_cat} as it can be more directly applied to the setting of pro-tensor network.
\end{remark}

Let $\cC=(\cC,\boxtimes,I,\alpha,\lambda,\rho)$ be a monoidal $\cV$-category. One can check that if $F\:\cA\arprof\cB,G\:\cB\arprof\cD$ admit right adjoint, then the right adjoint of $G\bullet F$ is given by $F^R\bullet G^R$; if $F\:\cA\arprof\cB$ and $K\:\cD\arprof\cE$ admit right adjoint, then the right adjoint of $K\os F$ is given by $K^R\os F^R$. Therefore, a pro-tensor formed by composition and tensor products of $\boxtimes_\ast,I_\ast$ and $\Id_\cC$ admit a right adjoint. In particular, the domains and codomains of $\alpha,\lambda,\rho$ all admit a right adjoint. Since the right adjoints of $\boxtimes_\ast,I_\ast$ and $\Id_\cC$ are all given by the ``mirror reflection'', the right adjoints of the composites of $\boxtimes_\ast,I_\ast$ and $\Id_\cC$ are all given by ``mirro reflections''. For example, the right adjoint of the codomain of $\alpha$ is given algebraicly by $(\Id_\cC\os\boxtimes^\ast)\bullet\boxtimes^\ast$, which should be graphically denoted as 
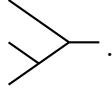
\[
    \ctikz{[scale=0.4,xscale=-1,yscale=-1]
        \draw (-1,0)--(0,0);
        \draw (0,0)--(2,1.4);
        \draw (0,0)--(2,-1.4);
        \draw (1,0.7)--(2,0);
    }.
\]
Thus we can form the transpose (see \eqref{eq.double_transpose})
\[
\alpha^R\:
    \ctikz{[scale=0.4,xscale=-1,yscale=-1]
        \draw (-1,0)--(0,0);
        \draw (0,0)--(2,1.4);
        \draw (0,0)--(2,-1.4);
        \draw (1,0.7)--(2,0);
    }
    \Rightarrow
    \ctikz{[scale=0.4,xscale=-1]
        \draw (-1,0)--(0,0);
        \draw (0,0)--(2,1.4);
        \draw (0,0)--(2,-1.4);
        \draw (1,0.7)--(2,0);
    },\quad
\lambda^R\:
        \diagram{
                \ctikz{[scale=0.4,xscale=-1]
            \draw[white] (1,0)circle[radius=0.1]; 
            \draw (-1,0)--(1,0);
        }
        \Rightarrow 
        \ctikz{[scale=0.4,xscale=-1]
            \draw (-1,0)--(0,0);
            \draw (0,0)--(1,0.7);
            \draw[fill=white] (1,0.7)circle[radius=0.25];
            \draw (0,0)--(1,-0.7);
        }
        }
        \quad\text{and}\quad
\rho^R\:
        \diagram{
        \ctikz{[scale=0.4]
            \draw[white] (1,0)circle[radius=0.1,xscale=-1]; 
            \draw (-1,0)--(1,0);
        }
        \Rightarrow 
        \ctikz{[scale=0.4,yscale=-1,xscale=-1]
            \draw (-1,0)--(0,0);
            \draw (0,0)--(1,0.7);
            \draw[fill=white] (1,0.7)circle[radius=0.25];
            \draw (0,0)--(1,-0.7);
        }
        }
\]
of the pro-tensor network maps $\alpha,\lambda$ and $\rho$ respectively. The definition of $\cV$-monoidal category, Theorem \ref{thm.local}, and eqs.~\eqref{eq.ex.conjoin_right_adj}, \eqref{eq.right_transpose}, and \eqref{eq2.right_transpose} together imply  that the pro-tensor network maps $\epsilon_2$, $\eta_2$, $\epsilon_0$, $\eta_0$, $\alpha$, $\lambda$, $\rho$, $\alpha^R$, $\lambda^R$ and $\rho^R$ fulfill the conditions recorded in Figure \ref{fig.property_monoidal}.
\begin{figure}[htp]
    \eqnn{\label{eq.pentagon_graph}
        \ctikz{[black,node distance=3cm, every node/.style={inner sep=0.5pt}]
            \node (A) at (162:2.25)  
                {$\ctikz{[scale=0.2]
                    \draw (-2,0)--(0,0);
                    \draw (0,0)--(3,2.1);
                    \draw (1,0.7)--(3,-0.7);
                    \draw (2,1.4)--(3,0.7);
                    \draw (0,0)--(3,-2.1);
                }$};
            \node (B) at (90:2.25)   
                {$\ctikz{[scale=0.2]
                    \draw (-2,0)--(0,0);
                    \draw (0,0)--(1,0.7)--(2,1.4)--(3,2.1);
                    \draw (2,-1.4)--(3,-0.7);
                    \draw (2,1.4)--(3,0.7);
                    \draw (0,0)--(3,-2.1);
                }$};
            \node (C) at (18:2.25)
                {$\ctikz{[scale=0.2,yscale=-1]
                    \draw (-2,0)--(0,0);
                    \draw (0,0)--(3,2.1);
                    \draw (1,0.7)--(3,-0.7);
                    \draw (2,1.4)--(3,0.7);
                    \draw (0,0)--(3,-2.1);
                }$};
            \node (D) at (234:2.25) 
                {$\ctikz{[scale=0.2]
                    \draw (-2,0)--(0,0);
                    \draw (0,0)--(3,2.1);
                    \draw (1,0.7)--(3,-0.7);
                    \draw (2,0)--(3,0.7);
                    \draw (0,0)--(3,-2.1);
                }$};
            \node (E) at (306:2.25)
                {$\ctikz{[scale=0.2,yscale=-1]
                    \draw (-2,0)--(0,0);
                    \draw (0,0)--(3,2.1);
                    \draw (1,0.7)--(3,-0.7);
                    \draw (2,0)--(3,0.7);
                    \draw (0,0)--(3,-2.1);
                }$};
            \draw[nattrans] (A) -- (B) node[midway, above=4pt, xshift=-3pt] {$\scriptstyle\alpha$};
            \draw[nattrans] (A) -- (D) node[midway, below=0pt, xshift=-10pt] {$\scriptstyle\alpha\star\id$};
            \draw[nattrans] (B) -- (C) node[midway, above=4pt, xshift=3pt] {$\scriptstyle\alpha$};
            \draw[nattrans] (D) -- (E) node[midway, below=5pt] {$\scriptstyle\alpha$};
            \draw[nattrans] (E) -- (C) node[midway, below=0pt, xshift=10pt] {$\scriptstyle\id\star\alpha$};
        }
        \qquad\qquad\qquad
        \ctikz{[black,every node/.style={inner sep=0.5pt}]
           \node (A) at (162:2.25)
                {$\ctikz{[scale=0.2,xscale=-1]
                    \draw (-2,0)--(0,0);
                    \draw (0,0)--(3,2.1);
                    \draw (1,0.7)--(3,-0.7);
                    \draw (2,1.4)--(3,0.7);
                    \draw (0,0)--(3,-2.1);
                }$};
            \node (B) at (90:2.25)
                {$\ctikz{[scale=0.2,xscale=-1]
                    \draw (-2,0)--(0,0);
                    \draw (0,0)--(1,0.7)--(2,1.4)--(3,2.1);
                    \draw (2,-1.4)--(3,-0.7);
                    \draw (2,1.4)--(3,0.7);
                    \draw (0,0)--(3,-2.1);
                }$};
            \node (C) at (18:2.25)
                {$\ctikz{[scale=0.2,yscale=-1,xscale=-1]
                    \draw (-2,0)--(0,0);
                    \draw (0,0)--(3,2.1);
                    \draw (1,0.7)--(3,-0.7);
                    \draw (2,1.4)--(3,0.7);
                    \draw (0,0)--(3,-2.1);
                }$};
            \node (D) at (234:2.25)  
                {$\ctikz{[scale=0.2,xscale=-1]
                    \draw (-2,0)--(0,0);
                    \draw (0,0)--(3,2.1);
                    \draw (1,0.7)--(3,-0.7);
                    \draw (2,0)--(3,0.7);
                    \draw (0,0)--(3,-2.1);
                }$};
            \node (E) at (306:2.25)
                {$\ctikz{[scale=0.2,yscale=-1,xscale=-1]
                    \draw (-2,0)--(0,0);
                    \draw (0,0)--(3,2.1);
                    \draw (1,0.7)--(3,-0.7);
                    \draw (2,0)--(3,0.7);
                    \draw (0,0)--(3,-2.1);
                }$};
            \draw[nattrans] (B) -- (A) node[midway, above=4pt, xshift=-3pt] {$\scriptstyle \alpha^R$};
            \draw[nattrans] (D) -- (A) node[midway, below=0pt, xshift=-11pt] {$\scriptstyle\alpha^R\star\id$};
            \draw[nattrans] (C) -- (B) node[midway, above=4pt, xshift=7pt] {$\scriptstyle\alpha^R$};
            \draw[nattrans] (E) -- (D) node[midway, below=5pt] {$\scriptstyle \alpha^R$};
            \draw[nattrans] (C) -- (E) node[midway, below=0pt, xshift=11pt] {$\scriptstyle\id\star\alpha^R$};
        }
    }
    \vspace{1pt}
    \eqnn{\label{eq.triangle_graph}
        \begin{array}{c}
            \diagram@C=1.5pc@R=1.5pc{
                &
                \ctikz{[scale=0.25,yscale=-1]
                    \draw (-1,0)--(0,0);
                    \draw (0,0)--(2,1.4);
                    \draw (0,0)--(2,-1.4);
                }
                \\
                \ctikz{[scale=0.25]
                    \draw (-1,0)--(0,0);
                    \draw (0,0)--(2,1.4);
                    \draw (0,0)--(2,-1.4);
                    \draw (1,0.7)--(2,0);
                    \draw[fill=white] (2,0)circle[radius=0.2];
                }
                \ar@{=>}[ru]^-{\rho} \ar@{=>}[rr]_-{\alpha}  & & 
                \ctikz{[scale=0.25,yscale=-1]
                    \draw (-1,0)--(0,0);
                    \draw (0,0)--(2,1.4);
                    \draw (0,0)--(2,-1.4);
                    \draw (1,0.7)--(2,0);
                    \draw[fill=white] (2,0)circle[radius=0.2];
                }
                \ar@{=>}[lu]_-{\lambda}
            }
            \qquad\qquad\qquad
            \diagram@C=1.5pc@R=1.5pc{
                &
                \ctikz{[scale=0.25,yscale=-1,xscale=-1]
                    \draw (-1,0)--(0,0);
                    \draw (0,0)--(2,1.4);
                    \draw (0,0)--(2,-1.4);
                }
                \\
                \ctikz{[scale=0.25,xscale=-1]
                    \draw (-1,0)--(0,0);
                    \draw (0,0)--(2,1.4);
                    \draw (0,0)--(2,-1.4);
                    \draw (1,0.7)--(2,0);
                    \draw[fill=white] (2,0)circle[radius=0.2];
                }
                \ar@{<=}[ru]^-{\rho^R} \ar@{<=}[rr]_-{\alpha^R}  & & 
                \ctikz{[scale=0.25,yscale=-1,xscale=-1]
                    \draw (-1,0)--(0,0);
                    \draw (0,0)--(2,1.4);
                    \draw (0,0)--(2,-1.4);
                    \draw (1,0.7)--(2,0);
                    \draw[fill=white] (2,0)circle[radius=0.2];
                }
                \ar@{<=}[lu]_-{\lambda^R}
            }
        \end{array}
    }
    \vspace{1pt}
    \eqnn{\label{eq.zig_zag_m}
        \hspace{1.75pc}
        \begin{array}{c}
            \diagram@=2.25pc{
                \ctikz{[scale=0.4]
                    \draw (-1,0)--(0,0);
                    \draw (0,0)--(1,0.7);
                    \draw (0,0)--(1,-0.7);
                }
                \ar@{=>}[d]_-1
                \ar@{=>}[r]^-1
                &
                \ctikz{[scale=0.4]
                    \draw (-1,0)--(0,0);
                    \draw (0,0)--(1,0.7)--(3,0.7);
                    \draw (0,0)--(1,-0.7)--(3,-0.7);
                }
                \ar@{=>}[d]^{\eta_2}
                \\
                \ctikz{[scale=0.4]
                    \draw (-1,0)--(0,0);
                    \draw (0,0)--(1,0.7);
                    \draw (0,0)--(1,-0.7);
                }
                & 
                \ctikz{[scale=0.4]
                    \draw (-1,0)--(0,0);
                    \draw (0,0)--(1,0.7)--(2,0);
                    \draw (0,0)--(1,-0.7)--(2,0);
                    \draw (2,0)--(3,0);
                    \draw (3,0)--(4,0.7);
                    \draw (3,0)--(4,-0.7);
                }
                \ar@{=>}[l]^-{\epsilon_2}
            }
            \qquad\qquad\qquad\quad
            \diagram@=2.25pc{
                \ctikz{[scale=0.4,xscale=-1]
                    \draw (-1,0)--(0,0);
                    \draw (0,0)--(1,0.7);
                    \draw (0,0)--(1,-0.7);
                }
                \ar@{=>}[d]_-1
                \ar@{=>}[r]^-1
                &
                \ctikz{[scale=0.4,xscale=-1]
                    \draw (-1,0)--(0,0);
                    \draw (0,0)--(1,0.7)--(3,0.7);
                    \draw (0,0)--(1,-0.7)--(3,-0.7);
                }
                \ar@{=>}[d]^{\eta_2}
                \\
                \ctikz{[scale=0.4,xscale=-1]
                    \draw (-1,0)--(0,0);
                    \draw (0,0)--(1,0.7);
                    \draw (0,0)--(1,-0.7);
                }
                & 
                \ctikz{[scale=0.4,xscale=-1]
                    \draw (-1,0)--(0,0);
                    \draw (0,0)--(1,0.7)--(2,0);
                    \draw (0,0)--(1,-0.7)--(2,0);
                    \draw (2,0)--(3,0);
                    \draw (3,0)--(4,0.7);
                    \draw (3,0)--(4,-0.7);
                }
                \ar@{=>}[l]^-{\epsilon_2}
            }
        \end{array}
    }
    \vspace{1pt}
    \eqnn{\label{eq.zig_zag_u}
        \hspace{0.75pc}
        \begin{array}{c}
            \diagram@R=2.75pc@C=2.5pc{
                \ctikz{
                    \draw (0,0)--(0.5,0);
                    \draw[fill=white] (0.5,0)circle[radius=0.1];
                }
                \ar@{=>}[r]^-{\eta_0} \ar@{=>}[rd]_-1 & 
                \ctikz{
                    \draw (0,0)--(0.5,0);
                    \draw[fill=white] (0.5,0)circle[radius=0.1];
                    \draw (1,0)--(1.5,0);
                    \draw[fill=white] (1,0)circle[radius=0.1];
                    \draw[fill=white] (1.5,0)circle[radius=0.1];
                }
                \ar@{=>}[d]^{\epsilon_0}
                \\
                & 
                \ctikz{
                    \draw (0,0)--(0.5,0);
                    \draw[fill=white] (0.5,0)circle[radius=0.1];
                }
            }
            \qquad\qquad\qquad\qquad\quad
            \diagram@R=2.75pc@C=2.5pc{
                \ctikz{[xscale=-1]
                    \draw (0,0)--(0.5,0);
                    \draw[fill=white] (0.5,0)circle[radius=0.1];
                }
                \ar@{=>}[r]^-{\eta_0} \ar@{=>}[rd]_-1& 
                \ctikz{[xscale=-1]
                    \draw (0,0)--(0.5,0);
                    \draw[fill=white] (1,0)circle[radius=0.1];
                    \draw (1,0)--(1.5,0);
                    \draw[fill=white] (1,0)circle[radius=0.1];
                    \draw[fill=white] (1.5,0)circle[radius=0.1];
                }
                \ar@{=>}[d]^{\epsilon_0}
                \\
                & 
                \ctikz{[xscale=-1]
                    \draw (0,0)--(0.5,0);
                    \draw[fill=white] (0.5,0)circle[radius=0.1];
                }
            }
        \end{array}
    }
    \vspace{1pt}
    \eqnn{\label{eq.right_dual_of_alpha}
        \begin{array}{c}
            \diagram@R=0.5pc@C=1.75pc{
                \ctikz{[scale=0.2,yscale=-1]
                    \draw (-3,0)--(-2,0);
                    \draw (-2,0)--(0,1.4)--(2,0);
                    \draw (-2,0)--(0,-1.4)--(2,0);
                    \draw (-1,-0.7)--(0,0)--(1,-0.7);
                    \draw (2,0)--(3,0);
                }
                \ar@{=>}[r]^-{\epsilon_2} \ar@{=>}[dd]_{\alpha\star(\alpha^R)^{-1}} & 
                \ctikz{[scale=0.4]
                    \draw (-2,0)--(-1,0);
                    \draw (-1,0)--(0,0.7)--(1,0);
                    \draw (-1,0)--(0,-0.7)--(1,0);
                    \draw (1,0)--(2,0);
                }\ar@{=>}[rd]^-{\epsilon_2}
                \\
                & &
                \ctikz{[scale=0.4,baseline=-0.5ex]
                    \draw (0,0)--(3,0);
                }
                \\
                \ctikz{[scale=0.2]
                    \draw (-3,0)--(-2,0);
                    \draw (-2,0)--(0,1.4)--(2,0);
                    \draw (-2,0)--(0,-1.4)--(2,0);
                    \draw (-1,-0.7)--(0,0)--(1,-0.7);
                    \draw (2,0)--(3,0);
                }
                \ar@{=>}[r]_-{\epsilon_2} & 
                \ctikz{[scale=0.4]
                    \draw (-2,0)--(-1,0);
                    \draw (-1,0)--(0,0.7)--(1,0);
                    \draw (-1,0)--(0,-0.7)--(1,0);
                    \draw (1,0)--(2,0);
                }
                \ar@{=>}[ru]_-{\epsilon_2}
            }
            \qquad
            \diagram@R=0.5pc@C=1.75pc{
                & 
                \ctikz{[scale=0.3]
                    \draw (-1.5,0.7)--(-0.5,0);
                    \draw (-1.5,-0.7)--(-0.5,0);
                    \draw (-0.5,0)--(0.5,0);
                    \draw (0.5,0)--(1.5,0.7);
                    \draw (0.5,0)--(1.5,-0.7);
                    \draw (-1.5,-1.05)--(1.5,-1.05);
                }
                \ar@{=>}[r]^-{\eta_2} &
                \ctikz{[scale=0.4]
                    \begin{scope}[xscale=-1]
                        \draw (-1,0.35)--(-0.5,0);
                        \draw (-1,-0.35)--(-0.5,0);
                        \draw (-0.5,0)--(0,0);
                        \draw (-1,-0.7)--(0,0);
                    \end{scope}
                    \begin{scope}[xshift=-0.5cm]
                        \draw (-1,0.35)--(-0.5,0);
                        \draw (-1,-0.35)--(-0.5,0);
                        \draw (-0.5,0)--(0,0);
                        \draw (-1,-0.7)--(0,0);
                    \end{scope}
                    \draw (-0.5,0)--(0,0);
                }
                \ar@{=>}[dd]_{(\alpha^R)^{-1}\star\alpha}\\
                \ctikz{[scale=0.3]
                    \draw (-1.5,0)--(1.5,0);
                    \draw (-1.5,0.875)--(1.5,0.875);
                    \draw (-1.5,-0.875)--(1.5,-0.875);
                }
                \ar@{=>}[ru]^-{\eta_2} \ar@{=>}[rd]_-{\eta_2} \\
                & \ar@{=>}[r]_-{\eta_2} 
                \ctikz{[scale=0.3,yscale=-1]
                    \draw (-1.5,0.7)--(-0.5,0);
                    \draw (-1.5,-0.7)--(-0.5,0);
                    \draw (-0.5,0)--(0.5,0);
                    \draw (0.5,0)--(1.5,0.7);
                    \draw (0.5,0)--(1.5,-0.7);
                    \draw (-1.5,-1.05)--(1.5,-1.05);
                }
                &
                \ctikz{[scale=0.4,yscale=-1]
                    \begin{scope}[xscale=-1]
                        \draw (-1,0.35)--(-0.5,0);
                        \draw (-1,-0.35)--(-0.5,0);
                        \draw (-0.5,0)--(0,0);
                        \draw (-1,-0.7)--(0,0);
                    \end{scope}
                    \begin{scope}[xshift=-0.5cm]
                        \draw (-1,0.35)--(-0.5,0);
                        \draw (-1,-0.35)--(-0.5,0);
                        \draw (-0.5,0)--(0,0);
                        \draw (-1,-0.7)--(0,0);
                    \end{scope}
                    \draw (-0.5,0)--(0,0);
                }
            }
        \end{array}
    }
    \vspace{1pt}
    \eqnn{\label{eq.right_dual_of_lambda}
        \begin{array}{c}
            \diagram@C=1.5pc@R=2.25pc{
                \ctikz{[scale=0.4]
                    \draw (-2,0)--(-1,0);
                    \draw (-1,0)--(-0.5,0.35);
                    \draw (-1,0)--(0,-0.7);
                    \draw (1,0)--(0,-0.7);
                    \draw (1,0)--(0.5,0.35);
                    \draw (2,0)--(1,0);
                    \draw[fill=white] (-0.5,0.35)circle[radius=0.175];
                    \draw[fill=white] (0.5,0.35)circle[radius=0.175];
                }
                \ar@{=>}[rd]^(0.4){\lambda\star(\lambda^R)^{-1}}
                \ar@{=>}[d]_{\epsilon_0} 
                & &
                \ctikz{[scale=0.4,yscale=-1]
                    \draw (-2,0)--(-1,0);
                    \draw (-1,0)--(-0.5,0.35);
                    \draw (-1,0)--(0,-0.7);
                    \draw (1,0)--(0,-0.7);
                    \draw (1,0)--(0.5,0.35);
                    \draw (2,0)--(1,0);
                    \draw[fill=white] (-0.5,0.35)circle[radius=0.175];
                    \draw[fill=white] (0.5,0.35)circle[radius=0.175];
                }
                \ar@{=>}[d]^{\epsilon_0}
                \ar@{=>}[ld]_(0.4){\rho\star(\rho^R)^{-1}}
                \\
                \ctikz{[scale=0.4]
                    \draw (-2,0)--(-1,0);
                    \draw (-1,0)--(0,0.7)--(1,0);
                    \draw (-1,0)--(0,-0.7)--(1,0);
                    \draw (1,0)--(2,0);
                }
                \ar@{=>}[r]_-{\epsilon_2}
                & 
                \ctikz{[scale=0.4,baseline=-0.5ex]
                    \draw (0,0)--(3,0);
                }
                &
                \ctikz{[scale=0.4]
                    \draw (-2,0)--(-1,0);
                    \draw (-1,0)--(0,0.7)--(1,0);
                    \draw (-1,0)--(0,-0.7)--(1,0);
                    \draw (1,0)--(2,0);
                }
                \ar@{=>}[l]^-{\epsilon_2}
            }
            \qquad
            \diagram@C=2pc@R=2.25pc{
            \ctikz{[scale=0.3]
                \draw (-1.5,-0.7)--(1.5,-0.7);
                \draw (-1.5,0.7)--(1.5,0.7);
                \draw[fill=white] (-1.5,0.7)circle[radius=0.25];
                \draw[fill=white] (1.5,0.7)circle[radius=0.25];
            }
            \ar@{=>}[d]_{\eta_2}
            & 
            \ctikz{[scale=0.3,baseline=-0.5ex]
                \draw (-1.5,0)--(1.5,0);
            }
            \ar@{=>}[rd]_(0.55){(\rho^R)^{-1}\star\rho} \ar@{=>}[r]^-{\eta_0} \ar@{=>}[ld]^(0.3){(\lambda^R)^{-1}\star\lambda}\ar@{=>}[l]_-{\eta_0} 
            &
            \ctikz{[scale=0.3,yscale=-1]
                \draw (-1.5,-0.7)--(1.5,-0.7);
                \draw (-1.5,0.7)--(1.5,0.7);
                \draw[fill=white] (-1.5,0.7)circle[radius=0.25];
                \draw[fill=white] (1.5,0.7)circle[radius=0.25];
            }
            \ar@{=>}[d]^{\eta_2}
            \\
            \ctikz{[scale=0.3]
                \draw (-1.5,0.7)--(-0.5,0);
                \draw (-1.5,-0.7)--(-0.5,0);
                \draw (-0.5,0)--(0.5,0);
                \draw (0.5,0)--(1.5,0.7);
                \draw (0.5,0)--(1.5,-0.7);
                \draw[fill=white] (-1.5,0.7)circle[radius=0.25];
                \draw[fill=white] (1.5,0.7)circle[radius=0.25];
            }
            &
            &
            \ctikz{[scale=0.3,yscale=-1]
                \draw (-1.5,0.7)--(-0.5,0);
                \draw (-1.5,-0.7)--(-0.5,0);
                \draw (-0.5,0)--(0.5,0);
                \draw (0.5,0)--(1.5,0.7);
                \draw (0.5,0)--(1.5,-0.7);
                \draw[fill=white] (-1.5,0.7)circle[radius=0.25];
                \draw[fill=white] (1.5,0.7)circle[radius=0.25];
            }
            }
        \end{array}
    }
    \caption{Pro-tensor network presentation of basic properties satisfied by $\cC$.} 
    \label{fig.property_monoidal}
\end{figure}


\begin{definition}
\label{def.moduleVcat}
Let $\cC=(\cC,\boxtimes,I,\alpha,\lambda,\rho)$ be a monoidal $\cV$-category. A \emph{left $\cC$-module} $(\cM,\boxdot,\alpha_\wr,\lambda_\wr)$ consists of the following data:
    \begin{itemize}
        \item A $\cV$-category $\cM$.
        \item A $\cV$-functor $\boxdot\:\cC\os\cM\to\cM$. The associated pro-tensors $\boxdot_\ast\:\cC\os\cM\arprof\cM,\boxdot^\ast\:\cM\arprof\cC\os\cM$ are denoted graphically by 
        \[
            \ctikz{[scale=0.4]
                \draw (1,0.7)--(0,0);
                \draw[color3] (-1,0)--(1,0);
            }
            \quad\text{and}\quad
            \ctikz{[scale=0.4,xscale=-1]
                \draw (1,0.7)--(0,0);
                \draw[color3] (-1,0)--(1,0);
            }
        \]
        respectively. The counit and unit of the adjunction $\boxdot_\ast\ladj\boxdot^\ast$ which exists by Example \ref{ex.conjoin_right_adj} are denoted by, respectively, 
        \[
        \epsilon_\wr\:
        \diagram{
        \ctikz{[scale=0.4]
            \draw (-1,0)--(0,0.7)--(1,0);
            \draw[color3] (-2,0)--(-1,0);
            \draw[color3] (-1,0)--(1,0);
            \draw[color3] (1,0)--(2,0);
        }
        \Rightarrow
        \ctikz{[scale=0.4]
            \draw [white] (0,0.7)circle[radius=0.1];
            \draw[color3] (0,0)--(3,0);
        }
        }
        \quad\text{and}\quad
        \eta_\wr\:
        \diagram{
            \ctikz{[scale=0.4]
                \draw (0,0.7)--(3,0.7);
                \draw[color3] (0,0)--(3,0);
            }
            \Rightarrow
            \ctikz{[scale=0.4]
                \draw (2,0)--(3,0.7);
                \draw (1,0)--(0,0.7);
                \draw[color3] (0,0)--(3,0);
            }
        }.
        \]
        $\epsilon_\wr$ and $\eta_\wr$ could be understood as part of the data of the left $\cC$-module $\cM$.
        \item Two invertible pro-tensor network maps

\[
    \alpha_\wr\:
    \diagram{
        \ctikz{[scale=0.36]
            \draw[color3] (-1,0)--(0,0);
            \draw[color3] (0,0)--(2.5,0);
            \draw (0,0)--(1.5,1.05);
            \draw (1.5,1.05)--(2.5,1.75);
            \draw (1.5,1.05)--(2.5,0.35);
        }
        \Rightarrow
        \ctikz{[scale=0.6]
            \draw[color3] (-1,0)--(1.5,0);
            \draw (0,0)--(1.5,1.05);
            \draw (1,0)--(1.5,0.35);
        }
    }
    \quad\text{and}\quad
    \lambda_\wr\:
    \diagram{
        \ctikz{[scale=0.4]
            \draw[color3] (-1,0)--(0,0);
            \draw[color3] (0,0)--(1,0);
            \draw (0,0)--(1,0.7);
            \draw[fill=white] (1,0.7)circle[radius=0.2];
        }
        \Rightarrow
        \ctikz{[scale=0.4]
            \draw[white] (1,0.7)circle[radius=0.1]; 
            \draw[color3] (-1,0)--(1,0);
        }
    }
\]
rendering the following diagrams commutative:
        \eqnn{\label{eq.pentagon_and_triangle_module}
            \ctikz{[black,node distance=3cm, every node/.style={inner sep=0.5pt},scale=0.9]
                \node (A) at (162:2.25)  
                    {$\ctikz{[scale=0.25]
                        \draw[color3] (-2,0)--(0,0);
                        \draw[color3] (0,0)--(3.5,0);
                        \draw (0,0)--(2,1.4);
                        \draw (2,1.4)--(3.5,2.45);
                        \draw (2,1.4)--(3.5,0.35);
                        \draw (2.75,1.925)--(3.5,1.4);
                    }$};
                \node (B) at (90:2.25)   
                    {$\ctikz{[scale=0.25]
                        \draw[color3] (-2,0)--(0,0);
                        \draw[color3] (0,0)--(3.5,0);
                        \draw (0,0)--(1.5,1.05);
                        \draw (1.5,1.05)--(2.5,1.75);
                        \draw (2.5,1.75)--(3.5,2.45);
                        \draw (2.5,1.75)--(3.5,1.05);
                        \draw (2.5,0)--(3.5,0.7);
                    }$};
                \node (C) at (18:2.25)
                    {$\ctikz{[scale=0.25]
                        \draw[color3] (-2,0)--(0,0);
                        \draw[color3] (0,0)--(3.5,0);
                        \draw (0,0)--(1.5,1.05);
                        \draw (1.5,1.05)--(2.5,1.75);
                        \draw (2.5,1.75)--(3.5,2.45);
                        \draw (1.25,0)--(3.5,1.575);
                        \draw (2.5,0)--(3.5,0.7);
                    }$};
                \node (D) at (234:2.25) 
                    {$\ctikz{[scale=0.25]
                        \draw[color3] (-2,0)--(0,0);
                        \draw[color3] (0,0)--(3.5,0);
                        \draw (0,0)--(2,1.4);
                        \draw (2,1.4)--(3.5,2.45);
                        \draw (2,1.4)--(2.75,0.875);
                        \draw (2.75,0.875)--(3.5,1.4);
                        \draw (2.75,0.875)--(3.5,0.35);
                    }$};
                \node (E) at (306:2.25)
                    {$\ctikz{[scale=0.25]
                        \draw[color3] (-2,0)--(0,0);
                        \draw[color3] (0,0)--(3.5,0);
                        \draw (0,0)--(2,1.4);
                        \draw (2,1.4)--(3.5,2.45);
                        \draw (1.5,0)--(2.75,0.875);
                        \draw (2.75,0.875)--(3.5,1.4);
                        \draw (2.75,0.875)--(3.5,0.35);
                    }$};
                \draw[nattrans] (A) -- (B) node[midway, above=4pt, xshift=-3pt] {$\scriptstyle\alpha_\wr$};
                \draw[nattrans] (A) -- (D) node[midway, below=0pt, xshift=-10pt] {$\scriptstyle\alpha\star\id$};
                \draw[nattrans] (B) -- (C) node[midway, above=4pt, xshift=3pt] {$\scriptstyle\alpha_\wr$};
                \draw[nattrans] (D) -- (E) node[midway, below=5pt] {$\scriptstyle\alpha_\wr$};
                \draw[nattrans] (E) -- (C) node[midway, below=0pt, xshift=10pt] {$\scriptstyle\id\star\alpha_\wr$};
            }
            \qquad\qquad
            \begin{array}{c}
                \diagram@C=1.5pc@R=1.5pc{
                    &
                    \ctikz{[scale=0.2]
                        \draw[color3] (-2,0)--(3,0);
                        \draw[color3] (0,0)--(3,0);
                        \draw (0,0)--(3,2.1);
                    }
                    \\
                    \ctikz{[scale=0.2]
                        \draw[color3] (-2,0)--(0,0);
                        \draw[color3] (0,0)--(3,0);
                        \draw (0,0)--(2,1.4);
                        \draw (2,1.4)--(3,2.1);
                        \draw (2,1.4)--(3,0.7);
                        \draw[fill=white] (3,0.7)circle[radius=0.25];
                    }
                    \ar@{=>}[ru]^-{\rho} \ar@{=>}[rr]_-{\alpha_\wr}  & & 
                    \ctikz{[scale=0.2]
                        \draw[color3] (-2,0)--(0,0);
                        \draw[color3] (0,0)--(3,0);
                        \draw (0,0)--(2,1.4);
                        \draw (2,1.4)--(3,2.1);
                        \draw (2,0)--(3,0.7);
                        \draw[fill=white] (3,0.7)circle[radius=0.25];
                    }
                    \ar@{=>}[lu]_-{\lambda_\wr}
                }
            \end{array}
            ,
        }
        where $\ctikz{[scale=0.5,xscale=-1]
        \draw (1,0)--(0,0);
        \draw (0,0)--(-1,0.7);
        \draw (0,0)--(-1,-0.7);
    }$ and $
    \ctikz{[xscale=-1]
        \draw (1,0)--(0,0);
        \draw[fill=white] (0,0)circle[radius=0.1];
    }$ stands for $\boxtimes_\ast$ and $I_\ast$ respectively as in Definition \ref{dfn.mon_cat}.
    \end{itemize}
\end{definition}
\begin{remark}
Given a monoidal $\cV$-category $\cC$, there is a canonically defined monoidal $\cV$-category $\cC^\rev$ satisfying $a\boxtimes^\rev b\defdtobe b\boxtimes a$. A right $\cC$-module is just a left $\cC^\rev$-module and a $\cC$-$\cC$-bimodule is just a left $\cC\ot\cC^\rev$-module. When ``left/right'' is omitted we mean left module by default.
\end{remark}

We can consider the transpose
\[
\alpha_\wr^R\:
    \diagram{
        \ctikz{[scale=0.6,xscale=-1]
            \draw[color3] (-1,0)--(1.5,0);
            \draw (0,0)--(1.5,1.05);
            \draw (1,0)--(1.5,0.35);
        }
        \Rightarrow
        \ctikz{[scale=0.36,xscale=-1]
            \draw[color3] (-1,0)--(0,0);
            \draw[color3] (0,0)--(2.5,0);
            \draw (0,0)--(1.5,1.05);
            \draw (1.5,1.05)--(2.5,1.75);
            \draw (1.5,1.05)--(2.5,0.35);
        }
    }
    \quad\text{and}\quad
    \lambda_\wr^R\:
    \diagram{
        \ctikz{[scale=0.4,xscale=-1]
            \draw[white] (1,0.7)circle[radius=0.1]; 
            \draw[color3] (-1,0)--(1,0);
        }
        \Rightarrow
        \ctikz{[scale=0.4,xscale=-1]
            \draw[color3] (-1,0)--(0,0);
            \draw[color3] (0,0)--(1,0);
            \draw (0,0)--(1,0.7);
            \draw[fill=white] (1,0.7)circle[radius=0.2];
        }
    }
\]
of $\alpha_\wr$ and $\lambda_\wr$ respectively.
The pro-tensor network maps $\epsilon_\wr,\eta_\wr,\alpha_\wr,\lambda_\wr,\alpha_\wr^R,\lambda_\wr^R$ satisfy the conditions recorded in Figure \ref{fig.property_module}, which are analogous to that in Figure \ref{fig.property_monoidal}. 

\begin{figure}[hbtp]
    \eqnn{\label{eq.pentagon_graph_module}
        \ctikz{[black,node distance=3cm, every node/.style={inner sep=0.5pt},scale=0.9]
                \node (A) at (162:2.25)  
                    {$\ctikz{[scale=0.25]
                        \draw[color3] (-2,0)--(0,0);
                        \draw[color3] (0,0)--(3.5,0);
                        \draw (0,0)--(2,1.4);
                        \draw (2,1.4)--(3.5,2.45);
                        \draw (2,1.4)--(3.5,0.35);
                        \draw (2.75,1.925)--(3.5,1.4);
                    }$};
                \node (B) at (90:2.25)   
                    {$\ctikz{[scale=0.25]
                        \draw[color3] (-2,0)--(0,0);
                        \draw[color3] (0,0)--(3.5,0);
                        \draw (0,0)--(1.5,1.05);
                        \draw (1.5,1.05)--(2.5,1.75);
                        \draw (2.5,1.75)--(3.5,2.45);
                        \draw (2.5,1.75)--(3.5,1.05);
                        \draw (2.5,0)--(3.5,0.7);
                    }$};
                \node (C) at (18:2.25)
                    {$\ctikz{[scale=0.25]
                        \draw[color3] (-2,0)--(0,0);
                        \draw[color3] (0,0)--(3.5,0);
                        \draw (0,0)--(1.5,1.05);
                        \draw (1.5,1.05)--(2.5,1.75);
                        \draw (2.5,1.75)--(3.5,2.45);
                        \draw (1.25,0)--(3.5,1.575);
                        \draw (2.5,0)--(3.5,0.7);
                    }$};
                \node (D) at (234:2.25) 
                    {$\ctikz{[scale=0.25]
                        \draw[color3] (-2,0)--(0,0);
                        \draw[color3] (0,0)--(3.5,0);
                        \draw (0,0)--(2,1.4);
                        \draw (2,1.4)--(3.5,2.45);
                        \draw (2,1.4)--(2.75,0.875);
                        \draw (2.75,0.875)--(3.5,1.4);
                        \draw (2.75,0.875)--(3.5,0.35);
                    }$};
                \node (E) at (306:2.25)
                    {$\ctikz{[scale=0.25]
                        \draw[color3] (-2,0)--(0,0);
                        \draw[color3] (0,0)--(3.5,0);
                        \draw (0,0)--(2,1.4);
                        \draw (2,1.4)--(3.5,2.45);
                        \draw (1.5,0)--(2.75,0.875);
                        \draw (2.75,0.875)--(3.5,1.4);
                        \draw (2.75,0.875)--(3.5,0.35);
                    }$};
                \draw[nattrans] (A) -- (B) node[midway, above=4pt, xshift=-3pt] {$\scriptstyle\alpha_\wr$};
                \draw[nattrans] (A) -- (D) node[midway, below=0pt, xshift=-10pt] {$\scriptstyle\alpha\star\id$};
                \draw[nattrans] (B) -- (C) node[midway, above=4pt, xshift=3pt] {$\scriptstyle\alpha_\wr$};
                \draw[nattrans] (D) -- (E) node[midway, below=5pt] {$\scriptstyle\alpha_\wr$};
                \draw[nattrans] (E) -- (C) node[midway, below=0pt, xshift=10pt] {$\scriptstyle\id\star\alpha_\wr$};
            }
        \qquad\qquad
        \ctikz{[black,every node/.style={inner sep=0.5pt},scale=0.9]
                \node (A) at (162:2.25)  
                    {$\ctikz{[scale=0.25,xscale=-1]
                        \draw[color3] (-1,0)--(0,0);
                        \draw[color3] (0,0)--(3.5,0);
                        \draw (0,0)--(2,1.4);
                        \draw (2,1.4)--(3.5,2.45);
                        \draw (2,1.4)--(3.5,0.35);
                        \draw (2.75,1.925)--(3.5,1.4);
                    }$};
                \node (B) at (90:2.25)   
                    {$\ctikz{[scale=0.25,xscale=-1]
                        \draw[color3] (-1,0)--(0,0);
                        \draw[color3] (0,0)--(3.5,0);
                        \draw (0,0)--(1.5,1.05);
                        \draw (1.5,1.05)--(2.5,1.75);
                        \draw (2.5,1.75)--(3.5,2.45);
                        \draw (2.5,1.75)--(3.5,1.05);
                        \draw (2.5,0)--(3.5,0.7);
                    }$};
                \node (C) at (18:2.25)
                    {$\ctikz{[scale=0.25,xscale=-1]
                        \draw[color3] (-1,0)--(0,0);
                        \draw[color3] (0,0)--(3.5,0);
                        \draw (0,0)--(1.5,1.05);
                        \draw (1.5,1.05)--(2.5,1.75);
                        \draw (2.5,1.75)--(3.5,2.45);
                        \draw (1.25,0)--(3.5,1.575);
                        \draw (2.5,0)--(3.5,0.7);
                    }$};
                \node (D) at (234:2.25) 
                    {$\ctikz{[scale=0.25,xscale=-1]
                        \draw[color3] (-1,0)--(0,0);
                        \draw[color3] (0,0)--(3.5,0);
                        \draw (0,0)--(2,1.4);
                        \draw (2,1.4)--(3.5,2.45);
                        \draw (2,1.4)--(2.75,0.875);
                        \draw (2.75,0.875)--(3.5,1.4);
                        \draw (2.75,0.875)--(3.5,0.35);
                    }$};
                \node (E) at (306:2.25)
                    {$\ctikz{[scale=0.25,xscale=-1]
                        \draw[color3] (-1,0)--(0,0);
                        \draw[color3] (0,0)--(3.5,0);
                        \draw (0,0)--(2,1.4);
                        \draw (2,1.4)--(3.5,2.45);
                        \draw (1.5,0)--(2.75,0.875);
                        \draw (2.75,0.875)--(3.5,1.4);
                        \draw (2.75,0.875)--(3.5,0.35);
                    }$};
            \draw[nattrans] (B) -- (A) node[midway, above=4pt, xshift=-3pt] {$\scriptstyle \alpha_\wr^R$};
            \draw[nattrans] (D) -- (A) node[midway, below=0pt, xshift=-11pt] {$\scriptstyle\alpha^R\star\id$};
            \draw[nattrans] (C) -- (B) node[midway, above=4pt, xshift=7pt] {$\scriptstyle\alpha_\wr^R$};
            \draw[nattrans] (E) -- (D) node[midway, below=5pt] {$\scriptstyle \alpha_\wr^R$};
            \draw[nattrans] (C) -- (E) node[midway, below=0pt, xshift=11pt] {$\scriptstyle\id\star\alpha_\wr^R$};
        }
    }
    \vspace{2pt}
    \eqnn{\label{eq.triangle_graph_module}
        \begin{array}{c}
            \diagram@C=1.5pc@R=1.5pc{
                &
                \ctikz{[scale=0.2]
                    \draw[color3] (-2,0)--(3,0);
                    \draw[color3] (0,0)--(3,0);
                    \draw (0,0)--(3,2.1);
                }
                \\
                \ctikz{[scale=0.2]
                    \draw[color3] (-2,0)--(0,0);
                    \draw[color3] (0,0)--(3,0);
                    \draw (0,0)--(2,1.4);
                    \draw (2,1.4)--(3,2.1);
                    \draw (2,1.4)--(3,0.7);
                    \draw[fill=white] (3,0.7)circle[radius=0.25];
                }
                \ar@{=>}[ru]^-{\rho} \ar@{=>}[rr]_-{\alpha_\wr}  & & 
                \ctikz{[scale=0.2]
                    \draw[color3] (-2,0)--(0,0);
                    \draw[color3] (0,0)--(3,0);
                    \draw (0,0)--(2,1.4);
                    \draw (2,1.4)--(3,2.1);
                    \draw (2,0)--(3,0.7);
                    \draw[fill=white] (3,0.7)circle[radius=0.25];
                }
                \ar@{=>}[lu]_-{\lambda_\wr}
            }
            \qquad\qquad
            \diagram@C=1.5pc@R=1.5pc{
                &
                \ctikz{[scale=0.2,xscale=-1]
                    \draw[color3] (-2,0)--(3,0);
                    \draw[color3] (0,0)--(3,0);
                    \draw (0,0)--(3,2.1);
                }
                \\
                \ctikz{[scale=0.2,xscale=-1]
                    \draw[color3] (-2,0)--(0,0);
                    \draw[color3] (0,0)--(3,0);
                    \draw (0,0)--(2,1.4);
                    \draw (2,1.4)--(3,2.1);
                    \draw (2,1.4)--(3,0.7);
                    \draw[fill=white] (3,0.7)circle[radius=0.25];
                }
                \ar@{<=}[ru]^-{{\rho^R}} \ar@{<=}[rr]_-{\alpha_\wr^R}  & & 
                \ctikz{[scale=0.2,xscale=-1]
                    \draw[color3] (-2,0)--(0,0);
                    \draw[color3] (0,0)--(3,0);
                    \draw (0,0)--(2,1.4);
                    \draw (2,1.4)--(3,2.1);
                    \draw (2,0)--(3,0.7);
                    \draw[fill=white] (3,0.7)circle[radius=0.25];
                }
                \ar@{<=}[lu]_-{\lambda_\wr^R}
            }
        \end{array}
    }
    \vspace{2pt}
    \eqnn{\label{eq.zig_zag_m_module}
        \begin{array}{c}
            \diagram{
                \ctikz{[scale=0.4]
                    \draw[color3] (-1,0)--(0,0);
                    \draw[color3] (0,0)--(1,0);
                    \draw (0,0)--(1,0.7);
                }
                \ar@{=>}[d]_-1
                \ar@{=>}[r]^-1
                &
                \ctikz{[scale=0.4]
                    \draw[color3] (-1,0)--(0,0)--(3,0);
                    \draw (0,0)--(1,0.7)--(3,0.7);
                }
                \ar@{=>}[d]^{\eta_\wr}
                \\
                \ctikz{[scale=0.4]
                    \draw[color3] (-1,0)--(0,0);
                    \draw[color3] (0,0)--(1,0);
                    \draw (0,0)--(1,0.7);
                }
                & 
                \ctikz{[scale=0.4]
                    \draw[color3] (-1,0)--(0,0)--(4,0);
                    \draw (0,0)--(1,0.7)--(2,0);
                    \draw (3,0)--(4,0.7);
                }
                \ar@{=>}[l]^-{\epsilon_\wr}
            }
            \qquad\qquad\qquad\qquad
            \diagram{
                \ctikz{[scale=0.4,xscale=-1]
                    \draw[color3] (-1,0)--(0,0);
                    \draw[color3] (0,0)--(1,0);
                    \draw (0,0)--(1,0.7);
                }
                \ar@{=>}[d]_-1
                \ar@{=>}[r]^-1
                &
                \ctikz{[scale=0.4,xscale=-1]
                    \draw[color3] (-1,0)--(0,0)--(3,0);
                    \draw (0,0)--(1,0.7)--(3,0.7);
                }
                \ar@{=>}[d]^{\eta_\wr}
                \\
                \ctikz{[scale=0.4,xscale=-1]
                    \draw[color3] (-1,0)--(0,0);
                    \draw[color3] (0,0)--(1,0);
                    \draw (0,0)--(1,0.7);
                }
                & 
                \ctikz{[scale=0.4,xscale=-1]
                    \draw[color3] (-1,0)--(0,0)--(4,0);
                    \draw (0,0)--(1,0.7)--(2,0);
                    \draw (3,0)--(4,0.7);
                }
                \ar@{=>}[l]^-{\epsilon_\wr}
            }
        \end{array}
    }
    \vspace{2pt}
    \eqnn{\label{eq.right_dual_of_alpha_module}
        \begin{array}{c}
            \diagram@R=0.3pc@C=1.5pc{
                \ctikz{[scale=0.2]
                    \draw[color3] (-3.5,0)--(-2.5,0);
                    \draw[color3] (-2.5,0)--(2.5,0);
                    \draw[color3] (2.5,0)--(3.5,0);
                    \draw (-2.5,0)--(-1,1.05);
                    \draw (-1,1.05)--(0,1.75)--(1,1.05);
                    \draw (-1,1.05)--(0,0.35)--(1,1.05);
                    \draw (1,1.05)--(2.5,0);
                }
                \ar@{=>}[r]^-{\epsilon_2} \ar@{=>}[dd]_{\alpha_\wr\star(\alpha_\wr^R)^{-1}} 
                & 
                \ctikz{[scale=0.4]
                    \draw[color3] (-2,0)--(2,0);
                    \draw (-1,0)--(0,0.7)--(1,0);
                }
                \ar@{=>}[rd]^-{\epsilon_\wr}
                \\
                & &
                \ctikz{[scale=0.4]
                    \draw[color3] (0,0)--(3,0);
                }
                \\
                \ctikz{[scale=0.2]
                    \draw[color3] (-3.5,0)--(-2.5,0);
                    \draw[color3] (-2.5,0)--(2.5,0);
                    \draw[color3] (2.5,0)--(3.5,0);
                    \draw (-2.5,0)--(-1,1.05);
                    \draw (-1,1.05)--(0,1.75)--(1,1.05);
                    \draw (-1,0)--(0,0.7)--(1,0);
                    \draw (1,1.05)--(2.5,0);
                }
                \ar@{=>}[r]_-{\epsilon_\wr} & 
                \ctikz{[scale=0.4]
                    \draw[color3] (-2,0)--(2,0);
                    \draw (-1,0)--(0,0.7)--(1,0);
                }
                \ar@{=>}[ru]_-{\epsilon_\wr}
            }
            \quad\quad
            \diagram@R=0.3pc@C=1.5pc{
                & 
                \ctikz{[scale=0.3]
                    \draw[color3] (-1.5,-1.05)--(1.5,-1.05);
                    \draw (-1.5,0.7)--(-0.5,0);
                    \draw (-1.5,-0.7)--(-0.5,0);
                    \draw (-0.5,0)--(0.5,0);
                    \draw (0.5,0)--(1.5,0.7);
                    \draw (0.5,0)--(1.5,-0.7);
                }
                \ar@{=>}[r]^-{\eta_\wr} &
                \ctikz{[scale=0.3]
                    \draw[color3] (0,0)--(3.5,0);
                    \draw (0,1.05)--(0.5,0.7);
                    \draw (0.5,0.7)--(1.5,0);
                    \draw (0,0.35)--(0.5,0.7);
                    \draw (2,0)--(3,0.7);
                    \draw (3,0.7)--(3.5,1.05);
                    \draw (3,0.7)--(3.5,0.35);
                }
                \ar@{=>}[dd]_{(\alpha_\wr^R)^{-1}\star\alpha_\wr}
                \\
                \ctikz{[scale=0.3]
                    \draw[color3] (-1.5,-0.875)--(1.5,-0.875);
                    \draw (-1.5,0)--(1.5,0);
                    \draw (-1.5,0.875)--(1.5,0.875);
                }
                \ar@{=>}[ru]^-{\eta_2} \ar@{=>}[rd]_-{\eta_\wr}
                \\
                & \ar@{=>}[r]_-{\eta_\wr} 
                \ctikz{[scale=0.3]
                    \draw[color3] (-1.5,0)--(1.5,0);
                    \draw (-1.5,0.7)--(-0.5,0);
                    \draw (0.5,0)--(1.5,0.7);
                    \draw (-1.5,1.4)--(1.5,1.4);
                }
                &
                \ctikz{[scale=0.3]
                    \draw[color3] (0,0)--(3.5,0);
                    \draw (0,1.05)--(1.5,0);
                    \draw (0,0.35)--(0.5,0);
                    \draw (2,0)--(3.5,1.05);
                    \draw (3,0)--(3.5,0.35);
                }
            }
        \end{array}
    }
    \vspace{2pt}
    \eqnn{\label{eq.right_dual_of_lambda_module}
        \begin{array}{c}
            \diagram@C=1.25pc{
                \ctikz{[scale=0.4]
                    \draw[color3] (-2,0)--(2,0);
                    \draw (-1,0)--(-0.3,0.49);
                    \draw (1,0)--(0.3,0.49);
                    \draw[fill=white] (-0.3,0.49)circle[radius=0.175];
                    \draw[fill=white] (0.3,0.49)circle[radius=0.175];
                }
                \ar@{=>}[rd]^(0.4){\lambda_\wr\star(\lambda_\wr^R)^{-1}}
                \ar@{=>}[d]_{\epsilon_0} 
                &
                \\
                \ctikz{[scale=0.4]
                    \draw[color3] (-2,0)--(2,0);
                    \draw (-1,0)--(0,0.7)--(1,0);
                }
                \ar@{=>}[r]_-{\epsilon_\wr}
                & 
                \ctikz{[scale=0.4]
                    \draw (0,0)--(3,0);
                }
            }
            \qquad\qquad\qquad\qquad
            \diagram@C=1.25pc{
            \ctikz{[scale=0.3]
                \draw[color3] (-1.5,-0.7)--(1.5,-0.7);
                \draw (-1.5,0.7)--(1.5,0.7);
                \draw[fill=white] (-1.5,0.7)circle[radius=0.25];
                \draw[fill=white] (1.5,0.7)circle[radius=0.25];
            }
            \ar@{=>}[d]_{\eta_\wr}
            & 
            \ctikz{[scale=0.3]
                \draw[color3] (-1.5,0)--(1.5,0);
            }
            \ar@{=>}[ld]^(0.5){(\lambda_\wr^R)^{-1}\star\lambda_\wr}\ar@{=>}[l]_-{\eta_0} 
            \\
            \ctikz{[scale=0.3]
                \draw[color3] (-1.5,0)--(1.5,0);
                \draw (-1.5,0.7)--(-0.5,0);
                \draw (0.5,0)--(1.5,0.7);
                \draw[fill=white] (-1.5,0.7)circle[radius=0.25];
                \draw[fill=white] (1.5,0.7)circle[radius=0.25];
            }
            }
        \end{array}
    }
    \caption{Pro-tensor network presentation of basic properties satisfied by $\cM$}.
    \label{fig.property_module}
\end{figure}
Moreover, applying the isomorphism in Proposition \ref{prp.transpose}.\ref{item1.prp.transpose} to the left equality in \eqref{eq.right_dual_of_lambda_module} and \eqref{eq.right_dual_of_alpha_module} respectively, one can prove the following two properties (see Appendix \ref{sub.two_module_transpose} for details).
\eqnn{\label{eq.unit_advanced}
    \diagram{
        \ctikz{[scale=0.3]
            \draw[color3] (-1.5,0)--(1.5,0);
            \draw (-1.5,1)--(1.5,1);
            \draw[fill=white] (1.5,1)circle[radius=0.25];
        }
        \ar@{=>}[rd]_{\lambda_\wr^R\star\lambda_\wr^{-1}\hspace{1ex}} \ar@{=>}[r]^-{\eta_\wr}
        &
        \ctikz{[scale=0.3]
            \draw[color3] (-1.5,0)--(1.5,0);
            \draw (-1.5,0.7)--(-0.5,0);
            \draw (0.5,0)--(1.5,0.7);
            \draw[fill=white] (1.5,0.7)circle[radius=0.25];
        }
        \\
        &
        \ctikz{[scale=0.3]
            \draw[color3] (-4.25,0)--(1.5,0);
            \draw (-1.5,0.7)--(-0.5,0);
            \draw (0.5,0)--(1.5,0.7);
            \draw (-2.75,1)--(-4.25,1);
            \draw[fill=white] (-2.75,1)circle[radius=0.25];
            \draw[fill=white] (-1.5,0.7)circle[radius=0.25];
            \draw[fill=white] (1.5,0.7)circle[radius=0.25];
        }
        \ar@{=>}[u]_{\epsilon_0}
    }
}
\eqnn{\label{eq.mult_advanced}
    \diagram{
        \ctikz{[scale=0.3]
            \draw[color3] (-1,0)--(1.5,0);
            \draw (-1,1.05)--(0.5,1.05);
            \draw (0.5,1.05)--(1.5,1.75);
            \draw (0.5,1.05)--(1.5,0.35);
        }
        \ar@{=>}[d]_{\eta_\wr}
        \ar@{=>}[r]^-{\eta_\wr}
        &
        \ctikz{[scale=0.3]
            \draw (0,0)--(1.5,1.05);
            \draw (-1,0)--(-2.5,1.05);
            \draw (-2.5,1.05)--(-1.5,1.75)--(1.5,1.75);
            \draw (-2.5,1.05)--(-3.5,1.05);
            \draw[color3] (-3.5,0)--(1.5,0);
        }
        \ar@{=>}[r]^-{\eta_\wr}
        &
        \ctikz{[scale=0.3]
            \draw (2.1,0)--(3.1,0.7);
            \draw (1.1,0)--(3.1,1.4);
            \draw (-1,0)--(-2.5,1.05);
            \draw (-2.5,1.05)--(-1.9,1.47);
            \draw (-1.9,1.47)--(0.1,0);
            \draw (-2.5,1.05)--(-3.5,1.05);
            \draw[color3] (-3.5,0)--(3.1,0);
        }
        \ar@{=>}[d]^-{\alpha_\wr^R\star\alpha_\wr^{-1}}
        \\
        \ctikz{[scale=0.3]
            \draw (0.5,1.05)--(1.5,1.75);
            \draw (0.5,1.05)--(1.5,0.35);
            \draw (-1,0)--(0.5,1.05);
            \draw (-2,0)--(-3,0.7);
            \draw[color3] (-3,0)--(1.5,0);
        }
        & &
        \ctikz{[scale=0.3]
            \draw (1.1,0)--(3.1,1.4);
            \draw (2.5,0.98)--(3.1,0.56);
            \draw (-1.3,1.05)--(-1.9,0.63)--(-2.5,1.05);
            \draw (-2.5,1.05)--(-1.9,1.47);
            \draw (-1.9,1.47)--(0.1,0);
            \draw (-2.5,1.05)--(-3.5,1.05);
            \draw[color3] (-3.5,0)--(3.1,0);
        }
        \ar@{=>}[ll]^-{\epsilon_2}
    }
}
Furthermore, the diagram in the left below can be proved commutative by a standard application of \eqref{eq.pentagon_and_triangle_module} in a similar way as \cite[Proposition 2.2.4]{Etingof_Gelaki_Nikshych_Ostrik_2015} or \cite[Proposition 8.1.3]{Marmolejo_1997}. The commutativity of the right diagram follows from the left.
\eqnn{\label{eq.triangle_other_module}
    \begin{array}{c}
        \diagram@C=1.5pc@R=1.5pc{
            &
            \ctikz{[scale=0.2]
                \draw[color3] (-2,0)--(3,0);
                \draw[color3] (0,0)--(3,0);
                \draw (0,0)--(3,2.1);
            }
            \\
            \ctikz{[scale=0.2]
                \draw[color3] (-2,0)--(0,0);
                \draw[color3] (0,0)--(3,0);
                \draw (0,0)--(2,1.4);
                \draw (2,1.4)--(3,2.1);
                \draw (2,1.4)--(3,0.7);
                \draw[fill=white] (3,2.1)circle[radius=0.4];
            }
            \ar@{=>}[ru]^-{\lambda_\wr} \ar@{=>}[rr]_-{\alpha_\wr}  & & 
            \ctikz{[scale=0.2]
                \draw[color3] (-2,0)--(0,0);
                \draw[color3] (0,0)--(3,0);
                \draw (0,0)--(2,1.4);
                \draw (2,1.4)--(3,2.1);
                \draw (2,0)--(3,0.7);
                \draw[fill=white] (3,2.1)circle[radius=0.4];
            }
            \ar@{=>}[lu]_-{\lambda_\wr}
        }
        \qquad\qquad
        \diagram@C=1.5pc@R=1.5pc{
            &
            \ctikz{[scale=0.2,xscale=-1]
                \draw[color3] (-2,0)--(3,0);
                \draw[color3] (0,0)--(3,0);
                \draw (0,0)--(3,2.1);
            }
            \\
            \ctikz{[scale=0.2,xscale=-1]
                \draw[color3] (-2,0)--(0,0);
                \draw[color3] (0,0)--(3,0);
                \draw (0,0)--(2,1.4);
                \draw (2,1.4)--(3,2.1);
                \draw (2,1.4)--(3,0.7);
                \draw[fill=white] (3,2.1)circle[radius=0.4];
            }
            \ar@{<=}[ru]^-{{\lambda_\wr^R}} \ar@{<=}[rr]_-{\alpha_\wr^R}  & & 
            \ctikz{[scale=0.2,xscale=-1]
                \draw[color3] (-2,0)--(0,0);
                \draw[color3] (0,0)--(3,0);
                \draw (0,0)--(2,1.4);
                \draw (2,1.4)--(3,2.1);
                \draw (2,0)--(3,0.7);
                \draw[fill=white] (3,2.1)circle[radius=0.4];
            }
            \ar@{<=}[lu]_-{\lambda_\wr^R}
        }
    \end{array}
}

We end this subsection by considering frobenius property and rigidity of monoidal $\cV$-categories.

First, notice that 
given a monoidal $\cV$-category $\cC=(\cC,\boxtimes,I,\alpha,\lambda,\rho,\epsilon_2,\eta_2,\epsilon_0,\eta_0)$, one can define a profunctor homomorphism $\alpha^\sharp$ as the unique profunctor homomorphism rendering the following ``candy wrapping'' diagram commutative:
\begin{equation}
\label{eq.Frob_bulk_1}
 \diagram{
    \ctikz{[x=0.75pt,y=0.75pt,yscale=0.6,xscale=0.6]
\draw     (120,87.33) -- (263,87.67) ;
\draw     (120.67,143.67) -- (263,144.33) ;
\draw    (159,144.33) -- (220.33,87.67) ;
    }
    \ar@{=>}[d]_{\alpha^\sharp}
    \ar@{=>}[rr]^-{\eta_2\star\eta_2}
    & &
    \ctikz{[x=0.75pt,y=0.75pt,yscale=0.6,xscale=0.6]
\draw    (153.67,87.67) -- (247.67,87.67) ;
\draw       (150.33,143.67) -- (249,144.33) ;
\draw       (182.33,144.33) -- (220.33,87.67) ;
\draw       (247.67,87.67) -- (271.67,114.33) ;
\draw       (249,144.33) -- (271.67,114.33) ;
\draw       (271.67,114.33) -- (307.67,114.33) ;
\draw       (307.67,114.33) -- (330.33,87.67) ;
\draw       (307.67,114.33) -- (331.67,144.33) ;
\draw       (129.67,117) -- (153.67,87.67) ;
\draw       (129.67,117) -- (150.33,143.67) ;
\draw       (91.67,117) -- (129.67,117) ;
\draw       (66.33,88.33) -- (91.67,117) ;
\draw       (91.67,117) -- (68.33,144.33) ;
    }
    \ar@{=>}[d]^{\alpha}
    \\
    \ctikz{
    [x=0.75pt,y=0.75pt,yscale=-0.8,xscale=0.8]
\draw    (242.33,88.33) -- (266.33,115.67) ;
\draw     (243.67,145) -- (266.33,115.67) ;
\draw     (266.33,115.67) -- (302.33,115.67) ;
\draw     (302.33,115.67) -- (325,89) ;
\draw     (302.33,115.67) -- (326.33,145.67) ;
    }
    & 
    \ctikz{[x=0.75pt,y=0.75pt,yscale=-0.6,xscale=0.6]
\draw     (153.67,87.67) -- (213.67,87) ;
\draw    (150.33,143.67) -- (215,143.67) ;
\draw     (213.67,87) -- (237.67,113.67) ;
\draw     (215,143.67) -- (237.67,113.67) ;
\draw     (237.67,113.67) -- (273.67,113.67) ;
\draw     (273.67,113.67) -- (296.33,87) ;
\draw     (273.67,113.67) -- (297.67,143.67) ;
\draw     (129.67,117) -- (153.67,87.67) ;
\draw     (129.67,117) -- (150.33,143.67) ;
\draw     (91.67,117) -- (129.67,117) ;
\draw     (66.33,88.33) -- (91.67,117) ;
\draw     (91.67,117) -- (68.33,144.33) ;
    }
    \ar@{=>}[l]^-{\epsilon_2}
    & 
    \ctikz{[x=0.75pt,y=0.75pt,yscale=-0.6,xscale=0.6]
\draw     (153.67,87.67) -- (247.67,87.67) ;
\draw       (150.33,143.67) -- (249,144.33) ;
\draw       (164.33,144.33) -- (199.73,112.33) ;
\draw       (247.67,87.67) -- (271.67,114.33) ;
\draw       (249,144.33) -- (271.67,114.33) ;
\draw       (271.67,114.33) -- (307.67,114.33) ;
\draw       (307.67,114.33) -- (330.33,87.67) ;
\draw       (307.67,114.33) -- (331.67,144.33) ;
\draw       (129.67,117) -- (153.67,87.67) ;
\draw       (129.67,117) -- (150.33,143.67) ;
\draw       (91.67,117) -- (129.67,117) ;
\draw       (66.33,88.33) -- (91.67,117) ;
\draw       (91.67,117) -- (68.33,144.33) ;
\draw       (199.73,112.33) -- (233.67,143.51) ;
    }
    \ar@{=>}[l]^-{\epsilon_2}.
    }
\end{equation}

\begin{remark}
$\alpha^\sharp$ is the image 
of $\alpha$ under a composition of two transposing isomorphisms in Proposition \ref{prp.transpose}:
\begin{align*}
    \Hom(\boxtimes_\ast\bullet(\boxtimes\os\Id_\cC)_\ast,\boxtimes_\ast\bullet(\Id_\cC\os\boxtimes)_\ast)
    &\cong\Hom
((\boxtimes\os\Id_\cC)_\ast,\boxtimes^\ast\bullet\boxtimes_\ast\bullet(\Id_\cC\os\boxtimes)_\ast)
    \\
    &\cong
\Hom((\boxtimes\os\Id_\cC)_\ast\bullet(\Id_\cC\os\boxtimes)^\ast,\boxtimes^\ast\bullet\boxtimes_\ast).
\end{align*}
\end{remark}
We will refer to $\alpha^\sharp$ as the (left) Frobenius structure.
Unpacking the definition of $\alpha^\sharp$, one can find its component is given by composition along the unique internal leg
\begin{equation}
\label{eq.alpha^sharp_component}
\begin{tikzcd}
{\int^{c\in\cC}\cC(x,w\boxtimes c)\ot\cC(c\boxtimes y,z)} && {\cC(x\boxtimes y,w\boxtimes z)} \\
	{\cC(x,w\boxtimes c)\ot\cC(c\boxtimes y,z)}
	\arrow["{\alpha^\sharp_{(x,y),(w,z)}}", from=1-1, to=1-3]
	\arrow["{\copr_c}", from=2-1, to=1-1]
	\arrow["{\text{composition along $c$}}"{description}, from=2-1, to=1-3].
\end{tikzcd}
\end{equation}

Similarly, we can define the (right) Frobenius structure
\begin{equation}
\label{eq.Frob_bulk_2}
     \alpha^\flat\:
    \ctikz{[scale=0.5]
        \draw (-1.5,0.525)--(1.5,0.525);
        \draw (-1.5,-0.525)--(1.5,-0.525);
        \draw (0.75,0.525)--(-0.75,-0.525);
    }
    \Rightarrow
    \ctikz{[scale=0.5]
        \draw (-1.5,0.7)--(-0.5,0)--(-1.5,-0.7);
        \draw (1.5,0.7)--(0.5,0)--(1.5,-0.7);
        \draw (-0.5,0)--(0.5,0);
    }.
\end{equation}

Analogously, if $(\cM,\boxdot,\alpha_\wr,\lambda_\wr)$ is a left $\cC$-module, then we can define module Frobenius structures
\eqnn{\label{eq.module_Frobeniusiator}
    \alpha_\wr^\sharp\:\ctikz{[scale=0.5,xscale=-1]
        \draw (-1.5,0.7)--(1.5,0.7);
        \draw[color3] (-1.5,0)--(1.5,0);
        \draw (0.5,0.7)--(-0.5,0);
    }
    \Rightarrow
    \ctikz{[scale=0.5]
        \draw (-0.5,0)--(-1.5,0.7);
        \draw (0.5,0)--(1.5,0.7);
        \draw[color3] (-1.5,0)--(1.5,0);
    }
\quad\text{and}\quad
    \alpha_\wr^\flat\:\ctikz{[scale=0.5]
        \draw (-1.5,0.7)--(1.5,0.7);
        \draw[color3] (-1.5,0)--(1.5,0);
        \draw (0.5,0.7)--(-0.5,0);
    }
    \Rightarrow
    \ctikz{[scale=0.5,xscale=-1]
        \draw (-0.5,0)--(-1.5,0.7);
        \draw (0.5,0)--(1.5,0.7);
        \draw[color3] (-1.5,0)--(1.5,0);
    }.
}
Although $\alpha$ is invertible, the Frobenius structures $\alpha^\sharp$ and $\alpha^\flat$ need not be invertible. We say $\cC$ is \emph{naturally Frobenius} \cite{Franco_Street_Wood_2011} if $\alpha^\sharp$ and $\alpha^\flat$ is invertible (in which case one can show that $\alpha_\wr^\sharp$ and $\alpha_\wr^\flat$ is automatically invertible regardless of the choice of $\cM$). Thus, if $\cC$ is naturally Frobenius and $\cM$ is a left $\cC$-module, then we have the following canonical isomorphisms induced from $\alpha_\wr^\sharp$ and $\alpha_\wr^\flat$:
\eqnn{\label{eq.Frob_intertible_unital}
    \ctikz{[scale=0.5,xscale=-1]
        \draw (-1.5,0.7)--(1.5,0.7);
        \draw[color3] (-1.5,0)--(1.5,0);
        \draw (0.5,0.7)--(-0.5,0);
        \draw[fill=white] (1.5,0.7)circle[radius=0.2];
    }
    \stackrel{\sim}{\Rightarrow}
    \ctikz{[scale=0.5]
        \draw (0.5,0)--(1.5,0.7);
        \draw[color3] (-1.5,0)--(1.5,0);
    }
    ,\qquad\qquad
    \ctikz{[scale=0.5]
        \draw (-1.5,0.7)--(1.5,0.7);
        \draw[color3] (-1.5,0)--(1.5,0);
        \draw (0.5,0.7)--(-0.5,0);
        \draw[fill=white] (1.5,0.7)circle[radius=0.2];
    }
    \stackrel{\sim}{\Rightarrow}
    \ctikz{[scale=0.5,xscale=-1]
        \draw (0.5,0)--(1.5,0.7);
        \draw[color3] (-1.5,0)--(1.5,0);
    }.
}

The purpose of this work is not to investigate the property of being naturally Frobenius in depth. We only record a sufficient condition for being naturally Frobenius:
\begin{definition}
     Let $\cC$ be a monoidal $\cV$-category. We say $\cC$ is \emph{rigid} if its underlying monoidal category $\underline{\cC}$ is rigid in the sense of \cite{Etingof_Gelaki_Nikshych_Ostrik_2015}, i.e., each object $x\in\underline{\cC}$ possesses a left dual $x^L$ and a right dual $x^R$.
\end{definition}
Left rigidity (resp. right rigidity) can be similarly defined when only left duals (resp. right duals) are present.
\begin{proposition}
\label{prp.rigit_to_NF}
    If $\cC$ is rigid, then $\cC$ is naturally Frobenius.
    \begin{proof}
        See Appendix~\ref{app.rigid_to_NF}.
    \end{proof}
\end{proposition}


\subsection{Promonads, probimonads and their modules}\label{sub.pmd}
While in the world of linear algebra, the notion of associative algebra is useful, we need a replacement of this notion in the world of categorified linear algebra or pro-tensor networks. Such a replacement is given by promonad. In this subsection, we introduce this notion. They're useful in the study of particle-like defects in string-net pro-tensor networks in Section \ref{sec.LW} and Section \ref{sec.main_thm}, and the discussion of topological holography in Section \ref{sec.TubeHolo}.


\begin{definition}
Let $\cC$ be a $\cV$-category. A \emph{promonad $T$ on $\cC$} consists of:
\begin{itemize}
  \item a pro-tensor $T\colon \cC \nrightarrow \cC$, denoted as 
  \[
          \onenodenolabeledge{T}
  \]
  \item a profunctor homomorphism
        \(
        \mu\colon T\bullet T \Longrightarrow T,
        \)
        called the \emph{multiplication}, denoted as 
        \[
            \begin{tikzpicture}
                [scale = 0.5 ,baseline=(current bounding box.center)]
                \node (A) at (-4,0){$\twonodenolabeledge{T}{T}$};
                \node (B) at (4,0){$\onenodenolabeledge{T}$};
                \draw[nat] (A)--(B) node[midway,above = 3pt]{\mu};
            \end{tikzpicture}
        \]
  \item a profunctor homomorphism
        \(
        \eta\colon \id_\cC \Longrightarrow T,
        \)
        called the \emph{unit}, denoted as 
        \[
            \begin{tikzpicture}
                [scale = 0.5 ,baseline=(current bounding box.center)]
                \node (A) at (-4,0){\begin{tikzpicture}[scale = 0.5 ,baseline=(current bounding box.center)]
                 \draw  (-2,0)--(2,0);
    \end{tikzpicture}};
          \node (B) at (4,0){$\onenodenolabeledge{T}$};
    \draw[nat] (A)--(B) node[midway,above = 3pt]{\eta};
            \end{tikzpicture}.
        \]
\end{itemize}
The profunctor homomorphisms $\mu$ and $\eta$ satisfy the usual associativity and unitality axioms:
\begin{equation}
\label{eq.pmd_ass}
  \begin{tikzpicture}
[scale = 0.5 ,baseline=(current bounding box.center)]
     \node (A) at (-4,0){$\threenodenolabeledge{T}{T}{T}$};
      \node (B) at (5,0){$\twonodenolabeledge{T}{T}$};
    \node (C) at (-4,-3){$\twonodenolabeledge{T}{T}$};
        \node (D) at (5,-3){$\onenodenolabeledge{T}$};
\draw[nat] (A)--(B) node[midway,above = 3pt]{\id_T \star \mu};
\draw[nat] (A)--(C) node[midway,left = 3pt]{\mu\star\id_T};
\draw[nat] (B)--(D) node[midway,right = 3pt]{\mu};
\draw[nat] (C)--(D) node[midway,below = 3pt]{\mu};
        \end{tikzpicture}
\end{equation}
\begin{equation}
\label{eq.pmd_unit}
\begin{tikzpicture}[baseline=(current bounding box.center)]
    \node (A) at (-5,0) {$\onenodenolabeledge{T}$};
\node (B) at (-5,-2)
{$\twonodenolabeledge{T}{T}$};
\node (C) at (-1,-2){$\onenodenolabeledge{T}$};
        \node (D) at (3,-2)
        {$\twonodenolabeledge{T}{T}$};
        \node (E) at (3,0) 
        {$\onenodenolabeledge{T}$};
\draw[nat] (A)--(B) node[midway, left = 3pt]{\eta\star\id_T};
\draw[nat] (B)--(C) node[midway, below = 3pt]{\mu};
\draw[nat] (A)--(C) node[midway, above = 3pt]{\id_T};
\draw[nat] (D)--(C)
node[midway, below = 3pt]
{\mu};
\draw[nat] (E)--(C)
node[midway, above = 3pt]
{\id_T};
\draw[nat] (E)--(D)
node[midway, right = 3pt]
{\id_T\star\eta};
\end{tikzpicture}.
\end{equation}
\end{definition}
\begin{definition}\label{dfn.promonad_module}
    Let $\cC$ and $\cA$ be $\cV$-categories and $(T,\mu,\eta)$ be a promonad on $\cC$. A \emph{a left $T$-module from $\cA$} is a pair $(P,\kappa)$ where  $P:\cA\nrightarrow \cC$ is a pro-tensor and  $\kappa: T\bullet P\Rightarrow P$ is a profunctor homomorphism called the \emph{$T$-action},
    denoted as  
    \[
        \begin{tikzpicture} [scale = 1 ,baseline=(current bounding box.center)]
          \draw  (-1,0)--(0,0);
          \draw[color4] (0,0)--(1,0);
          \mynode{0}{0}{0.3}{0.3}{P}
      \end{tikzpicture}
    \quad\text{and}\quad  
        \begin{tikzpicture}
            [scale = 0.5 ,baseline=(current bounding box.center)]
            \node (A) at (-4,0){\begin{tikzpicture} [scale = 1 ,baseline=(current bounding box.center)]
          \draw  (-2,0)--(0,0);
          \draw[color4] (0,0)--(1,0);
          \mynode{-1}{0}{0.3}{0.3}{T}
          \mynode{0}{0}{0.3}{0.3}{P}
      \end{tikzpicture}};
      \node (B) at (4,0){\begin{tikzpicture} [scale = 1 ,baseline=(current bounding box.center)]
          \draw  (-1,0)--(0,0);
          \draw[color4] (0,0)--(1,0);
          \mynode{0}{0}{0.3}{0.3}{P}
      \end{tikzpicture}};
\draw[nat] (A)--(B) node[midway,above = 3pt]{\kappa};
        \end{tikzpicture}
    \]
        respectively, satisfying the associativity and unitality axioms:
     \begin{equation}
     \label{eq.ModuleoOverMonad1}
       \begin{tikzpicture}
[scale = 0.5 ,baseline=(current bounding box.center)]
     \node (A) at (-4,0){\begin{tikzpicture} [scale = 1 ,baseline=(current bounding box.center)]
          \draw  (-3,0)--(0,0);
          \draw[color4] (0,0)--(1,0);
          \mynode{-2}{0}{0.3}{0.3}{T}
          \mynode{-1}{0}{0.3}{0.3}{T}
          \mynode{0}{0}{0.3}{0.3}{P}
      \end{tikzpicture}};
      \node (B) at (5,0){\begin{tikzpicture} [scale = 1 ,baseline=(current bounding box.center)]
          \draw  (-2,0)--(0,0);
          \draw[color4] (0,0)--(1,0);
          \mynode{-1}{0}{0.3}{0.3}{T}
          \mynode{0}{0}{0.3}{0.3}{P}
      \end{tikzpicture}};
    \node (C) at (-4,-3){\begin{tikzpicture} [scale = 1 ,baseline=(current bounding box.center)]
          \draw  (-2,0)--(0,0);
          \draw[color4] (0,0)--(1,0);
          \mynode{-1}{0}{0.3}{0.3}{T}
          \mynode{0}{0}{0.3}{0.3}{P}
      \end{tikzpicture}};
        \node (D) at (5,-3){\begin{tikzpicture} [scale = 1 ,baseline=(current bounding box.center)]
          \draw  (-1,0)--(0,0);
          \draw[color4] (0,0)--(1,0);
          \mynode{0}{0}{0.3}{0.3}{P}
      \end{tikzpicture}};
\draw[nat] (A)--(B) node[midway,above = 3pt]{\id_T\star\kappa};
\draw[nat] (A)--(C) node[midway,left = 3pt]{\mu\star\id_P};
\draw[nat] (B)--(D) node[midway,right = 3pt]{\kappa};
\draw[nat] (C)--(D) node[midway,below = 3pt]{\kappa};
        \end{tikzpicture}
\end{equation}
    \begin{equation}
            \label{eq.ModuleoOverMonad2}
 \begin{tikzpicture}[baseline=(current bounding box.center)]]
    \node (A) at (-5,0) {\begin{tikzpicture} [scale = 1 ,baseline=(current bounding box.center)]
          \draw  (-1,0)--(0,0);
          \draw[color4] (0,0)--(1,0);
          \mynode{0}{0}{0.3}{0.3}{P}
      \end{tikzpicture}};
\node (B) at (-5,-2)
{\begin{tikzpicture} [scale = 1 ,baseline=(current bounding box.center)]
          \draw  (-2,0)--(0,0);
          \draw[color4] (0,0)--(1,0);
          \mynode{-1}{0}{0.3}{0.3}{T}
          \mynode{0}{0}{0.3}{0.3}{P}
      \end{tikzpicture}};
\node (C) at (-1,-2){\begin{tikzpicture} [scale = 1 ,baseline=(current bounding box.center)]
          \draw  (-1,0)--(0,0);
          \draw[color4] (0,0)--(1,0);
          \mynode{0}{0}{0.3}{0.3}{P}
      \end{tikzpicture}};
\draw[nat] (A)--(B) node[midway, left = 3pt]{\eta\star\id_P};
\draw[nat] (B)--(C) node[midway, below = 3pt]{\kappa};
\draw[nat] (A)--(C) node[midway, above = 3pt]{\id_P};
\end{tikzpicture}.
\end{equation}
A \emph{$T$-module homomorphism} $(P,\kappa)\Rightarrow(Q,\psi)$ between left $T$-modules from $\cA$ is a pro-tensor network map $\phi\:P\Rightarrow Q$ preserving the $T$-actions.
\end{definition}
Right $T$-modules can be similarly defined.


Left $T$-modules from $\cA$ and $T$-module homomorphisms form a category, which we denote by $\Lmd_T(\cA)$. We'll be primarily concerned with the case $\cA=\ast$, in which case the $T$-module $P\:\cA\arprof\cC$ in Definition \ref{dfn.promonad_module} can be simply drawn as
\[
    \ctikz{
        \draw (-1,0)--(0,0);
        \mynode{0}{0}{0.3}{0.3}{P}
    }.
\]
An important aspect of promonad theory is the existence of an Eilenberg-Moore category of a promonad, which controls the representation theory of the latter. This is discussed in Section \ref{sec.EM}.

We have introduced the analogue of ``algebras'' in the theory of pro-tensor network. Bialgebras or Hopf algebras, on the other hand, have also been used ubiquitously in theoretical physics; to name but a few, \cite{Majid_1993, Schomerus_1995, Nill_Szlachanyi_1997, Connes_Kreimer_1998, Kitaev_2006, Kitaev_Kong_2012}. We now introduce probimonad, which is an analogue of bialgebras in the pro-tensor network theory. Note that probimonad is a well-known mathematical concept \cite{Lopez_Franco_2007, Chikhladze_Lack_Street_2010}, although no explicit name is given.

\begin{definition}\label{dfn.probimonad}
    Let $\cC$ be a $\cV$-category. A \emph{probimonad on $\cC$} consists of the following data:
    \begin{itemize}
        \item A promonad $(T,\mu,\eta)$ on $\cC$.
        \item Two profunctors $M\:\cC\os\cC\arprof\cC$ and $U\:\ast\arprof\cC$, denoted by 
        \[
        \ctikz{[scale=0.5,xscale=-1]
            \draw (1,0)--(0,0);
            \draw (0,0)--(-1,0.7);
            \draw (0,0)--(-1,-0.7);
        }
        \quad\text{and}\quad
        \ctikz{[xscale=-1]
            \draw (1,0)--(0,0);
            \draw[fill=white] (0,0)circle[radius=0.1];
        }
        \]
        respectively.
        \item Two maps of pro-tensor network
        \[
            \tau\:
            \ctikz{[scale=0.5]
                \draw (-1.5,0)--(1.5,0);
                \mynode{0}{0}{0.6}{0.6}{T} 
                \draw (1.5,0)--(3,1.05);
                \draw (1.5,0)--(3,-1.05);
            }
            \Rightarrow
            \ctikz{[scale=0.5]
                \draw (-1,0)--(1,0);
                \draw (1,0)--(3.5,1.75);
                \draw (1,0)--(3.5,-1.75);
                \mynode{2.25}{0.875}{0.6}{0.6}{T} 
                \mynode{2.25}{-0.875}{0.6}{0.6}{T} 
            }
            \quad\text{and}\quad
            \nu\:
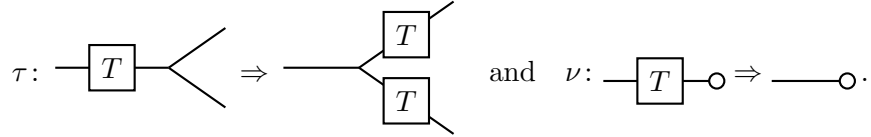
            \begin{tikzpicture}[scale=0.5,baseline = (current bounding box.center)]
                \draw (-1.5,0)--(1.5,0);
                \draw[fill=white] (1.5,0)circle[radius=0.2];
                \mynode{0}{0}{0.6}{0.6}{T}
            \end{tikzpicture}
            \Rightarrow
            \begin{tikzpicture}[scale=0.5,baseline = (current bounding box.center)]
                \draw (-1,0)--(1,0);
                \draw[fill=white] (1,0)circle[radius=0.2];
            \end{tikzpicture}
            \,.
        \]
        \item Three invertible maps of pro-tensor network
        \[
        \alpha^\bullet\:
        
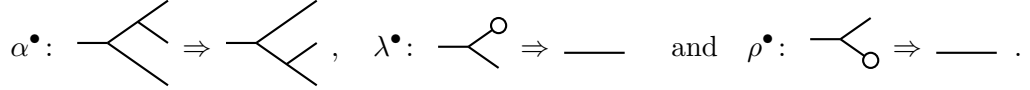
\diagram{
        \ctikz{[scale=0.4]
            \draw (-1,0)--(0,0);
            \draw (0,0)--(2,1.4);
            \draw (0,0)--(2,-1.4);
            \draw (1,0.7)--(2,0);
        }
        \Rightarrow 
        \ctikz{[scale=0.4,yscale=-1]
            \draw (-1,0)--(0,0);
            \draw (0,0)--(2,1.4);
            \draw (0,0)--(2,-1.4);
            \draw (1,0.7)--(2,0);
        }
        },\quad
        \lambda^\bullet\:
        \diagram{
        \ctikz{[scale=0.4]
            \draw (-1,0)--(0,0);
            \draw (0,0)--(1,0.7);
            \draw[fill=white] (1,0.7)circle[radius=0.25];
            \draw (0,0)--(1,-0.7);
        }
        \Rightarrow 
        \ctikz{[scale=0.4]
            \draw[white] (1,0)circle[radius=0.1]; 
            \draw (-1,0)--(1,0);
        }
        }
        \quad\text{and}\quad
        \rho^\bullet\:
        \diagram{
        \ctikz{[scale=0.4,yscale=-1]
            \draw (-1,0)--(0,0);
            \draw (0,0)--(1,0.7);
            \draw[fill=white] (1,0.7)circle[radius=0.25];
            \draw (0,0)--(1,-0.7);
        }
        \Rightarrow 
        \ctikz{[scale=0.4]
            \draw[white] (1,0)circle[radius=0.1]; 
            \draw (-1,0)--(1,0);
        }
        }.
        \]
    \end{itemize}
    They're required to satisfy the axioms listed in Figure \ref{fig.bimonad_axiom} in Appendix \ref{sec.graph.app}.
\end{definition}

It is well-known that given a bialgebra $A$, the category of left $A$-modules has a natural monoidal structure. This has the following analogy for probimonads.
\begin{proposition}\label{prp.bimonad_mod}
    Let $(T=(T,\mu,\eta),M,U,\tau,\nu,\alpha^\bullet,\lambda^\bullet,\rho^\bullet)$ be a probimonad on $\cC$. Then the category 
    $\Lmd_T(\ast)$ of $T$-modules from $\ast$ has a canonical structure of a monoidal category.
\end{proposition}
We now prove Proposition \ref{prp.bimonad_mod} by listing the complete monoidal structure of $\Lmd_T(\ast)$, which follow from a leisurely application of Axioms \eqref{eq.tau_intertwine}-\eqref{eq.pentagon_and_triangle_promonoidal}. For $(P,\kappa_P),(Q,\kappa_Q)\in\Lmd_T(\ast)$, we define their tensor product $P\conv Q$ as the pro-tensor
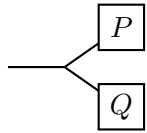
\[
    \ctikz{[scale=0.5]
        \draw (-1.5,0)--(0,0);
        \draw (0,0)--(1.5,1.05);
        \draw (0,0)--(1.5,-1.05);
        \mynode{1.5}{1.05}{0.6}{0.6}{P}
        \mynode{1.5}{-1.05}{0.6}{0.6}{Q}
    }
\]
carrying a $T$-action
\[
\kappa_{P\os Q}\defdtobe(
\diagram{
\ctikz{[scale=0.5]
        \draw (-3,0)--(0,0);
        \draw (0,0)--(2,1.4);
        \draw (0,0)--(2,-1.4);
        \mynode{-1.5}{0}{0.6}{0.6}{T}
        \mynode{2}{1.4}{0.6}{0.6}{P}
        \mynode{2}{-1.4}{0.6}{0.6}{Q}
    }
\ar@{=>}[r]^-\tau
&
\ctikz{[scale=0.5]
        \draw (-2,0)--(0,0);
        \draw (0,0)--(3,2.1);
        \draw (0,0)--(3,-2.1);
        \mynode{1.5}{1.05}{0.6}{0.6}{T}
        \mynode{1.5}{-1.05}{0.6}{0.6}{T}
        \mynode{3}{2.1}{0.6}{0.6}{P}
        \mynode{3}{-2.1}{0.6}{0.6}{Q}
    }
\ar@{=>}[r]^-{\kappa_P\star\kappa_Q}
&
\ctikz{[scale=0.5]
        \draw (-1.5,0)--(0,0);
        \draw (0,0)--(1.5,1.05);
        \draw (0,0)--(1.5,-1.05);
        \mynode{1.5}{1.05}{0.6}{0.6}{P}
        \mynode{1.5}{-1.05}{0.6}{0.6}{Q}
    }
}.
\]
One can check that $\kappa_{P\os Q}$ is a well-defined $T$-action by \eqref{eq.tau_intertwine}. $\conv$ naturally extends to a functor by locality of pro-tensor network maps (Theorem \ref{thm.local}). The tensor unit of $\Lmd_T(\ast)$ is given by $U$; indeed $U$ is a $T$-module from $\ast$ precisely by \eqref{eq.nu_intertwine}. Finally, the associators and unitors in $\Lmd_T(\ast)$ are induced from $\alpha^\bullet$ and $\lambda^\bullet,\rho^\bullet$ respectively, which are well-defined and satisfy the pentagon and triangle equalities by \eqref{eq.alpha_is_monad_2mor}-\eqref{eq.pentagon_and_triangle_promonoidal}.

\begin{remark}
    A promonad is just a monad in the 2-category $\vprof$. The axioms satisfied by a probimonad may be hard to grasp except for their consequence (Proposition \ref{prp.bimonad_mod}) and our concrete examples (see Section \ref{sub.proof}), for which we now introduce the abstract nonsense understanding.
    For any 2-category $\mathsf{K}$, \cite{Street_1972} defines a 2-category $\Mnd(\mathsf{K})$ of monads in $\mathsf{K}$. It can be verified that when $\mathsf{K}$ is a monoidal 2-cateory, then so is $\Mnd(\mathsf{K})$. Now we can define a bimonad as an pseudoalgebra object in $\Mnd(\mathsf{K})$ \cite{Chikhladze_Lack_Street_2010}. A probimonad is then a bimonad in the monoidal 2-category $\mathsf{K}=(\vprof,\os,\ast)$. For a more comprehensive graphical treatment of bimonads in general monoidal 2-categories, see \cite{Willerton_2008}.
\end{remark}
\six{
 with morphisms being the pro-tensor homomorphisms preserving the $T$-actions. 
}

\section{ Pro-tensor network realization and generalization of Levin-Wen model}\label{sec.LW}

In this section, we introduce a trivalent branching network equipped with a “uniform” pro-tensor assignment, which recovers and generalizes the Levin–Wen model. We describe the pro-tensor network realization and generalization of the Levin–Wen model in Section~\ref{sub.lw}, referring to this construction as the string-net pro-tensor network. We then analyze particle-like defects within the string-net pro-tensor network in Sections~\ref{sec.particle_is_module}.




\subsection{String-net pro-tensor network}
\label{sub.lw}
The input of the ordinary Levin-Wen model is a unitary fusion category $\cC$, we would like to generalize $\cC$ to be an arbitrary rigid monoidal category which is enriched in a general cosmos $\cV$. Such a generalization not only helps reveal the underlying mathematical structure of the Levin-Wen model, but also has profound implications for understanding new phases and the categorical description of symmetries that are not necessarily finite or semisimple.

\begin{figure}
    \centering
    \begin{tikzpicture}
[x=0.75pt,y=0.75pt,yscale=-1,xscale=1]

\draw   (257.5,145.5) -- (243.75,169.32) -- (216.25,169.32) -- (202.5,145.5) -- (216.25,121.68) -- (243.75,121.68) -- cycle ;
\draw   (175,145.5) -- (161.25,169.32) -- (133.75,169.32) -- (120,145.5) -- (133.75,121.68) -- (161.25,121.68) -- cycle ;
\draw   (216.25,121.68) -- (202.5,145.5) -- (175,145.5) -- (161.25,121.68) -- (175,97.87) -- (202.5,97.87) -- cycle ;
\draw   (257.5,193.13) -- (243.75,216.95) -- (216.25,216.95) -- (202.5,193.13) -- (216.25,169.32) -- (243.75,169.32) -- cycle ;
\draw   (216.25,169.32) -- (202.5,193.13) -- (175,193.13) -- (161.25,169.32) -- (175,145.5) -- (202.5,145.5) -- cycle ;
\draw   (298.75,169.32) -- (285,193.13) -- (257.5,193.13) -- (243.75,169.32) -- (257.5,145.5) -- (285,145.5) -- cycle ;
\draw   (175,193.13) -- (161.25,216.95) -- (133.75,216.95) -- (120,193.13) -- (133.75,169.32) -- (161.25,169.32) -- cycle ;
\draw   (133.75,169.32) -- (120,193.13) -- (92.5,193.13) -- (78.75,169.32) -- (92.5,145.5) -- (120,145.5) -- cycle ;
\draw   (216.25,216.95) -- (202.5,240.76) -- (175,240.76) -- (161.25,216.95) -- (175,193.13) -- (202.5,193.13) -- cycle ;
\draw    (119,99) -- (133.75,121.68) ;
\draw    (160.25,75.18) -- (175,97.87) ;
\draw    (77.75,122.82) -- (92.5,145.5) ;
\draw    (217,75) -- (202.5,97.87) ;
\draw    (258.25,98.82) -- (243.75,121.68) ;
\draw    (299.5,122.63) -- (285,145.5) ;
\draw    (202.5,240.76) -- (217.25,263.45) ;
\draw    (243.75,216.95) -- (258.5,239.63) ;
\draw    (285,193.13) -- (299.75,215.82) ;
\draw    (175,240.76) -- (160.5,263.63) ;
\draw    (133.75,216.95) -- (119.25,239.82) ;
\draw    (92.5,193.13) -- (78,216) ;
\draw    (298.75,169.32) -- (329,169) ;
\draw    (48.5,169.63) -- (78.75,169.32) ;

\draw (287,81.4) node [anchor=north west][inner sep=0.75pt]    {$\cC$};
\end{tikzpicture}
    \caption{$\cC$ string-net pro-tensor network }
    \label{fig.string_net}
\end{figure}
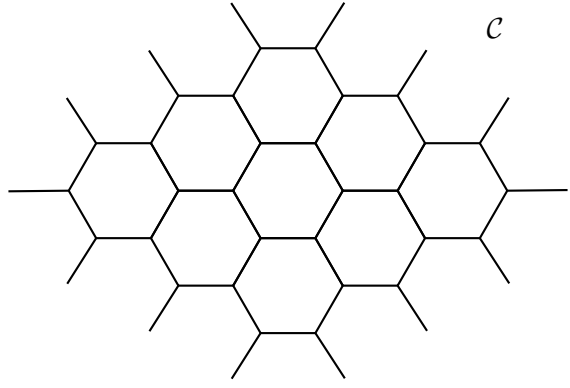

Let $\cC$ be a monoidal $\Vect$-category. Consider a finite trivalent pro-tensor network made of basic pro-tensors 
    \begin{tikzpicture}
[scale=0.4,baseline = (current bounding box.center)]
    \draw  (0,0)--(1.6,0);
    \draw  (-1,0.9)--(0,0);
    \draw  (-1,-0.9)--(0,0);
\end{tikzpicture} and 
    \begin{tikzpicture}
[scale=0.4,baseline = (current bounding box.center)]
    \draw  (0.2,0)--(1.8,0);
    \draw  (3,0.9)--(1.8,0);
    \draw  (3,-0.9)--(1.8,0);
\end{tikzpicture}, for example a honeycomb network, on an open disk, as shown in Figure~\ref{fig.string_net}.

If we assign objects in $\cC$ to all external edges, the network gives a potentially infinite dimensional vector space. We can view such a space as the analogue of the space of Levin-Wen model with all vertex terms $Q_I=1$~\cite{Levin_2005}, i.e. the space of allowed string-configuration, since back to finite semisimple case, the coend $\int^{c\in \cC}...\cC(a\boxtimes b,c)\ot\cC(c,d\boxtimes e)...\cong \oplus_{c\in \irr{\cC}}...\cC(a\boxtimes b,c)\otimes \cC(c,d\boxtimes e)...$ only takes care of the the strings that are compatible with each other.

The honeycomb pro-tensor can be transformed to the ``tree-bubble-tree" shape of pro-tensor by using $\alpha^\sharp$~\eqref{eq.Frob_bulk_1} and $\alpha^\flat$~\eqref{eq.Frob_bulk_2} as follows

\begin{equation}
    \begin{tikzpicture}
        \node (A) at (0,0) 
        {\begin{tikzpicture}[x=0.75pt,y=0.75pt,yscale=-0.4,xscale=0.4,baseline=(current bounding box.center)]

\draw   (259,145) -- (228.5,197.83) -- (167.5,197.83) -- (137,145) -- (167.5,92.17) -- (228.5,92.17) -- cycle ;
\draw    (134,42.5) -- (167.5,92.17) ;
\draw    (262,41.5) -- (228.5,92.17) ;
\draw    (64,144.5) -- (137,145) ;
\draw    (259,145) -- (338,145.5) ;
\draw    (167.5,197.83) -- (134,248.5) ;
\draw    (228.5,197.83) -- (262,247.5) ;
\draw  [dash pattern={on 0.84pt off 2.51pt}] (193.22,72.67) .. controls (208.23,53.66) and (243.96,56.86) .. (273.04,79.82) .. controls (302.11,102.77) and (313.52,136.79) .. (298.52,155.79) .. controls (283.51,174.8) and (247.78,171.6) .. (218.7,148.64) .. controls (189.63,125.69) and (178.22,91.67) .. (193.22,72.67) -- cycle ;

\end{tikzpicture}
};

\node (B) at (5,0) {
\begin{tikzpicture}[x=0.75pt,y=0.75pt,yscale=-0.4,xscale=0.4]

\draw    (134,42.5) -- (167.5,92.17) ;
\draw    (64,144.5) -- (137,145) ;
\draw    (167.5,197.83) -- (134,248.5) ;
\draw    (228.5,197.83) -- (262,247.5) ;
\draw    (137,145) -- (167.5,197.83) ;
\draw    (167.5,197.83) -- (228.5,197.83) ;
\draw    (228.5,197.83) -- (271,140.5) ;
\draw    (167.5,92.17) -- (271,140.5) ;
\draw    (271,140.5) -- (294,140.5) -- (328,140.5) ;
\draw    (328,140.5) -- (371,86.5) ;
\draw  [dash pattern={on 0.84pt off 2.51pt}] (119.45,165.08) .. controls (102.19,152.26) and (103.34,121.48) .. (122.01,96.33) .. controls (140.69,71.18) and (169.82,61.18) .. (187.09,74) .. controls (204.35,86.82) and (203.2,117.6) .. (184.52,142.75) .. controls (165.84,167.9) and (136.71,177.9) .. (119.45,165.08) -- cycle ;
\draw    (328,140.5) -- (368,199.5) ;
\draw    (137,145) -- (167.5,92.17) ;

\end{tikzpicture}}
;

\node (C) at (10,-0.2){
\begin{tikzpicture}[x=0.75pt,y=0.75pt,yscale=-0.4,xscale=0.4,baseline=(current bounding box.center)]

\draw    (64,144.5) -- (137,145) ;
\draw    (167.5,197.83) -- (134,248.5) ;
\draw    (228.5,197.83) -- (262,247.5) ;
\draw    (137,145) -- (167.5,197.83) ;
\draw    (167.5,197.83) -- (228.5,197.83) ;
\draw    (228.5,197.83) -- (271,140.5) ;
\draw    (197,105.5) -- (271,140.5) ;
\draw    (271,140.5) -- (294,140.5) -- (328,140.5) ;
\draw    (328,140.5) -- (371,86.5) ;
\draw    (328,140.5) -- (368,199.5) ;
\draw    (137,145) -- (197,105.5) ;
\draw  [dash pattern={on 0.84pt off 2.51pt}] (110.09,132.39) .. controls (122.61,117.87) and (151.65,122.38) .. (174.95,142.46) .. controls (198.26,162.54) and (207.01,190.6) .. (194.5,205.12) .. controls (181.98,219.64) and (152.95,215.13) .. (129.64,195.05) .. controls (106.33,174.97) and (97.58,146.91) .. (110.09,132.39) -- cycle ;
\draw    (24,85.5) -- (64,144.5) ;
\draw    (21,198.5) -- (64,144.5) ;
\end{tikzpicture}
};

\node (D) at (10,-3){
\begin{tikzpicture}[x=0.75pt,y=0.75pt,yscale=-0.4,xscale=0.4]

\draw    (127.33,35) -- (189.67,93) ;
\draw    (129.67,159) -- (189.67,93) ;
\draw    (127,108.33) -- (163,69) ;
\draw    (189.67,93) -- (261.67,93) ;
\draw   (334,50.17) -- (406.33,93) -- (334,135.83) -- (261.67,93) -- cycle ;
\draw    (540.24,150.74) -- (477.68,92.98) ;
\draw    (530.33,41) -- (477.68,92.98) ;
\draw    (334,135.83) -- (348.94,149.78) -- (354.2,154.69) -- (368.99,168.51) ;
\draw    (477.68,92.98) -- (405.68,93.26) ;
\draw  [dash pattern={on 0.84pt off 2.51pt}] (302.09,137.42) .. controls (292.38,117.06) and (309.76,88.51) .. (340.91,73.65) .. controls (372.07,58.78) and (405.19,63.23) .. (414.91,83.58) .. controls (424.62,103.94) and (407.24,132.49) .. (376.09,147.35) .. controls (344.93,162.22) and (311.81,157.77) .. (302.09,137.42) -- cycle ;

\end{tikzpicture}

};

\node (E) at (4,-3){
\begin{tikzpicture}[x=0.75pt,y=0.75pt,yscale=-0.4,xscale=0.4]

\draw    (127.33,35) -- (189.67,93) ;
\draw    (129.67,159) -- (189.67,93) ;
\draw    (127,108.33) -- (163,69) ;
\draw    (189.67,93) -- (261.67,93) ;
\draw   (334,50.17) -- (406.33,93) -- (334,135.83) -- (261.67,93) -- cycle ;
\draw    (540.24,150.74) -- (477.68,92.98) ;
\draw    (530.33,41) -- (477.68,92.98) ;
\draw    (501.34,70.32) -- (521.54,89.18) -- (536.33,103) ;
\draw    (477.68,92.98) -- (405.68,93.26) ;

\end{tikzpicture}
};
\draw[nattrans] (A)--(B);
\draw[nattrans] (B)--(C);
\draw[nattrans] (C)--(D);
\draw[nattrans] (D)--(E);
    \end{tikzpicture}.
\end{equation}

The above process gives a map $\mathfrak F$. The bubble in the last step can be further moved away by $\epsilon_2$, and we define the evaluation map 
\begin{equation}
    \begin{tikzpicture}
        \node (A) at (0,0){\begin{tikzpicture}[x=0.75pt,y=0.75pt,yscale=-0.6,xscale=0.6]

\draw   (152.33,112.5) -- (130.92,149.59) -- (88.08,149.59) -- (66.67,112.5) -- (88.08,75.41) -- (130.92,75.41) -- cycle ;
\draw     (66.67,38.31) -- (88.08,75.41) ;
\draw       (130.92,149.59) -- (152.33,186.69) ;
\draw       (130.92,75.41) -- (151,41) ;
\draw       (68.33,187) -- (88.08,149.59) ;
\draw       (152.33,112.5) -- (199.08,112.09) ;
\draw       (19.92,112.91) -- (66.67,112.5) ;

\end{tikzpicture}};
\node (B) at (6,0){\begin{tikzpicture}[x=0.75pt,y=0.75pt,yscale=-0.57,xscale=0.57]

\draw     (105.51,46.47) -- (166.37,112.21) ;
\draw        (109.67,185.67) -- (166.37,112.21) ;
\draw        (111.67,121) -- (140.84,85.09) ;
\draw        (166.37,112.21) -- (269.67,113) ;
\draw        (269.7,112.87) -- (329,187) ;
\draw        (269.7,112.87) -- (325.67,45.67) ;
\draw        (325.67,121.67) -- (294.17,84.43) ;

\end{tikzpicture}

};

\draw[nattrans](A)--(B) node [midway, above = 4pt]{$\ev:=\epsilon_2\cdot \mathfrak F$};
    \end{tikzpicture}
\end{equation}
Note that $\epsilon_2: \boxtimes_\ast\bullet \boxtimes^\ast \Rightarrow \cC(-,-)$ is the counit of the adjoint pair of profunctors induced by the functor $\boxtimes: \cC\times \cC \rightarrow \cC$. Its component  ${\epsilon_2}_{c,c'}: \int^{a,b}\cC(c,a\boxtimes b)\cC(a\boxtimes b,c')\rightarrow \cC(c,c')$ is induced by composition along $a\boxtimes b$ (see ~\eqref{eq.counit}). The Frobenius structure also essentially encode composition (see ~\eqref{eq.alpha^sharp_component}); together with $\epsilon_2$, this shows that the evaluation map $\ev$ is precisely the composition map. For $\cC$ a unitary fusion category, the dagger structure on morphism gives hom-space the structure of a finite dimensional Hilbert space, where the inner product of $f$ and $g\in \cC(a,b) $ is defined as $\Tr(f^\dagger g)$ up to a proper normalization. $\ev$ is then a map between finite dimensional Hilbert spaces, and $\ev^\dagger \ev$ matches the interpretation of the $B_p$ term in the Levin-Wen model~\cite{Kong_2012,Lan_2014,Green_2024} up to a normalization factor $1/\cD$, where $\cD$ is the total quantum dimension of $\cC$. The effective Levin-Wen Hamiltonian restricted on the subspace of allowed string configurations is then \begin{equation}
\label{eq.Levin-Wen Hamil}
    H = -\sum_{p\in\mathrm{plaquattes}}\ev_p^\dagger\ev_p/\cD.
\end{equation}
\begin{remark}
 There seems to be multiple ways to define the evaluation map, as we can apply the associator and the Frobenius structures at different positions in different orders. However, since the evaluation map essentially corresponds to composition, different ways of composing will always yield the same result up to canonical isomorphism. Hence, there is no ambiguity defining the evaluation map $\ev$.
\end{remark}
\begin{remark}
For a bosonic system with input unitary fusion category $\cC$ as discussed above, $\cC$ can be viewed as a $\hilbfd$-category, where $\hilbfd$ denotes the category of finite dimensional Hilbert spaces. In this finite-dimensional setting, the coends are finite-dimensional, ensuring $\mathrm{ev}^\dagger$ is always well-defined. The resulting Hamiltonian consists of commuting projectors — a consequence of the locality of the pro-tensor homomorphism — and describes a gapped phase. However, if the input category is non-finite, say $\cC=\Rep(G)$ for a compact Lie group $G$,  the coends may become infinite-dimensional, and the background category $\cV$ must be taken as the category of (possibly infinite-dimensional) Hilbert spaces, $\hilb$\footnote[1]{The category $\hilbfd$ is symmetric monoidal closed and finitely bi-complete: the internal hom $[U,W]$ can be identified with  the space of all linear maps from $U$ to $W$, endowed with the Hilbert-Schmidt inner product. In contrast, the category $\hilb$ is symmetric monoidal bicomplete, but  not closed in a naive way, since the space of bounded operators is merely a Banach space and does not carry a canonical Hilbert space structure. However, if one restricts morphisms to Hilbert-Schmidt operators, the resulting category $\hilb_{\mathrm{HS}}$ is symmetric monoidal closed, with internal hom the restricted hom, while failing to be complete or cocomplete. }.  If we discard the Hilbert space structure and consider $\cC$ as a $\Vect$-category, the operator $\ev$ may be unbounded, and the Hamiltonian may fail to be a well-defined, self-adjoint operator. Even when $\ev^\dagger$ exists,  $B_p \sim \ev_p^\dagger \ev_p$ remains not well-normalized. When \(B_p\) is ill-defined, the usual local plaquette dynamics implementing free string fluctuations is absent. Nevertheless, particle-like defects in the system still exhibit topological behavior, as will be explained in the next section.

For fermionic system, with  input unitary super fusion category or more generally a rigid monoidal $\mathbf{s}\hilbfd$-category $\cC$, the pro-tensor network can also recover fermionic string-net models.
\end{remark}

We refer to the above construction as a $\cC$ string-net pro-tensor network. 

\begin{remark}
A \(\cC\) string-net pro-tensor network adopts the perspective of a \(2+1\)D discrete quantum field theory, as explained in Remark~\ref{rmk.ptn_as_qft}. The same construction can also be interpreted from the \(1+1\)D operator perspective, where it may be named as a \(\cC\)-symmetric pro-tensor network.

Let \(\cC\) be a \(\Vect\)-enriched monoidal category. If we regard \(\cC\) as the charge category of a certain \(1+1\)D generalized symmetry, then the hom-space \(\cC(a,b)\) is the space of symmetric operators from \(a\) to \(b\). Consider a network as in Figure~\ref{fig.string_net}. If we assign to each link an object of \(\cC\), and to each vertex a morphism in the corresponding hom-space, then we obtain a \(\cC\)-symmetric tensor network. In a \(\cC\)-symmetric pro-tensor network, however, the data assigned to the vertices are the pro-tensors
$\boxtimes_*$ and $\boxtimes^*$, 
which already encode all possible symmetic local tensors that can be placed at the corresponding vertices. Thus, studying \(\cC\)-symmetric pro-tensor networks amounts to studying all \(\cC\)-symmetric tensor networks at once. In particular, once the objects on the external links are fixed, the contracted pro-tensor associated with a \(\cC\)-symmetric pro-tensor network produces a vector space. This vector space is precisely the space of all possible contracted \(\cC\)-symmetric tensors with those fixed boundary labels. In this sense, pro-tensor networks provide a unified framework for studying many many-body theories.
\end{remark}

For a rigid monoidal $\cV$-category $\cC$ and  a left $\cC$-module $\cV$-category $\cM$, we can consider the  $\cC$ string-net pro-tensor network with boundary $\cM$, as illustrated in Figure~\ref{fig:string_net_bdy}. The boundary pro-tensors \(
    \ctikz{[scale=0.4]
        \draw[color=color3  ,draw opacity=1, thick] (-1,0)--(0,0);
        \draw[color=color3  ,draw opacity=1, thick] (0,0)--(1,0);
        \draw (0,0)--(1,0.7);
    }
\),\(
    \ctikz{[scale=0.4,xscale=-1]
            \draw[color=color3  ,draw opacity=1, thick] (-1,0)--(0,0);
            \draw[color=color3  ,draw opacity=1, thick] (0,0)--(1,0);
            \draw (0,0)--(1,0.7);
    } \) are induced by the module action on $\cM$.

    \begin{figure}[htbp] 
    \centering  
   \tikzset{every picture/.style={line width=0.75pt}} 

\begin{tikzpicture}[x=0.75pt,y=0.75pt,yscale=-1,xscale=1]

\draw   (176,163) -- (161.5,188.11) -- (132.5,188.11) -- (118,163) -- (132.5,137.89) -- (161.5,137.89) -- cycle ;
\draw   (219.5,137.89) -- (205,163) -- (176,163) -- (161.5,137.89) -- (176,112.77) -- (205,112.77) -- cycle ;
\draw   (263,112.77) -- (248.5,137.89) -- (219.5,137.89) -- (205,112.77) -- (219.5,87.66) -- (248.5,87.66) -- cycle ;
\draw   (306.5,137.89) -- (292,163) -- (263,163) -- (248.5,137.89) -- (263,112.77) -- (292,112.77) -- cycle ;
\draw   (350,163) -- (335.5,188.11) -- (306.5,188.11) -- (292,163) -- (306.5,137.89) -- (335.5,137.89) -- cycle ;
\draw   (263,163) -- (248.5,188.11) -- (219.5,188.11) -- (205,163) -- (219.5,137.89) -- (248.5,137.89) -- cycle ;
\draw    (89,163) -- (118,163) ;
\draw    (350,163) -- (379,163) ;
\draw    (119,117) -- (132.5,137.89) ;
\draw    (163,93) -- (176,112.77) ;
\draw    (208,69) -- (219.5,87.66) ;
\draw    (248.5,87.66) -- (259,70) ;
\draw    (292,112.77) -- (304,92) ;
\draw    (335.5,137.89) -- (349,117) ;
\draw [color=color3  ,draw opacity=1 ][line width=1.25]    (79.5,188.11) -- (401.5,188.11) ;

\draw (392,199) node [anchor=north west][inner sep=0.75pt]   [align=left] {$\cM$};
\draw (340,83.4) node [anchor=north west][inner sep=0.75pt]    {$\cC$};
\end{tikzpicture}

    \caption{$\cC$ string-net pro-tensor network with boundary condition $\cM$}
    \label{fig:string_net_bdy}
\end{figure}

Similar to the bulk standalone case, such a pro-tensor network defines the space of compatible bulk–boundary string configurations. To obtain the ground state subspace, we again employ the evaluation map: we first reshape the network into the form of “tree–bubble–tree” using the associativity and Frobenius structure of both $\cC$  and $\cM$, then contract all bubbles via $\epsilon_2$ and $\epsilon_\wr$, as illustrated in
~\eqref{eq.bulk_boundary_ev}.
\begin{equation}
\label{eq.bulk_boundary_ev}
    \begin{tikzpicture}
        \node (A) at (-4,0){\begin{tikzpicture}[x=0.75pt,y=0.75pt,yscale=-0.7,xscale=0.7]

\draw   (188.79,122.75) -- (172.32,152.52) -- (139.37,152.52) -- (122.9,122.75) -- (139.37,92.98) -- (172.32,92.98) -- cycle ;
\draw    (122.9,122.75) -- (87.31,122.75) ;
\draw    (121.58,65) -- (139.37,92.98) ;
\draw    (172.32,92.98) -- (188.79,65) ;
\draw    (223.06,122.75) -- (188.79,122.75) ;
\draw    (223.06,122.75) -- (239.73,152.08) ;
\draw    (223.06,122.75) -- (238.87,96.63) ;
\draw [color=color3  ,draw opacity=1 ] [line width=1]   (57,152.76) -- (285,152.76) ;

\end{tikzpicture}};
\node (B) at (6,0){\begin{tikzpicture}[x=0.75pt,y=0.75pt,yscale=-0.7,xscale=0.7]

\draw [color=color3  ,draw opacity=1 ][line width=1]   (109.02,145.9) -- (242.02,145.39) -- (345,145) ;
\draw    (143,123) -- (168.02,144.9) ;
\draw   (113.75,116.69) -- (104.25,88.64) -- (133.5,94.95) -- (143,123) -- cycle ;
\draw    (84.02,70.9) -- (104.25,88.64) ;
\draw    (56.02,66.9) -- (84.02,70.9) ;
\draw    (75.02,46.9) -- (84.02,70.9) ;
\draw    (186.51,145.2) .. controls (207.02,106.9) and (243.02,68.9) .. (286.51,145.2) ;
\draw    (325.1,127.48) -- (305.05,145.42) ;
\draw    (332.45,100.17) -- (325.1,127.48) ;
\draw    (354.22,117.91) -- (325.1,127.48) ;

\end{tikzpicture}};
\node (C) at (1,-3){\begin{tikzpicture}[x=0.75pt,y=0.75pt,yscale=-1,xscale=1]

\draw    (221.1,128.48) -- (201.05,146.42) ;
\draw    (228.45,101.17) -- (221.1,128.48) ;
\draw    (250.22,118.91) -- (221.1,128.48) ;
\draw    (153.61,128) -- (172.87,146.79) ;
\draw    (126,120.89) -- (153.61,128) ;
\draw    (142.09,99.6) -- (153.61,128) ;
\draw [color=color3  ,draw opacity=1 ] [line width=1]  (119,146.89) -- (197,146.89) -- (250,146.89) ;

\end{tikzpicture}
};
\draw[nattrans] (A)--(C) node [midway, below left]{$\ev:=$};
\draw [nattrans] (A) -- (B) node
[midway, above = 2pt, align=center] {(module) Frobenius structures\\ and associators};
\draw[nattrans] (B)--(C) node [midway, below right]{$\epsilon_2\star\epsilon_\wr$};

    \end{tikzpicture}
\end{equation}

When $\cC$ is a unitary fusion category and $\cM$ is an indecomposable left $\cC$-module, the evaluation map $\ev$ projects onto the ground state subspace of Kitaev-Kong construction for the $\cC$ string-net with a gapped boundary labeled by $\cM$.

\subsection{Particle-like defects and modules over the renormalization algebra}
\label{sec.particle_is_module}
 
In this section, we discuss particle-like  defects in a $\cC$ string-net pro-tensor network. 
For the conventional Levin-Wen model, two primary approaches are employed to investigate such particle-like defects:
\begin{itemize}
\item One way treats particle-like defects as point-like excitations above the Levin-Wen ground state. In this picture, particle-like excitations are created in pairs from the vacuum via non-local string operators.  In the original paper~\cite{Levin_2005} of Levin-Wen model and a generalization~\cite{Lin_2021}, the authors explicitly constructed string operators on the lattice by deriving a set of consistency equations. A valid solution to these equations yields a string operator and, correspondingly, a particle excitation. While physically intuitive, this approach is computationally restrictive since the consistency equations are highly nonlinear and generally difficult to solve. More importantly, string operators are not naturally organized into a transparent algebraic object from which the classification of topological excitations can be directly extracted. This approach is therefore constructive rather than intrinsic, making a full classification practically challenging.
\item An alternative, systematically algebraic approach defines a particle-like defect as a localized region of elevated energy density embedded within the vacuum. We say a particle-like defect is at \emph{fixed-point} if it is scale-invariant, meaning that coarse-graining the surrounding region preserves its ``superselection sector".
For fixed-point models, such as the Levin-Wen models, any localized defect can be systematically promoted to a fixed-point defect by surrounding it with a region of the ground state. Because the system is at a fixed point, radially enlarging this surrounding ground-state region leaves the defect type unchanged. This implies that the  ground states space defined on the region surrounding the defect is equipped with a natural algebra structure: gluing two such concentric regions together corresponds to an associative multiplication operation. We define this as the \emph{renormalization algebra}. In this framework, a particle-like fixed-point defect is mathematically formalized as a module over the renormalization algebra, while the act of surrounding an arbitrary, unrefined defect with the ground state naturally yields a free module over this algebra.
For the ordinary Levin-Wen model with input unitary fusion category $\cC$, this algebra is the tube algebra $\tube(\cC)$~\cite{Lan_2014}, and the category of its representations is equivalent to the Drinfeld center of $\cC$~\cite{Ocneanu1994Chirality,Izumi:2000qa,Mueger:2001crc}:
\[
\Rep(\tube(\cC)) \cong \cZ(\cC),
\]
which is the unitary modular tensor category of anyons of the corresponding topological order.
According to~\cite{Kitaev_Kong_2012}, for the string-net model defined with input fusion category $\cC$, an indecomposable left $\cC$-module $\cM$ labels a gapped boundary condition. Associated to such a boundary \(\cM\) is an algebra  $A_{\cM,\cM}^\cC$, and its representation category is equivalent to the category of $\cC$-module functors from $\cM$ to $\cM$,
\[
\Rep(A_{\cM,\cM}^\cC) \cong {}_\cC[\cM,\cM]
\]
which is then the fusion category of particle-like topological defects on the boundary $\cM$~\cite{Kitaev_Kong_2012,Bai_Zhang_2025}. See also~\cite{Jia_2024,cordova2024,cordova2024representationtheorysolitons,Barter_2022,konechny2024fusingmatricesassociatedconformal,choi2026generalizedtubealgebrassymmetryresolved} for the study of defects from this perspective.
\end{itemize}
Note that the first (string operator) approach relies on the explicit form of the system’s Hamiltonian. String operators are constructed to commute with all local terms of the Hamiltonian everywhere except at the endpoints of the string. In contrast, the second (algebraic) approach characterizes particles intrinsically via the renormalization invariance of the surrounding ground states, which can be naturally generalized within our pro-tensor network framework, even when the Hamiltonian~\eqref{eq.Levin-Wen Hamil} is not well-defined.

\begin{figure}
    \centering
    \begin{subfigure}
        [b]{0.8\textwidth}
        \centering
        \begin{tikzpicture}[x=0.75pt,y=0.75pt,yscale=-1,xscale=1]

\draw  [fill=white,fill opacity=1 ][line width=0.75]  (210,155) -- (236,112) -- (287.31,112) -- (314.69,155) -- cycle ;
\draw    (212,81) -- (236,112) ;
\draw    (287.31,112) -- (308,82) ;
\draw [color=color3  ,draw opacity=1 ][line width=1]    (161,155) -- (372,155) ;
\draw  [fill=Mygray  ,fill opacity=1 ] (243.25,151.75) .. controls (243.25,141.67) and (251.64,133.5) .. (262,133.5) .. controls (272.36,133.5) and (280.75,141.67) .. (280.75,151.75) .. controls (280.75,161.83) and (272.36,170) .. (262,170) .. controls (251.64,170) and (243.25,161.83) .. (243.25,151.75) -- cycle ;

\draw (159,114) node [anchor=north west][inner sep=0.75pt]   [align=left] {...};
\draw (339,111) node [anchor=north west][inner sep=0.75pt]   [align=left] {...};
\draw (254,77) node [anchor=north west][inner sep=0.75pt]   [align=left] {...};
\draw (331,56) node [anchor=north west][inner sep=0.75pt]   [align=left] {$\cC$};
\draw (361,162) node [anchor=north west][inner sep=0.75pt]   [align=left] {$\cM$};
\end{tikzpicture}
    \caption{A particle-like  defect (shaded area) on the boundary $\cM$ of a $\cC$ string-net pro-tensor network.}
    \label{fig.boudary_defect}
    \end{subfigure}
    \begin{subfigure}
        [b]{0.8\textwidth}
        \centering 
            \begin{tikzpicture}[x=0.75pt,y=0.75pt,yscale=-1,xscale=1]

\draw [color=color3  ,draw opacity=1 ] [line width = 1]  (129.02,165.9) -- (262.02,165.39) -- (365,165) ;
\draw    (159,140) -- (179.25,165.64) ;
\draw    (206.51,165.2) .. controls (229,123) and (271,106) .. (306.51,165.2) ;
\draw    (350,138) -- (328.05,165.42) ;
\draw  [fill=white,fill opacity=1 ][dash pattern={on 0.84pt off 2.51pt}] (239,166.5) .. controls (239,157.39) and (246.39,150) .. (255.5,150) .. controls (264.61,150) and (272,157.39) .. (272,166.5) .. controls (272,175.61) and (264.61,183) .. (255.5,183) .. controls (246.39,183) and (239,175.61) .. (239,166.5) -- cycle ;

\draw (133,108) node [anchor=north west][inner sep=0.75pt]   [align=left] {...};
\draw (344,105) node [anchor=north west][inner sep=0.75pt]   [align=left] {...};
\draw (246,95) node [anchor=north west][inner sep=0.75pt]   [align=left] {...};

\end{tikzpicture}
\caption{The ground state on the surrounding area}
\label{fig.surr}
    \end{subfigure}
    \caption{}
\label{Fig.bdydefect&surroundingGS}
\end{figure}
We now study particle-like defects on the boundary $\cM$ of a $\cC$ string-net pro-tensor network using the second approach. Particle-like defects in the bulk are a special case via the folding trick, i.e., taking $\cM$ to be $\cC$ itself as a $\cC$-$\cC$-bimodule. As illustrated in Figure~\ref{fig.boudary_defect}, let the shaded area denote a particle-like defect on the boundary $\cM$. We can apply the evaluation map $\ev$ to obtain the ground state subspace on the area surrounding the defect, as shown in Figure~\ref{fig.surr}. Following the idea introduced above, this space should form a certain renormalization algebra. In the pro-tensor network, the algebraic structure is encoded in the pro-tensor network
surrounding the defects, which is given by
\begin{equation}
\cM\Omega_\cC \cM=:\label{eq.renormalization_alg}
    \begin{tikzpicture}[x=0.75pt,y=0.75pt,yscale=-1,xscale=1, baseline=-117 ]

\draw    (206.51,165.2) .. controls (229,123) and (271,106) .. (306.51,165.2) ;
\draw [color=color3  ,draw opacity=1 ]   (177,166) -- (235.01,165.7) ;
\draw [color=color3  ,draw opacity=1 ]   (277.51,165.35) -- (335.52,165.04) ;
\end{tikzpicture}.
\end{equation}
The multiplication is the renormalization of the surrounding ground state area by ``gluing" two such pro-tensors together and get back to itself. 
\begin{equation}
\label{eq.intuition_multiplication}
    \begin{tikzpicture}
        \node(A) at (-5.5,0){\begin{tikzpicture}[x=0.75pt,y=0.75pt,yscale=-0.7,xscale=0.7]

\draw    (101.51,162.2) .. controls (124,120) and (166,103) .. (201.51,162.2) ;
\draw [color=color3  ,draw opacity=1 ]   (44,163) -- (130.01,162.7) ;
\draw [color=color3  ,draw opacity=1 ]   (172.51,162.35) -- (264,162) ;
\draw    (68,163) .. controls (100,77) and (194,68) .. (233,163) ;
\end{tikzpicture}};
\node (B) at (0,0){\begin{tikzpicture}[x=0.75pt,y=0.75pt,yscale=-1,xscale=1]

\draw    (101.51,162.2) .. controls (124,120) and (166,103) .. (201.51,162.2) ;
\draw [color=color3  ,draw opacity=1 ]   (76,162) -- (130.01,162.7) ;
\draw [color=color3  ,draw opacity=1 ]   (174.26,162.37) -- (228.76,162.02) ;
\draw    (113.55,144.34) -- (188,144) ;

\end{tikzpicture}

};
\node (C) at (5.5,0){\begin{tikzpicture}[x=0.75pt,y=0.75pt,yscale=-1,xscale=1]

\draw    (206.51,165.2) .. controls (229,123) and (271,106) .. (306.51,165.2) ;
\draw [color=color3  ,draw opacity=1 ]   (177,166) -- (235.01,165.7) ;
\draw [color=color3  ,draw opacity=1 ]   (277.51,165.35) -- (335.52,165.04) ;
\end{tikzpicture}};
\draw[nattrans] (A)--(B)node[midway,above = 2pt]{$\alpha_\wr^{-1}\star\alpha_\wr^R$};
\draw[nattrans] (B)--(C) node [midway, above = 2pt]{$\epsilon_2$};
    \end{tikzpicture},
\end{equation}
The unit is given by growing a surrounding ground state area out of nothing
\begin{equation}
\label{eq.intuition_unit}
    \begin{tikzpicture}
        \node (A) at (-5.5,0){\begin{tikzpicture}[x=0.75pt,y=0.75pt,yscale=-1,xscale=1,baseline=-99]

\draw [color=color3  ,draw opacity=1 ]   (177,166) -- (235.01,165.7) ;
\draw [color=color3  ,draw opacity=1 ]   (277.51,165.35) -- (335.52,165.04) ;
\end{tikzpicture}};
\node (B) at (0,0){\begin{tikzpicture}[x=0.75pt,y=0.75pt,yscale=-1,xscale=1,baseline=(current bounding box.center)]

\draw   (206,166)--(236,136);
\draw [fill = white] (236,136) circle (4.5);
\draw    (306,166)--(276,136);
\draw [fill = white] (276,136) circle (4.5);
\draw [color=color3  ,draw opacity=1 ]   (177,166) -- (235.01,165.7) ;
\draw [color=color3  ,draw opacity=1 ]   (277.51,165.35) -- (335.52,165.04) ;
\end{tikzpicture}};
\node (C) at (5.5,0){\begin{tikzpicture}[x=0.75pt,y=0.75pt,yscale=-1,xscale=1,baseline=-145]

\draw    (206.51,165.2) .. controls (229,123) and (271,106) .. (306.51,165.2) ;
\draw [color=color3  ,draw opacity=1 ]   (177,166) -- (235.01,165.7) ;
\draw [color=color3  ,draw opacity=1 ]   (277.51,165.35) -- (335.52,165.04) ;
\end{tikzpicture}};
\draw[nattrans] (A)--(B) node [midway, above = 2pt]{$\lambda_\wr^{-1}\star\lambda_\wr^R$};
\draw[nattrans] (B)--(C) node [midway, above = 2pt]{$\epsilon_0$};
    \end{tikzpicture}.
\end{equation}

We will show in Appendix~\ref{sub.is_probimonad} the associativity and unitality, which makes~\eqref{eq.renormalization_alg} indeed an algebra.

If the shaded area is a particle-like fixed-point defect, it would be a module over such algebra, which is a pro-tensor from $\cM$ to $\cM$  with a certain module action
\begin{equation}
\label{eq.Intuition_Module_Action}
    \begin{tikzpicture}
        \node (A) at (-3,0){\begin{tikzpicture}[x=0.75pt,y=0.75pt,yscale=-1,xscale=1]

\draw [color=color3  ,draw opacity=1 ]   (179,165) -- (262.02,165.39) -- (333,165) ;
\draw    (206.51,165.2) .. controls (229,123) and (271,106) .. (306.51,165.2) ;
\draw  [fill=Mygray,fill opacity=1 ] (239,166.5) .. controls (239,157.39) and (246.39,150) .. (255.5,150) .. controls (264.61,150) and (272,157.39) .. (272,166.5) .. controls (272,175.61) and (264.61,183) .. (255.5,183) .. controls (246.39,183) and (239,175.61) .. (239,166.5) -- cycle ;

\end{tikzpicture}
};
\node(B) at (3,0){\begin{tikzpicture}
[x=0.75pt,y=0.75pt,yscale=-0.9,xscale=0.9,baseline = -3.2cm]

\draw [color=color3  ,draw opacity=1 ]   (180,165) -- (262.02,165.39) -- (322,165) ;
\draw  [fill=Mygray  ,fill opacity=1 ] (239,166.5) .. controls (239,157.39) and (246.39,150) .. (255.5,150) .. controls (264.61,150) and (272,157.39) .. (272,166.5) .. controls (272,175.61) and (264.61,183) .. (255.5,183) .. controls (246.39,183) and (239,175.61) .. (239,166.5) -- cycle ;
\end{tikzpicture}};
\draw[nattrans] (A)--(B) node [midway, above = 2pt]{$\kappa$};
    \end{tikzpicture},
\end{equation}
satisfying associativity
    \begin{equation}
\label{eq.intuition_asso}
    \begin{tikzpicture}
        \node(A) at (-5,0){\begin{tikzpicture}[x=0.75pt,y=0.75pt,yscale=-0.7,xscale=0.7]

\draw    (101.51,162.2) .. controls (124,120) and (166,103) .. (201.51,162.2) ;
\draw [color=color3  ,draw opacity=1 ]   (44,163) -- (130.01,162.7) ;
\draw [color=color3  ,draw opacity=1 ]   (172.51,162.35) -- (264,162) ;
\draw [fill = Mygray,fill opacity =1] (151.26,162.525) circle (21.25);

\draw    (68,163) .. controls (100,77) and (194,68) .. (233,163) ;
\end{tikzpicture}};
\node (B) at (0,0){\begin{tikzpicture}[x=0.75pt,y=0.75pt,yscale=-1,xscale=1]

\draw    (101.51,162.2) .. controls (124,120) and (166,103) .. (201.51,162.2) ;
\draw [color=color3  ,draw opacity=1 ]   (76,162) -- (138.01,162.7) ;
\draw [color=color3  ,draw opacity=1 ]   (166.26,162.37) -- (228.76,162.02) ;
\draw [fill = Mygray,fill opacity = 1] (152.635,162.535) circle (13.93);
\draw    (113.55,144.34) -- (188,144) ;

\end{tikzpicture}

};
\node (C) at (5,0){\begin{tikzpicture}[x=0.75pt,y=0.75pt,yscale=-1,xscale=1]

\draw    (206.51,165.2) .. controls (229,123) and (271,106) .. (306.51,165.2) ;
\draw [color=color3  ,draw opacity=1 ]   (177,166) -- (250.01,165.7) ;
\draw [color=color3  ,draw opacity=1 ]   (265.51,165.35) -- (335.52,165.04) ;
\draw [fill = Mygray,fill opacity = 1] (257.76,165.525) circle (13.93);
\end{tikzpicture}};
\node (D) at (-5,-4){\begin{tikzpicture}[x=0.75pt,y=0.75pt,yscale=-1,xscale=1]

\draw    (206.51,165.2) .. controls (229,123) and (271,106) .. (306.51,165.2) ;
\draw [color=color3  ,draw opacity=1 ]   (177,166) -- (250.01,165.7) ;
\draw [color=color3  ,draw opacity=1 ]   (265.51,165.35) -- (335.52,165.04) ;
\draw [fill = Mygray,fill opacity = 1] (257.76,165.525) circle (13.93);
\end{tikzpicture}};
\node (E) at (5,-4){\begin{tikzpicture}[x=0.75pt,y=0.75pt,yscale=-0.9,xscale=0.9,baseline = -3.2cm]

\draw [color=color3  ,draw opacity=1 ]   (180,165) -- (262.02,165.39) -- (322,165) ;
\draw  [fill=Mygray  ,fill opacity=1 ] (239,166.5) .. controls (239,157.39) and (246.39,150) .. (255.5,150) .. controls (264.61,150) and (272,157.39) .. (272,166.5) .. controls (272,175.61) and (264.61,183) .. (255.5,183) .. controls (246.39,183) and (239,175.61) .. (239,166.5) -- cycle ;
\end{tikzpicture}};
\draw[nattrans] (A)--(B)node[midway,above = 2pt]{$\alpha_\wr^{-1}\star\alpha_\wr^R$};
\draw[nattrans] (B)--(C) node [midway, above = 2pt]{$\epsilon_2$};
\draw[nattrans] (A)--(D) node [midway, right = 2pt]{$\kappa$};
\draw[nattrans] (D)--(E) node [midway, above = 2pt]{$\kappa$};
\draw[nattrans] (C)--(E) node [midway, left = 2pt]{$\kappa$};
    \end{tikzpicture},
\end{equation}
and unitality
\begin{equation}
\label{eq.intuition_uni}
      \begin{tikzpicture}
        \node (A) at (-4,0){\begin{tikzpicture}[x=0.75pt,y=0.75pt,yscale=-1,xscale=1,baseline=-99]
\draw [color=color3  ,draw opacity=1 ]   (177,166) -- (255.01,165.7) ;
\draw [color=color3  ,draw opacity=1 ]   (257.51,165.35) -- (335.52,165.04) ;
\draw [fill = Mygray,fill opacity =1] (256,165) circle (15);
\end{tikzpicture}};
\node (B) at (4,0){\begin{tikzpicture}[x=0.75pt,y=0.75pt,yscale=-1,xscale=1,baseline=(current bounding box.center)]

\draw   (206,166)--(236,136);
\draw [fill = white] (236,136) circle (4.5);
\draw    (306,166)--(276,136);
\draw [fill = white] (276,136) circle (4.5);
\draw [color=color3  ,draw opacity=1 ]   (177,166) -- (255.01,165.7) ;
\draw [color=color3  ,draw opacity=1 ]   (257.51,165.35) -- (335.52,165.04) ;
\draw [fill = Mygray,fill opacity =1] (256,165) circle (15);
\end{tikzpicture}};
\node (C) at (4,-3){\begin{tikzpicture}[x=0.75pt,y=0.75pt,yscale=-1,xscale=1,baseline=-145]

\draw    (206.51,165.2) .. controls (229,123) and (271,106) .. (306.51,165.2) ;
\draw [color=color3  ,draw opacity=1 ]   (177,166) -- (255.01,165.7) ;
\draw [color=color3  ,draw opacity=1 ]   (257.51,165.35) -- (335.52,165.04) ;
\draw [fill = Mygray,fill opacity =1] (256,165) circle (15);
\end{tikzpicture}};
\node (D) at (-4,-3){\begin{tikzpicture}[x=0.75pt,y=0.75pt,yscale=-1,xscale=1,baseline=-99]
\draw [color=color3  ,draw opacity=1 ]   (177,166) -- (255.01,165.7) ;
\draw [color=color3  ,draw opacity=1 ]   (257.51,165.35) -- (335.52,165.04) ;
\draw [fill = Mygray,fill opacity =1] (256,165) circle (15.25);
\end{tikzpicture}};
\draw[nattrans] (A)--(B) node [midway, above = 2pt]{$\lambda_\wr^{-1}\star\lambda_\wr^R$};
\draw[nattrans] (B)--(C) node [midway, left = 2pt]{$\epsilon_0$};
\draw[nattrans] (C)--(D) node [midway, above = 2pt]
{$\kappa$};
\draw[nattrans] (A) -- (D) node [midway, right = 2pt ]{$1$};
\end{tikzpicture}.
\end{equation}
Especially, the left hand side of~\eqref{eq.right_dual_of_alpha_module} and~\eqref{eq.right_dual_of_lambda_module} show that vacuum, i.e. the identity profunctor $1_\cM$ from $\cM$ to $\cM$, is an $\MnCM$-module.

For two $\MnCM$-modules, a morphism between them is  a module map, defined as a pro-tensor homomorphism between the underlying pro-tensors that intertwines the module actions. We denote the category of modules over $\cM\Omega_\cC\cM$ as $\Lmd_{\cM\Omega_\cC\cM}$.  $\Lmd_{\MnCM}$ is then the category of particle-like fixed-point defects on the $\cM$ boundary of a $\cC$ string-net pro-tensor network. Also, similar to the category of modules over promonad,
there is a canonical way to upgrade $\Lmd_{\cM\Omega_\cC\cM}$ to be a $\cV$-category denoted by $\bLmd_{\cM\Omega_\cC\cM}$ (see Appendix~\ref{app.enriched_structure}).


The ground state space in the presence of a boundary particle-like fixed-point defect $F\in \Lmd_{\cM\Omega_\cC\cM}$ is again the image under the evaluation map $\ev$.  In this setting, apart from transformations \eqref{eq.bulk_boundary_ev},   $\ev$ additionally incorporates  the module action $\kappa$ (see \eqref{eq.Intuition_Module_Action}). The equation
\begin{equation}
    \begin{tikzpicture}
        \node (A) at (0,0){\begin{tikzpicture}[x=0.75pt,y=0.75pt,yscale=-0.8,xscale=0.8]

\draw  [fill=white,fill opacity=1 ][line width=0.75]  (87,106) -- (113,63) -- (164.31,63) -- (191.69,106) -- cycle ;
\draw    (89,32) -- (113,63) ;
\draw    (164.31,63) -- (185,33) ;
\draw [color=color3  ,draw opacity=1 ][line width=1.25]    (38,106) -- (249,106) ;
\draw  [fill=Mygray,fill opacity=1 ] (139,102) circle (20); 

\draw (130,94.4) node [anchor=north west][inner sep=0.75pt]    {$F$};

\end{tikzpicture}};
\node (B) at (6,0) {\begin{tikzpicture}[x=0.75pt,y=0.75pt,yscale=-0.8,xscale=0.8]

\draw    (67,95) -- (91,126) ;
\draw    (231.31,125) -- (240.51,111.66) -- (252,95) ;
\draw [color=color3, draw opacity=1 ][line width=1.25]    (58,126) -- (269,126) ;
\draw  [fill=Mygray,fill opacity=1 ] (159,122) circle (20);

\draw (150,114.4) node [anchor=north west][inner sep=0.75pt]    {$F$};
\end{tikzpicture}};
\draw[nattrans] (A)--(B) node [midway, above = 4pt]{$\ev$};
    \end{tikzpicture}
\end{equation}
implements the projection of the total Hilbert space (in the presence of a boundary particle 
$F$
) onto the corresponding ground state subspace carrying $F$.

Consider inserting an arbitrary  profunctor $G:\cM\nrightarrow \cM$ (not necessarily a module over $\cM\Omega_\cC \cM$) into the boundary as a defect,  we may say surrounding $G$ by a ground-state region and then doing evaluation as a step of \emph{renormalization} 
\begin{equation}
    \begin{tikzpicture}
        \node (A) at (0,0){\begin{tikzpicture}[x=0.75pt,y=0.75pt,yscale=-0.8,xscale=0.8]

\draw  [fill=white,fill opacity=1 ][line width=0.75]  (87,106) -- (113,63) -- (164.31,63) -- (191.69,106) -- cycle ;
\draw    (89,32) -- (113,63) ;
\draw    (164.31,63) -- (185,33) ;
\draw [color=color3  ,draw opacity=1 ][line width=1.25]    (38,106) -- (249,106) ;
\draw  [fill=Mygray,fill opacity=1 ] (139,102) circle (18);

\draw (130,94.4) node [anchor=north west][inner sep=0.75pt]    {$G$};

\end{tikzpicture}};
\node (B) at (6,0) {\begin{tikzpicture}[x=0.75pt,y=0.75pt,yscale=-0.8,xscale=0.8]

\draw    (67,95) -- (91,126) ;
\draw    (231.31,125) -- (240.51,111.66) -- (252,95) ;
\draw [color=color3  ,draw opacity=1 ][line width=1.25]    (58,126) -- (269,126) ;
\draw  [fill=Mygray,fill opacity=1 ] (158,122) circle (18); 
\draw    (106,125) .. controls (127,80) and (177,62) .. (211,125) ;

\draw (150,114.4) node [anchor=north west][inner sep=0.75pt]    {$G$};
\end{tikzpicture}};
\draw[nattrans] (A)--(B) node [midway, above = 4pt]{$\ev$};
    \end{tikzpicture}.
\end{equation}
This is analogous to the usual real-space
renormalization procedure, which consists of coarse-graining followed by
rescaling. Note that while $G$ itself may not be a module, the composite
\begin{equation}
    (\boxdot_{\cM})_\ast\bullet (\Id_\cC\os G) \bullet (\boxdot_{\cM})^\ast=\begin{tikzpicture}[x=0.75pt,y=0.75pt,yscale=-1,xscale=1, baseline=-84.75]

\draw [color=color3  ,draw opacity=1 ][line width=1.25]    (85,125) -- (229,125) ;
\draw  [fill=Mygray,fill opacity=1 ] (159,122) circle (16); 
\draw    (106,125) .. controls (127,80) and (177,62) .. (211,125) ;

\draw (151,116.4) node [anchor=north west][inner sep=0.75pt]    {$G$};

\end{tikzpicture}
\label{eq.freemodule}
\end{equation}
is indeed a  free  $\cM\Omega_\cC\cM$ module with action  

\begin{equation}
    \begin{tikzpicture}
        \node (A) at (0,0){\begin{tikzpicture}[x=0.75pt,y=0.75pt,yscale=-0.9,xscale=0.9]

\draw [color=color3  ,draw opacity=1 ][line width=1.25]    (55,125) -- (270,126) ;
\draw  [fill=Mygray,fill opacity=1 ] (159,122) circle (18); 
\draw    (106,125) .. controls (133,70) and (177,62) .. (211,125) ;
\draw    (76,125) .. controls (100,35) and (211,35) .. (238,125) ;

\draw (151,113.4) node [anchor=north west][inner sep=0.75pt]    {$G$};

\end{tikzpicture}
};

\node (B) at (6,0) {\begin{tikzpicture}[x=0.75pt,y=0.75pt,yscale=-1,xscale=1, baseline=-84.75]

\draw [color=color3  ,draw opacity=1 ][line width=1.25]    (85,125) -- (229,125) ;
\draw  [fill=Mygray,fill opacity=1 ] (159,122) circle (16); 
\draw    (106,125) .. controls (127,80) and (177,62) .. (211,125) ;

\draw (151,116.4) node [anchor=north west][inner sep=0.75pt]    {$G$};

\end{tikzpicture}};
\draw[nattrans] (A)--(B) node [midway, above = 4pt]{$\epsilon_2\cdot(\alpha_\wr^{-1}\star\alpha_\wr^R)$};
    \end{tikzpicture},
\end{equation}

and hence a particle-like fixed defect. 
Therefore, inserting an arbitrary defect profunctor $G$ is physically
equivalent to inserting the
particle-like fixed-point
defect~\eqref{eq.freemodule}. In this sense, the defect $G$ flows to the fixed point~\eqref{eq.freemodule} under one step of renormalization. We may likewise immediately generalize this picture  to  defects on the junction of two boundaries $\cM$ and $\cN$, by considering modules over
\begin{equation}
\cM\Omega_\cC \cN=:\label{eq.MNrenormalization_alg}
    \begin{tikzpicture}[x=0.75pt,y=0.75pt,yscale=-1,xscale=1, baseline=-117 ]

\draw    (206.51,165.2) .. controls (229,123) and (271,106) .. (306.51,165.2) ;
\draw [color=color4  ,draw opacity=1 ]   (177,166) -- (235.01,165.7) ;
\draw [color=color3  ,draw opacity=1 ]   (277.51,165.35) -- (335.52,165.04) ;
\draw node at (185,175){$\cN$};
\draw node at (325,175){$\cM$};
\end{tikzpicture}.
\end{equation}
Our framework implements an intrinsic and natural mechanism such that an arbitrary pro-tensor insertion automatically becomes a particle-like fixed-point defect in a string-net pro-tensor network.

Note that the category $\Lmd_{\cM\Omega\cM}$ is a monoidal category, where the tensor product of two modules over $\cM\Omega_\cC \cM$  exactly corresponds to the contraction of the underlying pro-tensors. Inspired by \cite[Section 3]{Kitaev_Kong_2012}, for two modules   $(F_1,\kappa_1)$ and $(F_2,\kappa_2)$, we define a canonical module structure $\kappa_{1,2}$ on $F_1\bullet F_2$
\begin{equation}
    \label{eq.tensor_product}
    \begin{tikzpicture}
        \node (A) at (-5.5,0){\begin{tikzpicture}[x=0.75pt,y=0.75pt,yscale=-0.7,xscale=0.7,baseline = -32]

\draw [color={rgb, 255:red, 245; green, 166; blue, 35 }  ,draw opacity=1 ]   (100,128) -- (331,126) ;
\draw  [fill={rgb, 255:red, 155; green, 155; blue, 155 }  ,fill opacity=1 ] (148.06,127.62) .. controls (148.3,116.03) and (157.89,106.83) .. (169.48,107.07) .. controls (181.07,107.31) and (190.27,116.9) .. (190.03,128.5) .. controls (189.79,140.09) and (180.2,149.29) .. (168.61,149.05) .. controls (157.02,148.81) and (147.81,139.22) .. (148.06,127.62) -- cycle ;
\draw  [fill={rgb, 255:red, 155; green, 155; blue, 155 }  ,fill opacity=1 ] (217,127) .. controls (217,115.95) and (225.95,107) .. (237,107) .. controls (248.05,107) and (257,115.95) .. (257,127) .. controls (257,138.05) and (248.05,147) .. (237,147) .. controls (225.95,147) and (217,138.05) .. (217,127) -- cycle ;
\draw    (124.5,127.5) .. controls (167,33) and (263,50) .. (292,127) ;

\draw (158,120.4) node [anchor=north west][inner sep=0.75pt]    {$F_{1}$};
\draw (225,118.4) node [anchor=north west][inner sep=0.75pt]    {$F_{2}$};

\end{tikzpicture}};
\node (B) at (0,0)
{\begin{tikzpicture}[x=0.75pt,y=0.75pt,yscale=-0.7,xscale=0.7, baseline = -32]

\draw    (137,148) .. controls (151.5,110.5) and (187,92) .. (216,148) ;
\draw    (235.5,147) .. controls (254,110.5) and (283,92) .. (314.5,147) ;
\draw [color={rgb, 255:red, 245; green, 166; blue, 35 }  ,draw opacity=1 ]   (120,148) -- (227,147.07) -- (351,146) ;
\draw  [fill={rgb, 255:red, 155; green, 155; blue, 155 }  ,fill opacity=1 ] (256,146) .. controls (256,134.95) and (264.95,126) .. (276,126) .. controls (287.05,126) and (296,134.95) .. (296,146) .. controls (296,157.05) and (287.05,166) .. (276,166) .. controls (264.95,166) and (256,157.05) .. (256,146) -- cycle ;
\draw  [fill={rgb, 255:red, 155; green, 155; blue, 155 }  ,fill opacity=1 ] (156.06,146.62) .. controls (156.3,135.03) and (165.89,125.83) .. (177.48,126.07) .. controls (189.07,126.31) and (198.27,135.9) .. (198.03,147.5) .. controls (197.79,159.09) and (188.2,168.29) .. (176.61,168.05) .. controls (165.02,167.81) and (155.81,158.22) .. (156.06,146.62) -- cycle ;

\draw (166,137.4) node [anchor=north west][inner sep=0.75pt]    {$F_{1}$};
\draw (265,136.4) node [anchor=north west][inner sep=0.75pt]    {$F_{2}$};

\end{tikzpicture}
};
\node (C) at (5.5,0){\begin{tikzpicture}[x=0.75pt,y=0.75pt,yscale=-0.7,xscale=0.7, baseline = -52]

\draw [color={rgb, 255:red, 245; green, 166; blue, 35 }  ,draw opacity=1 ]   (120,148) -- (227,147.07) -- (351,146) ;
\draw  [fill={rgb, 255:red, 155; green, 155; blue, 155 }  ,fill opacity=1 ] (256,146) .. controls (256,134.95) and (264.95,126) .. (276,126) .. controls (287.05,126) and (296,134.95) .. (296,146) .. controls (296,157.05) and (287.05,166) .. (276,166) .. controls (264.95,166) and (256,157.05) .. (256,146) -- cycle ;
\draw  [fill={rgb, 255:red, 155; green, 155; blue, 155 }  ,fill opacity=1 ] (156.06,146.62) .. controls (156.3,135.03) and (165.89,125.83) .. (177.48,126.07) .. controls (189.07,126.31) and (198.27,135.9) .. (198.03,147.5) .. controls (197.79,159.09) and (188.2,168.29) .. (176.61,168.05) .. controls (165.02,167.81) and (155.81,158.22) .. (156.06,146.62) -- cycle ;

\draw (166,137.4) node [anchor=north west][inner sep=0.75pt]    {$F_{1}$};
\draw (265,136.4) node [anchor=north west][inner sep=0.75pt]    {$F_{2}$};

\end{tikzpicture}};
\draw[nattrans] (A) -- (B) node [midway, above = 2pt]{$\eta_\wr$};
\draw[nattrans] (B) -- (C) node [midway, above = 2pt]{$\kappa_1\star \kappa_2$};
\end{tikzpicture}.
    \end{equation}
    This tensor product is denoted as $\conv$, and then
    \begin{equation}
    \label{eq.tensor_of_Omega_module}
        (F_1,\kappa_1)\conv(F_2,\kappa_2):= (F_1\bullet F_2, \kappa_{1,2}).
    \end{equation}
We elaborate on the monoidal structure in greater detail in Appendix~\ref{sub.is_probimonad}, in which~\eqref{eq.omega_bimonad1} and~\eqref{eq.omega_bimonad2} are associativity and unitality of the module action~\eqref{eq.tensor_product}, respectively;~\eqref{eq.omega_bimonad_ass} and~\eqref{eq.omega_bimonad_unit} are the associativity and unitality of the tensor product $\conv$, respectively.  

We now consider the fusion of boundary particle-like fixed-point defects. 

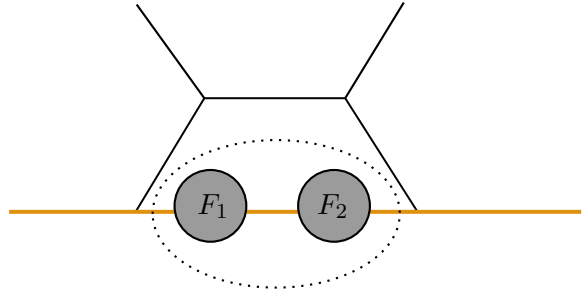
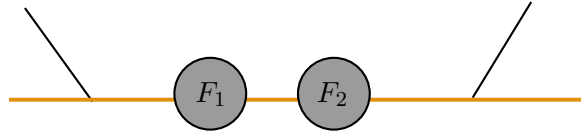
\begin{figure}
    \centering
    \begin{subfigure}[b]{0.8\textwidth}
    \centering
        \begin{tikzpicture}[x=0.75pt,y=0.75pt,yscale=-1,xscale=1]
\draw  [fill=white,fill opacity=1 ][line width=0.75]  (73,126) -- (99,83) -- (150.31,83) -- (177.69,126) -- cycle ;
\draw    (75,52) -- (99,83) ;
\draw    (150.31,83) -- (171,53) ;
\draw [color=color3  ,draw opacity=1 ][line width=1.5]    (24,126) -- (235,126) ;
\draw  [fill=Mygray,fill opacity=1 ] (125,122) circle  (18);
\draw  [fill=white,fill opacity=1 ][line width=0.75]  (373,126) -- (399,83) -- (450.31,83) -- (477.69,126) -- cycle ;
\draw    (375,52) -- (399,83) ;
\draw    (450.31,83) -- (471,53) ;
\draw [color=color3  ,draw opacity=1 ][line width=1.5]    (324,126) -- (535,126) ;
\draw  [fill=Mygray,fill opacity=1 ] (425,122) circle (18); 
\draw [color=color3  ,draw opacity=1 ] [dash pattern={on 4.5pt off 4.5pt}]  (235,126) -- (324,126) ;

\draw (117,114.4) node [anchor=north west][inner sep=0.75pt]    {$F_{1}$};
\draw (417,114.4) node [anchor=north west][inner sep=0.75pt]    {$F_{2}$};
\draw (224,71.4) node [anchor=north west][inner sep=0.75pt]    {$...$};
\draw (304,71.4) node [anchor=north west][inner sep=0.75pt]    {$...$};

\end{tikzpicture}
 \caption{Two particle-like fixed-point defects $F_1$ and $F_2$ on the boundary $\cM$}
    \label{fig.two_bdy_particle}
    \end{subfigure}
 \vspace{1cm}

\begin{subfigure}[b]{0.8\textwidth}
\centering
    \begin{tikzpicture}[x=0.75pt,y=0.75pt,yscale=-1,xscale=1]

\draw  [fill=white,fill opacity=1 ][line width=0.75]  (107.46,145) -- (141.93,88) -- (212.62,88) -- (248.92,145) -- cycle ;
\draw [color=color3  ,draw opacity=1 ][line width=1.5]    (44,145) -- (331,145) ;
\draw  [fill=Mygray,fill opacity=1 ] (145,142) circle (18); 
\draw  [fill=Mygray,fill opacity=1 ] (207,142) circle (18); 
\draw    (108,41) -- (141.93,88) ;
\draw    (242,40) -- (212.62,88) ;
\draw  [dash pattern={on 0.84pt off 2.51pt}] (116,146) .. controls (116,125.57) and (143.76,109) .. (178,109) .. controls (212.24,109) and (240,125.57) .. (240,146) .. controls (240,166.43) and (212.24,183) .. (178,183) .. controls (143.76,183) and (116,166.43) .. (116,146) -- cycle ;

\draw (137,134.4) node [anchor=north west][inner sep=0.75pt]    {$F_{1}$};
\draw (197,134.4) node [anchor=north west][inner sep=0.75pt]    {$F_{2}$};
\end{tikzpicture}
    \caption{A single particle-like fixed-point defect $F_1\bullet F_2$ on the boundary $\cM$}
    \label{fig.fused_particle}
    \end{subfigure}
 \vspace{1cm}
    
    \begin{subfigure}[b]{0.8\textwidth}
    \centering
    \begin{tikzpicture}[x=0.75pt,y=0.75pt,yscale=-1,xscale=1]

\draw    (72,119) -- (95.07,150.95) -- (105.93,166) ;
\draw    (326,116) -- (308.35,144.83) -- (296.62,164) ;
\draw [color=color3  ,draw opacity=1 ][line width=1.5]    (64,165) -- (351,165) ;
\draw  [fill=Mygray,fill opacity=1 ] (227,162) circle (18); 
\draw  [fill=Mygray,fill opacity=1 ] (165,162) circle (18);

\draw (156,154.4) node [anchor=north west][inner sep=0.75pt]    {$F_{1}$};
\draw (217,154.4) node [anchor=north west][inner sep=0.75pt]    {$F_{2}$};

\end{tikzpicture}
\caption{Ground state subspace of both \ref{fig.two_bdy_particle} and \ref{fig.fused_particle}}
\label{fig.gss_two_particles}
    \end{subfigure}
    \caption{Fusion of two particles on the boundary $\cM$}
    \label{fig.fusion_of_bdy_particles}
\end{figure}

We study two distinct boundary configurations: one where two fixed-point defects 
$F_1$
 and 
$F_2$ are inserted on the boundary (Figure~\ref{fig.two_bdy_particle}), and the other where a single particle $F_1\bullet F_2$
is inserted (Figure~\ref{fig.fused_particle}). After applying the evaluation map, the ground state subspace of both configurations is identical, as shown in Figure~\ref{fig.gss_two_particles}. This observation manifests that the fusion of two boundary particles corresponds precisely to the composition of underlying profunctors, which is the monoidal structure of $\Lmd_{\cM\Omega_\cC\cM}$. Moreover, the fact that the ground state subspace remains the same whether the boundary fixed-point defects
$F_1$ and $F_2$
 are placed close together or far apart implies that separating these particles incurs no energy cost, revealing the topological nature of them.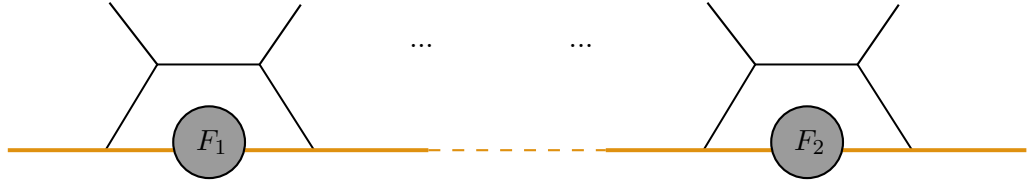

In Figure~\ref{fig.boudary_defect}, the boundary defects are depicted without any attached bulk strings.  More generally, one may also allow bulk strings to end on boundary defects microscopically. For instance, 
Figure~\ref{fig:1-bulk_string_bdy_defect} shows a boundary defect with a single bulk string attached on it. In this case, the corresponding renormalization algebra is given by
\begin{equation}
\label{eq.1_bulk_string_monad}
\begin{tikzpicture}[x=0.75pt,y=0.75pt,yscale=-1,xscale=1]

\draw    (101.51,162.2) .. controls (124,120) and (166,103) .. (201.51,162.2) ;
\draw [color={rgb, 255:red, 245; green, 166; blue, 35 }  ,draw opacity=1 ]   (76,162) -- (130.01,162.7) ;
\draw [color={rgb, 255:red, 245; green, 166; blue, 35 }  ,draw opacity=1 ]   (165.38,162.7) -- (219.88,162.35) ;
\draw    (158,124) -- (177,100) ;
\draw    (176,133) -- (161,152) ;
\end{tikzpicture}
\end{equation}
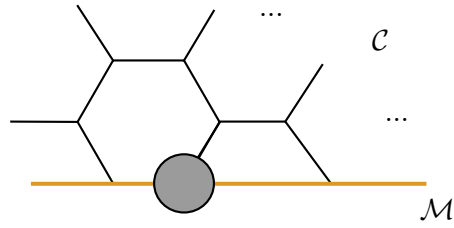
\begin{figure}
    \centering
    \begin{tikzpicture}[x=0.75pt,y=0.75pt,yscale=-0.8,xscale=0.8]

\draw   (257.25,135.21) -- (235,173.75) -- (190.5,173.75) -- (168.25,135.21) -- (190.5,96.67) -- (235,96.67) -- cycle ;
\draw   (233.5,174) -- (257.67,135) -- (298.51,135) -- (327.3,174) -- cycle ;
\draw    (126,135) -- (168.25,135.21) ;
\draw [color={rgb, 255:red, 221; green, 152; blue, 34 }  ,draw opacity=1 ][line width=1.5]    (139,174) -- (308,174) -- (387,174) ;
\draw  [fill={rgb, 255:red, 155; green, 155; blue, 155 }  ,fill opacity=1 ] (216.25,173.75) .. controls (216.25,163.67) and (224.64,155.5) .. (235,155.5) .. controls (245.36,155.5) and (253.75,163.67) .. (253.75,173.75) .. controls (253.75,183.83) and (245.36,192) .. (235,192) .. controls (224.64,192) and (216.25,183.83) .. (216.25,173.75) -- cycle ;
\draw    (168,62) -- (190.5,96.67) ;
\draw    (235,96.67) -- (254,65) ;
\draw    (298.51,135) -- (322,99) ;

\draw (280.27,65.04) node [anchor=north west][inner sep=0.75pt]  [rotate=-0.38] [align=left] {...};
\draw (359,131) node [anchor=north west][inner sep=0.75pt]   [align=left] {...};
\draw (351,76) node [anchor=north west][inner sep=0.75pt]   [align=left] {$\cC$};
\draw (381,182) node [anchor=north west][inner sep=0.75pt]   [align=left] {$\cM$};
\end{tikzpicture}
    \caption{A particle-like defect on the boundary $\cM$ of a $\cC$ string-net pro-tensor network, with a bulk string attached on it.}
    \label{fig:1-bulk_string_bdy_defect}
\end{figure}
with multiplication 
\begin{equation}
\label{eq.1_bulk_string_multi}
    \begin{tikzpicture}
        \node (A) at (0,0){\begin{tikzpicture}[x=0.75pt,y=0.75pt,yscale=-0.6,xscale=0.6]

\draw    (101.51,162.2) .. controls (124,120) and (166,103) .. (201.51,162.2) ;
\draw [color={rgb, 255:red, 245; green, 166; blue, 35 }  ,draw opacity=1 ]   (38,162) -- (130.01,162.7) ;
\draw [color={rgb, 255:red, 245; green, 166; blue, 35 }  ,draw opacity=1 ]   (165.38,162.7) -- (281,163) ;
\draw    (158,124) -- (177,100) ;
\draw    (176,133) -- (161,152) ;
\draw    (69.51,163.2) .. controls (103.65,94.82) and (164.63,63.45) .. (220.04,136.97) .. controls (225.76,144.57) and (231.43,153.28) .. (237,163.2) ;
\draw    (158.12,93.91) -- (179,63) ;
\end{tikzpicture}
};
\node (B) at (5,0){\begin{tikzpicture}[x=0.75pt,y=0.75pt,yscale=-0.6,xscale=0.6]

\draw [color={rgb, 255:red, 245; green, 166; blue, 35 }  ,draw opacity=1 ]   (38,162) -- (130.01,162.7) ;
\draw [color={rgb, 255:red, 245; green, 166; blue, 35 }  ,draw opacity=1 ]   (165.38,162.7) -- (281,163) ;
\draw    (69.51,163.2) .. controls (103.65,94.82) and (164.63,63.45) .. (220.04,136.97) .. controls (225.76,144.57) and (231.43,153.28) .. (237,163.2) ;
\draw    (209.12,122.91) -- (219.29,107.85) -- (230,92) ;
\draw  [fill={rgb, 255:red, 255; green, 255; blue, 255 }  ,fill opacity=1 ] (106.92,103.84) -- (127.29,100.98) -- (118.58,119.61) -- (98.21,122.47) -- cycle ;
\draw  [fill={rgb, 255:red, 255; green, 255; blue, 255 }  ,fill opacity=1 ] (179.03,91.42) -- (192.03,107.92) -- (171.44,112.09) -- (158.44,95.59) -- cycle ;
\draw    (127.12,125.91) -- (137.29,110.85) -- (148,95) ;
\end{tikzpicture}
};
\node (C) at (10,0){\begin{tikzpicture}[x=0.75pt,y=0.75pt,yscale=-1,xscale=1]

\draw    (101.51,162.2) .. controls (124,120) and (166,103) .. (201.51,162.2) ;
\draw [color={rgb, 255:red, 245; green, 166; blue, 35 }  ,draw opacity=1 ]   (76,162) -- (130.01,162.7) ;
\draw [color={rgb, 255:red, 245; green, 166; blue, 35 }  ,draw opacity=1 ]   (165.38,162.7) -- (219.88,162.35) ;
\draw    (158,124) -- (177,100) ;
\draw    (176,133) -- (161,152) ;
\end{tikzpicture}};
\draw[nattrans] (A) -- (B)node
[midway, above = 2pt, align=center] {Frobenius structure\\ and associativity};
\draw[nattrans] (B) -- (C) node [midway, above = 4pt, align=center]{$\epsilon_2\star \epsilon_2$ \\ and rigidity};
    \end{tikzpicture}.
\end{equation}
A particle-like fixed defect of this type is therefore described by a module over the algebra~\eqref{eq.1_bulk_string_monad}. Nevertheless, we have checked\footnote{The proof is omitted from the present paper and will be provided in a forthcoming work.} that the renormalization algebra~\eqref{eq.1_bulk_string_monad} is actually Morita equivalent to $\MnCM$. Consequently, the corresponding categories of modules are  equivalent. Furthermore, we believe that the renormalization algebra obtained by attaching any number of bulk strings are all Morita equivalent to one another~\cite{Kong_2012,Lan_2014}, and in particular Morita equivalent to $\MnCM$,
\begin{equation}
    \begin{tikzpicture}[x=0.75pt,y=0.75pt,yscale=-0.7,xscale=0.7, baseline=(current bounding box.center)  ]

\draw [color={rgb, 255:red, 245; green, 166; blue, 35 }  ,draw opacity=1 ]   (38,162) -- (130.01,162.7) ;
\draw [color={rgb, 255:red, 245; green, 166; blue, 35 }  ,draw opacity=1 ]   (165.38,162.7) -- (281,163) ;
\draw    (69.51,163.2) .. controls (103.65,94.82) and (164.63,63.45) .. (220.04,136.97) .. controls (225.76,144.57) and (231.43,153.28) .. (237,163.2) ;
\draw    (209.12,122.91) -- (219.29,107.85) -- (230,92) ;
\draw    (127.12,125.91) -- (137.29,110.85) -- (148,95) ;
\draw    (167.12,136.91) -- (177.29,121.85) -- (188,106) ;
\draw    (174.12,98.91) -- (184.29,83.85) -- (195,68) ;

\draw (151,110) node [anchor=north west][inner sep=0.75pt]   [align=left] {...};
\draw (193,95) node [anchor=north west][inner sep=0.75pt]   [align=left] {...};
\end{tikzpicture}\simeq_{\mathrm{Morita}}
 \begin{tikzpicture}[x=0.75pt,y=0.75pt,yscale=-1,xscale=1,baseline=(current bounding box.center)]

\draw    (206.51,165.2) .. controls (229,123) and (271,106) .. (306.51,165.2) ;
\draw [color=color3  ,draw opacity=1 ]   (177,166) -- (235.01,165.7) ;
\draw [color=color3  ,draw opacity=1 ]   (277.51,165.35) -- (335.52,165.04) ;
\end{tikzpicture}.
\end{equation}
 Hence the choice of representative renormalization algebra is an immaterial microscopic detail, any of these Morita equivalent algebras yields the equivalent category of particle-like fixed-point defects on the boundary.

\begin{remark}
The rigidity of the input category $\cC$ for a string-net pro-tensor network serves 
as a physical assumption that underlies the spatial ``homogeneity'' of the system.  For example, as shown in Figure~\ref{fig.string_net}, bulk strings may occur in three different orientations: one horizontal and two inclined. Inserting a defect on links of these three types therefore leads to three distinct local configurations. When $\cC$ is rigid, the invertibility of $\alpha^\sharp$ and $\alpha^\flat$ allow one to rearrange the links appropriately, so that all three configurations give rise to the same renormalization algebra, $\cC\Omega_{\cC\ot \cC^\rev}\cC$. Without rigidity, this argument break down: certain links fail to match properly, and defects on the three types of links may no longer be treated on an equal footing.
The rigidity of $\cC$ also ensures 
that the algebra multiplication \eqref{eq.1_bulk_string_multi}, as well as its 
multi-bulk-string generalizations, is well defined, thereby endowing the theory 
with macroscopic robustness.
In addition, in the discussion of topological holography in Section~\ref{sec.TubeHolo}, 
our physical interpretation of the SymTFT inherently requires the input category 
$\cC$ of symmetry charges to be rigid. Nevertheless, from a purely mathematical 
standpoint, our theorems concerning the representation of $\MnCN$ in 
Section~\ref{sec.main_thm} do not require $\cC$ to be rigid.
\end{remark}


\section{
The generalized Kitaev-Kong theorem
}\label{sec.main_thm}
In Section \ref{sec.particle_is_module}, we have observed that a particle-like defect living at the $\cM$-$\cN$-junction of a $\cC$ string-net pro-tensor network is a  module over $\cM\Omega_\cC \cN$.
Although this description is both physically illuminating and mathematically rigorous, we will show that the category of defects also admits two alternative mathematical characterizations, both offering conceptual insight and/or computational convenience\footnote{The computational convenience of the point of view in \ref{item1.the_five} is discussed in Section~\ref{sub.recovery}. However, we do not know whether the second characterization is computationally convenient, i.e., if there are efficient methods of computing lax module profunctors independent from \ref{item1.the_five}.}:
\begin{enumerate}
    \item \label{item1.the_five} as the category of left modules over the promonad $\MCN$ which will be defined  in Section \ref{sub.MCN}.
    \item \label{item2.the_five} as the category of lax $\cC$-module profunctors $\cM\arprof\cN$.
\end{enumerate}

In particular, \six{we will use the graphical calculus in pro-tensor network to prove} the categories in \ref{item1.the_five} and \ref{item2.the_five} are themselves equivalent. This equivalence is first proved in a more restrictive case and a different language by Kitaev and Kong \cite{Kitaev_Kong_2012}, where they prove in order to study topological defects in the ordinary Levin-Wen model. Our results vastly generalize Kitaev-Kong's theorem by disposing the requirements that $\cC,\cM,\cN$ be finite semsimple and $\Vect$-enriched, and that $\cC$ be rigid.

Moreover, using the same method, one can prove that there are yet two other equivalent characterizations of the category of defects (Remark \ref{rmk.Tambara}):
\begin{enumerate}
    \setcounter{enumi}{2}
    \item \label{item3.the_five} as the category of left Tambara modules $\cM\arprof\cN$.
    \item \label{item4.the_five} as the category of left oplax $\cC$-comodule profunctors $\cM\arprof\cN$.
\end{enumerate}
In mathematical terms, the five categories (the category of $\MnCN$-modules
 and the categories in \ref{item1.the_five}, \ref{item2.the_five}, \ref{item3.the_five} and \ref{item4.the_five} above) are canonically equivalent. The reasoning we based on is Section \ref{sub.duality} which is not sophisticated from a mathematical point of view, however, we have not seen literature pointing out the equivalence of these five structures even in the case $\cV=\set$ or $\cV=\vect$.

In \ref{sub.MCN} and \ref{sub.module_prof}, we introduce the promonad $\MCN$ and the concept of module profunctors, respectively. In \ref{sub.proof}, we provide the above mutually equivalent characterizations. In \ref{sub.recovery}, we show how to recover Kitaev-Kong's theorem from our results. 
\subsection{The promonad $\MCN$}
\label{sub.MCN}


Let $\cV$ be a cosmos. Let $\cC=(\cC,\boxtimes,I)$ be a monoidal $\cV$-category, and $\cM,\cN$ be left $\cC$-modules, with $\cC$-actions $\boxdot_\cM:\cC\times \cM\rightarrow \cM$ and $\boxdot_\cN:\cC\times \cN\rightarrow \cN$ respectively. 
We denote the structure pro-tensor network maps associated to $\cC$, which are introduced in Section \ref{sub.mon_VcatVModule}, by $(\epsilon_2,\eta_2,\epsilon_0,\eta_0,\alpha,\lambda,\rho,\alpha^R,\lambda^R,\rho^R)$. For a left $\cC$-module $i\in\{\cM,\cN\}$, we use $(\epsilon_{\wr,i},\eta_{\wr,i},\alpha_{\wr,i},\lambda_{\wr,i},\alpha_{\wr,i}^R,\lambda_{\wr,i}^R)$ to denote the structure pro-tensor network maps associated to $i$.

Our graphical notation also agrees with Section \ref{sub.mon_VcatVModule} (and hence also Section \ref{sec.LW}). In particular, the pro-tensors $(\boxdot_\cM)_*,$ $ (\boxdot_\cM)^*,$ $ (\boxdot_\cN)_*,$ $ (\boxdot_\cN)^*$ induced by the module actions are denoted by \(
    \ctikz{[scale=0.4]
        \draw[color3] (-1,0)--(0,0);
        \draw[color3] (0,0)--(1,0);
        \draw  (0,0)--(1,0.7);
    }
\),\(
    \ctikz{[scale=0.4,xscale=-1]
            \draw[color3] (-1,0)--(0,0);
            \draw[color3] (0,0)--(1,0);
            \draw  (0,0)--(1,0.7);
    } \), \(
    \ctikz{[scale=0.4]
        \draw[color4] (-1,0)--(0,0);
        \draw[color4] (0,0)--(1,0);
        \draw  (0,0)--(1,0.7);
    }
\),\(
    \ctikz{[scale=0.4,xscale=-1]
            \draw[color4] (-1,0)--(0,0);
            \draw[color4] (0,0)--(1,0);
            \draw  (0,0)--(1,0.7);
    } \), respectively.

We define the following pro-tensor:
 \begin{equation}
 \label{eq.MCN}
     \cM\bbH_\cC\cN \defdtobe
     \begin{tikzpicture}[scale = 0.7 ,baseline=(current bounding box.center)]
         \draw[color3,thick] (0.5,1.5)--(2,1.5);
         
         \draw[color3,thick] (.9,0.5)--(2.,0.5);
         \draw[color3,thick] (.9,-0.5)--(2.,-0.5);
         
         \draw[color3,thick] (2.,0.5) arc (-90:90:0.5cm);
         \draw[color3,thick] (.9,0.5) arc (90:270:0.5cm);
         \draw[color4,thick] (-1.5,-1)--(2.5,-1); 
    
         \draw[black,thick] (-0.7,-1) .. controls (0,1) .. (1,0.5);
         \node  at (-0.3,0.8) {$\cC$};
         \node  at (2.5,-0.2) {$\cM^\op$};
         \node at  (0.5,1.8){$\cM^\op$};
         \node  at (-1.5,-1.3) {$\cN$};
         \node  at (2.5,-1.3) {$
         \cN$};
     \end{tikzpicture}
     \:
     \cM^\op\ot\cN\nrightarrow\cM^\op\ot \cN. 
 \end{equation}

Algebraically, one can check that its effect on objects are given by
\[
     \cM\bbH_\cC\cN((m_1,n_1),(m_2,n_2))=\int^{x\in \cC}\cN(n_1,x\boxdot_\cN n_2)\ot\cM(x\boxdot_\cM m_2,m_1)
\]
 for $m_1,m_2\in\cM,n_1,n_2\in\cN$.
 

\begin{remark}
     Note that, by bending the $\cM^\op$ legs in \eqref{eq.MCN} to $\cM$, one exactly obtains the renormalization algebra $\cM\Omega_\cC \cN$ in~\eqref{eq.MNrenormalization_alg}. The similarity comes as no surprise, as both the renormalization algebra $\MnCN$ and the promonad $\MCN$ are motivated by constructions in \cite{Kitaev_Kong_2012}. In fact, if we view~\eqref{eq.MCN} as an H-shape graph
 
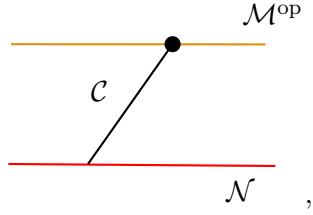
\begin{equation*}
     \begin{tikzpicture}[x=0.75pt,y=0.75pt,yscale=-1,xscale=1]
\draw [color = color4  ,draw opacity=1 ]   (166,166) -- (300,167) ;
\draw [color=color3,draw opacity=1 ]   (167.42,105.9) -- (295,106) ;
\draw    (206,166) -- (248.42,105.9) ;
\draw  [fill=black,fill opacity=1 ] (244.92,105.9) .. controls (244.92,103.96) and (246.48,102.4) .. (248.42,102.4) .. controls (250.35,102.4) and (251.92,103.96) .. (251.92,105.9) .. controls (251.92,107.83) and (250.35,109.4) .. (248.42,109.4) .. controls (246.48,109.4) and (244.92,107.83) .. (244.92,105.9) -- cycle ;
\draw (282,84.4) node [anchor=north west][inner sep=0.75pt]    {$\cM^\op$};
\draw (206,123) node [anchor=north west][inner sep=0.75pt]   [align=left] {$\cC$};
\draw (274,173) node [anchor=north west][inner sep=0.75pt]   [align=left] {$\cN$};
\end{tikzpicture},
 \end{equation*}
 then it agrees with those in \cite[Eq.(17)]{Kitaev_Kong_2012} (as geometric shapes) and also explains our notation for the promonad. More precise relationships between these algebraic constructions are given in Sections \ref{sub.proof} and \ref{sub.recovery}.
 \end{remark}

The pro-tensor $\cM\bbH_\cC \cN$ can be equipped with a promonad structure, with multiplication and unit defined in an analogous way as ~\eqref{eq.intuition_multiplication} and~\eqref{eq.intuition_unit}, respectively:
 \begin{equation}
     \mu: \begin{tikzpicture}[scale = 0.7 ,baseline=(current bounding box.center)]
         \draw[color3,thick] (0.5,1.5)--(2.,1.5);
         
         \draw[color3,thick] (0.5+0.2,0.5)--(2.,0.5);
         \draw[color3,thick] (0.5+0.2,-0.5)--(2.,-0.5);
         
         \draw[color3,thick] (2.,0.5) arc (-90:90:0.5cm);
         \draw[color3,thick] (0.5+0.2,0.5) arc (90:270:0.5cm);
       \draw[black,thick] (0.8,0.5)
  .. controls (0,0.8) ..
  (-0.6,-1);

\draw  (1.35,0.5)
  .. controls (-0.1,1.3) ..
  (-0.9,-1);
         \draw[color4,thick] (-1.5,-1)--(2.5,-1);
    
     \end{tikzpicture}\xRightarrow{\alpha_{\wr,\cN}^{-1}\star\alpha_{\wr,\cM}^R} \begin{tikzpicture}[scale = 0.7 ,baseline=(current bounding box.center)]
         \draw[color3,thick] (0.5,1.5)--(2.,1.5);
         
         \draw[color3,thick] (1.1,0.5)--(2.,0.5);
         \draw[color3,thick] (1.1,-0.5)--(2.,-0.5);
         
         \draw[color3,thick] (2.,0.5) arc (-90:90:0.5cm);
         \draw[color3,thick] (1.1,0.5) arc (90:270:0.5cm);
         \draw[color4,thick] (-1.5,-1)--(2.5,-1);
       \draw  (1.5,0.5)
  .. controls (0.3,1.2) ..
  (-1.2,-1);

\draw  (0.9,0.85)--(-0.5,0);
         \end{tikzpicture}\xRightarrow{\epsilon_2}
         \begin{tikzpicture}[scale = 0.7 ,baseline=(current bounding box.center)]
         \draw[color3,thick] (0.5,1.5)--(2,1.5);
         
         \draw[color3,thick] (1.1,0.5)--(2.,0.5);
         \draw[color3,thick] (1.1,-0.5)--(2.,-0.5);
         
         \draw[color3,thick] (2.,0.5) arc (-90:90:0.5cm);
         \draw[color3,thick] (1.1,0.5) arc (90:270:0.5cm);
         \draw[color4,thick] (-1.5,-1)--(2.5,-1); 
          \draw[black,thick] (1,0.5)
  .. controls (0,1) ..
  (-0.7,-1);
         \end{tikzpicture},
 \end{equation}
\begin{equation} 
    \eta:
    \begin{tikzpicture}[scale = 0.7 ,baseline=(current bounding box.center)]
         \draw[color3,thick] (0.5,1.5)--(2.,1.5);
         
         \draw[color3,thick] (0.5,0.5)--(2.,0.5);
         \draw[color3,thick] (0.5,-0.5)--(2.,-0.5);
         
         \draw[color3,thick] (2.,0.5) arc (-90:90:0.5cm);
         \draw[color3,thick] (0.5,0.5) arc (90:270:0.5cm);
         \draw[color4,thick] (-1.5,-1)--(2.5,-1); 
         \end{tikzpicture}\xRightarrow{\lambda_{\wr,\cN}^{-1}\star\lambda_{\wr,\cM}^R}
\begin{tikzpicture}[scale = 0.7 ,baseline=(current bounding box.center)]
         \draw[color3,thick] (0.5,1.5)--(2.,1.5);
         
         \draw[color3,thick] (0.5,0.5)--(2.,0.5);
         \draw[color3,thick] (0.5,-0.5)--(2.,-0.5);
         
         \draw[color3,thick] (2.,0.5) arc (-90:90:0.5cm);
         \draw[color3,thick] (0.5,0.5) arc (90:270:0.5cm);
         \draw[color4,thick] (-1.5,-1)--(2.5,-1);
         \draw  (-1,-1)--(-0.8,-0.3);
         \draw (1,0.5)--(0,1);
          \draw[fill=white] (0,1)circle[radius=0.1];
           \draw[fill=white] (-0.8,-0.3)circle[radius=0.1];
         \end{tikzpicture}\xRightarrow{\epsilon_0}
         \begin{tikzpicture}[scale = 0.7 ,baseline=(current bounding box.center)]
         \draw[color3,thick] (0.5,1.5)--(2.,1.5);
         
         \draw[color3,thick] (0.9,0.5)--(2.,0.5);
         \draw[color3,thick] (0.9,-0.5)--(2.,-0.5);
         
         \draw[color3,thick] (2.,0.5) arc (-90:90:0.5cm);
         \draw[color3,thick] (0.9,0.5) arc (90:270:0.5cm);
         \draw[color4,thick] (-1.5,-1)--(2.5,-1); 
       \draw[black,thick] (1,0.5)
  .. controls (0,1) ..
  (-0.7,-1);
         \end{tikzpicture}.
\end{equation}
The verification of the associativity and unitality of $\mu$, $\eta$ is the same as that of \eqref{eq.omega_ass} and \eqref{eq.omega_unit} for the renormalization algebra $\MnCN$, up to applications of Theorem \ref{thm.local} and \eqref{eq.zigzagger}. Thus we can conclude that $(\cM\bbH_\cC \cN,\mu,\eta)$ is a promonad. 

\begin{remark}\label{rmk.mcn}
 The promonad $\cM\bbH_\cC \cN$ or its variants has been extensively studied in the literature, see for example~\cite{pastro2007doublesmonoidalcategories,Lopez_Franco_2007,Clarke_2024}. In the study of so called Tambara modules, it is showed in \cite{pastro2007doublesmonoidalcategories,Tambara2006DistributorsOA} (see also \cite{Clarke_2024,boisseau2020} in the computer science community where $\cM\bbH_\cC \cN$ is named ``profunctor optics") that $\cM\bbH_\cC \cN$ is induced by a monad $\Phi$ on the functor $\cV$-category $\enf[]{\cN^\op\ot\cM}{\bcV}$  defined by 
\begin{equation}
    \Phi(S)(n,m)=\int^{x\in\cC,p\in\cM,q\in\cN} \cN(n,x\boxdot_\cN q)\ot\cM(x\boxdot_\cM p,m)\ot S(q,p).
\end{equation}
The monad $\Phi$ is precisely $\MnCN$~\eqref{eq.MNrenormalization_alg}.
$\Phi$ is the left adjoint of the endofunctor $\Theta$ on $[\cN^\op\ot\cM, \bcV]$, where 
\begin{equation}
    \Theta(T)(n,m) = \int_{x\in \cC}T(x\boxdot_\cN n,x\boxdot_\cM m).
\end{equation}
$\Theta$ has an obvious comonad structure induced by the monad structure on $\Phi$. Let $Y:\cM^\op\ot \cN\rightarrow \enf[]{\cN^\op\ot \cM}{\bcV}$ be the Yoneda embedding,  $\Phi Y$ then has a canonical promonad structure~\cite{Clarke_2024}, which  coincides with  $(\cM\bbH_\cC\cN,\mu,\eta)$. 
\end{remark}


\subsection{Module profunctors}\label{sub.module_prof}
Let $\cC,\cM,\cN$ be as in Section \ref{sub.MCN}.
\begin{definition}
    A \emph{lax $\cC$-module profunctor} $(F,\beta)$ from $\cM$ to $\cN$ is a profunctor $F:\cM\nrightarrow \cN$ together with a profunctor homomorphism $\beta\:(\boxdot_\cN)_\ast\bullet(\Id_\cC\os F)\Rightarrow F\bullet (\boxdot_\cM)_\ast$, denoted as 
\begin{equation}\label{eq.ModuleFuncMor}
\begin{tikzpicture}[scale = 0.5,baseline=0]
    \draw[color4] (-2,0)--(0,0);
    \draw[color3] (0,0)--(2,0);
    \mynode{0}{0}{0.4}{0.4}{F}
\end{tikzpicture}
\quad\text{and}\quad
\begin{tikzpicture}[scale=0.5,baseline=0]
    \draw[color3] (2,0)--(0,0);
    \draw[color4] (0,0)--(-2,0);
    \mynode{0}{0}{0.4}{0.4}{F}
    \draw (2,1.65) parabola (-1.35,0); 
\end{tikzpicture}
\stackrel{\beta}{\Rightarrow}
\begin{tikzpicture}[scale=0.5,baseline=0]
    \draw[color3] (2.5,0)--(0,0);
    \draw[color4] (0,0)--(-2,0);
    \mynode{0}{0}{0.4}{0.4}{F}
    \draw (2.5,1.01) parabola (1.15,0);
\end{tikzpicture}
\end{equation}
respectively, satisfying the associativity axiom
\begin{equation}\label{eq.ModuleProfAss}
    \begin{tikzpicture}[scale = 1.2,xscale=1.5]
        \node (A) at (0,2)
        {\begin{tikzpicture}
            [scale = 0.5, baseline = (current bounding box.center)]
            \mygrid{ 
                \draw[gray!30, step=0.5, opacity = 0.5] (-5,-5) grid (5,5);
                \foreach \x in {-3,-2,-1,0,1,2,3}
                    \draw (\x,3) node[above] {\small $\x$};
                \foreach \y in {-1,0,1,2,3}
                    \draw (3,\y) node[right] {\small $\y$};
            }
            \draw[color3] (2.35,0)--(0,0);
            \draw[color4] (0,0)--(-2.5,0);
            \mynode{0}{0}{0.4}{0.4}{F}
            \draw (0.5,1.15) parabola (-1.35,0);
            \draw (0.5,1.65) parabola (-2,0); 
        \end{tikzpicture}};
        \node (B) at (2,2){\begin{tikzpicture}[scale = 0.5, baseline = (current bounding box.center)]
            \mygrid{ 
                \draw[gray!30, step=0.5, opacity = 0.5] (-5,-5) grid (5,5);
                \foreach \x in {-3,-2,-1,0,1,2,3}
                    \draw (\x,3) node[above] {\small $\x$};
                \foreach \y in {-1,0,1,2,3}
                    \draw (3,\y) node[right] {\small $\y$};
            }
            \draw[color3] (2.85,0)--(0,0);
            \draw[color4] (0,0)--(-2,0);
            \mynode{0}{0}{0.4}{0.4}{F}
            \draw (2.85,1.01) parabola (1.5,0);
            \draw (0.5,1.35) parabola (-1.35,0); 
        \end{tikzpicture}};
        \node (D) at (4,2)
            {\begin{tikzpicture}[scale = 0.5, baseline = (current bounding box.center)]
            \mygrid{ 
                \draw[gray!30, step=0.5, opacity = 0.5] (-5,-5) grid (5,5);
                \foreach \x in {-3,-2,-1,0,1,2,3}
                    \draw (\x,3) node[above] {\small $\x$};
                \foreach \y in {-1,0,1,2,3}
                    \draw (3,\y) node[right] {\small $\y$};
            }
            \draw[color3] (4.35,0)--(1,0);
            \draw[color4] (1,0)--(-0.5,0);
            \mynode{1}{0}{0.4}{0.4}{F}
            \draw (4.35,1.01) parabola (3,0);
            \draw (4.35,1.4) parabola (2.5,0);
        \end{tikzpicture}};  
        \node (C) at (0,0)
         {\begin{tikzpicture}[scale = 0.5, baseline = (current bounding box.center)]
            \draw (0.5,1.4) parabola (-1.65,0);
            \draw[white,fill=white] (-0.95,0.1)rectangle(0.55,1.75);
            \draw (0.5,1.15) parabola (-0.5,0.8);
            \draw (0.5,1.65) parabola (-0.75,1.2); 
            \draw (-1,0.74) to[out=90, in=220](-0.75,1.2);
            \draw (-1,0.74) to[out=-30, in=220](-0.5,0.8);
            \draw[color3] (2.35,0)--(0,0);
            \draw[color4] (0,0)--(-2.5,0);
            \mynode{0}{0}{0.4}{0.4}{F}
        \end{tikzpicture}};   
        \node (E) at (4,0) {\begin{tikzpicture}[scale = 0.5, baseline = (current bounding box.center)]
            \mygrid{ 
                \draw[gray!30, step=0.5, opacity = 0.5] (-5,-5) grid (5,5);
                \foreach \x in {-3,-2,-1,0,1,2,3}
                    \draw (\x,3) node[above] {\small $\x$};
                \foreach \y in {-1,0,1,2,3}
                    \draw (3,\y) node[right] {\small $\y$};
            }
            \draw[color3] (4.35,0)--(1,0);
            \draw[color4] (1,0)--(-0.5,0);
            \mynode{1}{0}{0.4}{0.4}{F}
            \begin{scope}[xshift=3.85cm]
                \draw (0.5,1.4) parabola (-1.65,0);
                \draw[white,fill=white] (-0.95,0.1)rectangle(0.55,1.75);
                \draw (0.5,1.15) parabola (-0.5,0.8);
                \draw (0.5,1.65) parabola (-0.75,1.2); 
                \draw (-1,0.74) to[out=90, in=220](-0.75,1.2);
                \draw (-1,0.74) to[out=-30, in=220](-0.5,0.8);
            \end{scope}
        \end{tikzpicture}};
        \draw[nat] (A)--(B) node[midway,above]{\beta};
        \draw[nat] (A)--(C) node[midway,left = 2 pt]{\alpha_{\wr,\cN}^{-1}};
        \draw[nat] (C)--(E) node[midway, below]{\beta};
        \draw[nat] (D)--(E) node[midway,right = 4pt]{\alpha_{\wr,\cM}^{-1}};
        \draw[nat] (B)--(D) node[midway,above]{\beta};
    \end{tikzpicture}
\end{equation}
and the unitality axiom
\begin{equation}
    \label{eq.ModuleProfUni}
    \diagram{
    \begin{tikzpicture}[scale=0.5,baseline=0]
        \mygrid{ 
            \draw[gray!30, step=0.5, opacity = 0.5] (-3,-3) grid (3,3);
            \foreach \x in {-3,-2,-1,0,1,2,3}
                \draw (\x,3) node[above] {\small $\x$};
            \foreach \y in {-1,0,1,2,3}
                \draw (3,\y) node[right] {\small $\y$};
        }
        \draw (2,1.45) parabola (-1.35,0); 
        \draw[white,fill=white] (0.5,0)rectangle(2,1.7);
        \draw[fill=white] (0.5,1.15)circle[radius=0.1]; 
        \draw[color3] (2,0)--(0,0);
        \draw[color4] (0,0)--(-2,0);
        \mynode{0}{0}{0.4}{0.4}{F}
    \end{tikzpicture}
    \ar@{=>}[rr]^-\beta  \ar@{=>}[rd]_-{\lambda_{\wr,\cN}} & &
    \begin{tikzpicture}[scale=0.5,baseline=0]
        \mygrid{ 
            \draw[gray!30, step=0.5, opacity = 0.5] (-3,-3) grid (3,3);
            \foreach \x in {-3,-2,-1,0,1,2,3}
                \draw (\x,3) node[above] {\small $\x$};
            \foreach \y in {-1,0,1,2,3}
                \draw (3,\y) node[right] {\small $\y$};
        }
        \draw[color3] (2,0)--(0,0);
        \draw[color4] (0,0)--(-2,0);
        \mynode{0}{0}{0.4}{0.4}{F}
        \draw (2.7,0.7) parabola (1.35,0);
        \draw[white,fill=white] (1.9,0.2)rectangle(2.7,1.1);
        \draw[fill=white] (1.9,0.45)circle[radius=0.1];
    \end{tikzpicture}
    \ar@{=>}[ld]^-{\lambda_{\wr,\cM}}
    \\
    &
    \begin{tikzpicture}[scale=0.5,baseline=0]
        \draw[color4] (-2,0)--(0,0);
        \draw[color3] (0,0)--(2,0);
        \mynode{0}{0}{0.4}{0.4}{F}
    \end{tikzpicture}
}.
\end{equation}
When $\beta$ is invertible, we call $F$ a \emph{strong} $\cC$-module profunctor, or simply a $\cC$-module profunctor.

\end{definition}

\begin{definition}
Let $F=(F,\beta_F),G=(G,\beta_G)$ be two lax $\cC$-module profunctors from $\cM$ to $\cN$. A \emph{module profunctor homomorphism} from $F$ to $G$ is a profunctor homomorphism $\phi:F\Rightarrow G$ such that the following diagram commutes:
\[
    \diagram{
    \begin{tikzpicture}[scale=0.5,baseline=0]
        \draw[color3] (2,0)--(0,0);
        \draw[color4] (0,0)--(-2,0);
        \mynode{0}{0}{0.4}{0.4}{F}
        \draw (2,1.65) parabola (-1.35,0); 
    \end{tikzpicture}
    \ar@{=>}[d]_{\phi}
    \ar@{=>}[r]^-{\beta_F}
    &
    \begin{tikzpicture}[scale=0.5,baseline=0]
        \draw[color3] (2.5,0)--(0,0);
        \draw[color4] (0,0)--(-2,0);
        \mynode{0}{0}{0.4}{0.4}{F}
        \draw (2.5,1.01) parabola (1.15,0);
    \end{tikzpicture}
    \ar@{=>}[d]^\phi
    \\
    \begin{tikzpicture}[scale=0.5,baseline=0]
        \draw[color3] (2,0)--(0,0);
        \draw[color4] (0,0)--(-2,0);
        \mynode{0}{0}{0.4}{0.4}{G}
        \draw (2,1.65) parabola (-1.35,0); 
    \end{tikzpicture}
    \ar@{=>}[r]_-{\beta_G}
    &
    \begin{tikzpicture}[scale=0.5,baseline=0]
        \draw[color3] (2.5,0)--(0,0);
        \draw[color4] (0,0)--(-2,0);
        \mynode{0}{0}{0.4}{0.4}{G}
        \draw (2.5,1.01) parabola (1.15,0);
    \end{tikzpicture}
    }.
\]
\end{definition}
Lax $\cC$-module profunctors $\cM\arprof\cN$ and module profunctor homomorphisms form a category, which we denote by $_\cC\vprof^\lax(\cM,\cN)$.

\begin{remark}
    \label{rmk.rigidCModuleProf}
    If $\cC$ is left rigid, a lax $\cC$-module profunctor $F:\cM\nrightarrow \cN$ is automatically strong. The key observation is that $\cC$ is (left) naturally Frobenius in this case (Proposition \ref{prp.rigit_to_NF}), so that $\alpha_\wr^\sharp$ defined in \eqref{eq.module_Frobeniusiator} is invertible. Now given a lax $\cC$-module profunctor $(F,\beta)$ in \eqref{eq.ModuleFuncMor}, the inverse of $\beta$ is given by 
    \begin{equation}
    \label{eq.inverseModuleProfMor}
        \begin{tikzpicture}
        \node (A) at (0,0) {\begin{tikzpicture}
        [scale = 0.5,baseline=(current bounding box.center)]     
         \draw[color4] (-0.3,0)--(2,0);
    \draw[color3] (2,0)--(5,0);
    \mynode{2}{0}{0.4}{0.4}{F}
    \draw  (3.1,0)--(4.3,1.8);
        \end{tikzpicture}};
       \node (B) at (0,-4) {\begin{tikzpicture}
           [scale = 0.5,baseline=(current bounding box.center)]
            \draw[color4] (-1,0)--(2,0);
    \draw[color3] (2,0)--(4,0);
    \mynode{2}{0}{0.4}{0.4}{F}
    \draw  (0,0)--(1.2,1.8);
       \end{tikzpicture}}; 
\node (C) at (5,0) {\begin{tikzpicture}
        [scale = 0.5,baseline=(current bounding box.center)]     
         \draw[color4] (0,0)--(2,0);
    \draw[color3] (2,0)--(4.6,0);
    \mynode{2}{0}{0.4}{0.4}{F}
    \draw  (3.9,0)--(2.7,0.9);
    \draw  (1.8,0.9)--(2.7,0.9);
    \draw[fill = white] (1.8,0.9) circle [radius = 0.1];
    \draw  (2.7,0.9)--(3.9,1.8);
     \end{tikzpicture}};
\node (D) at (10,0) {\begin{tikzpicture}
        [scale = 0.5,baseline=(current bounding box.center)]     
         \draw[color4] (-1,0)--(2,0);
    \draw[color3] (2,0)--(4,0);
    \mynode{2}{0}{0.4}{0.4}{F}
    \draw  (0.2,0)--(-.4,0.9);
    \draw  (-0.9,0.9)--(-.4,0.9);
    \draw[fill = white] (-0.9,0.9) circle [radius = 0.1];
    \draw  (-0.4,0.9)--(0.2,1.8);
    \myparabola{1.5}{0.5}{0}{3.5}{0}
     \end{tikzpicture}};

     \node (E) at (10,-4) {\begin{tikzpicture}
        [scale = 0.5,baseline=(current bounding box.center)]     
         \draw[color4] (-1,0)--(1.4,0);
    \draw[color3] (1.4,0)--(6,0);
    \mynode{1.4}{0}{0.4}{0.4}{F}
    \draw  (0.2,0)--(-.4,0.9);
    \draw  (-0.9,0.9)--(-.4,0.9);
    \draw[fill = white] (-0.9,0.9) circle [radius = 0.1];
    \draw  (-0.4,0.9)--(0.2,1.8);
    \myparabola{1.5}{2.7}{0}{5.7}{0}
     \end{tikzpicture}};

\node (F) at (5,-4){\begin{tikzpicture}
        [scale = 0.5,baseline=(current bounding box.center)]     
         \draw[color4] (-1,0)--(1.4,0);
    \draw[color3] (1.4,0)--(4,0);
    \mynode{1.4}{0}{0.4}{0.4}{F}
    \draw  (0.2,0)--(-.4,0.9);
    \draw  (-0.9,0.9)--(-.4,0.9);
    \draw[fill = white] (-0.9,0.9) circle [radius = 0.1];
    \draw  (-0.4,0.9)--(0.2,1.8);
     \end{tikzpicture}};
\draw[nattrans] (A)--(B) node [midway, right = 4pt] {$\scriptstyle :=
\beta^{-1}$};
\draw[nattrans] (A)--(C) node [midway, above = 4pt] {$\scriptstyle (\alpha_{\wr,\cM}^\sharp)^{-1}$};
\draw[nattrans] (C)--(D) node [midway, above = 4pt] {$\scriptstyle \eta_{\wr,\cN}$};
\draw[nattrans] (D)--(E) node [midway, right = 4pt] {$\scriptstyle \beta$};
\draw[nattrans] (E)--(F) node [midway, above = 4pt] {$\scriptstyle \epsilon_{\wr,\cM}$};
\draw[nattrans] (F)--(B) node [midway, above = 4pt] {$\scriptstyle \alpha_{\wr,\cN}^\sharp$};
\end{tikzpicture}
    \end{equation}
\end{remark}

\begin{remark}
\label{rmk.AdjModuleFun}
    If $(F\:\cM\arprof\cN,\beta\:(\boxdot_\cN)_\ast\bullet (\Id_\cC\os F)\Rightarrow F\bullet (\boxdot_\cM)_\ast)$ is a strong $\cC$-module profunctor and the profunctor part $F$ admits a right adjoint $F^R$, then $(F^R,\beta')$ is a lax $\cC$-module profunctor with $\beta'$ defined as the image of $\beta^{-1}$ under the transposing isomorphism
    \[
        \Hom(F\bullet (\boxdot_\cN)_\ast,(\boxdot_\cM)_\ast\bullet (\Id_\cC\os F))\cong \Hom((\boxdot_\cN)_\ast\bullet (\Id_\cC\os F^R),F^R\bullet (\boxdot_\cM)_\ast)
    \]
    in Proposition \ref{prp.transpose}.\footnote{Clearly the requirement on $F$ can relaxed to being only an oplax $\cC$-module profunctor.} Moreover, $(F^R,\beta')$ serves as the right adjoint of $(F,\beta)$ as lax $\cC$-module profunctors. Combined with Remark \ref{rmk.rigidCModuleProf}, if $\cC$ is rigid, given a $\cC$-module profunctor $(F,\beta)$, it has a right adjoint as a $\cC$-module profunctor iff the profunctor part $F$ admits a right adjoint.
\end{remark}

\begin{remark}
    Left $\cC$-modules, (lax) module profunctors and module profunctor homomorphisms form an ordinary 2-category ${}_\cC\vprof$ (${}_\cC\vprof^\lax$). Module functors are similarly defined but we omit here. Left $\cC$-modules, (lax) module functors and natural transformations form an ordinary 2-category ${}_\cC\Vcat$ (${}_\cC\Vcat^{\lax}$). The hom categories ${}_\cC\vprof(\cM,\cN)$ (${}_\cC\vprof^{\lax}(\cM,\cN)$) and ${}_\cC\Vcat(\cM,\cN)$ (${}_\cC\Vcat^{\lax}(\cM,\cN)$) in these 2-categories can be canonically enriched over $\cV$, which are denoted by ${}_\cC\bvprof(\cM,\cN)$ (${}_\cC\bvprof^{\lax}(\cM,\cN)$) and $\enf[\cC]\cM\cN$ ($\enf[\cC]\cM\cN^{\lax}$). They are essentially variants of internal homs; see Appendix \ref{app.enriched_structure}. When considering right $\cC$-modules or $\cC$-$\cC$-bimodules, we change the position of subscript $\cC$, for example, $\Vcat_\cC$ or $_\cC\Vcat_\cC$.
\end{remark}
\begin{remark}\label{rmk.relationship_module_fun}
    (Lax) $\cC$-module profunctors bear a close kinship with (lax) $\cC$-module functors. First of all, it can be verified that a (lax) $\cC$-module functor $H\:\cM\to\cN$ gives rise to a (lax) $\cC$-module profunctor $H_\ast\:\cM\arprof\cN$. On the other hand, any (lax) $\cC$-module profunctor $\cM\arprof\cN$ is precisely a (lax) $\cC$-module functor $\cM\arprof\enf[]{\cN^\op}{\bcV}$, if $\enf[]{\cN^\op}{\bcV}$ is equipped with the following $\cC$-module action:
\[
 \begin{split}
      \odot:\cC\ot\enf[]{\cN^\op}{\bcV}&\rightarrow \enf[]{\cN^\op}{\bcV}\\
      (x,F)&\mapsto x\odot F:= \cN(-,x\boxdot_\cN -)\bullet F\equiv\int^{n\in \cN}\cN(-,x\boxdot_\cN n)F(n).
 \end{split}
\]
More precisely, there is an equivalence between ordinary categories
  \begin{equation}
  \label{eq.moduleprof-prof}
        {}_\cC\vprof^\lax(\cM,\cN)\cong {}_\cC\Vcat^\lax({\cM},{\enf{\cN^\op}{\bcV}}),
    \end{equation}
which can be further upgraded to an equivalence between $\cV$-categories (see Lemma~\ref{lem.enriched_structure})
    \begin{equation*}
        {}_\cC\bvprof^\lax(\cM,\cN)\cong {}_\cC[\cM,[\cN^\op,\bcV]]^\lax.
    \end{equation*}
\end{remark} 

\subsection{The representations of $\MCN$ and $\MCM$}\label{sub.proof}
\begin{theorem}\label{thm.unenriched_Kitaev_Kong}
    There exist canonical equivalences among the following three categories:
\begin{enumerate}
    \item \label{item1.thm.unenriched_Kitaev_Kong} The category $\Lmd_{\MnCN}$ of $\MnCN$-modules and $\MnCN$-module maps.
    \item \label{item2.thm.unenriched_Kitaev_Kong} The category $\Lmd_{\MCN}(\ast)$ of left $\MCN$-modules and $\MCN$-module maps.
    \item \label{item3.thm.unenriched_Kitaev_Kong} The category ${}_\cC\vprof^\lax(\cM,\cN)$ of lax $\cC$-module profunctors $\cM\arprof\cN$.
\end{enumerate}
\begin{proof}
Let us first establish an equivalence between \ref{item1.thm.unenriched_Kitaev_Kong} and \ref{item3.thm.unenriched_Kitaev_Kong} by constructing a pair of mutually inverse functors. We focus on objects' assignments. Given a lax $\cC$-module profunctor
\[
(\begin{tikzpicture}[scale=0.5,baseline=0]
    \draw[color3] (2,0)--(0,0);
    \draw[color4] (0,0)--(-2,0);
    \mynode{0}{0}{0.4}{0.4}{F}
\end{tikzpicture}
,
\begin{tikzpicture}[scale=0.5,baseline=0]
    \draw[color3] (2,0)--(0,0);
    \draw[color4] (0,0)--(-2,0);
    \mynode{0}{0}{0.4}{0.4}{F}
    \draw (2,1.65) parabola (-1.35,0); 
\end{tikzpicture}
\stackrel{\beta}{\Rightarrow}
\begin{tikzpicture}[scale=0.5,baseline=0]
    \mygrid{ 
        \draw[gray!30, step=0.5, opacity = 0.5] (-3,-3) grid (3,3);
        \foreach \x in {-3,-2,-1,0,1,2,3}
            \draw (\x,3) node[above] {\small $\x$};
        \foreach \y in {-1,0,1,2,3}
            \draw (3,\y) node[right] {\small $\y$};
    }
    \draw[color3] (2.5,0)--(0,0);
    \draw[color4] (0,0)--(-2,0);
    \mynode{0}{0}{0.4}{0.4}{F}
    \draw (2.5,1.01) parabola (1.15,0);
\end{tikzpicture}
),
\]
we define an $\MnCN$-module with underlying pro-tensor $F$ and module action
\[
\omega\defdtobe(
\begin{tikzpicture}[scale=0.5,baseline=0]
    \draw[color3] (2,0)--(0,0);
    \draw[color4] (0,0)--(-2,0);
    \mynode{0}{0}{0.4}{0.4}{F}
    \myparabola{1.5}{-1.35}{0}{1.35}{0}
\end{tikzpicture}
\stackrel{\beta}{\Rightarrow}
\begin{tikzpicture}[scale=0.5,baseline=0]
    \mygrid{ 
        \draw[gray!30, step=0.5, opacity = 0.5] (-3,-3) grid (3,3);
        \foreach \x in {-3,-2,-1,0,1,2,3}
            \draw (\x,3) node[above] {\small $\x$};
        \foreach \y in {-1,0,1,2,3}
            \draw (3,\y) node[right] {\small $\y$};
    }
    \draw[color3] (4.2,0)--(0,0);
    \draw[color4] (0,0)--(-2,0);
    \mynode{0}{0}{0.4}{0.4}{F}
    \myparabola{1.5}{1}{0}{3.7}{0}
\end{tikzpicture}
\stackrel{\epsilon_{\wr,\cM}}{\Rightarrow}
\begin{tikzpicture}[scale=0.5,baseline=0]
    \draw[color3] (2,0)--(0,0);
    \draw[color4] (0,0)--(-2,0);
    \mynode{0}{0}{0.4}{0.4}{F}
\end{tikzpicture}
).
\]
We remark that it is possible to realize $\omega$ as the image of $\beta$ under an isomorphism given in Proposition \ref{prp.transpose}.\ref{item2.prp.transpose}. That $(F,\omega)$ is an $\MnCN$-module is verified in Fig \ref{fig.modprof_to_omega}. The assignment $(F,\beta)\mapsto (F,\omega)$ easily extends to a functor $\Phi\:{}_\cC\vprof^\lax(\cM,\cN)\to\Lmd_{\MnCN}$.
\begin{figure}[htbp]
\[
    \ctikz{[scale=1.25,xscale=1.5]
        \mygrid{
            \draw[gray!30, step=0.5, opacity = 0.5] (0,0) grid (6,8);
            \foreach \x in {0,1,2,3,4,5,6}
                \draw (\x,-1) node[below] {\small $\x$};
            \foreach \y in {0,1,2,3,4,5,6,7,8}
                \draw (-1,\y) node[left] {\small $\y$};
        }
        \node (1_1) at (0,4){$
        \ctikz{[scale=0.5]
            \mygrid{ 
                \draw[gray!30, step=0.5, opacity = 0.5] (-3,-3) grid (3,3);
                \foreach \x in {-3,-2,-1,0,1,2,3}
                    \draw (\x,3) node[above] {\small $\x$};
                \foreach \y in {-1,0,1,2,3}
                    \draw (3,\y) node[right] {\small $\y$};
            }
            \myparabola{1.5}{-1.35}{0}{1.35}{0}
            \draw[white, fill=white] (-0.8,0)rectangle(0.8,1.1);
            \draw[fill=white](-0.8,0.6)circle[radius=0.1];
            \draw[fill=white](0.8,0.6)circle[radius=0.1];
            \draw[color3] (2,0)--(0,0);
            \draw[color4] (0,0)--(-2,0);
            \mynode{0}{0}{0.4}{0.4}{F}
        }$};
        \node (1_2) at (2,4){$
        \ctikz{[scale=0.5]
            \mygrid{ 
                \draw[gray!30, step=0.5, opacity = 0.5] (-3,-3) grid (3,3);
                \foreach \x in {-3,-2,-1,0,1,2,3}
                    \draw (\x,3) node[above] {\small $\x$};
                \foreach \y in {-1,0,1,2,3}
                    \draw (3,\y) node[right] {\small $\y$};
            }
            \myparabola{1.5}{-1.35}{0}{1.35}{0}
            \draw[color3] (2,0)--(0,0);
            \draw[color4] (0,0)--(-2,0);
            \mynode{0}{0}{0.4}{0.4}{F}
        }$};
        \node (1_3) at (4,4){$
        \ctikz{[scale=0.5]
            \mygrid{ 
                \draw[gray!30, step=0.5, opacity = 0.5] (-3,-3) grid (3,3);
                \foreach \x in {-3,-2,-1,0,1,2,3}
                    \draw (\x,3) node[above] {\small $\x$};
                \foreach \y in {-1,0,1,2,3}
                    \draw (3,\y) node[right] {\small $\y$};
            }
            \myparabola{1.5}{1}{0}{3.7}{0}     
            \draw[color3] (4.2,0)--(0,0);
            \draw[color4] (0,0)--(-2,0);
            \mynode{0}{0}{0.4}{0.4}{F}
        }$};
        \node (2_2) at (2,2){$
        \ctikz{[scale=0.5]
            \mygrid{ 
                \draw[gray!30, step=0.5, opacity = 0.5] (-3,-3) grid (3,3);
                \foreach \x in {-3,-2,-1,0,1,2,3}
                    \draw (\x,3) node[above] {\small $\x$};
                \foreach \y in {-1,0,1,2,3}
                    \draw (3,\y) node[right] {\small $\y$};
            }
            \begin{scope}[xshift=2.35cm]
                \myparabola{1.5}{-1.35}{0}{1.35}{0}
                \draw[white, fill=white] (-0.8,0)rectangle(0.8,1.1);
                \draw[fill=white](-0.8,0.6)circle[radius=0.1];
                \draw[fill=white](0.8,0.6)circle[radius=0.1];
            \end{scope}
            \draw[color3] (4.2,0)--(0,0);
            \draw[color4] (0,0)--(-2,0);
            \mynode{0}{0}{0.4}{0.4}{F}
        }$};
        \node (3_1) at (0,0){$
        \ctikz{[scale=0.5]
            \mygrid{ 
                \draw[gray!30, step=0.5, opacity = 0.5] (-3,-3) grid (3,3);
                \foreach \x in {-3,-2,-1,0,1,2,3}
                    \draw (\x,3) node[above] {\small $\x$};
                \foreach \y in {-1,0,1,2,3}
                    \draw (3,\y) node[right] {\small $\y$};
            }
            \begin{scope}[xshift=1cm]
                \draw (0,1.01) parabola (1.35,0);
                \draw[white, fill=white] (-0.8,0)rectangle(0.8,1.1);
                \draw[fill=white](0.8,0.6)circle[radius=0.1];
            \end{scope}
            \draw[color3] (3,0)--(0,0);
            \draw[color4] (0,0)--(-2,0);
            \mynode{0}{0}{0.4}{0.4}{F}
        }$};
        \node (3_3) at (4,0){$
        \begin{tikzpicture}[scale=0.5,baseline=0]
            \draw[color3] (2,0)--(0,0);
            \draw[color4] (0,0)--(-2,0);
            \mynode{0}{0}{0.4}{0.4}{F}
        \end{tikzpicture}$};
        \draw[nat] (2_2)--(3_1)node[midway, above left]{\lambda_{\wr,\cM}};
        \draw[nat] (3_1)--(3_3)node[midway, below]{(\lambda_{\wr,\cM}^R)^{-1}};
        \draw[nat] (1_1)--(3_1)node[midway, left]{\lambda_{\wr,\cN}};
        \draw[nat] (1_1)--(1_2)node[midway, above]{\epsilon_0};
        \draw[nat] (1_2)--(1_3)node[midway, above]{\beta};
        \draw[nat] (1_3)--(3_3)node[midway, right]{\epsilon_{\wr,\cM}};
        \draw[nat] (2_2)--(1_3)node[midway, below right]{\epsilon_0};
        \draw[nat] (2_2)--(3_3)node[midway, above right, yshift=0.3cm, xshift=-0.6cm]{\lambda_{\wr,\cM}\star(\lambda_{\wr,\cM}^R)^{-1}};
        \draw[nat] (1_1)--(2_2)node[midway, below left]{\beta};
        \draw (2,3.1) node{\tiny(Theorem \ref{thm.local})};
        \draw (0.65,2)node{\tiny(\eqref{eq.ModuleProfUni})};
        \draw (3.35,2)node{\tiny(\eqref{eq.right_dual_of_lambda_module})};
    }
\]
\[
    \ctikz{[scale=1.5,xscale=2] 
        \mygrid{
            \draw[gray!30, step=0.5, opacity = 0.5] (0,0) grid (6,8);
            \foreach \x in {0,1,2,3,4,5,6}
                \draw (\x,-1) node[below] {\small $\x$};
            \foreach \y in {0,1,2,3,4,5,6,7,8}
                \draw (-1,\y) node[left] {\small $\y$};
        }
        \node (1_1) at (0,6){$
        \begin{tikzpicture}[scale=0.5,baseline=0]
            \mygrid{ 
            \draw[gray!30, step=0.5, opacity = 0.5] (-3,-3) grid (3,3);
            \foreach \x in {-3,-2,-1,0,1,2,3}
                \draw (\x,3) node[above] {\small $\x$};
            \foreach \y in {-1,0,1,2,3}
                \draw (3,\y) node[right] {\small $\y$};
            }
            \draw[color3] (3,0)--(0,0);
            \draw[color4] (0,0)--(-3,0);
            \mynode{0}{0}{0.4}{0.4}{F}
            \myparabola{1.5}{-1.15}{0}{1.15}{0}
            \myparabola{1.5}{-2.4}{0}{2.4}{0}
        \end{tikzpicture}$};
        \node (1_2) at (2,6){$
        \begin{tikzpicture}[scale=0.5,baseline=0]
            \mygrid{ 
            \draw[gray!30, step=0.5, opacity = 0.5] (-3,-3) grid (3,3);
            \foreach \x in {-3,-2,-1,0,1,2,3}
                \draw (\x,3) node[above] {\small $\x$};
            \foreach \y in {-1,0,1,2,3}
                \draw (3,\y) node[right] {\small $\y$};
            }
            \draw[color3] (4.3,0)--(0,0);
            \draw[color4] (0,0)--(-2,0);
            \mynode{0}{0}{0.4}{0.4}{F}
            \myparabola{1.5}{1}{0}{3.3}{0}
            \myparabola{1.5}{-1}{0}{3.8}{0}
        \end{tikzpicture}$};
        \node (1_3) at (4,6){$
        \begin{tikzpicture}[scale=0.5,baseline=0]
            \draw[color3] (3,0)--(0,0);
            \draw[color4] (0,0)--(-3,0);
            \mynode{0}{0}{0.4}{0.4}{F}
            \myparabola{1.5}{-1.35}{0}{1.35}{0}
        \end{tikzpicture}$};
        \node (2_2) at (2,4){$
        \begin{tikzpicture}[scale=0.5,baseline=0]
            \mygrid{ 
            \draw[gray!30, step=0.5, opacity = 0.5] (-3,-3) grid (3,3);
            \foreach \x in {-3,-2,-1,0,1,2,3}
                \draw (\x,3) node[above] {\small $\x$};
            \foreach \y in {-1,0,1,2,3}
                \draw (3,\y) node[right] {\small $\y$};
            }
            \draw[color3] (6.3,0)--(0,0);
            \draw[color4] (0,0)--(-2,0);
            \mynode{0}{0}{0.4}{0.4}{F}
            \myparabola{1.5}{1}{0}{5.8}{0}
            \myparabola{1.5}{2.25}{0}{4.55}{0}
        \end{tikzpicture}$};
        \node (2_3) at (4,4){$
        \begin{tikzpicture}[scale=0.5,baseline=0]
            \mygrid{ 
                \draw[gray!30, step=0.5, opacity = 0.5] (-3,-3) grid (3,3);
                \foreach \x in {-3,-2,-1,0,1,2,3}
                    \draw (\x,3) node[above] {\small $\x$};
                \foreach \y in {-1,0,1,2,3}
                    \draw (3,\y) node[right] {\small $\y$};
            }   
            \draw[color3] (4.2,0)--(0,0);
            \draw[color4] (0,0)--(-2,0);
            \mynode{0}{0}{0.4}{0.4}{F}
            \myparabola{1.5}{1}{0}{3.7}{0}
        \end{tikzpicture}$};
        \node (3_1) at (0,2){
        $
        \begin{tikzpicture}[scale=0.5,baseline=0]
            \myparabola{1.5}{-1.75}{0}{1.75}{0}
            \draw[white,fill=white] (-1.15,0)rectangle(1.15,1.5);
            \draw (-1.2,0.725) to[out=90, in=230](-1,1.3);
            \draw (-1.2,0.725) to[out=-20, in = 220](-0.7,0.8);
            \begin{scope}[xscale=-1]
                \draw (-1.2,0.725) to[out=90, in=230](-1,1.3);
                \draw (-1.2,0.725) to[out=-20, in = 220](-0.7,0.8);
            \end{scope}
            \myparabola{1}{-1}{1.3}{1}{1.3}
            \myparabola{1.2}{-0.7}{0.8}{0.7}{0.8}
            \draw[color3] (3,0)--(0,0);
            \draw[color4] (0,0)--(-3,0);
            \mynode{0}{0}{0.4}{0.4}{F}
            \mygrid{ 
                \draw[gray!30, step=0.5, opacity = 0.5] (-3,-3) grid (3,3);
                \foreach \x in {-3,-2,-1,0,1,2,3}
                    \draw (\x,3) node[above] {\small $\x$};
                \foreach \y in {-1,0,1,2,3}
                    \draw (3,\y) node[right] {\small $\y$};
            }
        \end{tikzpicture}$
        };
        \node (3_2) at (2,2){$
        \begin{tikzpicture}[scale=0.5,baseline=0]
            \mygrid{ 
                \draw[gray!30, step=0.5, opacity = 0.5] (-3,-3) grid (3,3);
                \foreach \x in {-3,-2,-1,0,1,2,3}
                    \draw (\x,3) node[above] {\small $\x$};
                \foreach \y in {-1,0,1,2,3}
                    \draw (3,\y) node[right] {\small $\y$};
            }   
            \draw[color3] (6.3,0)--(0,0);
            \draw[color4] (0,0)--(-2,0);
            \mynode{0}{0}{0.4}{0.4}{F}
           \begin{scope}[xshift=3.4cm]
               \myparabola{1.5}{-1.75}{0}{1.75}{0}
                \draw[white,fill=white] (-1.15,0.1)rectangle(1.15,1.5);
                \draw (-1.2,0.725) to[out=90, in=230](-1,1.3);
                \draw (-1.2,0.725) to[out=-20, in = 220](-0.7,0.8);
                \begin{scope}[xscale=-1]
                    \draw (-1.2,0.725) to[out=90, in=230](-1,1.3);
                    \draw (-1.2,0.725) to[out=-20, in = 220](-0.7,0.8);
                \end{scope}
                \myparabola{1}{-1}{1.3}{1}{1.3}
                \myparabola{1.2}{-0.7}{0.8}{0.7}{0.8}
           \end{scope}
        \end{tikzpicture}$};
        \node (4_1) at (0,0){$
        \begin{tikzpicture}[scale=0.5,baseline=0]
            \draw[color3] (3,0)--(0,0);
            \draw[color4] (0,0)--(-3,0);
            \mynode{0}{0}{0.4}{0.4}{F}
            \myparabola{1.5}{-1.35}{0}{1.35}{0}
        \end{tikzpicture}$};
        \node (4_2) at (2,0){$
        \begin{tikzpicture}[scale=0.5,baseline=0]
            \mygrid{ 
                \draw[gray!30, step=0.5, opacity = 0.5] (-3,-3) grid (3,3);
                \foreach \x in {-3,-2,-1,0,1,2,3}
                    \draw (\x,3) node[above] {\small $\x$};
                \foreach \y in {-1,0,1,2,3}
                    \draw (3,\y) node[right] {\small $\y$};
            }   
            \draw[color3] (4.2,0)--(0,0);
            \draw[color4] (0,0)--(-2,0);
            \mynode{0}{0}{0.4}{0.4}{F}
            \myparabola{1.5}{1}{0}{3.7}{0}
        \end{tikzpicture}$};
        \node (4_3) at (4,0){$
        \begin{tikzpicture}[scale=0.5,baseline=0]
            \draw[color3] (3,0)--(0,0);
            \draw[color4] (0,0)--(-3,0);
            \mynode{0}{0}{0.4}{0.4}{F}
        \end{tikzpicture}$};
        \draw[nat] (1_1)--(1_2)node[midway, above]{\beta};
        \draw[nat] (1_2)--(1_3)node[midway, above]{\epsilon_{\wr,\cM}};
        \draw[nat] (1_3)--(2_3)node[midway, right]{\beta};
        \draw[nat] (2_3)--(4_3)node[midway, right]{\epsilon_{\wr,\cM}};
        \draw[nat] (1_1)--(3_1)node[midway, left]{\alpha_{\wr,\cN}^{-1}\star\alpha_{\wr,\cM}^R};
        \draw[nat] (3_1)--(4_1)node[midway, left]{\epsilon_2};
        \draw[nat] (4_1)--(4_2)node[midway, below]{\beta};
        \draw[nat] (4_2)--(4_3)node[midway, below]{\epsilon_{\wr,\cM}};
        \draw[nat] (3_1)--(3_2)node[midway, above]{\beta};
        \draw[nat] (3_2)--(4_2)node[midway, right]{\epsilon_2};
        \draw[nat] (2_2)--(3_2)node[midway, right]{\alpha_{\wr,\cM}^{-1}\star\alpha_{\wr,\cM}^R};
        \draw[nat] (1_2)--(2_2)node[midway, right]{\beta};
        \draw[nat] (2_2)--(2_3)node[midway, below]{\epsilon_{\wr,\cM}};
        \draw (3,5)node{\tiny(Theorem \ref{thm.local})};
        \draw (1,1)node{\tiny(Theorem \ref{thm.local})};
        \draw (1,4.5)node{\tiny(\eqref{eq.ModuleProfAss})};
        \draw (3,2)node{\tiny(\eqref{eq.right_dual_of_alpha_module})};
    }
\]
\caption{Verification of the $\cM\Omega_\cC\cN$-module structure of $(F,\omega)$.}
\label{fig.modprof_to_omega}
\end{figure}

Conversely, given an $\MnCN$-module
\[
(
\begin{tikzpicture}[scale=0.5,baseline=0]
    \draw[color3] (2,0)--(0,0);
    \draw[color4] (0,0)--(-2,0);
    \mynode{0}{0}{0.4}{0.4}{F}
\end{tikzpicture}
,
\begin{tikzpicture}[scale=0.5,baseline=0]
    \draw[color3] (2,0)--(0,0);
    \draw[color4] (0,0)--(-2,0);
    \mynode{0}{0}{0.4}{0.4}{F}
    \myparabola{1.5}{-1.35}{0}{1.35}{0}
\end{tikzpicture}
\stackrel{\omega}{\Rightarrow}
\begin{tikzpicture}[scale=0.5,baseline=0]
    \draw[color3] (2,0)--(0,0);
    \draw[color4] (0,0)--(-2,0);
    \mynode{0}{0}{0.4}{0.4}{F}
\end{tikzpicture}
),
\]
we define a lax $\cC$-module profunctor with underlying profunctor $F$ and lax $\cC$-module data
\[
\beta\defdtobe(
\begin{tikzpicture}[scale=0.5,baseline=0]
    \draw[color3] (2,0)--(0,0);
    \draw[color4] (0,0)--(-2,0);
    \mynode{0}{0}{0.4}{0.4}{F}
    \draw (2,1.65) parabola (-1.35,0); 
\end{tikzpicture}
\stackrel{\eta_{\wr,\cM}}{\Rightarrow}
\begin{tikzpicture}[scale=0.5,baseline=0]
    \mygrid{ 
        \draw[gray!30, step=0.5, opacity = 0.5] (-3,-3) grid (3,3);
        \foreach \x in {-3,-2,-1,0,1,2,3}
            \draw (\x,3) node[above] {\small $\x$};
        \foreach \y in {-1,0,1,2,3}
            \draw (3,\y) node[right] {\small $\y$};
    }
    \draw[color3] (3.35,0)--(0,0);
    \draw[color4] (0,0)--(-2,0);
    \mynode{0}{0}{0.4}{0.4}{F}
    \myparabola{1.5}{-1.35}{0}{1.35}{0}
    \draw (3.35,1.01) parabola (2,0);
\end{tikzpicture}
\stackrel{\omega}{\Rightarrow}
\begin{tikzpicture}[scale=0.5,baseline=0]
    \mygrid{ 
        \draw[gray!30, step=0.5, opacity = 0.5] (-3,-3) grid (3,3);
        \foreach \x in {-3,-2,-1,0,1,2,3}
            \draw (\x,3) node[above] {\small $\x$};
        \foreach \y in {-1,0,1,2,3}
            \draw (3,\y) node[right] {\small $\y$};
    }
    \draw[color3] (2.5,0)--(0,0);
    \draw[color4] (0,0)--(-2,0);
    \mynode{0}{0}{0.4}{0.4}{F}
    \draw (2.5,1.01) parabola (1.15,0);
\end{tikzpicture}
).
\]
It is possible to realize $\beta$ as the image of $\omega$ under an isomorphism given in Proposition \ref{prp.transpose}.\ref{item2.prp.transpose}. That $(F,\beta)$ is indeed a lax $\cC$-module profunctor is verified in Figure \ref{fig.omega_to_modprof}. The assignment $(F,\omega)\mapsto (F,\beta)$ easily extends to a functor $\Psi\:\Lmd_{\MnCN}\to{}_\cC\vprof^\lax(\cM,\cN)$.
\begin{figure}[htbp]
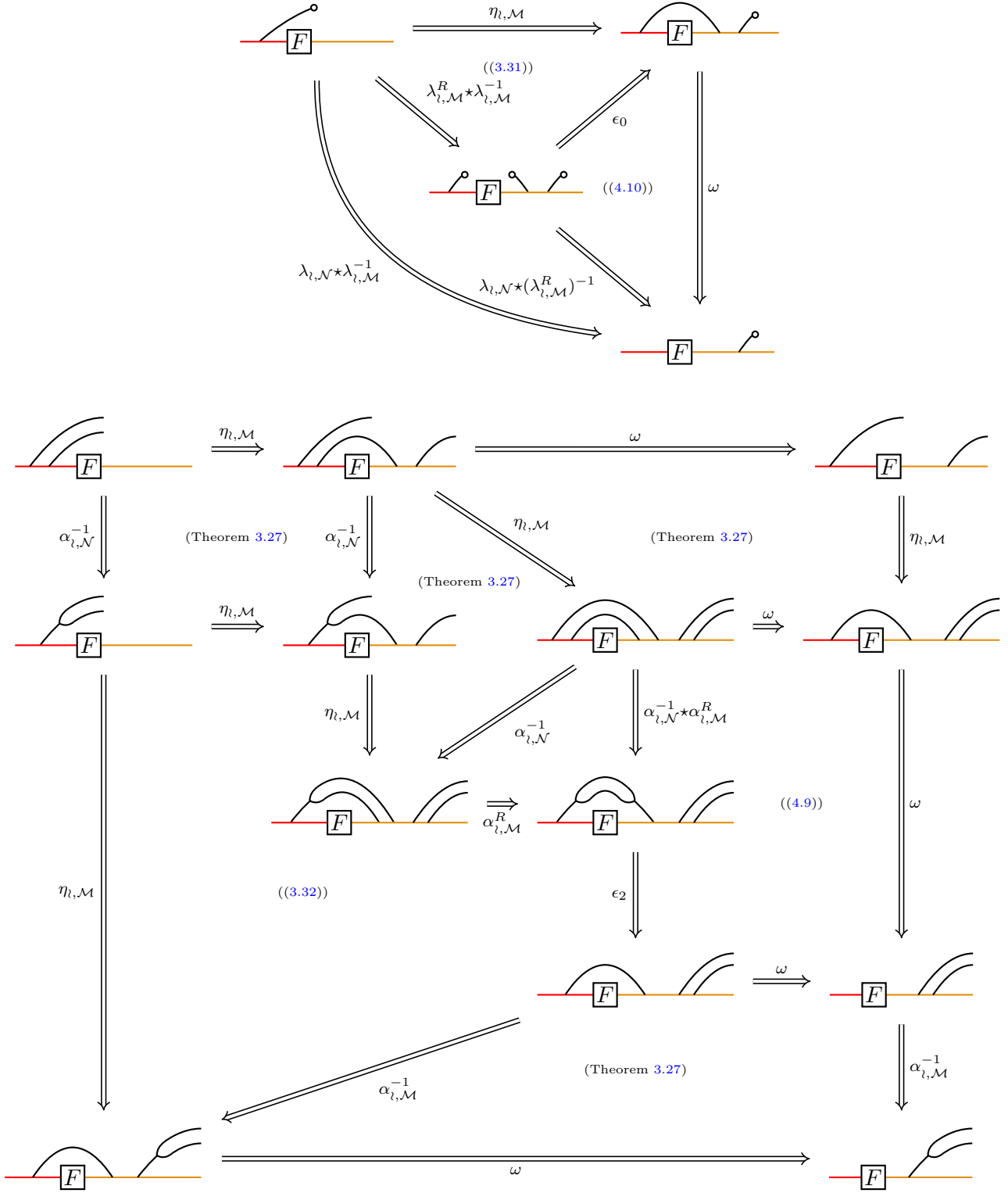

\[
    \ctikz{[scale=1.35, xscale=1.2]
        \mygrid{
            \draw[gray!30, step=0.5, opacity = 0.5] (0,0) grid (6,6);
            \foreach \x in {0,1,2,3,4,5,6}
                \draw (\x,6) node[above] {\small $\x$};
            \foreach \y in {0,1,2,3,4,5,6}
                \draw (6,\y) node[right] {\small $\y$};
        }
        \node (1_1) at (0,4){$
        \ctikz{[scale=0.5]
            \mygrid{ 
                \draw[gray!30, step=0.5, opacity = 0.5] (-3,-3) grid (3,3);
                \foreach \x in {-3,-2,-1,0,1,2,3}
                    \draw (\x,3) node[above] {\small $\x$};
                \foreach \y in {-1,0,1,2,3}
                    \draw (3,\y) node[right] {\small $\y$};
            }
            \draw (2,1.45) parabola (-1.35,0); 
            \draw[white,fill=white] (0.5,0)rectangle(2,1.7);
            \draw[fill=white] (0.5,1.15)circle[radius=0.1]; 
            \draw[color3] (3.2,0)--(0,0);
            \draw[color4] (0,0)--(-2,0);
            \mynode{0}{0}{0.4}{0.4}{F}
        }$};
        \node (1_3) at (4,4){$
        \ctikz{[scale=0.5]
            \mygrid{ 
                \draw[gray!30, step=0.5, opacity = 0.5] (-3,-3) grid (3,3);
                \foreach \x in {-3,-2,-1,0,1,2,3}
                    \draw (\x,3) node[above] {\small $\x$};
                \foreach \y in {-1,0,1,2,3}
                    \draw (3,\y) node[right] {\small $\y$};
            }
            \draw[color3] (3.35,0)--(0,0);
            \draw[color4] (0,0)--(-2,0);
            \mynode{0}{0}{0.4}{0.4}{F}
            \myparabola{1.5}{-1.35}{0}{1.35}{0}
            \draw (3.35,1.01) parabola (2,0);
            \draw[white,fill=white] (2.55,0.2)rectangle(3.35,1.1);
            \draw[fill=white] (2.55,0.6)circle[radius=0.1];      
        }$};
        \node (2_2) at (2,2){$
        \ctikz{[scale=0.5]
            \mygrid{ 
                \draw[gray!30, step=0.5, opacity = 0.5] (-3,-3) grid (3,3);
                \foreach \x in {-3,-2,-1,0,1,2,3}
                    \draw (\x,3) node[above] {\small $\x$};
                \foreach \y in {-1,0,1,2,3}
                    \draw (3,\y) node[right] {\small $\y$};
            }
            \myparabola{1.5}{-1.35}{0}{1.35}{0}
            \draw[white, fill=white] (-0.8,0)rectangle(0.8,1.1);
            \draw[fill=white](-0.8,0.6)circle[radius=0.1];
            \draw[fill=white](0.8,0.6)circle[radius=0.1];
            \draw (3.35,1.01) parabola (2,0);
            \draw[white,fill=white] (2.55,0.2)rectangle(3.35,1.1);
            \draw[fill=white] (2.55,0.6)circle[radius=0.1];   
            \draw[color3] (3.2,0)--(0,0);
            \draw[color4] (0,0)--(-2,0);
            \mynode{0}{0}{0.4}{0.4}{F}
        }$};
        \node (3_3) at (4,0){$
        \ctikz{[scale=0.5]
            \mygrid{ 
                \draw[gray!30, step=0.5, opacity = 0.5] (-3,-3) grid (3,3);
                \foreach \x in {-3,-2,-1,0,1,2,3}
                    \draw (\x,3) node[above] {\small $\x$};
                \foreach \y in {-1,0,1,2,3}
                    \draw (3,\y) node[right] {\small $\y$};
            }
            \draw[color3] (3.2,0)--(0,0);
            \draw[color4] (0,0)--(-2,0);
            \mynode{0}{0}{0.4}{0.4}{F}
            \draw (3.35,1.01) parabola (2,0);
            \draw[white,fill=white] (2.55,0.2)rectangle(3.35,1.1);
            \draw[fill=white] (2.55,0.6)circle[radius=0.1];   
        }$};
        \draw[nat] (1_1)--(1_3)node[midway, above]{\eta_{\wr,\cM}};
        \draw[nat] (1_3)--(3_3)node[midway, right]{\omega};
        \draw[nat] (2_2)--(3_3)node[midway, below left]{\lambda_{\wr,\cN}\star(\lambda_{\wr,\cM}^R)^{-1}};
        \draw[nat] (1_1)--(2_2)node[midway, above right]{\lambda_{\wr,\cM}^R\star\lambda_{\wr,\cM}^{-1}};
        \draw[nat] (1_1)..controls (0,1.5) and (1,0.5)..(3_3)node[midway, below left]{\lambda_{\wr,\cN}\star\lambda_{\wr,\cM}^{-1}};
        \draw[nat] (2_2)--(1_3)node[midway, below right]{\epsilon_0};
        \draw (3.25,2)node{\tiny(\eqref{eq.intuition_uni})};
        \draw (2,3.5)node{\tiny(\eqref{eq.unit_advanced})};
    }
\]
\[
    \hspace{-3pc}
    \ctikz{[scale=1.5,xscale=1.5]
        \mygrid{
            \draw[gray!30, step=0.5, opacity = 0.5] (0,0) grid (6,8);
            \foreach \x in {0,1,2,3,4,5,6}
                \draw (\x,-1) node[below] {\small $\x$};
            \foreach \y in {0,1,2,3,4,5,6,7,8}
                \draw (-1,\y) node[left] {\small $\y$};
        }
        \node (1_1) at (0,8){$
        \ctikz{[scale=0.5,baseline=0]
            \mygrid{ 
                \draw[gray!30, step=0.5, opacity = 0.5] (-5,-5) grid (5,5);
                \foreach \x in {-3,-2,-1,0,1,2,3}
                    \draw (\x,3) node[above] {\small $\x$};
                \foreach \y in {-1,0,1,2,3}
                    \draw (3,\y) node[right] {\small $\y$};
            }
            \draw[color3] (3.5,0)--(0,0);
            \draw[color4] (0,0)--(-2.5,0);
            \mynode{0}{0}{0.4}{0.4}{F}
            \draw (0.5,1.15) parabola (-1.35,0);
            \draw (0.5,1.65) parabola (-2,0); 
        }$};
        \node (1_2) at (2,8){$
        \ctikz{[scale=0.5,baseline=0]
            \mygrid{ 
                \draw[gray!30, step=0.5, opacity = 0.5] (-5,-5) grid (5,5);
                \foreach \x in {-3,-2,-1,0,1,2,3}
                    \draw (\x,3) node[above] {\small $\x$};
                \foreach \y in {-1,0,1,2,3}
                    \draw (3,\y) node[right] {\small $\y$};
            }
            \draw[color3] (3.35,0)--(0,0);
            \draw[color4] (0,0)--(-2.5,0);
            \mynode{0}{0}{0.4}{0.4}{F}
            \myparabola{1.5}{-1.35}{0}{1.35}{0}
            \draw (3.35,1.01) parabola (2,0);
            \draw (0.5,1.65) parabola (-2,0); 
        }$};
        \node (1_4) at (6,8){$
        \ctikz{[scale=0.5,baseline=0]
            \mygrid{ 
                \draw[gray!30, step=0.5, opacity = 0.5] (-5,-5) grid (5,5);
                \foreach \x in {-3,-2,-1,0,1,2,3}
                    \draw (\x,3) node[above] {\small $\x$};
                \foreach \y in {-1,0,1,2,3}
                    \draw (3,\y) node[right] {\small $\y$};
            }
            \draw[color3] (3.35,0)--(0,0);
            \draw[color4] (0,0)--(-2.5,0);
            \mynode{0}{0}{0.4}{0.4}{F}
            \draw (3.35,1.01) parabola (2,0);
            \draw (0.5,1.65) parabola (-2,0); 
        }$};
        \node (2_1) at (0,6){$
        \ctikz{[scale=0.5,baseline=0]
            \mygrid{ 
                \draw[gray!30, step=0.5, opacity = 0.5] (-5,-5) grid (5,5);
                \foreach \x in {-3,-2,-1,0,1,2,3}
                    \draw (\x,3) node[above] {\small $\x$};
                \foreach \y in {-1,0,1,2,3}
                    \draw (3,\y) node[right] {\small $\y$};
            }
            \draw (0.5,1.4) parabola (-1.65,0);
            \draw[white,fill=white] (-0.95,0.1)rectangle(0.55,1.75);
            \draw (0.5,1.15) parabola (-0.5,0.8);
            \draw (0.5,1.65) parabola (-0.75,1.2); 
            \draw (-1,0.74) to[out=90, in=220](-0.75,1.2);
            \draw (-1,0.74) to[out=-30, in=220](-0.5,0.8);
            \draw[color3] (3.5,0)--(0,0);
            \draw[color4] (0,0)--(-2.5,0);
            \mynode{0}{0}{0.4}{0.4}{F}
        }$};
        \node (2_2) at (2,6){$
        \ctikz{[scale=0.5,baseline=0]
            \mygrid{ 
                \draw[gray!30, step=0.5, opacity = 0.5] (-5,-5) grid (5,5);
                \foreach \x in {-3,-2,-1,0,1,2,3}
                    \draw (\x,3) node[above] {\small $\x$};
                \foreach \y in {-1,0,1,2,3}
                    \draw (3,\y) node[right] {\small $\y$};
            }
            \draw (0.5,1.4) parabola (-1.65,0);
            \draw[white,fill=white] (-0.95,0.1)rectangle(0.55,1.75);
            \draw (0.5,1.65) parabola (-0.75,1.2);
            \myparabola{1.2}{-0.5}{0.8}{1.35}{0}
            \draw (-1,0.74) to[out=90, in=220](-0.75,1.2);
            \draw (-1,0.74) to[out=-30, in=220](-0.5,0.8);
            \draw[color3] (3.35,0)--(0,0);
            \draw[color4] (0,0)--(-2.5,0);
            \mynode{0}{0}{0.4}{0.4}{F}
            \draw (3.35,1.01) parabola (2,0);
        }$};
        \node (2_3) at (4,6){$
        \ctikz{[scale=0.5,baseline=0]
            \mygrid{ 
                \draw[gray!30, step=0.5, opacity = 0.5] (-5,-5) grid (5,5);
                \foreach \x in {-3,-2,-1,0,1,2,3}
                    \draw (\x,3) node[above] {\small $\x$};
                \foreach \y in {-1,0,1,2,3}
                    \draw (3,\y) node[right] {\small $\y$};
            }
            \draw[color3] (4.35,0)--(0,0);
            \draw[color4] (0,0)--(-2.3,0);
            \mynode{0}{0}{0.4}{0.4}{F}
            \myparabola{1.5}{-1.15}{0}{1.15}{0}
            \myparabola{1.5}{-1.8}{0}{1.8}{0}
            \draw (4.35,1.01) parabola (3,0);
            \draw (4.35,1.4) parabola (2.5,0);
        }$};
        \node (2_4) at (6,6){$
        \ctikz{[scale=0.5,baseline=0]
            \mygrid{ 
                \draw[gray!30, step=0.5, opacity = 0.5] (-5,-5) grid (5,5);
                \foreach \x in {-3,-2,-1,0,1,2,3}
                    \draw (\x,3) node[above] {\small $\x$};
                \foreach \y in {-1,0,1,2,3}
                    \draw (3,\y) node[right] {\small $\y$};
            }
            \draw[color3] (4.35,0)--(0,0);
            \draw[color4] (0,0)--(-2.3,0);
            \mynode{0}{0}{0.4}{0.4}{F}
            \myparabola{1.5}{-1.35}{0}{1.35}{0}
            \draw (4.35,1.01) parabola (3,0);
            \draw (4.35,1.4) parabola (2.5,0);
        }$};
        \node (3_2) at (2,4){$
        \ctikz{[scale=0.5,baseline=0]
            \mygrid{ 
                \draw[gray!30, step=0.5, opacity = 0.5] (-5,-5) grid (5,5);
                \foreach \x in {-3,-2,-1,0,1,2,3}
                    \draw (\x,3) node[above] {\small $\x$};
                \foreach \y in {-1,0,1,2,3}
                    \draw (3,\y) node[right] {\small $\y$};
            }
            \draw (0.5,1.4) parabola (-1.65,0);
            \draw[white,fill=white] (-0.95,0.1)rectangle(0.55,1.75);
            \myparabola{1.2}{-0.75}{1.2}{1.85}{0}
            \myparabola{1.2}{-0.5}{0.8}{1.35}{0}
            \draw (-1,0.74) to[out=90, in=220](-0.75,1.2);
            \draw (-1,0.74) to[out=-30, in=220](-0.5,0.8);
            \draw[color3] (4.35,0)--(0,0);
            \draw[color4] (0,0)--(-2.3,0);
            \mynode{0}{0}{0.4}{0.4}{F}
            \draw (4.35,1.01) parabola (3,0);
            \draw (4.35,1.4) parabola (2.5,0);
        }$};
        \node (3_3) at (4,4){$
        \ctikz{[scale=0.5,baseline=0]
            \mygrid{ 
                \draw[gray!30, step=0.5, opacity = 0.5] (-5,-5) grid (5,5);
                \foreach \x in {-3,-2,-1,0,1,2,3}
                    \draw (\x,3) node[above] {\small $\x$};
                \foreach \y in {-1,0,1,2,3}
                    \draw (3,\y) node[right] {\small $\y$};
            }
            \draw (0.5,1.4) parabola (-1.65,0);
            \begin{scope}[xscale=-1]
                \draw (0.5,1.4) parabola (-1.65,0); 
            \end{scope}
            \draw[white,fill=white] (-0.95,0.1)rectangle(0.95,1.75);
            \myparabola{0.9}{-0.75}{1.2}{0.75}{1.2}
            \myparabola{1.1}{-0.5}{0.8}{0.5}{0.8}
            \draw (-1,0.74) to[out=90, in=220](-0.75,1.2);
            \draw (-1,0.74) to[out=-30, in=220](-0.5,0.8);
            \begin{scope}[xscale=-1]
                \draw (-1,0.74) to[out=90, in=220](-0.75,1.2);
                \draw (-1,0.74) to[out=-30, in=220](-0.5,0.8);
            \end{scope}
            \draw[color3] (4.35,0)--(0,0);
            \draw[color4] (0,0)--(-2.3,0);
            \mynode{0}{0}{0.4}{0.4}{F}
            \draw (4.35,1.01) parabola (3,0);
            \draw (4.35,1.4) parabola (2.5,0);
        }$};
        \node (4_3) at (4,2){$
        \ctikz{[scale=0.5,baseline=0]
            \mygrid{ 
                \draw[gray!30, step=0.5, opacity = 0.5] (-5,-5) grid (5,5);
                \foreach \x in {-3,-2,-1,0,1,2,3}
                    \draw (\x,3) node[above] {\small $\x$};
                \foreach \y in {-1,0,1,2,3}
                    \draw (3,\y) node[right] {\small $\y$};
            }
            \draw[color3] (4.35,0)--(0,0);
            \draw[color4] (0,0)--(-2.3,0);
            \mynode{0}{0}{0.4}{0.4}{F}
            \myparabola{1.5}{-1.35}{0}{1.35}{0}
            \draw (4.35,1.01) parabola (3,0);
            \draw (4.35,1.4) parabola (2.5,0);
        }$};
        \node (4_4) at (6,2){$
        \ctikz{[scale=0.5,baseline=0]
            \mygrid{ 
                \draw[gray!30, step=0.5, opacity = 0.5] (-5,-5) grid (5,5);
                \foreach \x in {-3,-2,-1,0,1,2,3}
                    \draw (\x,3) node[above] {\small $\x$};
                \foreach \y in {-1,0,1,2,3}
                    \draw (3,\y) node[right] {\small $\y$};
            }
            \draw[color3] (4.35,0)--(1,0);
            \draw[color4] (1,0)--(-0.5,0);
            \mynode{1}{0}{0.4}{0.4}{F}
            \draw (4.35,1.01) parabola (3,0);
            \draw (4.35,1.4) parabola (2.5,0);
        }$};
        \node (5_1) at (0,0){$
        \ctikz{[scale=0.5,baseline=0]
            \mygrid{ 
                \draw[gray!30, step=0.5, opacity = 0.5] (-5,-5) grid (5,5);
                \foreach \x in {-3,-2,-1,0,1,2,3}
                    \draw (\x,3) node[above] {\small $\x$};
                \foreach \y in {-1,0,1,2,3}
                    \draw (3,\y) node[right] {\small $\y$};
            }
            \draw[color3] (4.35,0)--(0,0);
            \draw[color4] (0,0)--(-2.3,0);
            \mynode{0}{0}{0.4}{0.4}{F}
            \myparabola{1.5}{-1.35}{0}{1.35}{0}
            \begin{scope}[xshift=3.85cm]
                \draw (0.5,1.4) parabola (-1.65,0);
                \draw[white,fill=white] (-0.95,0.1)rectangle(0.55,1.75);
                \draw (0.5,1.15) parabola (-0.5,0.8);
                \draw (0.5,1.65) parabola (-0.75,1.2); 
                \draw (-1,0.74) to[out=90, in=220](-0.75,1.2);
                \draw (-1,0.74) to[out=-30, in=220](-0.5,0.8);
            \end{scope}
        }$};
        \node (5_4) at (6,0){$
        \ctikz{[scale=0.5,baseline=0]
            \mygrid{ 
                \draw[gray!30, step=0.5, opacity = 0.5] (-5,-5) grid (5,5);
                \foreach \x in {-3,-2,-1,0,1,2,3}
                    \draw (\x,3) node[above] {\small $\x$};
                \foreach \y in {-1,0,1,2,3}
                    \draw (3,\y) node[right] {\small $\y$};
            }
            \draw[color3] (4.35,0)--(1,0);
            \draw[color4] (1,0)--(-0.5,0);
            \mynode{1}{0}{0.4}{0.4}{F}
            \begin{scope}[xshift=3.85cm]
                \draw (0.5,1.4) parabola (-1.65,0);
                \draw[white,fill=white] (-0.95,0.1)rectangle(0.55,1.75);
                \draw (0.5,1.15) parabola (-0.5,0.8);
                \draw (0.5,1.65) parabola (-0.75,1.2); 
                \draw (-1,0.74) to[out=90, in=220](-0.75,1.2);
                \draw (-1,0.74) to[out=-30, in=220](-0.5,0.8);
            \end{scope}
        }$};
        \draw[nat] (1_1)--(1_2)node[midway, above]{\eta_{\wr,\cM}};
        \draw[nat] (1_2)--(1_4)node[midway, above]{\omega};
        \draw[nat] (1_4)--(2_4)node[midway, right]{\eta_{\wr,\cM}};
        \draw[nat] (2_4)--(4_4)node[midway, right]{\omega};
        \draw[nat] (4_4)--(5_4)node[midway, right]{\alpha_{\wr,\cM}^{-1}};
        \draw[nat] (5_1)--(5_4)node[midway, below]{\omega};
        \draw[nat] (1_1)--(2_1)node[midway, left]{\alpha_{\wr,\cN}^{-1}};
        \draw[nat] (2_1)--(5_1)node[midway, left]{\eta_{\wr,\cM}};
        \draw[nat] (2_1)--(2_2)node[midway, above]{\eta_{\wr,\cM}};
        \draw[nat] (2_2)--(3_2)node[midway, left]{\eta_{\wr,\cM}};
        \draw[nat] (3_2)--(3_3)node[midway, below]{\alpha_{\wr,\cM}^R};
        \draw[nat] (1_2)--(2_2)node[midway, left]{\alpha_{\wr,\cN}^{-1}};
        \draw[nat] (1_2)--(2_3)node[midway, above right]{\eta_{\wr,\cM}};
        \draw[nat] (2_3)--(2_4)node[midway, above]{\omega};
        \draw[nat] (2_3)--(3_2)node[midway, below right]{\alpha_{\wr,\cN}^{-1}};
        \draw[nat] (2_3)--(3_3)node[midway, right]{\alpha_{\wr,\cN}^{-1}\star\alpha_{\wr,\cM}^R};
        \draw[nat] (3_3)--(4_3)node[midway, left]{\epsilon_2};
        \draw[nat] (4_3)--(4_4)node[midway, above]{\omega};
        \draw[nat] (4_3)--(5_1)node[midway, below right]{\alpha_{\wr,\cM}^{-1}};
        \draw (1,7)node{\tiny (Theorem \ref{thm.local})};
        \draw (2.75,6.5)node{\tiny (Theorem \ref{thm.local})};
        \draw (1.5,3)node{\tiny (\eqref{eq.mult_advanced})};
        \draw (4,1)node{\tiny (Theorem \ref{thm.local})};
        \draw (4.5,7)node{\tiny (Theorem \ref{thm.local})};
        \draw (5.25,4)node{\tiny(\eqref{eq.intuition_asso})};
    }
\]
\caption{Verification of the lax $\cC$-module profunctor structure of $(F,\beta)$.}
\label{fig.omega_to_modprof}
\end{figure}

The proof that the functors $\Phi$ and $\Psi$ are mutually inverses to each other assembles that of Proposition \ref{prp.transpose} and is omitted.

The equivalence between \ref{item1.thm.unenriched_Kitaev_Kong} and \ref{item2.thm.unenriched_Kitaev_Kong} can be similarly given as above, making use of Theorem \ref{thm.higher_transpose}; a full proof is provided in Appendix \ref{sec.graph.app}.
\end{proof}
\end{theorem}

To take full advantage of the strength of enriched category theory, we record the following result.
\begin{theorem}[Enriched version of Theorem \ref{thm.unenriched_Kitaev_Kong}]\label{thm.gen_Kitaev_Kong}
    The three categories in Theorem \ref{thm.unenriched_Kitaev_Kong} are all naturally $\cV$-enriched. Denote the corresponding $\cV$-category by 
    \[
        \bLmd_{\MnCN},\;\bLmd_{\MCN}(\ast) \quad\text{and}\quad {}_\cC\bvprof^\lax(\cM,\cN)
    \]
    respectively. Then these $\cV$-categories are all $\cV$-equivalent:
    \[
        \bLmd_{\MnCN}\cong\bLmd_{\MCN}(*)\cong  {}_\cC\bvprof^\lax(\cM,\cN).
    \]
\end{theorem}
See Section \ref{app.enriched_structure} for a sketch proof of Theorem \ref{thm.gen_Kitaev_Kong}.



\begin{remark}\label{rmk.Tambara}
    The method employed in the proof of Theorem \ref{thm.unenriched_Kitaev_Kong} can be readily applied to prove the equivalences among the following five categories, including the three in Theorem \ref{thm.unenriched_Kitaev_Kong}:
    \begin{enumerate}
        \item \label{item1.rmk.Tambara} The category $\Lmd_{\MCN}(\ast)$ of $\MCN$-modules from $\ast$.
        \item \label{item2.rmk.Tambara} The category $\Lmd_{\MnCN}$ of $\MnCN$-modules.
        \item \label{item3.rmk.Tambara} The category ${}_\cC\vprof^\lax(\cM,\cN)$ of lax $\cC$-module profunctors $\cM\arprof\cN$.
        \item \label{item4.rmk.Tambara} The category ${}_\cC\vprof^{\co,\oplax}(\cM,\cN)$ of \emph{oplax $\cC$-comodule profunctors from $\cM$ to $\cN$}.
        \item \label{item5.rmk.Tambara} The category ${}_\cC\LTamb(\cM,\cN)$ of \emph{left Tambara modules from $\cM$ to $\cN$}.
\end{enumerate}
\begin{figure}[htbp]
\eqnn{\label{eq.oplax_1}
\diagram{
    \begin{tikzpicture}[scale=0.5,xscale=-1,baseline=0]
        \mygrid{ 
            \draw[gray!30, step=0.5, opacity = 0.5] (-3,-3) grid (3,3);
            \foreach \x in {-3,-2,-1,0,1,2,3}
                \draw (\x,3) node[above] {\small $\x$};
            \foreach \y in {-1,0,1,2,3}
                \draw (3,\y) node[right] {\small $\y$};
        }
        \draw (2,1.45) parabola (-1.35,0); 
        \draw[white,fill=white] (0.5,0)rectangle(2,1.7);
        \draw[fill=white] (0.5,1.15)circle[radius=0.1]; 
        \draw[color4] (2,0)--(0,0);
        \draw[color3] (0,0)--(-2,0);
        \mynode{0}{0}{0.4}{0.4}{F}
    \end{tikzpicture}
    \ar@{=>}[rr]^-\gamma  \ar@{=>}[rd]_-{(\lambda_{\wr,\cM}^R)^{-1}\hspace{0.5em}} & &
    \begin{tikzpicture}[scale=0.5,xscale=-1,baseline=0]
        \mygrid{ 
            \draw[gray!30, step=0.5, opacity = 0.5] (-3,-3) grid (3,3);
            \foreach \x in {-3,-2,-1,0,1,2,3}
                \draw (\x,3) node[above] {\small $\x$};
            \foreach \y in {-1,0,1,2,3}
                \draw (3,\y) node[right] {\small $\y$};
        }
        \draw[color4] (2,0)--(0,0);
        \draw[color3] (0,0)--(-2,0);
        \mynode{0}{0}{0.4}{0.4}{F}
        \draw (2.7,0.7) parabola (1.35,0);
        \draw[white,fill=white] (1.9,0.2)rectangle(2.7,1.1);
        \draw[fill=white] (1.9,0.45)circle[radius=0.1];
    \end{tikzpicture}
    \ar@{=>}[ld]^-{(\lambda_{\wr,\cN}^R)^{-1}}
    \\
    &
    \begin{tikzpicture}[scale=0.5,baseline=0]
        \draw[color4] (-2,0)--(0,0);
        \draw[color3] (0,0)--(2,0);
        \mynode{0}{0}{0.4}{0.4}{F}
    \end{tikzpicture}
}}
\eqnn{\label{eq.oplax_2}
\begin{tikzpicture}[scale = 1.2,xscale=1.5]
    \node (A) at (0,2)
        {\begin{tikzpicture}
            [scale = 0.5, xscale=-1, baseline = (current bounding box.center)]
            \mygrid{ 
                \draw[gray!30, step=0.5, opacity = 0.5] (-5,-5) grid (5,5);
                \foreach \x in {-3,-2,-1,0,1,2,3}
                    \draw (\x,3) node[above] {\small $\x$};
                \foreach \y in {-1,0,1,2,3}
                    \draw (3,\y) node[right] {\small $\y$};
            }
            \draw[color4] (2.35,0)--(0,0);
            \draw[color3] (0,0)--(-2.5,0);
            \mynode{0}{0}{0.4}{0.4}{F}
            \draw (0.5,1.15) parabola (-1.35,0);
            \draw (0.5,1.65) parabola (-2,0); 
        \end{tikzpicture}};
    \node (B) at (2,2){\begin{tikzpicture}[scale = 0.5, xscale=-1, baseline = (current bounding box.center)]
            \mygrid{ 
                \draw[gray!30, step=0.5, opacity = 0.5] (-5,-5) grid (5,5);
                \foreach \x in {-3,-2,-1,0,1,2,3}
                    \draw (\x,3) node[above] {\small $\x$};
                \foreach \y in {-1,0,1,2,3}
                    \draw (3,\y) node[right] {\small $\y$};
            }
            \draw[color4] (2.85,0)--(0,0);
            \draw[color3] (0,0)--(-2,0);
            \mynode{0}{0}{0.4}{0.4}{F}
            \draw (2.85,1.01) parabola (1.5,0);
            \draw (0.5,1.35) parabola (-1.35,0); 
        \end{tikzpicture}};
    \node (D) at (4,2)
            {\begin{tikzpicture}[scale = 0.5, xscale=-1, baseline = (current bounding box.center)]
            \mygrid{ 
                \draw[gray!30, step=0.5, opacity = 0.5] (-5,-5) grid (5,5);
                \foreach \x in {-3,-2,-1,0,1,2,3}
                    \draw (\x,3) node[above] {\small $\x$};
                \foreach \y in {-1,0,1,2,3}
                    \draw (3,\y) node[right] {\small $\y$};
            }
            \draw[color4] (4.35,0)--(1,0);
            \draw[color3] (1,0)--(-0.5,0);
            \mynode{1}{0}{0.4}{0.4}{F}
            \draw (4.35,1.01) parabola (3,0);
            \draw (4.35,1.4) parabola (2.5,0);
        \end{tikzpicture}};  
    \node (C) at (0,0)
         {\begin{tikzpicture}[scale = 0.5, xscale=-1, baseline = (current bounding box.center)]
            \draw (0.5,1.4) parabola (-1.65,0);
            \draw[white,fill=white] (-0.95,0.1)rectangle(0.55,1.75);
            \draw (0.5,1.15) parabola (-0.5,0.8);
            \draw (0.5,1.65) parabola (-0.75,1.2); 
            \draw (-1,0.74) to[out=90, in=220](-0.75,1.2);
            \draw (-1,0.74) to[out=-30, in=220](-0.5,0.8);
            \draw[color4] (2.35,0)--(0,0);
            \draw[color3] (0,0)--(-2.5,0);
            \mynode{0}{0}{0.4}{0.4}{F}
        \end{tikzpicture}};   
    \node (E) at (4,0) {\begin{tikzpicture}[scale = 0.5, xscale=-1, baseline = (current bounding box.center)]
            \mygrid{ 
                \draw[gray!30, step=0.5, opacity = 0.5] (-5,-5) grid (5,5);
                \foreach \x in {-3,-2,-1,0,1,2,3}
                    \draw (\x,3) node[above] {\small $\x$};
                \foreach \y in {-1,0,1,2,3}
                    \draw (3,\y) node[right] {\small $\y$};
            }
            \draw[color4] (4.35,0)--(1,0);
            \draw[color3] (1,0)--(-0.5,0);
            \mynode{1}{0}{0.4}{0.4}{F}
            \begin{scope}[xshift=3.85cm]
                \draw (0.5,1.4) parabola (-1.65,0);
                \draw[white,fill=white] (-0.95,0.1)rectangle(0.55,1.75);
                \draw (0.5,1.15) parabola (-0.5,0.8);
                \draw (0.5,1.65) parabola (-0.75,1.2); 
                \draw (-1,0.74) to[out=90, in=220](-0.75,1.2);
                \draw (-1,0.74) to[out=-30, in=220](-0.5,0.8);
            \end{scope}
        \end{tikzpicture}};
        \draw[nat] (A)--(B) node[midway, above]{\gamma};
        \draw[nat] (A)--(C) node[midway, left]{(\alpha_{\wr,\cM}^R)^{-1}};
        \draw[nat] (C)--(E) node[midway, below]{\gamma};
        \draw[nat] (D)--(E) node[midway, right]{(\alpha_{\wr,\cN}^R)^{-1}};
        \draw[nat] (B)--(D) node[midway, above]{\gamma};
    \end{tikzpicture}
}
    \eqnn{\label{eq.Tambara_1}
    \diagram{
        \ctikz{[scale=0.5]
            \draw[color4] (-2,0)--(0,0);
            \draw[color3] (0,0)--(2,0);
            \mynode{0}{0}{0.4}{0.4}{F}
        }
        \ar@{=>}[rr]^-{\eta_0} \ar@{=>}[rd]_-{\lambda_{\wr,\cN}^R\star\lambda_{\wr,\cM}^{-1}}
        & &
        \ctikz{[scale=0.5]
            \draw[color4] (-2,0)--(0,0);
            \draw[color3] (0,0)--(2,0);
            \mynode{0}{0}{0.4}{0.4}{F}
            \draw (-2,1.01)--(2,1.01);
            \draw[fill=white] (-2,1.01)circle[radius=0.1];
            \draw[fill=white] (2,1.01)circle[radius=0.1];
        }
        \ar@{=>}[ld]^-{\theta}
        \\
        &
        \ctikz{[scale=0.5]
            \draw (2.5,1.01) parabola (1.15,0);
            \draw (-2.5,1.01) parabola (-1.15,0);
            \draw[white, fill=white] (1.7,-0.1)rectangle(2.5,1.1);
            \draw[white, fill=white] (-1.7,-0.1)rectangle(-2.5,1.1);
            \draw[fill=white] (1.7,0.67)circle[radius=0.1];
            \draw[fill=white] (-1.7,0.67)circle[radius=0.1];
            \draw[color4] (-2,0)--(0,0);
            \draw[color3] (0,0)--(2,0);
            \mynode{0}{0}{0.4}{0.4}{F}
        }
    }
    }
    \eqnn{\label{eq.Tambara_2}
        \ctikz{[scale=1.25,xscale=1.5]
            \node (1_1) at (0,2){$
            \ctikz{[scale=0.5]
                \draw[color3] (2.5,0)--(0,0);
                \draw[color4] (0,0)--(-2.5,0);
                \draw (2.5,1.45)--(-2.5,1.45);
                \draw (2.5,0.72)--(-2.5,0.72);
                \mynode{0}{0}{0.4}{0.4}{F}
            }$};
            \node (1_2) at (2,2){$
            \ctikz{[scale=0.5]
                \draw[color3] (2.5,0)--(0,0);
                \draw[color4] (0,0)--(-2.5,0);
                \draw (2.5,1.45)--(-2.5,1.45);
                \draw (2.5,0.81) parabola (1.35,0);
                \draw (-2.5,0.81) parabola (-1.35,0);
                \mynode{0}{0}{0.4}{0.4}{F}
            }$};
            \node (1_3) at (4,2){$
            \ctikz{[scale=0.5]
                \draw[color3] (2.5,0)--(0,0);
                \draw[color4] (0,0)--(-2.5,0);
                \draw (2.5,0.81) parabola (1.35,0);
                \draw (-2.5,0.81) parabola (-1.35,0);
                \draw (2.5,1.35) parabola (0.8,0);
                \draw (-2.5,1.35) parabola (-0.8,0);
                \mynode{0}{0}{0.4}{0.4}{F}
            }$};
            \node (2_1) at (0,0){$
            \ctikz{[scale=0.5]
                \mygrid{ 
                    \draw[gray!30, step=0.5, opacity = 0.5] (-5,-5) grid (5,5);
                    \foreach \x in {-3,-2,-1,0,1,2,3}
                        \draw (\x,3) node[above] {\small $\x$};
                    \foreach \y in {-1,0,1,2,3}
                        \draw (3,\y) node[right] {\small $\y$};
                }
                \draw[color3] (2.5,0)--(0,0);
                \draw[color4] (0,0)--(-2.5,0);
                \draw (1.25,1.23)--(-1.25,1.23);
                \draw (1.25,1.23)to[out = -50, in = 180](2.5,0.91);
                \draw (1.25,1.23)to[out = 50, in = 180](2.5,1.55);
                \begin{scope}[xscale=-1]
                    \draw (1.25,1.23)to[out = -50, in = 180](2.5,0.91);
                    \draw (1.25,1.23)to[out = 50, in = 180](2.5,1.55);                    
                \end{scope}
                \mynode{0}{0}{0.4}{0.4}{F}
            }$};
            \node (2_3) at (4,0){$
            \ctikz{[scale=0.5]
                \begin{scope}[xshift=2.5cm]
                    \draw (0,1.3) parabola (-1.75,0);
                    \draw[white,fill=white] (-1.15,0)rectangle(0.1,1.5);
                    \draw (-1.2,0.725) to[out=90, in=230](-1,1.3);
                    \draw (-1.2,0.725) to[out=-20, in = 220](-0.7,0.8);
                    \draw (0,1.2) parabola (-0.7,0.8);
                    \draw (0,1.8) parabola (-1,1.3);
                \end{scope}
                \begin{scope}[xshift=-2.5cm,xscale=-1]
                    \draw (0,1.3) parabola (-1.75,0);
                    \draw[white,fill=white] (-1.15,0)rectangle(0.1,1.5);
                    \draw (-1.2,0.725) to[out=90, in=230](-1,1.3);
                    \draw (-1.2,0.725) to[out=-20, in = 220](-0.7,0.8);
                    \draw (0,1.2) parabola (-0.7,0.8);
                    \draw (0,1.8) parabola (-1,1.3);
                \end{scope}
                \draw[color3] (2.5,0)--(0,0);
                \draw[color4] (0,0)--(-2.5,0);
                \mynode{0}{0}{0.4}{0.4}{F}
            }$};
            \draw[nat] (1_1)--(1_2)node[midway, above]{\theta};
            \draw[nat] (1_2)--(1_3)node[midway, above]{\theta};
            \draw[nat] (2_1)--(2_3)node[midway, below]{\theta};
            \draw[nat] (1_3)--(2_3)node[midway, right]{\alpha_{\wr,\cN}^R\star\alpha_{\wr,\cM}^{-1}};
            \draw[nat] (1_1)--(2_1)node[midway, left]{\eta_2};
        }
    }
    \caption{Axioms satisfied by an oplax $\cC$-module profunctor and a left Tambara module}
    \label{fig.oplax_co_and_Tambara}
\end{figure}
In above, an oplax $\cC$-comodule profunctor from $\cM$ to $\cN$ is a profunctor $F\:\cM\arprof\cN$ equipped with a profunctor homomorphism
        \[
\gamma\:
\begin{tikzpicture}[scale=0.5,baseline=0]
    \draw[color3] (2,0)--(0,0);
    \draw[color4] (0,0)--(-2,0);
    \mynode{0}{0}{0.4}{0.4}{F}
    \draw (-2,1.65) parabola (1.35,0); 
\end{tikzpicture}
\Rightarrow
\begin{tikzpicture}[scale=0.5,baseline=0]
    \mygrid{ 
        \draw[gray!30, step=0.5, opacity = 0.5] (-3,-3) grid (3,3);
        \foreach \x in {-3,-2,-1,0,1,2,3}
            \draw (\x,3) node[above] {\small $\x$};
        \foreach \y in {-1,0,1,2,3}
            \draw (3,\y) node[right] {\small $\y$};
    }
    \draw (-2.5,1.01) parabola (-1.15,0);
    \draw[color3] (2,0)--(0,0);
    \draw[color4] (0,0)--(-2.5,0);
    \mynode{0}{0}{0.4}{0.4}{F}
\end{tikzpicture}
        \]
        satisfying conditions \eqref{eq.oplax_1} and \eqref{eq.oplax_2} in Figure \ref{fig.oplax_co_and_Tambara}. A morphism between two oplax $\cC$-comodule profunctors is evidently defined.
        A left Tambara module from $\cM$ to $\cN$ is a profunctor $F\:\cM\arprof\cN$ equipped with a profunctor homomorphism
        \[
            \theta\:\ctikz{[scale=0.5]
                \draw[color4] (-2,0)--(0,0);
                \draw[color3] (0,0)--(2,0);
                \mynode{0}{0}{0.4}{0.4}{F}
                \draw (-2,1.01)--(2,1.01);
            }
            \Rightarrow
            \ctikz{[scale=0.5]
                \draw (2.5,1.01) parabola (1.15,0);
                \draw (-2.5,1.01) parabola (-1.15,0);
                \draw[color4] (-2.4,0)--(0,0);
                \draw[color3] (0,0)--(2.4,0);
                \mynode{0}{0}{0.4}{0.4}{F}
            }
        \]
        satisfying the ``crab-like'' conditions \eqref{eq.Tambara_1} and \eqref{eq.Tambara_2} in Figure \ref{fig.oplax_co_and_Tambara}.\footnote{Note that the more familiar definition of Tambara module in \cite[Definition 4.1]{Clarke_2024}(see also \cite[\S 2]{Tambara2006DistributorsOA} for the original $\Vect$-enriched version) differs from ours, however, the two are equivalent by an application of Yoneda lemma.} A morphism between two left Tambara modules is evidently defined.

    The equivalences among the five categories are given briefly as follows. By Proposition \ref{prp.transpose} we have canonical transposing isomorphisms
    \eqnn{\label{eq.transport.rmk.Tambara}
    \begin{split}
        & \phantom{\cong}\Hom((\boxdot_\cN)_\ast\bullet(\Id_\cC\os F),F\bullet (\boxdot_\cM)_\ast)
        \cong \Hom((\boxdot_\cN)_\ast\bullet(\Id_\cC\os F)\bullet (\boxdot_\cM)^\ast,F)\\
        & \cong \Hom((\Id_\cC\os F)\bullet(\boxdot_\cM)^\ast,(\boxdot_\cN)^\ast\bullet F) 
        \cong\Hom(\Id_\cC\os F,(\boxdot_\cN)^\ast\bullet F\bullet (\boxdot_\cM)_\ast)
    \end{split}
    }
    for any profunctor $\cM\arprof\cN$. Given $(F,\beta)\in{}_\cC\vprof^\lax(\cM,\cN)$, transporting $\beta$ across the isomorphisms in \eqref{eq.transport.rmk.Tambara} (note that $\beta$ lies in first hom set) give the corresponding the corresponding $\MnCN$-module, oplax $\cC$-module profunctor, and left Tambara module, and conversely. This establish the equivalences among categories \ref{item2.rmk.Tambara}-\ref{item4.rmk.Tambara} listed in the beginning of this remark, hence by Theorem \ref{thm.unenriched_Kitaev_Kong} all the five categories are equivalent.
    
    Moreover, one can check that the above equivalences lift to $\cV$-equivalences 
    \eqnn{\label{eq.five_equiv.rmk.Tambara}
        \begin{multlined}
            \bLmd_{\MCN}(\ast)\cong\bLmd_{\MnCN}\cong{}_\cC\bvprof^\lax(\cM,\cN)\\
            \cong{}_\cC\bvprof^{\co,\oplax}(\cM,\cN)\cong{}_\cC\bLTamb(\cM,\cN),
        \end{multlined}
    }
    where ${}_\cC\bvprof^{\co,\oplax}(\cM,\cN)$ and $_\cC\bLTamb(\cM,\cN)$ are respectively the $\cV$-enriched version of ${}_\cC\vprof^{\co,\oplax}$ and ${}_\cC\LTamb(\cM,\cN)$, constructed similarly as ${}_\cC\bvprof^\lax(\cM,\cN)$ and $\bLmd_{\MnCN}$. We leave it to the reader to fill in the missing details.

    

\end{remark}
\begin{remark}
Below are more comments on the literature on Tambara modules:
    \begin{itemize}
        \item  \eqref{eq.five_equiv.rmk.Tambara} recovers the result of \cite{pastro2007doublesmonoidalcategories, Tambara2006DistributorsOA}, that ${}_{\cC\os\cC^\rev}\bLTamb(\cC,\cC)$ is monoidally equivalent to the center of $\hat{\cC}$ when $\cC$ is rigid. (See also \eqref{eq.equiBetweeenVProf&Z}, Theorem~\ref{thm.mon_Kitaev_Kong} and Remark~\ref{rmk.rigidCModuleProf}.) The graphical proof provided here complements the original algebraic proofs provided by \cite{pastro2007doublesmonoidalcategories, Tambara2006DistributorsOA}.
        \item \cite{Tambara2006DistributorsOA} obtains an equivalence between the third and the fifth $\cV$-categories in \eqref{eq.five_equiv.rmk.Tambara} under the assumption that $\cC$ is left rigid.
        \item \cite{Clarke_2024} obtains the equivalence between the first and the fifth $\cV$-categories in \eqref{eq.five_equiv.rmk.Tambara}.
    \end{itemize}
\end{remark}

Let us now consider the three players in Theorem \ref{thm.unenriched_Kitaev_Kong} in the case $\cN=\cM$. $\Lmd_{\MnCM}$ is a monoidal category,  
with the tensor product $\conv$~\eqref{eq.tensor_of_Omega_module} encoding the fusion of defects. It is tempting to ask whether ${}_\cC\vprof^\lax(\cM,\cM)$ and $\Lmd_{\MCM}(\ast)$ also has a natural monoidal structure making the equivalence in Theorem \ref{thm.unenriched_Kitaev_Kong} is also monoidal, so that the tensor product and the unit of ${}_\cC\vprof^\lax(\cM,\cM)$ and $\Lmd_{\MCM}(\ast)$ captures the fusion of defects and vacuum defects, respectively. The answer turn out to be yes. Let us begin with the monoidal structures on ${}_\cC\vprof^\lax(\cM,\cM)$ and $\Lmd_{\MCM}(\ast)$:
\begin{itemize}
    \item ${}_\cC\vprof^\lax(\cM,\cM)$ has a monoidal structure with the tensor product $\circ$ given by composition of lax $\cC$-module profunctors, and tensor unit given by the identity profunctor $\Id_\cM$.
    \item The category $\Lmd_{\MCM}(\ast)$ is also monoidal with monoidal structure given by Proposition \ref{prp.bimonad_mod}. Indeed, define 
\eqnn{\label{eq.probimonad_of_MCM_1}
    M\defdtobe\ctikz{[scale=0.667]
        \draw[color3] (0,0)--(-1,0);
        \draw[color3] (-1,0) arc(270:90:0.5);
        \draw[color3] (-1,1)--(0,1);
        \draw[color3] (0,1.5)--(-2,1.5);
        \draw[color3] (0,-0.5)--(-2,-0.5);
    },\quad
    \tau\defdtobe(
    \diagram@C=2pc{
    \ctikz{[scale=0.333]
        \mygrid{ 
            \draw[gray!30, step=0.5, opacity = 0.5] (-6,-4) grid (6,4);
            \foreach \x in {-6,-5,-4,-3,-2,-1,0,1,2,3,4,5,6}
                \draw (\x,3) node[above] {\small $\x$};
            \foreach \y in {-1,0,1,2,3,4}
                \draw (3,\y) node[right] {\small $\y$};
        }
        \draw[color3] (0,0)--(-1,0);
        \draw[color3] (-1,0) arc(270:90:0.5);
        \draw[color3] (-1,1)--(0,1);
        \draw[color3] (0,2)--(-0.5,2);
        \draw[color3] (0,-1)--(-5.5,-1);
        \draw[color3] (-4,4)--(-5.5,4);
        \begin{scope}[yshift=2cm,xshift=-0.5cm]
            \basetwoturn
        \end{scope}
        \myparabola{3}{-5}{-1}{-2.75}{3}
    }
    \ar@{=>}[r]^-{\eta_{\wr,\cM}}
    &
    \ctikz{[scale=0.333]
        \mygrid{ 
            \draw[gray!30, step=0.5, opacity = 0.5] (-6,-1) grid (1,4);
            \foreach \x in {-6,-5,-4,-3,-2,-1,0,1,2,3,4,5,6}
                \draw (\x,3) node[above] {\small $\x$};
            \foreach \y in {-1,0,1,2,3,4}
                \draw (3,\y) node[right] {\small $\y$};
        }
        \draw[color3] (-3.75,0)--(0,0);
        \draw[color3] (-3.75,0) arc(270:90:0.5);
        \draw[color3] (-3.75,1)--(0,1);
        \draw[color3] (0,2)--(-2,2);
        \draw[color3] (0,-1)--(-5.5,-1);
        \draw[color3] (-3.5,4)--(-5.5,4);
        \begin{scope}[yshift=2cm,xshift=0.5cm]
                \draw[color3] (-0.5,0)--(-2.5,0);
                \draw[color3] (-2.5,0) arc (270:90:0.5);
                \draw[color3] (-2.5,1)--(-1.5,1);
                \draw[color3] (-1.5,1) arc (-90:90:0.5);
                \draw[color3] (-1.5,2)--(-4,2);
        \end{scope}
        \myparabola{2.75}{-5}{-1}{-3.6}{1}
        \myparabola{2.75}{-3.25}{1}{-1.85}{3}
    }
    \ar@{=>}[r]^-{\chi_\cM^{-1}}
    &
    \ctikz{[scale=0.238]
        \mygrid{ 
            \draw[gray!30, step=0.5, opacity = 0.5] (-6,-1) grid (1,8);
            \foreach \x in {-6,-5,-4,-3,-2,-1,0,1}
                \draw (\x,8) node[above] {\small $\x$};
            \foreach \y in {-1,0,1,2,3,4,5,6,7,8}
                \draw (1,\y) node[right] {\small $\y$};
        }
        \basetwoturn
        \draw[color3] (-4,2)arc(270:90:0.5);
        \draw[color3] (-4,3)--(0,3);
        \begin{scope}[yshift=4cm]
            \basetwoturn
        \end{scope}
        \draw[color3] (-4,6)--(-5,6);
        \draw[color3] (-5,-1)--(0,-1);
        \myparabola{2.75}{-3.75}{-1}{-2.35}{1}
        \myparabola{2.75}{-3.75}{3}{-2.35}{5}
    }
    }
    )
}
\eqnn{\label{eq.probimonad_of_MCM_2}
    U\defdtobe\ctikz{[scale=0.667,xscale=-1]
        \draw[color3] (0,0)--(-1,0);
        \draw[color3] (-1,0) arc(270:90:0.5);
        \draw[color3] (-1,1)--(0,1);
    },\quad
    \nu\defdtobe(
    \diagram@C=2pc{
    \ctikz{[scale=0.222]
        \mygrid{ 
            \draw[gray!30, step=0.5, opacity = 0.5] (-6,-1) grid (1,4);
            \foreach \x in {-6,-5,-4,-3,-2,-1,0,1}
                \draw (\x,4) node[above] {\small $\x$};
            \foreach \y in {-1,0,1,2,3,4,5,6,7,8}
                \draw (1,\y) node[right] {\small $\y$};
        }
        \begin{scope}[scale=1,xscale=-1]
            \basethreeturn
        \end{scope}
        \draw[color3] (0,0)--(-3.5,0);
        \draw[color3] (-3,3)--(-3.5,3);
        \myparabola{2.75}{-2.75}{0}{-1.35}{2}
    }
    \ar@{=>}[r]^-{\chi_\cM}
    &
    \ctikz{[scale=0.333]
        \draw[color3] (2,0)--(0,0);
        \draw[color3] (-2,0)--(0,0);
        \draw[color3] (2,0) arc (-90:90:1);
        \draw[color3] (2,2)--(-2,2);
        \myparabola{1.5}{-1.35}{0}{1.35}{0}
    }
    \ar@{=>}[r]^-{\epsilon_2}
    &
    \ctikz{[scale=0.667,xscale=-1]
        \draw[color3] (0,0)--(-1,0);
        \draw[color3] (-1,0) arc(270:90:0.5);
        \draw[color3] (-1,1)--(0,1);
    }
    }
).
}
  Then $(M,\tau,U,\nu)$ fits into a probimonad structure on the promonad $\MCM\:\cM^\op\os\cM\arprof\cM^\op\os\cM$ (this is proved in Proposition \ref{prp.is_bimonad}).  
\end{itemize}
Now we can prove that the monoidal structure on $\Lmd_{\MnCM}$ are well-captured by both $_\cC\vprof^\lax(\cM,\cM)$ and $\Lmd_{\MCM}(\ast)$ as follows.
\begin{theorem}[Monoidal version of Theorem \ref{thm.unenriched_Kitaev_Kong}]\label{thm.unenriched_mon_Kitaev_Kong}
    There exist canonical monoidal equivalences among the following three monoidal categories:
    \[
    \diagram@C=1.5pc{
    {}_\cC\vprof^\lax(\cM,\cM)
    \ar@<-0.5ex>[r]
    &
    \Lmd_{\MnCM}
    \ar@<-0.5ex>[l]
    \ar@<-0.5ex>[r]
    &
    \Lmd_{\MCM}(\ast)
    \ar@<-0.5ex>[l]
    }
    \]
\begin{proof}
    Let us first show that the equivalence $\Psi\:\Lmd_{\MnCM}\to{}_\cC\vprof^\lax(\cM,\cM)$ constructed in the proof of Theorem \ref{thm.unenriched_Kitaev_Kong} is monoidal. It suffices to prove that $\Psi$ is a strict monoidal functor. To this end, let
    \[
    (
\begin{tikzpicture}[scale=0.5,baseline=0]
    \draw[color3] (2,0)--(0,0);
    \draw[color3] (0,0)--(-2,0);
    \mynode{0}{0}{0.4}{0.4}{F}
\end{tikzpicture}
,
\begin{tikzpicture}[scale=0.5,baseline=0]
    \draw[color3] (2,0)--(0,0);
    \draw[color3] (0,0)--(-2,0);
    \mynode{0}{0}{0.4}{0.4}{F}
    \myparabola{1.5}{-1.35}{0}{1.35}{0}
\end{tikzpicture}
\stackrel{\omega}{\Rightarrow}
\begin{tikzpicture}[scale=0.5,baseline=0]
    \draw[color3] (2,0)--(0,0);
    \draw[color3] (0,0)--(-2,0);
    \mynode{0}{0}{0.4}{0.4}{F}
\end{tikzpicture}
)
\quad\text{and}\quad
(
\begin{tikzpicture}[scale=0.5,baseline=0]
    \draw[color3] (2,0)--(0,0);
    \draw[color3] (0,0)--(-2,0);
    \mynode{0}{0}{0.4}{0.4}{G}
\end{tikzpicture}
,
\begin{tikzpicture}[scale=0.5,baseline=0]
    \draw[color3] (2,0)--(0,0);
    \draw[color3] (0,0)--(-2,0);
    \mynode{0}{0}{0.4}{0.4}{G}
    \myparabola{1.5}{-1.35}{0}{1.35}{0}
\end{tikzpicture}
\stackrel{\omega'}{\Rightarrow}
\begin{tikzpicture}[scale=0.5,baseline=0]
    \draw[color3] (2,0)--(0,0);
    \draw[color3] (0,0)--(-2,0);
    \mynode{0}{0}{0.4}{0.4}{G}
\end{tikzpicture}
)\]
be objects in $\Lmd_{\MnCM}$. Then one can see that $\Psi(G)\circ\Psi(F)=\Psi(G\conv F)$ as the diagram
\[
    \diagram{
        \begin{tikzpicture}[scale=0.5,baseline=0]
            \draw[color3] (-2,0)--(5,0);
            \draw (1,1.45) parabola (-1.65,0);
            \mynode{0}{0}{0.4}{0.4}{G}
            \mynode{3}{0}{0.4}{0.4}{F}
        \end{tikzpicture}
        \ar@{=>}[r]^-{\eta_\wr}
        \ar@{=>}[d]+< 0em,-2em >_{\eta_\wr}
        \ar@{}[rdd]|{\text{\tiny Theorem \ref{thm.local}}}
        &
        \begin{tikzpicture}[scale=0.5,baseline=0]
            \draw [color3] (-2,0)--(5,0);
            \myparabola{1.5}{-1.15}{0}{1.15}{0}
            \draw (3.5,1.15) parabola (1.85,0);
            \mynode{0}{0}{0.4}{0.4}{G}
            \mynode{3}{0}{0.4}{0.4}{F}
        \end{tikzpicture}
        \ar@{=>}[r]^-{\omega'}
        \ar@{=>}[dd]^{\eta_\wr}
        \ar@{}[rdd]|(0.35){\hspace{5em}\text{\tiny Theorem \ref{thm.local}}}
        &
        \begin{tikzpicture}[scale=0.5,baseline=0]
            \draw [color3] (-2,0)--(5,0);
            \draw (3.5,1.15) parabola (1.85,0);
            \mynode{0}{0}{0.4}{0.4}{G}
            \mynode{3}{0}{0.4}{0.4}{F}
        \end{tikzpicture}
        \ar@{=>}[d]^{\eta_\wr}
        \\
        &&
        \begin{tikzpicture}[scale=0.5,baseline=0]
            \draw [color3] (-2,0)--(6,0);
            \myparabola{1.5}{1.85}{0}{4.15}{0}
            \draw (6,0.85) parabola (4.65,0);
            \mynode{0}{0}{0.4}{0.4}{G}
            \mynode{3}{0}{0.4}{0.4}{F}
        \end{tikzpicture}
        \ar@{=>}[d]^{\omega}
        \\
        \begin{tikzpicture}[scale=0.5,baseline=0]
            \mygrid{ 
                \draw[gray!30, step=0.5, opacity = 0.5] (-3,-4) grid (6,3);
                \foreach \x in {-3,-2,-1,0,1,2,3,4,5,6}
                    \draw (\x,3) node[above] {\small $\x$};
                \foreach \y in {-1,0,1,2,3}
                    \draw (3,\y) node[right] {\small $\y$};
            }
            \draw [color3] (-2,0)--(6,0);
            \draw (6,0.85) parabola (4.65,0);
            \mynode{0}{0}{0.4}{0.4}{G}
            \mynode{3}{0}{0.4}{0.4}{F}
            \draw (-1.15,0) parabola bend (1.5,1.7)  (4.15,0);
        \end{tikzpicture}
        \ar@{=>}[r]_-{\eta_\wr}
        &
        \begin{tikzpicture}[scale=0.5,baseline=0]
            \draw [color3] (-2,0)--(6,0);
            \myparabola{1.5}{-1.15}{0}{1.15}{0}
            \myparabola{1.5}{1.85}{0}{4.15}{0}
            \draw (6,0.85) parabola (4.65,0);
            \mynode{0}{0}{0.4}{0.4}{G}
            \mynode{3}{0}{0.4}{0.4}{F}
        \end{tikzpicture}
        \ar@{=>}[r]_-{\omega'\star\omega}
        \ar@{=>}[ur]^-{\omega'}
        &
        \begin{tikzpicture}[scale=0.5,baseline=0]
            \draw [color3] (-2,0)--(6,0);
            \draw (6,0.85) parabola (4.65,0);
            \mynode{0}{0}{0.4}{0.4}{G}
            \mynode{3}{0}{0.4}{0.4}{F}
        \end{tikzpicture}
    }
\]
is commutative. The naturality in $G$ and $F$ is clear. Moreover, $\Psi$ sends the tensor unit of $\Lmd_{\MnCM}$ strictly to the tensor unit of ${}_\cC\vprof^\lax(\cM,\cM)$ by the left diagram in \eqref{eq.zig_zag_m_module}. This concludes the proof that $\Psi$ is a strict monoidal functor, hence a monoidal equivalence.

The proof of the existence of a monoidal equivalence $\Lmd_{\MnCM}\cong\Lmd_{\MCM}(\ast)$ is given in Appendix \ref{sec.graph.app}.
\end{proof}
\end{theorem}

The $\cV$-enriched analogue of Theorem \ref{thm.unenriched_mon_Kitaev_Kong} is also true:
\begin{theorem}[Enriched monoidal version of Theorem \ref{thm.unenriched_Kitaev_Kong}]\label{thm.mon_Kitaev_Kong}
    There exists canonical equivalences among the following monoidal $\cV$-categories:
    $\bLmd_{\MnCM}$, $\bLmd_{\MCM}(\ast)$ and ${}_\cC\bvprof^\lax(\cM,\cM)$.
\end{theorem}
Again we give a sketch proof in Appendix \ref{app.enriched_structure}.


\subsection{Recovery of the classical Kitaev-Kong theorem}\label{sub.recovery}
We consider the following $\cV$-category $\widetilde{\MCN}$ with 
\[
    \ob(\widetilde{\MCN})\defdtobe\ob(\cM^\op\os\cN)
\]
and 
\[
    (\widetilde{\MCN})((m,n),(m',n'))\defdtobe\int^{c\in\cC}\cM(m',c\boxdot_\cM m)\os\cN(c\boxdot_\cN n,n'),
\]
which is the Eilenberg-Moore cateogry of the promonad $\MCN$ (see Appendix~\ref{sec.EM}). Then we have:
\begin{theorem}\label{thm.presheaf_Kitaev_Kong}
    The three $\cV$-categories in Theorem \ref{thm.gen_Kitaev_Kong} are also canonically equivalent to 
    \eqnn{\label{eq.thm.presheaf_Kitaev_Kong}
        \bvprof(\ast,\widetilde{\MCN})\equiv[\widetilde{\MCN}^\op,\bcV].
    }
\end{theorem}
Theorem \ref{thm.presheaf_Kitaev_Kong} offers computational convenience for computing lax $\cC$-module profunctors or $\MnCN$-modules. This computational convenience can be illustrated, for example, in the following consideration of a special case of Theorem \ref{thm.presheaf_Kitaev_Kong}, namely when $\cV=\vect$ and $\cC,\cM,\cN$ are both finite semisimple $\vect$-categories \cite{Etingof_Gelaki_Nikshych_Ostrik_2015}. In this case, if $m_1,m_2\in\cM,n_1,n_2\in\cN$ and $\oplus$ denotes the direct sum in $\cM$ and $\cN$, then we have
\[
\begin{multlined}
\widetilde{\MCN}((m,n),(m_1\oplus m_2,n_1\oplus n_2))\cong \bigoplus_{c\in\irr(\cC)}\cM(m_1\oplus m_2,c\boxdot_\cM m)\os\cN(c\boxdot_\cN n,n_1\oplus n_2)\\
\cong \bigoplus_{c\in\irr(\cC)}\bigoplus_{i,j=1,2}\cM(m_i,c\boxdot_\cM m)\os\cN(c\boxdot_\cN n,n_j)
\cong \bigoplus_{i,j=1,2}\widetilde{\MCN}((m,n),(m_i,n_j))
\end{multlined}
\]
which establishes $(m_1\oplus m_2,n_1\oplus n_2)$ as the direct sum of the four objects $(m_1,n_1),(m_1,n_2),(m_2,n_1)$ and $(m_2,n_2)$ in $\widetilde{\MCN}$. By inspection, a linear functor $\widetilde{\MCN}^\op\to\overline{\vect}$ is determined by its action on the full subcategory $\widetilde{\widetilde{\MCN}}$ of $\widetilde{\MCN}$ spanned by objects from $\irr(\cM)\times\irr(\cN)$. Define the Kitaev-Kong algebra $$A_{\cM,\cN}^\cC= \bigoplus_{\substack{c\in\irr(\cC)\\m_1,m_2\in\irr(\cM)\\n_1,n_2\in \irr(\cN)}}\cN(n_1,c\boxdot_\cN n_2)\ot\cM(c\boxdot_\cM m_2,m_1)$$ with multiplication defined in a similar way as \cite[Section 2.1]{Bai_Zhang_2025}. Then we have an $\C$-linear equivalence
    \eqnn{\label{eq.pre_cor.Kitaev_Kong}
    [\widetilde{\MCN}^\op,\overline{\vect}]\cong[(\widetilde{\widetilde{\MCN}})^\op,\overline{\vect}]\cong\rep(A_{\cM,\cN}^\cC),
    }
    where $\rep(A_{\cM,\cN}^\cC)$ is the category of finite-dimensional left $A_{\cM,\cN}^\cC$-modules.
\begin{corollary}[Kitaev-Kong]
    There is a $\C$-linear equivalence
    \[
        {}_\cC[\cM,\cN]^\lax\cong\rep(A^\cC_{\cM,\cN}).
    \]
    When $\cC$ is rigid, there is a $\C$-linear equivalence
    \[
        {}_\cC[\cM,\cN]\cong\rep(A_{\cM,\cN}^\cC).
    \]
    \begin{proof}
        We have 
        \[
    \rep(A_{\cM,\cN}^\cC)\cong[\widetilde{\MCN}^\op,\overline{\vect}]\cong {}_\cC\bvprof^\lax(\cM,\cN)\cong {}_\cC[\cM,[\cN^\op,\overline{\vect}]]^\lax,
        \]
        where the first, second and third isomorphism follows from \eqref{eq.pre_cor.Kitaev_Kong}, Theorem \ref{thm.presheaf_Kitaev_Kong} and Remark \ref{rmk.relationship_module_fun}, respectively. Since $\cN$ is finite semisimple, one can further show that $\cN\cong[\cN^\op,\overline{\vect}]$ as $\cC$-modules, hence the result. The second statement follows from that when $\cC$ is rigid, any lax $\cC$-module functor is automatically strong.
    \end{proof}
\end{corollary}

\section{Tube promonad and topological holography}
\label{sec.TubeHolo} 
In this section, we briefly talk about the topological holography from the perspective of pro-tensor network. We will not distinguish $\bLmd_{\MnCN}$ and $\bLmd_{\MCN}(\ast)$, owing to Theorem \ref{thm.mon_Kitaev_Kong}.
    
  Suppose $\cC$ is a rigid monoidal $\cV$-category, then $\cC$ is a left $\cC\ot \cC^\rev$ module, with the module action $(x\boxtimes y) \boxdot z:= x\boxtimes z \boxtimes y$. By Theorem~\ref{thm.gen_Kitaev_Kong}, we have 
    \begin{equation}
        \bLmd_{\cC\bbH_{\cC\ot \cC^\rev}\cC}(*)
        \cong {}_\cC\bvprof_\cC(\cC,\cC).
    \end{equation}

The category ${}_\cC\bvprof_\cC(\cC,\cC)$ can be understood as the category of particle-like fixed-point defects on $\cC$ itself as a (trivial) defect between two $\cC$ string-net pro-tensor network.  Equivalently, it is the category of particle-like
fixed-point defects in the bulk of the \(\cC\)-string-net pro-tensor
network. 
        
   Reference~\cite{Lan_2025} introduces the tube category \(\X\cC\) associated 
to a rigid monoidal \(\cV\)-category \(\cC\), which has the same objects as 
\(\cC\) and hom-object is given by \(\X\cC(a,b) = \int^x \cC(a x^{RR}, x b)\). When $\cC$ is pivotal (i.e, there is a monoidal isomorphism $\delta:x^{RR}\cong x$),  $\X\cC(a,b)$ is interpreted as the space of fixed point local tensors, in which the physical bonds of these ``fixed point tensors" are summed over or renormalized, and $a,b$ are virtual bonds.   $\X\cC$ can be viewed as a pro-tensor 
   \begin{equation}
       \X\cC =  \begin{tikzpicture}[scale = 0.6,baseline=(current bounding box.center)]
    \draw  (0,0)--(3.5,0);
    \draw  (4.5,0)--(8,0);
    \draw  (2,0)
      .. controls (4,1.5) ..
      (6,0);
    \draw  (1,0)
      .. controls (4,-1.5) ..
      (7,0);
    \draw[fill = white] (4.5,0) circle [radius = 0.1];
    \draw[fill = white] (8,0) circle [radius = 0.1];
    \end{tikzpicture} \cong 
    \begin{tikzpicture}[scale = 0.6,baseline=(current bounding box.center)]
    \draw  (0,0)--(3.5,0);
    \draw (6,0)--(7.5,0);
    \draw  (2,0)
      .. controls (4,1.5) ..
      (6,0);
    \draw  (1,0)
      .. controls (4,-1.5) ..
      (6,0);
    \draw[fill = white] (7.5,0) circle [radius = 0.1];
    \end{tikzpicture}
    :\cC\nrightarrow \cC,
\end{equation}
    it inherits the structures of promonad and probimonad from  $\cC\bbH_{\cC\ot \cC^\rev}\cC$, and we will call $\X\cC$ as the tube promonad.
    \begin{remark}
         The construction of the tube promonad appeared earlier in~\cite{Lopez_Franco_2007} under a different name and in a more general 2-categorical context. 
    \end{remark}

Now we consider the left modules over $\X\cC$ from $*$. 
Let $P\in \bLmd_{\X\cC}(\ast)$, it is a profunctor  $P:\ast\nrightarrow \cC$ with $\X\cC$ action $\X\cC\bullet P\rightarrow P$. 
Since $\X\cC(a,b)$ is the space of fixed point tensors with virtual bond $a$ and $b$, for a module $P:\ast\nrightarrow \cC$, the component $P(b)$ can be physically interpreted as \emph{the space of potentially non-local tensors with a virtual bond $b$ that can be locally deformed by symmetric fixed point tensors}. Notably, this physical intuition aligns perfectly with the DHR bimodule formalism for the SymTFT, recently developed by Jones and collaborators \cite{Jones_2024,evans2026} within the framework of operator algebras, where a DHR bimodule of the quasi-local algebra given by the fusion spin chain with input unitary fusion category $\cC$ is interpreted as a sector of movable localized non-local operators. Our idea is also inspired by the topological sectors of operators interpretation of SymTFT studied in~\cite{Kong_2022,XuZhang_2024}.

Let $\cC$ be a pivotal monoidal $\C$-linear ($\Vect$-enriched) category, there are some physically illustrative  examples of modules over $\X\cC$. \begin{itemize}
    \item Let $(z,\gamma)\in Z(\cC)$, then $\cC(-,z):\ast \nrightarrow \cC$ is a $\X\cC$ module with module action 
    \begin{equation}
\label{eqn.XC_action_on_C(-,z)}
\begin{split}
\X\cC(a,b)\ot\cC(b,z)&\rightarrow \cC(a,z)\\
\begin{tikzpicture}
    [scale = 0.6,baseline=(current bounding box.center)]
    \draw (-0.6,-0.3)--(1,0);
    \draw  node at (-0.1,0.15){$a$};
    \draw (1,0.5)rectangle (2,-0.5);
    \draw  node at (1.5,0) {$f$}; 
    \draw  (2,0)--(3.6,0.3);
    \draw  node at (3,0.55){$b$};
    \draw  (1.5,-0.5)--(1.5,-2.1);
    \draw  node at (1.9,-1.6){$x$};
    \draw (1.5,0.5)--(1.5,2.1);
    \draw  node at (1.9,1.6){$x$};
\end{tikzpicture}  \ot 
\begin{tikzpicture}
    [scale = 0.6,baseline=(current bounding box.center)]
     \draw (-0.6,-0.3)--(1,0);
     \draw  node at (-0.1,0.15){$b$};
    \draw (1,0.5)rectangle (2,-0.5);
    \draw  node at (1.5,0) {$g$}; 
    \draw  (2,0)--(3.6,0.3);
    \draw  node at (3,0.55){$z$};
\end{tikzpicture}&\mapsto
\begin{tikzpicture}
    [scale = 0.6,baseline=(current bounding box.center)]
     \draw (-0.6,-0.3)--(1,0);
    \draw  node at (-0.1,0.15){$a$};
    \draw (1,0.5)rectangle (2,-0.5);
    \draw  node at (1.5,0) {$f$}; 
    \draw  node at (1.4,1.7 ){$x$};
    \draw  (2,0)--(3.4,0.3);
    \draw  node at (2.8,0.55){$b$};
    \draw  (3.4,0.8) rectangle (4.4,-0.2);
    \draw  node at (3.9,0.3){$g$};
    \draw  (4.4,0.3)--(6.6,0.7125);
    \draw  (1.5,0.5) .. controls (2,3.6) and (4.9,3.6) .. (5.6,0.72);
    \draw  (1.5,-0.5).. controls (2.2,-3.5) and (5.6,-3.5) .. (5.6,0.41);
    \draw  node at (6.2,1) {$z$};
    \draw  node at (5.2,0){$\gamma$};
    \end{tikzpicture}.
\end{split}
    \end{equation}
     Physically, $(z,\gamma)$ represents a quantum current~\cite{Lan_2024}, which can be viewed as a non-local operator defined by the half-braiding of $z$ with the lattice sites. Consequently, $\cC(b,z)$ can be interpreted as the space of endpoints of the quantum current $(z,\gamma)$. In the specific case where $\cC$ is a fusion category, a quantum  current is holographically dual to a string operator in the $2+1$D topological order $Z(\cC)$, and its endpoints correspond to anyon excitations.
    \item
 $P_\cC:=\int^x\cC(-x,x)$, or more generally $P_{\cM}=\int^{m}\cM(-m,m)$ for a left $\cC$ module $\cM$, $\X\cC$ canonically acts on them by contraction of tensors
\begin{equation}
\label{eqn.XC_action_on_PM}
\begin{split}
\X\cC(a,b)\ot P_\cM(b)&\rightarrow P_\cM(a)\\
\begin{tikzpicture}
[scale = 0.6,baseline=(current bounding box.center)]
    \draw (-0.6,0)--(1,0);
    \draw  node at (-0.1,0.3){$a$};
    \draw (1,0.5)rectangle (2,-0.5);
    \draw  node at (1.5,0) {$f$}; 
    \draw  (2,0)--(3.6,0);
    \draw  node at (3,0.35){$b$};
    \draw  (1.5,-0.5)--(1.5,-2.1);
    \draw  node at (1.9,-1.6){$x$};
    \draw (1.5,0.5)--(1.5,2.1);
    \draw  node at (1.9,1.6){$x$};
\end{tikzpicture}   
\ot\begin{tikzpicture}
[scale = 0.6,baseline=(current bounding box.center)]
     \draw (-0.6,0)--(1,0);
     \draw  node at (-0.1,0.35){$b$};
    \draw (1,0.5)rectangle (2,-0.5);
    \draw  node at (1.5,0) {$g$}; 
    \draw  (1.5,-0.5)--(1.5,-2.1);
    \draw (1.5,0.5)--(1.5,2.1);
    \draw  node at (1.9,1.6){$m$};
    \draw  node at (1.9,-1.6){$m$};
\end{tikzpicture}&\mapsto 
\begin{tikzpicture}
[scale = 0.6,baseline=(current bounding box.center)]
    \draw (-0.6,0)--(1,0);
    \draw  node at (-0.1,0.3){$a$};
    \draw (1,0.5)rectangle (2,-0.5);
    \draw  node at (1.5,0) {$f$}; 
    \draw  (2,0)--(3.6,0);
    \draw  (1.5,-0.5)--(1.5,-2.5);
    \draw (1.5,0.5)--(1.5,2.5);
    \draw (3.6,0.5)rectangle (4.6,-0.5);
    \draw  node at (4.1,0) {$g$}; 
    \draw  (4.1,-0.5)--(4.1,-2.5);
    \draw (4.1,0.5)--(4.1,2.5);
    \draw  node at (1.9,1.6){$x$};
    \draw  node at (1.9,-1.6){$x$};
    \draw  node at (4.6,1.6){$m$};
    \draw  node at (4.6,-1.6){$m$};
\end{tikzpicture}.
\end{split}
\end{equation}
Here, $P_\cM(b)$ is the space of boundary fixed-point tensors with a virtual bond $b$.
Note that $m\in \cM$ in~\eqref{eqn.XC_action_on_PM} may represent tensor product of lattice sites within an arbitrary-length interval with physical boundaries included, which makes $g$ a potentially non-local operator.
\end{itemize}

It's proved that in an algebraic way~\cite{Lopez_Franco_2007} (see also~\cite{Lan_2025}) that there is an equivalence between $\cV$-categories  
\begin{equation}
    \bLmd_{\X\cC}(*)\cong Z(Y:\cC\xhookrightarrow{}[\cC^\op,\bcV] ),
\end{equation}
where $Y$ is the Yoneda embedding and $Z(Y)$ is the  relative center $\cV$-category of $Y$. The objects in $Z(Y)$ are functors 
``commuting" with representable functors $\cC(-,a) = Y(a)$ under convolution, i.e. $F:\cC^\op\rightarrow \bcV$ (or pro-tensors  $*\nrightarrow\cC$) together with a half braiding pro-tensor isomorphism 
$\gamma_{-,F}$ \eqref{eq.halfbraiding} satisfying certain coherent conditions.  
\begin{remark} One can check that \begin{equation}
\label{ZYZ}
    Z(Y)\cong Z(\enf{\cC^\op}{\bcV})
\end{equation}  as braided $\cV$-categories, using the fact 
that $\enf{\cC^\op}{\bcV}$-$\enf{\cC^\op}{\bcV}$-bimodule functors out of $\enf{\cC^\op}{\bcV}$ automatically preserves colimits, and $\enf{\cC^\op}{\bcV}$ is the free cocompletion of $\cC$.
\end{remark}
When $\cC$ is a fusion category, we have $\cC\cong \enf{\cC^\op}{\bcV}$ and thus $Z(Y)\cong Z(\cC)$. 
Since $Z(\cC)$ is believed to be the SymTFT of the fusion category $\cC$ of symmetry charge~\cite{Kong_2020,Ji_2020, KongZheng_2022,Jones_2024,Lan_2024,XuZhang_2024,KongZheng_2024,LanYueWang_2024},
we then propose that $\bLmd_{\X\cC}({\ast})$ is a possible microscopic mechanism of SymTFT. In our approach, the category of symmetry charge $\cC$ can be generalized, not necessarily being finite or semisimple, 

In terms of pro-tensors,
an object in $\bLmd_{\X\cC}(*)$ is a pro-tensor 
\begin{equation*}
\begin{tikzpicture}
[scale = 0.6,baseline=(current bounding box.center)]
    \draw  (-0.5,0)--(2,0);
    \draw  node at (0.6,0.5){$\cC$};
    \draw[fill=white] (2,0.5)rectangle(3,-.5);
    \draw  node at (2.5,0) {$P$};
\end{tikzpicture}    
\end{equation*}
with module action 

\begin{equation}
    \begin{tikzpicture}
        \node (A) at (0,0){ \begin{tikzpicture}[scale = 0.6,baseline=(current bounding box.center)]
    \draw  (0,0)--(3.5,0);
    \draw[fill = white] (3.5,0.5)rectangle(4.5,-0.5);
    \draw node at (4,0)
    {$P$};
    \draw  (6,0)--(7.5,0);
    \draw (2,0)--(4,1.5)--(6,0);
    \draw (1,0)--(4,-1.5)--(6,0);
    \draw[fill = white] (7.5,0) circle [radius = 0.1];
    \end{tikzpicture}};
    \node (B) at (6,0){\begin{tikzpicture}
[scale = 0.6,baseline=(current bounding box.center)]
    \draw  (-0.5,0)--(2,0);
    \draw[fill=white] (2,0.5)rectangle(3,-.5);
    \draw  node at (2.5,0) {$P$};
\end{tikzpicture}    };
\draw[nattrans] (2.5,0)--(4.5,0) node [midway, above = 4pt]{$\kappa$};
    \end{tikzpicture}
\end{equation}
satisfying associativity and unitality; and an object in $Z(Y)$ is a pro-tensor 
\begin{equation*}
\begin{tikzpicture}
[scale = 0.6,baseline=(current bounding box.center)]
    \draw  (-0.5,0)--(2,0);
    \draw  node at (0.6,0.5){$\cC$};
    \draw[fill=white] (2,0.5)rectangle(3,-.5);
    \draw  node at (2.5,0) {$F$};
\end{tikzpicture}    
\end{equation*}
together with a half braiding 

\begin{equation}
    \begin{tikzpicture}
        \node (A) at (0,0) {\begin{tikzpicture}
[scale = 0.6,baseline=(current bounding box.center)]
    \draw  (-0.5,0)--(2,0);
    \draw[fill=white] (2,0.5)rectangle(3,-.5);
    \draw  node at (2.5,0) {$F$};
    \draw  (0.5,0)--(2.5,1.2);
\end{tikzpicture}   };
\node (B) at (4,0)  {\begin{tikzpicture}
[scale = 0.6]
    \draw  (-0.5,0)--(2,0);
    \draw[fill=white] (2,0.5)rectangle(3,-.5);
    \draw  node at (2.5,0) {$F$};
    \draw  (0.5,0)--(2.5,-1.2);
\end{tikzpicture} };
\draw[nattrans] (1.5,0)--(3,0) node[midway, above=4pt]{$\gamma_{-,F}$};
    \end{tikzpicture}
    \label{eq.halfbraiding}
\end{equation}
satisfying braiding hexagon equation.

The equivalence between them can be  presented graphically as follows.
Given $(P,\kappa)\in \bLmd_{\X\cC}(*)$, the module action gives a canonical half braiding 
\begin{equation}
         \begin{tikzpicture}
        \node (A) at (0,0) {\begin{tikzpicture}
[scale = 0.7,baseline=(current bounding box.center)]
    \draw  (-0.5,0)--(2,0);
    \draw[fill=white] (2,0.5)rectangle(3,-.5);
    \draw  node at (2.5,0) {$P$};
    \draw  (0.5,0)--(2.5,1.2);
\end{tikzpicture}};




\node (C) at (11,0) {\begin{tikzpicture}
[scale = 0.7,baseline=(current bounding box.center)]
    \draw  (-2.9,0)--(2,0);
    \draw[fill=white] (2,0.5)rectangle(3,-.5);
    \draw  node at (2.5,0) {$P$};

       \draw  (0.6,0)
      to[out=60,in=180]
      (2.5,1);
    \draw  (2.5,1)
      to[out=0,in=120]
      (4.4,0);
    \draw  (4.4,0)
      to[out=-120,in=0]
      (2.5,-1);

    \draw  (0,-1)--(2.5,-1);
    \draw  (0,-1)
    to[out=180,in=60] (-1.9,-2) to[out=-60,in=180] (0,-3);
    \draw  (-2.9,-2)--(-1.9,-2);
    \draw[fill = white] (-2.9,-2) circle [radius = 0.1];
    \draw  (5.4,0)--(4.4,0);
    \draw[fill = white] (5.4,0) circle [radius = 0.1];
\end{tikzpicture}};

\node (D) at (11,-4) {\begin{tikzpicture}
[scale = 0.7,baseline=(current bounding box.center)]
    \draw  (-2.3,0)--(2,0);
    \draw[fill=white] (2,0.5)rectangle(3,-.5);
    \draw  node at (2.5,0) {$P$};
    \draw  (0.6,0)
      to[out=60,in=180]
      (2.5,1);
    \draw  (2.5,1)
      to[out=0,in=120]
      (4.4,0);
    \draw  (4.4,0)
      to[out=-120,in=0]
      (2.5,-1);

    \draw  (2.5,-1)to[out=180,in=-60](-0.2,0);

    \draw  (-1,0)--(-1.5,-1);
    \draw  (-1.5,-1)--(-2.3,-1);
    \draw[fill = white] (-2.3,-1) circle [radius = 0.1];
    \draw  (-1.5,-1)--(-0.7,-2);
     \draw  (5.4,0)--(4.4,0);
    \draw[fill = white] (5.4,0) circle [radius = 0.1];    
\end{tikzpicture}};

\node (E) at (5,-4) {\begin{tikzpicture}
[scale = 0.7,baseline=(current bounding box.center)]
    \draw  (-2,0)--(1.5,0);
    \draw[fill=white] (0.5,0.5)rectangle(1.5,-.5);
    \draw  node at (1,0) {$P$};
    \draw  (-1,0)--(-1.5,-1);
    \draw  (-1.5,-1)--(-2.3,-1);
    \draw[fill = white] (-2.3,-1) circle [radius = 0.1];
    \draw  (-1.5,-1)--(-0.7,-2);  
\end{tikzpicture}};

\node (F) at (0,-4){\begin{tikzpicture}
[scale = 0.7, baseline = (current bounding box.center)]
    \draw  (-0.5,0)--(2,0);
    \draw[fill=white] (2,0.5)rectangle(3,-.5);
    \draw  node at (2.5,0) {$P$};
    \draw  (0.5,0)--(2.5,-1.2);
\end{tikzpicture}};

\draw[nattrans] (A)--(C) node [midway, above =4pt]{\eqref{eq.Frob_intertible_unital}};
\draw[nattrans] (C)--(D) node [midway, right =4pt]{$\eta_2$};
\draw[nattrans] (D)--(E) node [midway, above =4pt]{$\kappa$};
\draw[nattrans] (E)--(F) node [midway, above =4pt]{\eqref{eq.Frob_intertible_unital}};
\draw[nattrans] (A)--(F) node [midway, right =4pt]{$=:\gamma_{-,P}$};
    \end{tikzpicture}
\end{equation}
Given $(F,\gamma_{-,F})\in Z(Y)$, half braiding induces a canonical module action on $F$
\begin{equation}
    \begin{tikzpicture}
         \node (A) at (0,0){ \begin{tikzpicture}[scale = 0.55,baseline=(current bounding box.center)]
            \draw  (0,0)--(3.5,0);
            \draw[fill = white] (3.5,0.5)rectangle(4.5,-0.5);
            \draw  node at (4,0) {$F$};
            \draw  (6,0)--(7.5,0);
            \draw  (2,0)
              .. controls (4,1.5) ..
              (6,0);
            \draw  (1,0)
              .. controls (4,-1.5) ..
              (6,0);
            \draw[fill = white] (7.5,0) circle [radius = 0.1];
         \end{tikzpicture}};
         
         \node (B) at (5.5,0) 
         {\begin{tikzpicture}[scale = 0.55,baseline=(current bounding box.center)]
            \draw  (0,0)--(3.5,0);
            \draw[fill = white] (3.5,0.5)rectangle(4.5,-0.5);
            \draw  node at (4,0) {$F$};
            \draw  (6,0)--(7.5,0);
            \draw  (2,0)--(4.5,-1.15);
            \draw  (1,0)
              .. controls (4,-1.7) ..
              (6,0);
            \draw[fill = white] (7.5,0) circle [radius = 0.1];    
         \end{tikzpicture}};

         \node (C) at (11,0)  {\begin{tikzpicture}[scale = 0.55,baseline=(current bounding box.center)]
            \draw  (0,0)--(3.5,0);
            \draw[fill = white] (3.5,0.5)rectangle(4.5,-0.5);
            \draw  node at (4,0) {$F$};
            \draw  (6,0)--(7.5,0);
            \draw  (2.5,-0.85)--(5,-.85);
            \draw  (1,0)
              .. controls (4,-1.7) ..
              (6,0);
            \draw[fill = white] (7.5,0) circle [radius = 0.1];    
         \end{tikzpicture}};

         \node (D) at (11,-4){\begin{tikzpicture}
            [scale = 0.55,baseline=(current bounding box.center)]
            \draw  (0,0)--(3.5,0);
            \draw[fill = white] (3.5,0.5)rectangle(4.5,-0.5);
            \draw  node at (4,0) {$F$};
            \draw  (1,0)--(4,-1.7);
            \draw[fill = white] (4,-1.7) circle [radius = 0.1];   
         \end{tikzpicture}};

         \node (E) at (0,-4) {\begin{tikzpicture}
            [scale =0.6,baseline=(current bounding box.center) ]
            \draw  (0,0)--(3.5,0);
            \draw[fill = white] (3.5,0.5)rectangle(4.5,-0.5);
            \draw  node at (4,0) {$F$};
         \end{tikzpicture}};
         
         \draw[nattrans] (A)--(B) node [midway, above = 4pt]{$\gamma_{-,F}$};
         \draw[nattrans] (B)--(C) node [midway, above = 4pt]{$\alpha$};
         \draw[nattrans] (C)--(D) node [midway, right = 4pt]{$\epsilon_2$};
         \draw[nattrans] (D)--(E) node [midway, above = 4pt]{$\rho$};
         \draw[nattrans] (A)--(E) node [midway, right = 4pt]{$=:\kappa$};
    \end{tikzpicture}
\end{equation}
Also, just like the equivalence $Z(\cD)\cong \Fun_{\cD|\cD}(\cD,\cD)$ in the case of ordinary monoidal category $\cD$, we have 
\begin{equation}
    Z(Y:\cC\hookrightarrow \enf{\cC^\op}{\overline{\cV}})\cong {}_\cC\bvprof_\cC(\cC,\cC),
\end{equation}
    where the equivalence (on the object level) is given by 
    \begin{equation}
    \label{eq.equiBetweeenVProf&Z}
        \begin{split}
        Z(Y)&\cong {}_\cC\bvprof_\cC(\cC,\cC)\\
        \begin{tikzpicture}
                    [scale =0.6, baseline=(current bounding box.center) ]
                    \draw  (-2,0)--(0,0);
                    \draw[fill = white] (0,0.5) rectangle (1,-0.5);
                    \draw  node at (0.5,0){$F$};
                \end{tikzpicture}&\mapsto
                \begin{tikzpicture}
                     [scale =0.6, baseline=(current bounding box.center) ]
                    \draw  (-2,0)--(0,0);
                    \draw[fill = white] (0,0.5) rectangle (1,-0.5);
                    \draw  node at (0.5,0){$F$};
                    \draw  (-1,0)--(0,-1);
                \end{tikzpicture}
            \\
            \begin{tikzpicture}
            [scale =0.6, baseline=(current bounding box.center) ]
                \draw  (-1.5,0)--(0,0);
                \draw[fill = white] (0,0.5) rectangle (1,-0.5);
                \draw (1,0)--(2,0);
                \draw[fill = white] (2,0) circle [radius = 0.1];
                \draw  node at (0.5,0) {$G$};
            \end{tikzpicture}&\mapsfrom
            \begin{tikzpicture}
                 [scale =0.6, baseline=(current bounding box.center) ]
                \draw  (-1.5,0)--(0,0);
                \draw[fill = white] (0,0.5) rectangle (1,-0.5);
                \draw (1,0)--(2.5,0);
                \draw  node at (0.5,0) {$G$};
            \end{tikzpicture},
        \end{split}
    \end{equation}
and can be verified easily. 

    In summary, we have
    \begin{equation}
    \label{eq.topo_holo}
        \bLmd_{\X\cC}(*)\cong    Z(Y) \cong {}_\cC\bvprof_\cC(\cC,\cC) \cong\bLmd_{\cC\bbH_{\cC\ot \cC^\rev} \cC}(*)\cong Z([\cC^\op, \bcV])
    \end{equation}
    the two promonads $\X\cC$ and $\cC\bbH_{\cC\ot \cC^\rev} \cC$ have  equivalent representation categories, thus they are Morita equivalent. 
    This suggests the validity of topological holography.
        The category of spaces of $1+1$D potentially non-local tensors that can be locally deformed by $\cC$ symmetric fixed point tensors is equivalent to the category of particle-like fixed-point defects in the $2+1$D $\cC$ string-net pro-tensor network.
\begin{example}
\label{ex.U(1)symmetry}
  Let us consider $1+1$D \(U(1)\)
symmetry on a lattice model with finite-dimensional local Hilbert spaces. We take the charge
category of the \(U(1)\) symmetry to be
$\Rep_{\fd}(U(1))$,
the category of finite-dimensional representations of \(U(1)\). It is rigid monoidal semisimple. We regard
$\Rep_\fd(U(1))$ as a category enriched over \(\Vect\), the category of possibly
infinite-dimensional vector spaces, rather than over \(\vect\).
   The reason we use $\Vect$ enrichment  is that although each hom-space of $\Rep_{\mathrm{f.d}}(U(1))$ is finite dimensional, it has infinite many isomorphism classes of simple objects labeled by $n\in \Z$.  Many natural constructions built from the charge
category involve summing over all possible charge labels. These sums are
generally infinite-dimensional. For example, 
 the space of $U(1)$ symmetric fixed point tensors $\X \Rep_\fd(U(1))(a,b)\cong \oplus_{n\in \Z}\Rep_{\fd}(a,b)$ is infinite dimensional for any $a,b\in \Rep_{\fd}(U(1))$ and thus only exists in $\Vect$. Also, from $\Rep_\fd(U(1))$ string-net pro-tensor network perspective, taking \(\Vect\) as the enrich-background
 guarantees that the relevant pro-tensor contractions well defined. 
On the contrast, if we are dealing with finite symmetry, for example fusion category symmetry,  $\vect$ is a sufficient choice of enrichment. Despite $\Vect$ enrichment is still mathematically allowed, it's not physical reasonable anymore since $\Vect$ enrichment is just an artificial extension of $\vect$ in this completely finite setting.

$\Rep_\fd(U(1))$ is equivalent to the category of finite dimensional $\Z$-graded vector spaces $\Vect_{\Z,\fd}$, and its free cocompletion, $[\Rep_\fd(U(1))^\op, \overline{\Vect}]$ is equivalent to the category of potentially infinite dimensional $\Z$-graded vector spaces $\Vect_\Z$ as monoidal categories.
According to~\eqref{eq.topo_holo}, the SymTFT of such symmetry is described by the category $Z(\Vect_\Z)$. Unpacking the definition of center, an object in $Z(\Vect_\Z)$ is a pair $(V,T)$, where $V$ is a potentially infinite dimensional $\Z$-graded vector space $V=\oplus_{n\in \Z}V_n$, and $T$ is a grading preserving $\Z$-action
\begin{equation}
    T:\Z\times V\rightarrow V
\end{equation}
such that  $T^0=\id_V$ and $T^{n+m}=T^nT^m$. A morphism between $(V,T)$ and $(W,S)$ is a $\Z$-grading preserving map $f:V\rightarrow W$ that intertwines the $\Z$-actions. Equivalently,
\[
Z(\Vect_{\Z})
\cong
\Lmd_{\C[t,t^{-1}]}(\Vect_{\Z}),
\]
where \(\C[t,t^{-1}]\cong \C[\mathbb Z]\) is the Laurent polynomial algebra placed in grade $0$. 
We note that the $\Z$-action can be given by any invertible matrix, and is in general not semisimple. Passing from $ \Rep_\fd(U(1)) $ to its fixed point tensors $Z(\Vect_\Z)$, we see a loss of semisimplicity. We view this as an algebraic version of ``divergence''. In fact, we see similar ``divergence'' also in other examples of continuous symmetries.  We conjecture that such algebraic divergence is related to analytical divergence associated with continuous symmetries in 1 and 2 spatial dimensions, such as the Mermin–Wagner theorem (continuous symmetries cannot be spontaneously broken), and the confinement of $U(1)$ gauge theory.
\end{example}

\begin{remark}
    It is helpful to know how $Z(\cC)$ is contained in $Z(Y)$. By the definition of relative center, $Z(\cC)$ is exactly the full subcategory of objects in $Z(Y)$ whose images under the forgetful functor $Z(Y)\to \enf{\cC^\op}{\cV}$ are representable (see also Ref.~\cite{Lan_2025}). In this remark we would like to give an alternative characterization in terms of dualizability. We claim that for Cauchy complete rigid $\cC$, the full subcategory of dualizable objects in ${}_\cC\bvprof_\cC(\cC,\cC)$ is equivalent to $Z(\cC)$. 

    To see such result, we combine the following basic general facts, regarding a profucntor $F:\cM\arprof \cN$ 
    \begin{enumerate}

        \item  Every $\cV$-category is equivalent to its Cauchy completion in $\vprof$ (see for e.g. \cite[\S 4]{Street_1983}). Thus without losing generality, we can suppose that $\cN$ is Cauchy complete, then there exists a profunctor $F^R:\cN\arprof\cM$ right adjoint to $F$ if and only if $F$ is representable;
        \item Suppose that $F$ is in addition an oplax left $\cD$-module profunctor, then its right adjoint $F^R$ has a canonical structure of a lax $\cD$-module profunctor (Remark~\ref{rmk.AdjModuleFun});
        \item Suppose that $\cD$ has left duals, then every lax left $\cD$-module functor is automatically strong (Remark~\ref{rmk.rigidCModuleProf}).
    \end{enumerate}
    Together, these mean that if $\cN$ is Cauchy complete and $\cD$ has left duals, left adjoints in $_{\cD}\bvprof(\cM,\cN)$ are the same as representables. Specializing to the case $\cM=\cN=\cC$ and $\cD=\cC\otimes \cC^\rev$, we know when $\cC$ is Cauchy complete and rigid (having both left and right duals, such that $\cC\otimes \cC^\rev$ has left duals), left adjoints in ${}_\cC\bvprof_\cC(\cC,\cC)$ are exactly representables. Together with the monoidal equivalence
    \[ {}_\cC\bvprof_\cC(\cC,\cC)\cong \enf[\cC]{\cC}{\enf{\cC^\op}{\bcV}}_\cC\cong Z(Y:\cC\to \enf{\cC^\op}{\bcV}),\]
    we know the full subcategory of left adjoints in ${}_\cC\bvprof_\cC(\cC,\cC)$ is equivalent to $Z(\cC)$. Moreover, when $\cC$ is rigid, it is easy to show that $Z(\cC)$ is also rigid, and that all adjoints in ${}_\cC\bvprof_\cC(\cC,\cC)$ are two-sided.
\end{remark}

\acknowledgments

TL is supported by start-up funding from The Chinese University of
Hong Kong, by funding from Research Grants Council, University Grants Committee
of Hong Kong (ECS No.~24304722, CRF No.~C7015-24GF), and also by Guangdong Provincial Quantum Science Strategic Initiative Project GDZX2501013.



\appendix

\section{Coends}\label{sec.coend}
\subsection{Definition and basic properties of coends}
In this subsection, we provide the definition of coends, discuss its existence, and present the proofs of some assertions in the main text concerning coends.

Let $\cV$ be a symmetric monoidal closed category (e.g. a cosmos or $\cV=\vect$).
\begin{definition}\label{dfn.coend}
    Let $F\:\cC\arprof\cC$ be a $\cV$-profunctor in the sense of Definition \ref{dfn.prof_unpack}. The \emph{coend (indexed by $\cC$)} of $F$ is given by an object $\int^{c\in\cC}F(c,c)\in \cV$ together with a family $\{\copr_c\:F(c,c)\to \int^{c\in\cC}F(c,c)\}_{c\in\cC}$ of morphisms in $\cV$ such that:
    \begin{itemize}
        \item $(\int^{c\in\cC}F(c,c),\{\copr_c\}_{c\in\cC})$ satisfies the \emph{$\cC$-balancing} property: for any $c,c'\in\cC$, the diagram
        \[
            \diagram@C=4pc{
                \cC(c,c')\os F(c',c) \ar[d]_{F_l^{c'cc'}} \ar[r]^-{\tau_{\cC(c,c'),F(c',c)}} & F(c',c)\os\cC(c,c') \ar[r]^-{F_r^{cc'c}} & F(c,c) \ar[d]^{\copr_c} \\
                F(c',c') \ar[rr]_-{\copr_{c'}} & & \int^{c\in\cC}F(c,c)
            }
        \]
        in $\cV$ is commutative.
        \item For any pair $(w,\{\theta_c\:F(c,c)\to w\}_{c\in\cC})$ satisfying the $\cC$-balancing property, there exists a unique morphism $\underline{\theta}\:\int^{c\in\cC}F(c,c)\to w$ such that
        \[
            \underline{\theta}\circ\copr_c=\theta_c
        \]
        holds for any $c\in\cC$.
    \end{itemize}
\end{definition}

Before we discuss the existence of coends in the general setting, let us consider the special case $\cV=\Vect$. In this case, a profunctor $F\:\cC\arprof\cC$ is also a linear functor $\cC^\op\os\cC\to\overline{\Vect}$ as in Definition \ref{dfn.profunctor}, and the $\cC$-balancing property of a pair $(w,\{\theta_c\:F(c,c)\to w\}_{c\in\cC})$ can be translated to the following condition: for any morphism $f\:c\to c'$ in $\cC$, the diagram
    \[
        \diagram{
            F(c',c) \ar[r]^-{F(1,f)} \ar[d]_{F(f,1)} & F(c',c') \ar[d]^{\theta_{c'}}\\
            F(c,c) \ar[r]_-{\theta_c} & w
        }
    \]
    commutes.
Using this translation, it is easy to have the following concrete computation of the coend of a $\Vect$-profunctor $F\:\cC\arprof\cC$:
\begin{lemma}
\label{lem.coend_Vect_prot}
      The coend of a $\Vect$-profunctor $F\:\cC\arprof\cC$ is given by the vector space
    \[
        (\coprod_{c\in\cC} F(c,c))/\mathrm{span}\langle F(1,f)(x)-F(f,1)(x)\mid f\:c\to c'\in\Mor(\cC),x\in F(c',c)\rangle.
    \]
\end{lemma}
Concerning the existence of coends, we have the following two results.
\begin{proposition}\label{prp.existence_of_coend_cosmos}
    If $\cV$ is cocomplete (e.g. a cosmos), then coend of any profunctor $F\:\cC\arprof\cC$ for $\cC$ being a small $\cV$-category exists.
    \begin{proof}
        It is possible to realize the coend of $F$ as certain functor in $\cV$ whose domain has set of objects $(\ob(\cC)\times\ob(\cC))\coprod\ob(\cC)$.
    \end{proof}
\end{proposition}
However, sometimes we work with non-cosmos $\cV$ such as $\vect$. In this case, it is helpful to invoke the notion of \emph{Cauchy completion} of a $\cV$-enriched category \cite{Lawvere_1973}. For $\cV=\set$, Cauchy completion corresponds to completion under retracts of idempotents; for $\cV=\vect$, Cauchy completion corresponds to completion under retracts and direct sums. More generally, the $\cV$-Cauchy completion $\cau(\cC)$ of a $\cV$-category $\cC$ is the full subcategory of the presheaf category $\hat{\cC}\defdtobe[\cC^\op,\bcV]$ consisting of $W\in\hat{\cC}$ which admits a right adjoint viewed as a profunctor $\ast\arprof\cC$.
\begin{proposition}\label{prp.finite_coend}
    If $\cV$ admits finite limits and colimits (e.g. $\cV=\vect$), and $\cC=\cau(\cD)$ for some $\cV$-category $\cD$ with $\ob(\cD)$ finite, then the coend of any profunctor $F\:\cC\arprof\cC$ exists.
\end{proposition}
Proving Proposition \ref{prp.finite_coend} is more involved and is deferred to the end of this subsection.

\begin{proposition}[Fubini's theorem for coends]\label{prp.Fubini}
    Let $\cC,\cD$ be $\cV$-categories and $F\:\cC\os\cD\arprof\cC\os\cD$ be a profunctor. Suppose for any $c,c'\in\cC$, the profunctor $F(c,-,c',-)\:\cD\arprof\cD$ admits coend. Then the assignment
    \[
        (c,c')\mapsto \int^{d\in\cD}F(c,d,c',d)
    \]
    extends to a profunctor $\cC\arprof\cC$. Suppose for any $d,d'\in\cD$, the profunctor $F(-,d,-,d')\:\cC\arprof\cC$ admits coend. Then the assignment 
    \[
        (d,d')\mapsto \int^{c\in\cC}F(c,d,c,d')
    \]
    extends to a profunctor $\cD\arprof\cD$.
    
    Moreover, under the two existence conditions above, the following objects are canonically isomorphic with each other\footnote{One exists iff the other exist; when they exist, they're canonically isomorphic.}:
    \[
        \int^{c\in\cC}\int^{d\in\cD}F(c,d,c,d)\cong\int^{(c,d)\in\cC\os\cD}F(c,d,c,d)\cong\int^{d\in\cD}\int^{c\in\cC}F(c,d,c,d).
    \]
\end{proposition}
\begin{lemma}[Co-Yoneda Lemma]\label{lem.co_Yoneda}
    For any $\cV$-functor $F\:\cC\to\bcV$ and $G\:\cC^\op\to\bcV$, there is 
    \[
    \int^{c\in\cC}\cC(c,d)\os F(c)=F(d),\qquad \int^{c\in\cC}G(c)\os\cC(d,c)=G(d).
    \]
\end{lemma}
\begin{proof}[Proof of Proposition \ref{prp.finite_coend}]
    Let $\cD$ be a category with $\ob(\cD)$ finite. Then any coend indexed by $\cD$ exists using the proof of Proposition \ref{prp.existence_of_coend_cosmos}. For $d\in\cD$, we use $d_\ast\:\ast\to\cD,\;d'\mapsto\cD(d',d)$ to denote the image of $d$ under the inclusion $\cD\to\cau(\cD)$. The key observation is that for $W,Z\in\cau(\cD)$, we have
    \[
        \cau(\cD)(W,Z)\cong\int^{d\in\cD}\cau(\cD)(W,d_\ast)\os \cau(\cD)(d_\ast,Z).
    \]
    This can be proved, for example, by writing $Z$ as a colimit of representables and then using the equivalence of (i) and (ii) in \cite[Proposition 6.14]{Kelly_Schmitt_2005}. Now assume $F\:\cau(\cD)\arprof\cau(\cD)$ is a profunctor. Then we have the following canonical isomorphisms
    \[
    \begin{split}
    \int^{W\in\cau(\cD)}F(W,W)&\cong \int^{W\in\cau(\cD)}\int^{Z\in\cau(\cD)}F(W,Z)\os\cau(\cD)(Z,W) \\
    &\cong\int^{W\in\cau(\cD)}\int^{Z\in\cau(\cD)}\int^{d\in\cD}F(W,Z)\os \cau(\cD)(Z,d_\ast)\os\cau(\cD)(d_\ast,W)\\
    &\cong    \int^{W\in\cau(\cD)}\int^{d\in\cD}\int^{Z\in\cau(\cD)}F(W,Z)\os \cau(\cD)(Z,d_\ast)\os\cau(\cD)(d_\ast,W)\\
    &\cong\int^{W\in\cau(\cD)}\int^{d\in\cD}F(W,d_\ast)\os\cau(\cD)(d_\ast,W)\\
    &\cong\int^{d\in\cD}\int^{W\in\cau(\cD)}F(W,d_\ast)\os\cau(\cD)(d_\ast,W)\\
    &\cong\int^{d\in\cD}F(d_\ast,d_\ast),
    \end{split}
    \]
    where in the third and fifth isomorphism we use Fubini's theorem (Proposition \ref{prp.Fubini}), and in the first, fourth and sixth isomorphism we use co-Yoneda lemma (Lemma \ref{lem.co_Yoneda}).
    As a consequence, the coend of $F$ is given by $\int^{d\in\cD}F(d_\ast,d_\ast)$, which in particular shows that the coend of $F$ exists.
\end{proof}
\begin{remark}
Proposition \ref{prp.finite_coend} is inspired by \cite[Lemma 2.5]{huang20242charactertheoryfinite2groups}, where the case $\cV=\vect$ is proved.    
\end{remark}
\subsection{Constructions related to the profunctor $G\bullet F$}\label{sub.compose_of_prof.app}
Let $\cV$ be a cosmos. Let $F\:\cC\arprof\cD,G\:\cD\arprof\cE$ be profunctors. We have already defined the effect on objects of $G\bullet F$ in \eqref{eq.compose_prof}. For $c,c'\in\cC,e\in\cE$, we define
\[
    (G\bullet F)_l^{ecc'}\:\cC(c,c')\os(\int^{d\in\cD}F(d,c)\os G(e,d))\to \int^{d\in\cD}F(d,c)\os G(e,d)
\]
to be the unique morphism rendering the diagram
\[
    \diagram@C=3pc{\cC(c,c')\os F(d,c)\os G(e,d) \ar[d]_{F_l^{dcc'}\os\id} \ar[r]^-{\copr_d\os \id} & \cC(c,c')\os(\int^{d\in\cD}F(d,c)\os G(e,d))
   \ar@{-->}[d]^{(G\bullet F)_l^{ecc'}} \\
    F(d,c')\os G(e,d) \ar[r]_-{\copr_d} & \int^{d\in\cD}F(d,c)\os G(e,d)}
\]
commutative. Similarly, for $c\in\cC,e,e'\in\cE$, we define
\[
    (G\bullet F)_r^{ee'c}\:(\int^{d\in\cD}F(d,c)\os G(e',d))\os E(e,e') \to \int^{d\in\cD}F(d,c)\os G(e,d)
\]
to be the unique morphism rendering the diagram
\[
    \diagram@C=3pc{
        F(d,c)\os G(e',d)\os \cE(e,e') \ar[d]_{\id\os G_r^{ee'd}} \ar[r]^-{\copr_d\os \id} &(\int^{d\in\cD}F(d,c)\os G(e',d))\os \cE(e,e') \ar@{-->}[d]^{ (G\bullet F)_r^{ee'c}}\\
        F(d,c)\os G(e,d) \ar[r]_-{\copr_d} & \int^{d\in\cD}F(d,c)\os G(e,d)
    }
\]
commutative. 

It can be verified that $G\bullet F$ with left $\cC$-action and right $\cE$-action defined above is a well-defined profunctor $\cC\arprof\cE$.

With respect to the composition of profunctors,  profunctor homomorphisms can also compose with each other, but in two different ways.

\begin{definition}[Horizontal and vertical compositions of profunctor homomorphisms]
\label{def.comp_profun_homo}
     Suppose $G,G'\:\cB\arprof\cC$ and $F,F',F''\:\cA\arprof\cB$ are profunctors, $\alpha: F\Rightarrow F'$, $\alpha':F'\Rightarrow F''$, $\beta: G\Rightarrow G'$ are profunctor homomorphisms, then the horizontal composition of $\beta$ and $\alpha$ is a profunctor homomorphism from $G\bullet F$ to $G'\bullet F'$, denoted as $\beta\bullet \alpha:G\bullet F\Rightarrow G'\bullet F'$, whose component $(\beta\bullet\alpha)_{c,a}$ for $c\in \cC, a\in\cA$ is defined through the universal property of coends
    \eqnn{
        \diagram@C=2.5pc{
            (G\bullet F)(c,a) \ar@{-->}[r]^-{(\beta\bullet\alpha)_{c,a}} & (G'\bullet F')(c,a) \\
            F(b,a)\os G(c,b) \ar[u]^{\copr_b} \ar[r]_-{\alpha_{b,a}\os\beta_{c,b}} & F'(b,a)\os G'(c,b) \ar[u]_{\copr_b}
        }
    }
    the vertical composition of $\alpha'$ and $\alpha$ is a profunctor homomorphism from $F$ to $F''$, denoted as $\alpha'\cdot \alpha:F\Rightarrow F''$, whose component $(\alpha'\cdot \alpha)_{b,a}$ for $b\in \cB, a\in \cA$ is defined as the following composition of morphisms in $\cV$
    \eqnn{
         (\alpha'\cdot \alpha)_{b,a}:F(b,a)\xrightarrow{\alpha_{b,a}}F'(b,a)\xrightarrow{\alpha'_{b,a}}F''(b,a).
     }
\end{definition}

\begin{proposition}
    [Interchanging law of profunctor homomorphisms]
    \label{Prop.interchangeing}
    Suppose $G,G',G''\:\cB\arprof\cC$ and $F,F',F''\:\cA\arprof\cB$ are profunctors, $\alpha: F\Rightarrow F'$, $\alpha':F'\Rightarrow F''$, $\beta: G\Rightarrow G'$, $\beta':G'\Rightarrow G''$ are profunctor homomorphisms, then we have 
    \begin{equation}
        (\beta'\cdot \beta)\bullet(\alpha'\cdot \alpha) = (\beta'\bullet \alpha')\cdot (\beta\bullet \alpha).
    \end{equation}
\end{proposition}

The proof can be done by showing both $((\beta'\cdot \beta)\bullet(\alpha'\cdot \alpha))_{c,a}$ and $((\beta'\bullet \alpha')\cdot (\beta\bullet \alpha))_{c,a}$ for $c\in\cC,a\in\cA$ are the unique morphism rendering the outer square of the following diagram commutative:
\eqnn{
    \diagram@C=3pc@R=1.25pc{
        (G\bullet F)(c,a) \ar@{-->}[rr] \ar[rd]_-{(\beta\bullet\alpha)_{c,a}} & & (G''\bullet F'')(c,a)  \\
        & (G'\bullet F')(c,a) \ar[ru]_-{(\beta'\bullet\alpha')_{c,a}} \\
        & F'(b,a)\os G'(c,b) \ar[rd]^-{\alpha'_{b,a}\os\beta'_{c,b}} \ar[u]_{\copr_b}\\
        F(b,a)\os G(c,b) \ar[uuu]^{\copr_b} \ar[rr]_-{(\alpha'\cdot\alpha)_{b,a}\os(\beta'\cdot\beta)_{c,b}} \ar[ru]^-{\alpha_{b,a}\os\beta_{c,b}} & & F''(b,a)\os G''(c,b) \ar[uuu]_{\copr_b}
    }
}


\subsection{Proof of Proposition~\ref{prp.rigit_to_NF}}
\label{app.rigid_to_NF}
\begin{proof}[Proof of Proposition~\ref{prp.rigit_to_NF}]
We sketch the proof that  $\alpha^\sharp$ is invertible if $\cC$ is rigid, and similar for $\alpha^\flat, \alpha^\sharp_\wr, \alpha^\flat_\wr$.

Let $\cC$ be a rigid monoidal $\cV$-category,  we have the following natural isomorphism of hom-objects
\begin{equation}
    \cC(x\boxtimes y,z)\cong \cC(x,z\boxtimes y^L).
\end{equation}
We then have the following commuting diagram
\begin{equation}
\begin{tikzcd}
	{\int^{c\in\cC}\cC(x,wc)\ot\cC(c y,z)} &&&& {\cC(xy,w z)} \\
	& {\int^{c\in \cC}\cC(x,wc)\ot\cC(c,z y^L)} & {\cC(x,wzy^L)} \\
	\\
	& {\cC(x,wc)\ot\cC(c,zy^L)} \\
	\\
	{\cC(x,wc)\ot\cC(cy,z)}
	\arrow["{\alpha^\sharp_{(x,y),(w,z)}}", from=1-1, to=1-5]
	\arrow["\cong", from=1-1, to=2-2]
	\arrow["\cong", from=2-2, to=2-3]
	\arrow["\cong", from=2-3, to=1-5]
	\arrow["{\copr_c}", from=4-2, to=2-2]
	\arrow["{\text{composition along $c$}}"{description}, from=4-2, to=2-3]
	\arrow["{\copr_c}", from=6-1, to=1-1]
	\arrow["{\text{composition along $c$}}"{description}, curve={height=70pt}, from=6-1, to=1-5]
	\arrow["\cong"', from=6-1, to=4-2]
\end{tikzcd}.
\end{equation}
Here we have omitted the tensor product symbols \(\boxtimes\) of $\cC$  for simplicity.
Due to the universal property of the coend $\int^{c\in \cC}\cC(x,wc)\os\cC(cy,z)$, $\alpha^\sharp_{(x,y),(w,z)}$ is equal to the isomorphism
\begin{equation}
\int^{c\in \cC}\cC(x,wc)\os\cC(cy,z)\cong\int^{c\in\cC}\cC(x,wc)\os\cC(c,zy^L)\cong \cC(x,wzy^L)\cong \cC(xy,wz).
\end{equation}

Since the isomorphism above is induced by the rigidity of \(\cC\), it is natural in all variables. Hence the components $\alpha^\sharp_{(x,y),(w,z)}$ assemble objectwise to give the 
isomorphism $\alpha^\sharp$  of profunctors. 
\end{proof}

\section{Several canonical $\cV$-enriched structures}
\label{app.enriched_structure}
Let $\cV$ be a cosmos.
For a $\cV$-category $\cC$, note that $\Vcat(*,\cC)$ is exactly the underlying category $\underline{\cC}$. Conversely, if $\cD$ is an ordinary category that can be canonically extended to a representable 2-functor $\Vcat(-,\overline\cD): \Vcat^{1\op} \to \cat$ such that $\Vcat(*,\overline\cD)\cong \cD$, we call the representing $\cV$-category $\overline\cD$ as the canonical $\cV$-enriched structure of $\cD$.

We list the canonical $\cV$-enriched structures of the important ordinary categories we encounter in the main text: 

\begin{lemma}\label{lem.enriched_structure}
\begin{enumerate}
    \item \label{item1.lem.enriched_structure} Let $\cC,\cD$ be $\cV$-categories. Then the 2-functor $\Vcat(\cC\ot -,\cD)$ is represented by the \emph{internal hom} $\enf{\cC}{\cD}$. $\underline{\enf{\cC}{\cD}}=\Vcat(\cC,\cD).$
    \item \label{item2.lem.enriched_structure} Let $\cA,\cB$ be $\cV$-categories. Then the 2-functor $\vprof(\cA\ot -,\cB)$ is represented by $\cV$-category $\enf{\cB^\op\ot\cA}{\bcV}=:\bvprof(\cA,\cB)$. $\underline{\bvprof(\cA,\cB)}=\vprof(\cA,\cB).$
    \item \label{item3.lem.enriched_structure} Let $\cC$ be a monoidal $\cV$-category, and $(\cM,\boxdot_\cM),(\cN,\boxdot_\cN)$ be left $\cC$-modules. The 2-functor ${}_\cC\Vcat^\lax(\cM\ot -,\cN)$ is representable, and we denote the representing object by $\enf[\cC]{\cM}{\cN}^\lax$, which is essentially also an internal hom ( $_\cC\Vcat$ is naturally a right module 2-category over $\Vcat$). The 2-functor ${}_\cC\vprof^\lax(\cM\ot -,\cN)$ is similarly represented by $\enf[\cC]{\cM}{\enf{\cN^\op}{\bcV}}^\lax=:{}_{\cC}\bvprof^\lax(\cM,\cN)$. $\underline{\enf[\cC]\cM\cN^\lax}={}_\cC\Vcat^\lax(\cM,\cN),$ $\underline{{}_\cC\bvprof^\lax(\cM,\cN)}={}_{\cC}\vprof^\lax(\cM,\cN).$
    \item \label{item4.lem.enriched_structure} Let $\cA,\cC$ be $\cV$-categories and $T$ be a promonad on $\cC$. The 2-functor $\Lmd_{\Id_{-^\op} \ot T}(\cA)\cong \Lmd_T(-\ot \cA)$ is representable, and we denote the representing object by $\bLmd_{T}(\cA)$. $\underline{\bLmd_{T}(\cA)}=\Lmd_{T}(\cA).$
\end{enumerate}
    \begin{proof}
     
     \begin{enumerate}
         \item 
            The internal hom is a standard construction in enriched category theory. Given $F,G\in\Vcat(\cC,\cD)$, we set their hom space $\enf[]{\cC}{\cD}(F,G)$ as the end $$\int_{c\in\cC}\cD(Fc,Gc)\in\cV.$$ Then $\Vcat(\cX,\enf{\cC}{\cD})\cong \Vcat(\cC\ot \cX,\cD)$. For similicity, later we denote $\nat(F,G)=\enf{-}{-}(F,G)$ and suppress the $\cV$-categories.
        \item It is straightforward to check that
        \begin{align*}
        \Vcat(\cX, \bvprof(\cA,\cB))&=\Vcat(\cX, \enf{\cB^\op\ot\cA}{\bcV})\\
        &\cong \Vcat(\cB^\op\ot\cA\ot\cX,\bcV)=\vprof(\cA\ot\cX,\cB).
        \end{align*}
        \item  
            The case regarding module (pro)functors or promonads are actually constructed using full subcategories of certain \emph{inserters} in $\Vcat$, which can be further expressed concretely using equalizers in $\cV$.

We give the construction for module functors. Suppose $(F,\beta),(G,\gamma)\in {}_\cC\Vcat^\lax(\cM,\cN)$. $\beta$ as a natural transformation can be identified with a morphism  \[
                \beta\:\one\to\nat(\boxdot_\cN  (\id_\cC\os F),F  \boxdot_\cM),
            \]
and similarly for $\gamma$. Then we define the hom-object ${}_\cC[\cM,\cN]^\lax((F,\beta),(G,\gamma))$ as the equalizer of 
 \[
                \diagram{
                    \nat(F,G) \ar@<0.7ex>[r]^-{\beta^\ast} \ar@<-0.7ex>[r]_-{\gamma_\ast} & \nat(\boxdot_\cN (\id_\cC\os F),G \boxdot_\cM)
                }
            \]
            in $\cV$. Here $\beta^\ast$ is defined as the morphism rending the diagram
\begin{equation*}
    \begin{tikzcd}
    [scale cd=0.8]
	{\nat(F,G)} && {\nat(F\boxdot_\cM,G\boxdot_\cM)} \\
	{\nat(\boxdot_\cN(\id_\cC\os F),G \boxdot_\cM)} && {\nat(F\boxdot_\cM,G\boxdot_\cM)\os \nat(\boxdot_\cN(\id_\cC\os F),F \boxdot_\cM)}
	\arrow["{- \boxdot_\cM}", from=1-1, to=1-3]
	\arrow["{\beta^\ast}"', from=1-1, to=2-1]
	\arrow["{\id\otimes \beta}", from=1-3, to=2-3]
	\arrow["\circ", from=2-3, to=2-1]
\end{tikzcd}
\end{equation*}

           is commutative. 
Similar for $\gamma_\ast$
\begin{equation*}
    \begin{tikzcd}
    [scale cd=0.8]
	{\nat(F,G)} && {\nat(\boxdot_\cN (\id_\cC \otimes F),\boxdot_\cN (\id_\cC\otimes G))} \\
	{\nat(\boxdot_\cN (\id_\cC\otimes F),G (\boxdot_{\cM}))} && {\nat(\boxdot_\cN ( \id_\cC\otimes G),G\boxdot_\cM)\otimes \nat(\boxdot_\cN (\id_\cC \otimes F),\boxdot_\cN (\id_\cC\otimes G))}
	\arrow["{\boxdot_{\cN}(\id_\cC\ot-)}", from=1-1, to=1-3]
	\arrow["{\gamma_\ast}"', from=1-1, to=2-1]
	\arrow["{\gamma\otimes \id}", from=1-3, to=2-3]
	\arrow["\circ", from=2-3, to=2-1].
\end{tikzcd}
\end{equation*}
For the representability,  we should consider the adjunction 
\begin{equation*}
    \Vcat(\cX,[\cM,\cN])\cong \Vcat(\cM\otimes \cX,\cN).
\end{equation*}

A natural transformation in $\Vcat(\cX,{}_\cC[\cM,\cN]^\lax)$ is exactly one in the left hand side $\Vcat(\cX,[\cM,\cN])$ satisfying certain equalizing condition, which translates to the condition of respecting lax left module functor structures under the above adjunction. Therefore, 
\begin{equation*}
    \Vcat(\cX,{}_{\cC}[\cM,\cN]^\lax)\cong {}_\cC\Vcat^\lax(\cM\otimes \cX,\cN).
\end{equation*}

For the case of module profunctors, one can similarly define the hom-object\linebreak
${}_\cC\bvprof^\lax(\cM,\cN)((F,\beta),(G,\gamma))$ as the equalizer
  \[
                \diagram{
                    \nat(F,G) \ar@<0.7ex>[r]^-{\beta^\ast} \ar@<-0.7ex>[r]_-{\gamma_\ast} & \nat((\boxdot_\cN)_\ast\bullet(\Id_\cC\os F),G\bullet (\boxdot_\cM)_\ast)
                }
            \]
            with certain canonical definition of $\beta^\ast$ and $\gamma_\ast$.
 Alternatively, due to the equivalence~\eqref{eq.moduleprof-prof}, we have 
\begin{equation*}
     {}_\cC\vprof^\lax(\cM\os\cX,\cN)\cong{}_{\cC}\Vcat^\lax(\cM\otimes \cX,[\cN^\op,\bcV])\cong \Vcat(\cX,{}_\cC[\cM,[\cN^\op,\bcV]]^\lax).
\end{equation*}
Therefore, ${}_\cC[\cM,[\cN^\op,\bcV]]^\lax$ represents the 2-functor ${}_\cC\vprof^\lax(\cM\os-,\cN)$ and we denote ${}_\cC\bvprof^\lax(\cM,\cN):= {}_\cC[\cM,[\cN^\op,\bcV]]^\lax$.

            \item
            For the case of promonad, suppose $(P,\lambda),(Q,\kappa)\in\Lmd_T(\cA)$. Then $\lambda$ can be identified with a morphism
            \[
                \one\to \nat(T\bullet P,P),
            \]
            and similar holds for $\kappa$. We set the hom space $\bLmd_T(\cA)((P,\lambda),(Q,\kappa))$ as the equalizer of 
            \[
                \diagram{
                \nat(P,Q)\ar@<0.7ex>[r]^-{\lambda^\ast} \ar@<-0.7ex>[r]_-{\kappa_\ast} & \nat(T\bullet P,Q)
                }.
            \]
            Here, $\lambda^\ast$ and $\kappa_\ast$ are defined by the commutative diagrams
            \[
                \diagram{
                    \nat(P,Q) \ar[r]^-{1\os\lambda} \ar[d]_{\lambda^\ast} & \nat(P,Q)\os\nat(T\bullet P,P) \ar[ld]^-\circ \\
                    \nat(T\bullet P,Q)
                }
            \]
            and
            \[
                \diagram{
                    \nat(P,Q) \ar[d]_{\kappa_\ast} \ar[r]^-{T\bullet-} & \nat(T\bullet P,T\bullet Q) \ar[d]_{\kappa\os 1} \\
                    \nat(T\bullet P,Q) & \nat(T\bullet Q,Q)\os\nat(T\bullet P,T\bullet Q) \ar[l]^-\circ
                }.
            \]
        We make use of Lemma~\ref{lem.EM} to show the representibility. Denote the Eilenberg-Moore object of $T$ by $\wtT$. It is also clear that the Eilenberg-Moore object of an identity promonad $\Id_\cX:\cX\arprof\cX$ is $\cX$ itself. Then
        \begin{align*}
            \Vcat(\cX,\bLmd_T(\cA))
            &\cong \Vcat(\cX,\bvprof(\cA,\wtT))
            \\&\cong \Vcat(\cX,[\wtT^\op\ot \cA,\bcV])
            \cong \Vcat(\cX\ot \wtT^\op\ot \cA,\bcV)
            \\&\cong \vprof(\cA,\cX^\op \ot \wtT)
            \cong\vprof(\cX\ot\cA,\wtT)
            \\&\cong \Lmd_{\Id_{\cX^\op}\ot T}(\cA)\cong \Lmd_T(\cX\ot\cA).
        \end{align*}
        Note that the above further implies that
        \[ \bLmd_{\Id_{\cX^\op}\ot T}(\cA)\cong \bLmd_T(\cX\ot\cA).\]
     \end{enumerate}
    \end{proof}
\end{lemma}

 We now sketch the proof of Theorem~\ref{thm.gen_Kitaev_Kong}. 
As in the construction of item~(4) in Lemma~\ref{lem.enriched_structure}, 
for the renormalization algebra \(\MnCN\), and for $(P,\lambda), (Q,\kappa)\in \Lmd_{\MnCN}$, we define the hom object $\bLmd_{\MnCN}((P,\lambda), (Q,\kappa))$ to be the equalizer 
  \[
                \diagram{
                \nat(P,Q)\ar@<0.7ex>[r]^-{\lambda^\ast} \ar@<-0.7ex>[r]_-{\kappa_\ast} & \nat((\boxdot_\cN)_* \bullet (1_\cC\ot P)\bullet (\boxdot_\cM)^*,Q)
                }.
            \]

 Let $(P,\lambda'), (Q,\kappa')$ be the corresponding left $\cC$ module profunctor under the equivalence $\Psi$ in the proof of Theorem~\ref{thm.unenriched_Kitaev_Kong}. Using the transpose isomorphism in Proposition~\ref{prp.transpose}.\ref{item2.prp.transpose}
 \[\nat\big(
(\boxdot_\cN)_* \bullet (1_\cC\otimes P)\bullet(\boxdot_\cM)^*,
Q
\big)
\cong
\nat\big(
(\boxdot_\cN)_* \bullet (1_\cC\otimes P),
Q\bullet(\boxdot_\cM)_*
\big),\]
      the above equalizer is identified with the equalizer    
 \[
                \diagram{
                    \nat(P,Q) \ar@<0.7ex>[r]^-{(\lambda')^\ast} \ar@<-0.7ex>[r]_-{\kappa'_\ast} & \nat((\boxdot_\cN)_\ast\bullet(\Id_\cC\os P),Q\bullet (\boxdot_\cM)_\ast)
                }
            \]
            which defines the hom-object $_\cC\bvprof^\lax(\cM,\cN)((P,\lambda'),(Q,\kappa'))$. Therefore, for any \((P,\lambda),(Q,\kappa)\in \Lmd_{\MnCN}\), we obtain a canonical isomorphism
\[
\bLmd_{\MnCN}\big((P,\lambda),(Q,\kappa)\big)
\cong
{}_\cC\bvprof^\lax(\cM,\cN)
\big((P,\lambda'),(Q,\kappa')\big),
\] 
            it upgrades     $\Psi$ to an equivalence between $\cV$-categories
            \begin{equation}
                \bLmd_{\MnCN} \cong {}_\cC\bvprof^{\lax}(\cM,\cN).
            \end{equation}
           $\bLmd_{\MnCN}$  then  represents the 2-functor $\Lmd_{(\cM\ot-) \Omega_\cC \cN}$.

For the equivalence with $\bLmd_{\MCN}(\ast)$, note that 
$\bLmd_{\MCN}(\ast)$ and ${}_\cC\bvprof^\lax(\cM,\cN)$ represent $\Lmd_{1_{(-)^\op} \ot \MCN}(\ast)$ and $ {}_\cC\vprof^\lax(\cM-,\cN)$, respectively. For  any $\cV$-category $\cX$,  Theorem~\ref{thm.unenriched_Kitaev_Kong} gives the equivalence
\begin{equation*}
   \Lmd_{1_{\cX^\op} \ot \MCN}(\ast)\cong \Lmd_{\cM\cX\bbH_\cC \cN}(\ast)\cong {}_\cC\vprof^\lax(\cM\cX,\cN).
\end{equation*}
Hence the two represented functors are naturally equivalent. By Yoneda lemma, we have 
\begin{equation}
    \bLmd_{\MCN}(\ast)\cong {}_\cC\bvprof^\lax(\cM,\cN).
\end{equation}

To further upgrade the above to a monoidal equivalence when $\cM=\cN$, the most convenient viewpoint is to consider pseudoalgebra objects, algebra homomorphisms and centralizers in the 2-categories $\vprof$ and $\Vcat$. The lax monoidal 2-functor
\begin{align*}
    \bvprof(*,-)=[-^\op, \bcV]: \vprof &\to \Vcat\\
    \cC &\mapsto \bvprof(*,\cC)=[\cC^\op,\bcV]
\end{align*}
preserves pseudoalgebra objects and algebra homomorphisms, and it is easy to verify that this functor also preserves centralizers by tautology. Then, by the duality of $\vprof$, the action $(\boxdot_\cM)_\ast: \cC\ot\cM\arprof\cM$ is equivalent to an algebra homomorphism $\hat\boxdot_\cM:\cC\arprof\cM\ot\cM^\op$. One can show that the centralizer of $\hat\boxdot_\cM$ in $\vprof$ is the Eilenberg-Moore category $\widetilde{\MCM}$ using Theorem~\ref{thm.unenriched_Kitaev_Kong}. On the other hand, the centralizer of $\bvprof(*,  \hat\boxdot_\cM)$ in $\Vcat$ is exactly $_\cC\bvprof^\lax(\cM,\cM)$, we therefore have the desired equivalence
\[ \bLmd_{\MCM}(*)\cong \bvprof(*,  \widetilde{\MCM})\cong {}_\cC\bvprof^\lax(\cM,\cM),\]
which is automatically monoidal by the universal property of centralizer. More details regarding centralizers will be given in our future work.

\section{The Eilenberg-Moore category of a promonad}\label{sec.EM}

Let $\cV$ be a cosmos. Let $\cC$ be a $\cV$-category and $T\:\cC\arprof\cC$ be a promonad with multiplication $\mu$ and unit $\eta$. We define the following $\cV$-category $\wtT$ \cite{Justesen_1968}:
\begin{itemize}
    \item $\mathrm{ob}(\wtT) \defdtobe \mathrm{ob}(\cC)$.
    \item For objects $a, b \in \wtT$, $\wtT(a,b) := T(a,b)\in\cV$ 
   \item For $a,b,c\in \cC$, the composition $\circ^{a,b,c}_\wtT:\wtT(b,c)\os\wtT(a,b)\rightarrow \wtT(a,c)$ is given by the unique morphism rendering the diagram
   \begin{equation}
       \begin{tikzcd}
	{T(b,c)\os T(a,b)} && {T(a,c)} \\
	{\int^bT(a,b)\os T(b,c)}
	\arrow["{\circ^{a,b,c}_\wtT}", from=1-1, to=1-3]
	\arrow["{\copr_b}"', from=1-1, to=2-1]
	\arrow["{\mu_{a,c}}"', from=2-1, to=1-3]
\end{tikzcd}
   \end{equation}
   commutative, where $\copr_b$ is the coprojection map associated with the coend.
   
\item For $a \in \wtT$, the unit $i^a_\wtT$ is defined as
\begin{equation}
\label{id}
i^a_\wtT : \one \xrightarrow{ i^a_\cC}\cC(a,a)\xrightarrow{\eta_{a,a}}T(a,a).
\end{equation}
\end{itemize}
The promonad axioms \eqref{eq.pmd_ass} and \eqref{eq.pmd_unit} guarantee that $\wtT$ is well-defined. There is also a canonical $\cV$-functor $H: \cC \to \wtT$ defined by $H(a)=a$ for $a\in\ob(\cC)$, and 
\[
    H_{a,b}:=\eta_{a,b}:\cC(a,b)\rightarrow T(a,b)
\]
for $a,b\in\cC$.
\begin{lemma}[\textnormal{\cite{Street_1981} or\cite[\S 10.1]{Lopez_Franco_2007}}]\label{lem.EM}
    For any $\cV$-category $\cA$, there is a $\cV$-equivalence
    \[
        \bLmd_T(\cA)\cong \bvprof(\cA,\wtT).
    \]
    In particular, we have $\bLmd_T(*)\cong\bvprof(*,\wtT)$.
    \begin{proof}
        It suffices to prove the first statement. Our aim is to construct a $\cV$-equivalence \[J\:\bLmd_T(\cA)\to\bvprof(\cA,\wtT).\]
        Given $(P,\lambda\:T\bullet P\Rightarrow P)\in\bLmd_T(\cA)$, define a morphism
        \[
           r_{dca}\:\diagram{
            P(c,a)\os\wtT(d,c)\ar[r]^-{\copr_c} & \int^{c\in\cC}\wtT(d,c)\os P(c,a) \ar[r]^-{\lambda_{d,a}} & P(d,a)
            }
        \]
        for $a\in\cA$ and $c,d\in\cC$. Then one can check that there is a profunctor $P'\:\cA\arprof\wtT$ defined as follows:
        \begin{itemize}
            \item  $P'(c,a)\defdtobe P(c,a)$ for $c\in\wtT,a\in\cA$.
            \item The left $\cA$-actions on $P'$ are inherited from the left $\cA$ action on $P$.
            \item The right $\wtT$-actions on $P'$ given by $\{r_{dca}\}_{d,c\in\cC,a\in\cA}$. 
        \end{itemize}
        We set $J(P)\defdtobe P'$.

        We need to define $J$'s action on hom spaces. To this end, for $P=(P,\lambda),Q=(Q,\kappa)\in\bLmd_T(\cA)$, we show that there exists a canonical isomorphism 
        \begin{equation}\label{eq.lem.EM}
            \bLmd_T(\cA)(P,Q)\isom \bvprof(\cA,\wtT)(P',Q').
        \end{equation}
        Note that the l.h.s. of \eqref{eq.lem.EM} is by definition the terminal object of the category $\cX$ of morphisms
        \begin{equation}\label{eq2.lem.EM}
            q\:L\to \nat(P,Q)
        \end{equation}
        in $\cV$ equalizing 
        $\diagram{
                \nat(P,Q)\ar@<0.7ex>[r]^-{\lambda^\ast} \ar@<-0.7ex>[r]_-{\kappa_\ast} & \nat(T\bullet P,Q)
                }$.
        On the other hand, the r.h.s. of \eqref{eq.lem.EM} is defined as the terminal object of the category $\cY$ of wedges
        \begin{equation}\label{eq3.lem.EM}
            (L,\{\theta_{c,a}\:L\to [P(c,a),Q(c,a)]\}_{c\in\wtT,a\in\cA}).
        \end{equation}
        It is enough to prove that $\cX$ and $\cY$ are canonically equivalent. This is indeed the case, as a morphism in \eqref{eq2.lem.EM} is precisely a wedge
        \[
            (L,\{\theta_{c,a}\:L\to [P(c,a),Q(c,a)]\}_{c\in\cC,a\in\cA})
        \]
        by universal property of the end $\nat(P,Q)$, while the equalizing condition on $q$ precisely states that $(L,\{\theta_{c,a}\}_{c\in\cC,a\in\cA})$ is in addition a wedge in the sense of \eqref{eq3.lem.EM}. Now $\cX\cong\cY$ canonically, yielding an isomorphism in \eqref{eq.lem.EM}, which we define as $J_{P,Q}$. One can verify that $\{J_{P,Q}\}_{P,Q\in\bLmd_T(\cA)}$ form a well-defined $\cV$-functor.

        Let us show that $J$ is a $\cV$-equivalence. Note that $J$ is $\cV$-fully-faithful by definition. To establish that $J$ is $\cV$-essentially surjective, assume $F\in\bvprof(\cA,\wtT)$. Then there is a family of morphisms
        \[
            r_{dca}\:F(c,a)\os\wtT(d,c)\to F(d,a)
        \]
        for $c,d\in\wtT,a\in\cA$. Fixing $d,a$, then the morphisms $\{r_{dca}\}_{c\in\wtT}$ is $\cC$-balancing, hence induces a morphism
        \[
            \lambda_{d,a}\:\int^{c\in\cC}\wtT(d,c)\os F(c,a)\to F(d,a).
        \]
        On the other hand, note that $F$ gives rise to a profunctor $\tilde{F}\:\cA\arprof\cC$ by
        \[
            \tilde{F}: \cA\overset{F}{\nrightarrow}\wtT\overset{H^\ast}{\nrightarrow}\cC.
        \]
        One can verify that $(\tilde{F},\{\lambda_{d,a}\}_{d\in\cC,a\in\cA})$ is a left $T$-module from $\cA$. The $T$-action $\widetilde{\lambda}:T\bullet \tilde{F}\Rightarrow \tilde{F}$ is induced by the universal property of coend
        \[
\begin{tikzcd}
	{\int^{c\in\cC} T(c',c)\os\tilde{F}(c,a)} && {T(c',c)\os F(H(c),a)} \\
	&& {\wtT(c',c)\os F(c,a)} \\
	&& {F(c,a)\os \wtT(c',c)} \\
	{\tilde{F}(c',a)} && {F(c',a)}
	\arrow["{\widetilde{\lambda_{c',a}}:=}"', from=1-1, to=4-1]
	\arrow["{\copr_{c}}"', from=1-3, to=1-1]
	\arrow["\id"', from=1-3, to=2-3]
	\arrow["{\tau_{\wtT(c',c),F(c,a)}}"', from=2-3, to=3-3]
	\arrow["{r_{c'ca}}"', from=3-3, to=4-3]
	\arrow["\id"', from=4-3, to=4-1]
\end{tikzcd}
        \]
        The module associativity follows from the $\wtT$-associativity of $F$, the module unitality follows from the definition of the right $\cC$ action on $\tilde{F}:\cA\nrightarrow\cC$.

        Moreover, $J(\tilde{F})\cong F$ as objects in $\bvprof(\cA,\wtT)$. Hence $J$ is $\cV$-essentially surjective, which completes the proof.
        
    \end{proof}
\end{lemma}
\begin{remark}
    Lemma \ref{lem.EM} implies the equivalence
    \[
        \Lmd_T(\cA)\cong\vprof(\cA,\wtT)
    \]
    of ordinary categories. One can furthermore check that this equivalence is natural in $\cA$ in a certain sense. These natural equivalence precisely dictates that $\wtT$ is the Eilenberg-Moore object of the monad $T$ in the 2-category $\vprof$ of $\cV$-profunctors \cite{Street_1972}. In fact, $\wtT$ is also the Kleisli object of $T$, that is, the Eilenberg-Moore object of the monad $T$ in the 2-category $\vprof^{1\op}$. This latter role played by $\wtT$ can be found in \cite{Street_1981} for example, and is more well-known, which actually brings $\wtT$ the name ``Kleisli category of $T$'' in some literature. That $\wtT$ is also the Eilenberg-Moore object must also be well-known since there is a 2-equivalence $\vprof\cong\vprof^{1\op}$; it is explicitly mentioned in \cite{Lopez_Franco_2007}.
\end{remark}

\section{Supplementary graphical axioms, properties and proofs}\label{sec.graph.app}

\subsection{Proof of Theorem \ref{thm.higher_transpose}}\label{sub.pf_of_higher_transpose}
\begin{proof}[Proof of Theorem \ref{thm.higher_transpose}]
    Let us construct a functor $J\:\vprof(\cC\os\cD,\cE)\to\vprof(\cD,\cC^\op\os\cE)$. The objects' assignment is designated by the statement of the theorem. For a profunctor homomorphism $\beta\:F\Rightarrow G\:\cC\os\cD\arprof\cE$, we set $J(F)$ to be the induced pro-tensor network map 
    \[
    \diagram{
    \ctikz{
        \draw (-1,0)--(0,0)node[pos=0.5,below]{$\cE$};
        \draw (0,0.15)--(0.5,0.15);
        \draw (0.5,0.15) arc (-90:90:0.25);
        \draw (0.5,0.65)--(-1,0.65);
        \draw (-0.4,0.9)node{$\cC^\op$};
        \draw (0,-0.15)--(1,-0.15) node[pos=0.6,below]{$\cD$};
        \mynode{0}{0}{0.3}{0.3}{F}
    }
    \ar@{=>}[r]^-\beta &
    \ctikz{
        \draw (-1,0)--(0,0)node[pos=0.5,below]{$\cE$};
        \draw (0,0.15)--(0.5,0.15);
        \draw (0.5,0.15) arc (-90:90:0.25);
        \draw (0.5,0.65)--(-1,0.65);
        \draw (-0.4,0.9)node{$\cC^\op$};
        \draw (0,-0.15)--(1,-0.15) node[pos=0.6,below]{$\cD$};
        \mynode{0}{0}{0.3}{0.3}{G}
    }
    }
    \]
    which is well-defined and moreover functorial by Theorem \ref{thm.locally_induce}. A functor $K\:\vprof(\cD,\cC^\op\os\cE)\to\vprof(\cC\os\cD,\cE)$ which also extends the objects' assignment in the statement of the theorem can be similarly constructed. Note that there exists a natural transformation $\theta\:KJ\Rightarrow \Id_{\vprof(\cC\os\cD,\cE)}$ whose component on $F\in\vprof(\cC\os\cD,\cE)$ is given by the pro-tensor network map
    \[
        \ctikz{[yscale=0.75]
            \draw (-1,0)--(0,0)node[pos=0.5,below]{$\cE$};
            \draw (0,0.15)--(0.5,0.15);
            \draw (0.5,0.15) arc (-90:90:0.25);
            \draw (0.5,0.65)--(-0.5,0.65);
            \draw (-0.5,0.65) arc (270:90:0.25);
            \draw (-0.5,1.15)--(1,1.15)node[pos=0.75,above]{$\cC$};
            \draw (0,-0.15)--(1,-0.15) node[pos=0.6,below]{$\cD$};
            \mynode{0}{0}{0.3}{0.4}{F}
        }
        \Rightarrow
        \ctikz{
            \draw (-1,0)--(0,0)node[pos=0.5,below]{$\cE$};
            \draw (0,0.15)--(1,0.15)node[pos=0.6,above]{$\cC$};
            \draw (0,-0.15)--(1,-0.15) node[pos=0.6,below]{$\cD$};
            \mynode{0}{0}{0.3}{0.3}{F}
        }
    \]
    induced by $\chi_\cC$. Indeed, the naturality follows from that the diagram
    \[
        \diagram@=3pc{
            \ctikz{[yscale=0.75]
                \draw (-1,0)--(0,0);
                \draw (0,0.15)--(0.5,0.15);
                \draw (0.5,0.15) arc (-90:90:0.25);
                \draw (0.5,0.65)--(-0.5,0.65);
                \draw (-0.5,0.65) arc (270:90:0.25);
                \draw (-0.5,1.15)--(1,1.15);
                \draw (0,-0.15)--(1,-0.15);
                \mynode{0}{0}{0.3}{0.4}{F}
            }
            \ar@{=>}[d]_{\chi_\cC} \ar@{=>}[r]^-{\beta}
            \ar@{}[rd]|{\text{(Theorem \ref{thm.local})}}
            &
            \ctikz{[yscale=0.75]
                \draw (-1,0)--(0,0);
                \draw (0,0.15)--(0.5,0.15);
                \draw (0.5,0.15) arc (-90:90:0.25);
                \draw (0.5,0.65)--(-0.5,0.65);
                \draw (-0.5,0.65) arc (270:90:0.25);
                \draw (-0.5,1.15)--(1,1.15);
                \draw (0,-0.15)--(1,-0.15);
                \mynode{0}{0}{0.3}{0.4}{G}
            }
            \ar@{=>}[d]^{\chi_\cC}
            \\
            \ctikz{
                \draw (-1,0)--(0,0);
                \draw (0,0.15)--(1,0.15);
                \draw (0,-0.15)--(1,-0.15);
                \mynode{0}{0}{0.3}{0.3}{F}
            }
            \ar@{=>}[r]_-\beta
            &
            \ctikz{
                \draw (-1,0)--(0,0);
                \draw (0,0.15)--(1,0.15);
                \draw (0,-0.15)--(1,-0.15);
                \mynode{0}{0}{0.3}{0.3}{G}
            }
        }
    \]
    of pro-tensor network maps commute for any $\beta\:F\Rightarrow G\:\cC\os\cD\arprof\cE$.
    Moreover, $\theta$ is a natural isomorphism by that $\chi_\cC$ is an isomorphism. A natural isomorphism $JK\Rightarrow\Id_{\vprof(\cD,\cC^\op\os\cE)}$ can be constructed from $\zeta_\cC$ in a similar way. As a result, $J$ and $K$ are mutually inverse equivalences.
\end{proof}
\subsection{Proof of \eqref{eq.unit_advanced} and \eqref{eq.mult_advanced}}\label{sub.two_module_transpose}
\begin{proof}[Proof of \eqref{eq.unit_advanced} and \eqref{eq.mult_advanced}]
    \eqref{eq.unit_advanced} is proved by 
\[
    \ctikz{[scale=1,xscale=1.6]
        \node (1_1) at (0,4){$
        \ctikz{[scale=0.3]
            \draw[color3] (-1.5,0)--(1.5,0);
            \draw (-1.5,1)--(1.5,1);
            \draw[fill=white] (1.5,1)circle[radius=0.25];
        }$};
        \node (1_3) at (4,4){$
        \ctikz{[scale=0.3]
            \draw[color3] (-4.25,0)--(1.5,0);
            \draw (-1.5,0.7)--(-0.5,0);
            \draw (0.5,0)--(1.5,0.7);
            \draw (-2.75,1)--(-4.25,1);
            \draw[fill=white] (-2.75,1)circle[radius=0.25];
            \draw[fill=white] (-1.5,0.7)circle[radius=0.25];
            \draw[fill=white] (1.5,0.7)circle[radius=0.25];
        }$};
        \node (2_2) at (2,2){$
        \ctikz{[scale=0.3]
            \draw[color3] (-2,0)--(3.5,0);
            \draw (-2,1)--(-0.5,1);
            \draw (1,1)--(3,1);
            \draw[fill=white] (3,1)circle[radius=0.25];
            \draw[fill=white] (1,1)circle[radius=0.25];
            \draw[fill=white] (-0.5,1)circle[radius=0.25];
        }$};
        \node (3_1) at (0,0){$
        \ctikz{[scale=0.3]
            \draw[color3] (-1.5,0)--(1.5,0);
            \draw (-1.5,1)--(1.5,1);
            \draw[fill=white] (1.5,1)circle[radius=0.25];
        }$};
        \node (3_3) at (4,0){$
        \ctikz{[scale=0.3]
            \draw[color3] (-1.5,0)--(1.5,0);
            \draw (-1.5,0.7)--(-0.5,0);
            \draw (0.5,0)--(1.5,0.7);
            \draw[fill=white] (1.5,0.7)circle[radius=0.25];
        }$};
        \draw[nat] (1_1)--(1_3)node[midway, above]{\lambda_\wr^R\star\lambda_\wr^{-1}};
        \draw[nat] (1_1)--(3_1)node[midway, left]{\id};
        \draw[nat] (1_1)--(2_2)node[midway, above right]{\eta_0};
        \draw[nat] (2_2)--(1_3)node[midway, above left]{\eta_\wr};
        \draw[nat] (2_2)--(3_1)node[midway, below right]{\epsilon_0};
        \draw[nat] (1_3)--(3_3)node[midway, right]{\epsilon_0};
        \draw[nat] (3_1)--(3_3)node[midway, below]{\eta_\wr};
        \draw (2,3.25)node{\tiny(\eqref{eq.right_dual_of_lambda_module})};
        \draw (3,1.35)node{\tiny(Theorem \ref{thm.local})};
        \draw (0.7,2)node{\tiny(\eqref{eq.zig_zag_u})};
    }.
\]
\eqref{eq.mult_advanced} is proved by
\[
    \ctikz{[scale=1.3,xscale=1.25]
        \node (1_1) at (0,4){$
        \ctikz{[scale=0.3]
            \draw[color3] (-1,0)--(1.5,0);
            \draw (-1,1.05)--(0.5,1.05);
            \draw (0.5,1.05)--(1.5,1.75);
            \draw (0.5,1.05)--(1.5,0.35);
        }$};
        \node (1_2) at (2,4){$
        \ctikz{[scale=0.3]
            \draw (0,0)--(1.5,1.05);
            \draw (-1,0)--(-2.5,1.05);
            \draw (-2.5,1.05)--(-1.5,1.75)--(1.5,1.75);
            \draw (-2.5,1.05)--(-3.5,1.05);
            \draw[color3] (-3.5,0)--(1.5,0);
        }$};
        \node (1_3) at (4,4){$
        \ctikz{[scale=0.3]
            \draw (2.1,0)--(3.1,0.7);
            \draw (1.1,0)--(3.1,1.4);
            \draw (-1,0)--(-2.5,1.05);
            \draw (-2.5,1.05)--(-1.9,1.47);
            \draw (-1.9,1.47)--(0.1,0);
            \draw (-2.5,1.05)--(-3.5,1.05);
            \draw[color3] (-3.5,0)--(3.1,0);
        }$};
        \node (1_4) at (6,4){$
        \ctikz{[scale=0.3]
            \draw (1.1,0)--(3.1,1.4);
            \draw (2.5,0.98)--(3.1,0.56);
            \draw (-1.3,1.05)--(-1.9,0.63)--(-2.5,1.05);
            \draw (-2.5,1.05)--(-1.9,1.47);
            \draw (-1.9,1.47)--(0.1,0);
            \draw (-2.5,1.05)--(-3.5,1.05);
            \draw[color3] (-3.5,0)--(3.1,0);
        }$};
        \node (2_2) at (2,2){$
        \ctikz{[scale=0.3]
            \draw[color3] (-3.5,0)--(1.5,0);
            \draw (-3.5,1.05)--(-2.5,1.05);
            \draw (-0.5,1.05)--(0.5,1.05);
            \draw (-2.5,1.05)--(-1.5,1.75)--(-0.5,1.05);
            \draw (-2.5,1.05)--(-1.5,0.35)--(-0.5,1.05);
            \draw (0.5,1.05)--(1.5,1.75);
            \draw (0.5,1.05)--(1.5,0.35);
        }$};
        \node (3_1) at (0,0){$
        \ctikz{[scale=0.3]
            \draw[color3] (-1,0)--(1.5,0);
            \draw (-1,1.05)--(0.5,1.05);
            \draw (0.5,1.05)--(1.5,1.75);
            \draw (0.5,1.05)--(1.5,0.35);
        }$};
        \node (3_4) at (6,0){$
        \ctikz{[scale=0.3]
            \draw (0.5,1.05)--(1.5,1.75);
            \draw (0.5,1.05)--(1.5,0.35);
            \draw (-1,0)--(0.5,1.05);
            \draw (-2,0)--(-3,0.7);
            \draw[color3] (-3,0)--(1.5,0);
        }$};
        \draw[nat] (1_1)--(1_2)node[midway, above]{\eta_\wr};
        \draw[nat] (1_2)--(1_3)node[midway, above]{\eta_\wr};
        \draw[nat] (1_3)--(1_4)node[midway, above]{\alpha_\wr^R};
        \draw[nat] (3_1)--(3_4)node[midway, below]{\eta_\wr};
        \draw[nat] (1_1)--(3_1)node[midway, left]{\id};
        \draw[nat] (1_4)--(3_4)node[midway, right]{\epsilon_2};
        \draw[nat] (1_1)--(2_2)node[midway, above right]{\eta_2};
        \draw[nat] (2_2)--(3_1)node[midway, below right]{\epsilon_2};
        \draw[nat] (2_2)--(1_4)node[midway, above left]{\eta_\wr};
        \draw (2.5,3.2) node{\tiny(\eqref{eq.right_dual_of_alpha_module})};
        \draw (0.6,2)node{\tiny(\eqref{eq.zig_zag_m})};
        \draw (4.2,1.5)node{\tiny(Theorem \ref{thm.local})};
    }.
\]
\end{proof}
\subsection{Axioms of a probimonad}
The axioms of a probimonad we need in Definition \ref{dfn.probimonad} are recorded in Figure \ref{fig.bimonad_axiom}.
\begin{figure}[htbp]
\eqnn{\label{eq.tau_intertwine}
\diagram@C=1pc@R=1.15pc{
    \ctikz{[scale=0.4]
        \draw (-3,0)--(1,0);
        \mynode{-1.5}{0}{0.5}{0.5}{T}
        \mynode{0}{0}{0.5}{0.5}{T} 
        \draw (1,0)--(2.5,1.05);
        \draw (1,0)--(2.5,-1.05);
    }
    \ar@{=>}[r]^-{\tau}
    \ar@{=>}[d]_{\mu}
    &
    \ctikz{[scale=0.4]
        \draw (-1.5,0)--(1,0);
        \mynode{0}{0}{0.5}{0.5}{T} 
        \draw (1,0)--(3.5,1.75);
        \draw (1,0)--(3.5,-1.75);
        \mynode{2.25}{0.875}{0.5}{0.5}{T} 
        \mynode{2.25}{-0.875}{0.5}{0.5}{T} 
    }
    \ar@{=>}[r]^-{\tau}
    &
    \ctikz{[scale=0.4]
        \draw (-1.5,0)--(1,0);
        \draw (1,0)--(5,2.8);
        \draw (1,0)--(5,-2.8);
        \mynode{2.25}{0.875}{0.5}{0.5}{T} 
        \mynode{2.25}{-0.875}{0.5}{0.5}{T} 
        \mynode{3.75}{1.925}{0.5}{0.5}{T} 
        \mynode{3.75}{-1.925}{0.5}{0.5}{T} 
    }
    \ar@{=>}[d]^-{\mu\star\mu}
    \\
    \ctikz{[scale=0.4]
        \draw (-1.5,0)--(1,0);
        \mynode{0}{0}{0.5}{0.5}{T} 
        \draw (1,0)--(2.5,1.05);
        \draw (1,0)--(2.5,-1.05);
    }
    \ar@{=>}[rr]_-{\tau}
    &
    &
    \ctikz{[scale=0.4]
        \draw (-1,0)--(1,0);
        \draw (1,0)--(3.5,1.75);
        \draw (1,0)--(3.5,-1.75);
        \mynode{2.25}{0.875}{0.5}{0.5}{T} 
        \mynode{2.25}{-0.875}{0.5}{0.5}{T} 
    }   
}
\qquad
\raisebox{1.5em}{
\diagram@R=1pc{
    \ctikz{[scale=0.4]
        \draw (-1,0)--(1,0);
        \draw (1,0)--(2,0.7);
        \draw (1,0)--(2,-0.7);
    }
    \ar@/^4pc/@{=>}[dd]^-{\eta\star\eta}
    \ar@{=>}[d]_-{\eta}
    \\
    \ctikz{[scale=0.4]
        \draw (-1.5,0)--(1,0);
        \mynode{0}{0}{0.5}{0.5}{T} 
        \draw (1,0)--(2.5,1.05);
        \draw (1,0)--(2.5,-1.05);
    }
    \ar@{=>}[d]_-{\tau} 
    \\
    \ctikz{[scale=0.4]
        \draw (-1,0)--(1,0);
        \draw (1,0)--(3.5,1.75);
        \draw (1,0)--(3.5,-1.75);
        \mynode{2.25}{0.875}{0.5}{0.5}{T} 
        \mynode{2.25}{-0.875}{0.5}{0.5}{T} 
    } 
}}}
\vskip -2ex
\eqnn{\label{eq.nu_intertwine}
\diagram@R=1.25pc{
    \ctikz{[scale=0.4]
        \draw (-3,0)--(1.5,0);
        \draw[fill=white] (1.5,0)circle[radius=0.2];
        \mynode{-1.5}{0}{0.5}{0.5}{T}
        \mynode{0}{0}{0.5}{0.5}{T}
    }
    \ar@{=>}[r]^-{\nu}
    \ar@{=>}[d]_{\mu}
    &
    \ctikz{[scale=0.4]
        \draw (-1.5,0)--(1.5,0);
        \draw[fill=white] (1.5,0)circle[radius=0.2];
        \mynode{0}{0}{0.5}{0.5}{T}
    }
    \ar@{=>}[d]^{\nu}
    \\
    \ctikz{[scale=0.4]
        \draw (-1.5,0)--(1.5,0);
        \draw[fill=white] (1.5,0)circle[radius=0.2];
        \mynode{0}{0}{0.5}{0.5}{T}
    }
    \ar@{=>}[r]_-\nu
    &
    \ctikz{[scale=0.4]
        \draw (-1,0)--(1,0);
        \draw[fill=white] (1,0)circle[radius=0.2];
    }
}
\qquad
\diagram@R=1.25pc{
    \ctikz{[scale=0.4]
        \draw (-1,0)--(1,0);
        \draw[fill=white] (1,0)circle[radius=0.2];
    }
    \ar@{=>}[d]_{\eta}
    \ar@{=>}[rd]^-{\id}
    \\
    \ctikz{[scale=0.4]
        \draw (-1.5,0)--(1.5,0);
        \draw[fill=white] (1.5,0)circle[radius=0.2];
        \mynode{0}{0}{0.5}{0.5}{T}
    }
    \ar@{=>}[r]_-{\nu}
    &
    \ctikz{[scale=0.4]
        \draw (-1,0)--(1,0);
        \draw[fill=white] (1,0)circle[radius=0.2];
    }
}}
\eqnn{\label{eq.alpha_is_monad_2mor}
\diagram@R=1.15pc{
    \ctikz{[scale=0.4]
        \draw (-1.75,0)--(1.75,0);
        \mynode{0}{0}{0.5}{0.5}{T}
        \draw (1.75,0)--(2.75,0.7);
        \draw (2.75,0.7)--(3.75,1.4);
        \draw (2.75,0.7)--(3.75,0);
        \draw (1.75,0)--(3.75,-1.4);
    }
    \ar@{=>}[r]^-\tau
    \ar@{=>}[d]_\alpha
    &
    \ctikz{[scale=0.4]
        \draw (-1.5,0)--(0,0);
        \draw (0,0)--(2,1.4);
        \mynode{1}{0.7}{0.5}{0.5}{T}
        \draw (2,1.4)--(3,2.1);
        \draw (2,1.4)--(3,0.7);
        \draw (0,0)--(3,-2.1);
        \mynode{1}{-0.7}{0.5}{0.5}{T}
    }
    \ar@{=>}[r]^-{\tau}
    &
    \ctikz{[scale=0.4]
        \draw (-1.5,0)--(0,0);
        \draw (0,0)--(2,1.4);
        \draw (2,1.4)--(4,2.8);
        \draw (2,1.4)--(4,0);
        \draw (0,0)--(4,-2.8);
        \mynode{3}{2.1}{0.5}{0.5}{T}
        \mynode{3}{0.7}{0.5}{0.5}{T}
        \mynode{3}{-2.1}{0.5}{0.5}{T}
    }
    \ar@{=>}[d]^-\alpha
    \\
    \ctikz{[scale=0.4,yscale=-1]
        \draw (-1.75,0)--(1.75,0);
        \mynode{0}{0}{0.5}{0.5}{T}
        \draw (1.75,0)--(2.75,0.7);
        \draw (2.75,0.7)--(3.75,1.4);
        \draw (2.75,0.7)--(3.75,0);
        \draw (1.75,0)--(3.75,-1.4);
    }
    \ar@{=>}[r]_-\tau
    &
    \ctikz{[scale=0.4,yscale=-1]
        \draw (-1.5,0)--(0,0);
        \draw (0,0)--(2,1.4);
        \mynode{1}{0.7}{0.5}{0.5}{T}
        \draw (2,1.4)--(3,2.1);
        \draw (2,1.4)--(3,0.7);
        \draw (0,0)--(3,-2.1);
        \mynode{1}{-0.7}{0.5}{0.5}{T}
    }
    \ar@{=>}[r]_-\tau
    &
    \ctikz{[scale=0.4,yscale=-1]
        \draw (-1.5,0)--(0,0);
        \draw (0,0)--(2,1.4);
        \draw (2,1.4)--(4,2.8);
        \draw (2,1.4)--(4,0);
        \draw (0,0)--(4,-2.8);
        \mynode{3}{2.1}{0.5}{0.5}{T}
        \mynode{3}{0.7}{0.5}{0.5}{T}
        \mynode{3}{-2.1}{0.5}{0.5}{T}
    }
}}
\vskip -1em
\eqnn{\label{eq.lambda_rho_are_monad_2mor}
\diagram@R=1.15pc{
    \ctikz{[scale=0.4]
        \draw (-1.5,0)--(1,0);
        \mynode{0}{0}{0.5}{0.5}{T} 
        \draw (1,0)--(2,0.7);
        \draw[fill=white] (2,0.7)circle[radius=0.2];
        \draw (1,0)--(2.5,-1.05);
    }
    \ar@{=>}[r]^-\tau
    \ar@{=>}[d]_-\lambda
    &
    \ctikz{[scale=0.4]
        \draw (-1,0)--(1,0);
        \draw[fill=white] (1,0)--(3.5,1.75);
        \draw[fill=white] (3.5,1.75)circle[radius=0.2];
        \draw (1,0)--(3.5,-1.75);
        \mynode{2.25}{0.875}{0.5}{0.5}{T} 
        \mynode{2.25}{-0.875}{0.5}{0.5}{T}
    }
    \ar@{=>}[d]^-{\nu}
    \\
    \ctikz{[scale=0.4]
        \draw (-1.5,0)--(1.5,0);
        \mynode{0}{0}{0.5}{0.5}{T}
    }
    &
    \ctikz{[scale=0.4]
        \draw (-1,0)--(1,0);
        \draw (1,0)--(2,0.7);
        \draw[fill=white] (2,0.7)circle[radius=0.2];
        \draw (1,0)--(3.5,-1.75);
        \mynode{2.25}{-0.875}{0.5}{0.5}{T}
    }
    \ar@{=>}[l]^-{\lambda}
}
\qquad
\diagram@R=1.15pc{
    \ctikz{[scale=0.4,yscale=-1]
        \draw (-1.5,0)--(1,0);
        \mynode{0}{0}{0.5}{0.5}{T} 
        \draw (1,0)--(2,0.7);
        \draw[fill=white] (2,0.7)circle[radius=0.2];
        \draw (1,0)--(2.5,-1.05);
    }
    \ar@{=>}[r]^-\tau
    \ar@{=>}[d]_-\rho
    &
    \ctikz{[scale=0.4,yscale=-1]
        \draw (-1,0)--(1,0);
        \draw[fill=white] (1,0)--(3.5,1.75);
        \draw[fill=white] (3.5,1.75)circle[radius=0.2];
        \draw (1,0)--(3.5,-1.75);
        \mynode{2.25}{0.875}{0.5}{0.5}{T} 
        \mynode{2.25}{-0.875}{0.5}{0.5}{T}
    }
    \ar@{=>}[d]^-{\nu}
    \\
    \ctikz{[scale=0.4,yscale=-1]
        \draw (-1.5,0)--(1.5,0);
        \mynode{0}{0}{0.5}{0.5}{T}
    }
    &
    \ctikz{[scale=0.4,yscale=-1]
        \draw (-1,0)--(1,0);
        \draw (1,0)--(2,0.7);
        \draw[fill=white] (2,0.7)circle[radius=0.2];
        \draw (1,0)--(3.5,-1.75);
        \mynode{2.25}{-0.875}{0.5}{0.5}{T}
    }
    \ar@{=>}[l]^-{\rho}
}}
\eqnn{\label{eq.pentagon_and_triangle_promonoidal}
    \ctikz{[black,node distance=3cm, every node/.style={inner sep=0.5pt},scale=0.9]
    \node (A) at (162:2.25)  
        {$\ctikz{[scale=0.2]
            \draw (-2,0)--(0,0);
            \draw (0,0)--(3,2.1);
            \draw (1,0.7)--(3,-0.7);
            \draw (2,1.4)--(3,0.7);
            \draw (0,0)--(3,-2.1);
        }$};
    \node (B) at (90:2.25)   
        {$\ctikz{[scale=0.2]
            \draw (-2,0)--(0,0);
            \draw (0,0)--(1,0.7)--(2,1.4)--(3,2.1);
            \draw (2,-1.4)--(3,-0.7);
            \draw (2,1.4)--(3,0.7);
            \draw (0,0)--(3,-2.1);
        }$};
    \node (C) at (18:2.25)
        {$\ctikz{[scale=0.2,yscale=-1]
            \draw (-2,0)--(0,0);
            \draw (0,0)--(3,2.1);
            \draw (1,0.7)--(3,-0.7);
            \draw (2,1.4)--(3,0.7);
            \draw (0,0)--(3,-2.1);
        }$};
    \node (D) at (234:2.25) 
        {$\ctikz{[scale=0.2]
            \draw (-2,0)--(0,0);
            \draw (0,0)--(3,2.1);
            \draw (1,0.7)--(3,-0.7);
            \draw (2,0)--(3,0.7);
            \draw (0,0)--(3,-2.1);
        }$};
    \node (E) at (306:2.25)
        {$\ctikz{[scale=0.2,yscale=-1]
            \draw (-2,0)--(0,0);
            \draw (0,0)--(3,2.1);
            \draw (1,0.7)--(3,-0.7);
            \draw (2,0)--(3,0.7);
            \draw (0,0)--(3,-2.1);
        }$};
    \draw[nattrans] (A) -- (B) node[midway, above=4pt, xshift=-3pt] {$\scriptstyle\alpha^\bullet$};
    \draw[nattrans] (A) -- (D) node[midway, below=0pt, xshift=-10pt] {$\scriptstyle\alpha^\bullet\star\id$};
    \draw[nattrans] (B) -- (C) node[midway, above=4pt, xshift=3pt] {$\scriptstyle\alpha^\bullet$};
    \draw[nattrans] (D) -- (E) node[midway, below=5pt] {$\scriptstyle\alpha^\bullet$};
    \draw[nattrans] (E) -- (C) node[midway, below=0pt, xshift=10pt] {$\scriptstyle\id\star\alpha^\bullet$};
}
    \qquad
    \begin{array}{c}
        \diagram@C=1.5pc@R=1.5pc{
        &
        \ctikz{[scale=0.25,yscale=-1]
            \draw (-1,0)--(0,0);
            \draw (0,0)--(2,1.4);
            \draw (0,0)--(2,-1.4);
        }
        \\
        \ctikz{[scale=0.25]
            \draw (-1,0)--(0,0);
            \draw (0,0)--(2,1.4);
            \draw (0,0)--(2,-1.4);
            \draw (1,0.7)--(2,0);
            \draw[fill=white] (2,0)circle[radius=0.32];
        }
        \ar@{=>}[ru]^-{\rho^\bullet} \ar@{=>}[rr]_-{\alpha^\bullet}  & & 
        \ctikz{[scale=0.25,yscale=-1]
            \draw (-1,0)--(0,0);
            \draw (0,0)--(2,1.4);
            \draw (0,0)--(2,-1.4);
            \draw (1,0.7)--(2,0);
            \draw[fill=white] (2,0)circle[radius=0.32];
        }
        \ar@{=>}[lu]_-{\lambda^\bullet}
    }
    \end{array}
}
\caption{Axioms for a probimonad (Definition \ref{dfn.probimonad})}
\label{fig.bimonad_axiom}
\end{figure}
\subsection{Properties of $\MnCN$ and the proof that $\MCM$ is a probimonad}\label{sub.is_probimonad}
Some important commutative diagrams of pro-tensor networks involving $\MnCN$ or $\MnCM$ are listed in Figure \ref{fig.MnCN} and \ref{fig.MnCM}. 
\begin{figure}[htbp]
\eqnn{\label{eq.omega_ass}
    \diagram@C=2.5pc{
        \begin{tikzpicture}[scale=0.4,baseline=0]
            \draw[color4] (-1.35-1.4,0)--(-1.35+0.7,0);
            \draw[color3] (1.35-0.7,0)--(1.35+1.4,0);
            \myparabola{1.5}{-1.75}{0}{1.75}{0}
            \myparabola{1.5}{-2.3}{0}{2.3}{0}
            \myparabola{1.5}{-1.2}{0}{1.2}{0}
        \end{tikzpicture}
        \ar@{=>}[r]^-{\alpha_{\wr,\cN}^{-1}\star\alpha_{\wr,\cM}^R}
        \ar@{=>}[d]_-{\alpha_{\wr,\cN}^{-1}\star\alpha_{\wr,\cM}^R}
        &
        \begin{tikzpicture}[scale=0.4,baseline=0]
            \draw[color4] (-1.35-1.4,0)--(-1.35+0.7,0);
            \draw[color3] (1.35-0.7,0)--(1.35+1.4,0);
            \myparabola{1.5}{-2.3}{0}{2.3}{0}
            \begin{scope}[scale=0.85]
                \draw (0.5,1.4) parabola (-1.65,0);
                \begin{scope}[xscale=-1]
                    \draw (0.5,1.4) parabola (-1.65,0); 
                \end{scope}
                \draw[white,fill=white] (-0.95,0.1)rectangle(0.95,1.75);
                \myparabola{0.9}{-0.75}{1.2}{0.75}{1.2}
                \myparabola{1.1}{-0.5}{0.8}{0.5}{0.8}
                \draw (-1,0.74) to[out=90, in=220](-0.75,1.2);
                \draw (-1,0.74) to[out=-30, in=220](-0.5,0.8);
                \begin{scope}[xscale=-1]
                    \draw (-1,0.74) to[out=90, in=220](-0.75,1.2);
                    \draw (-1,0.74) to[out=-30, in=220](-0.5,0.8);
                \end{scope}
            \end{scope}
        \end{tikzpicture}
        \ar@{=>}[r]^-{\epsilon_2}
        &
        \begin{tikzpicture}[scale=0.4,baseline=0]
            \draw[color4] (-1.35-1.4,0)--(-1.35+0.7,0);
            \draw[color3] (1.35-0.7,0)--(1.35+1.4,0);
            \myparabola{1.5}{-2}{0}{2}{0}
            \myparabola{1.5}{-1.2}{0}{1.2}{0}
        \end{tikzpicture}
        \ar@{=>}[d]^-{\alpha_{\wr,\cN}^{-1}\star\alpha_{\wr,\cM}^R}
        \\
        \begin{tikzpicture}[scale=0.4,baseline=0]
            \draw[color4] (-1.35-1.4,0)--(-1.35+0.7,0);
            \draw[color3] (1.35-0.7,0)--(1.35+1.4,0);
            \begin{scope}[scale=1.35]
                \draw (0.5,1.4) parabola (-1.65,0);
                \begin{scope}[xscale=-1]
                    \draw (0.5,1.4) parabola (-1.65,0); 
                \end{scope}
                \draw[white,fill=white] (-0.95,0.1)rectangle(0.95,1.75);
                \myparabola{0.9}{-0.75}{1.2}{0.75}{1.2}
                \myparabola{1.1}{-0.5}{0.8}{0.5}{0.8}
                \draw (-1,0.74) to[out=90, in=220](-0.75,1.2);
                \draw (-1,0.74) to[out=-30, in=220](-0.5,0.8);
                \begin{scope}[xscale=-1]
                    \draw (-1,0.74) to[out=90, in=220](-0.75,1.2);
                    \draw (-1,0.74) to[out=-30, in=220](-0.5,0.8);
                \end{scope}
            \end{scope}
            \myparabola{1.5}{-1.35}{0}{1.35}{0}
        \end{tikzpicture}
        \ar@{=>}[d]_-{\epsilon_2}   
        & &
        \begin{tikzpicture}[scale=0.4,baseline=0]
            \draw[color4] (-1.35-1.4,0)--(-1.35+0.7,0);
            \draw[color3] (1.35-0.7,0)--(1.35+1.4,0);
            \begin{scope}[scale=1.03]
                \draw (0.5,1.4) parabola (-1.65,0);
                \begin{scope}[xscale=-1]
                    \draw (0.5,1.4) parabola (-1.65,0); 
                \end{scope}
                \draw[white,fill=white] (-0.95,0.1)rectangle(0.95,1.75);
                \myparabola{0.9}{-0.75}{1.2}{0.75}{1.2}
                \myparabola{1.1}{-0.5}{0.8}{0.5}{0.8}
                \draw (-1,0.74) to[out=90, in=220](-0.75,1.2);
                \draw (-1,0.74) to[out=-30, in=220](-0.5,0.8);
                \begin{scope}[xscale=-1]
                    \draw (-1,0.74) to[out=90, in=220](-0.75,1.2);
                    \draw (-1,0.74) to[out=-30, in=220](-0.5,0.8);
                \end{scope}
            \end{scope}
        \end{tikzpicture}
        \ar@{=>}[d]^-{\epsilon_2}
        \\
        \begin{tikzpicture}[scale=0.4,baseline=0]
            \draw[color4] (-1.35-1.4,0)--(-1.35+0.7,0);
            \draw[color3] (1.35-0.7,0)--(1.35+1.4,0);
            \myparabola{1.5}{-2.3}{0}{2.3}{0}
            \myparabola{1.5}{-1.5}{0}{1.5}{0}
        \end{tikzpicture}
        \ar@{=>}[r]_-{\alpha_{\wr,\cN}^{-1}\star\alpha_{\wr,\cM}^R}
        &
        \begin{tikzpicture}[scale=0.4,baseline=0]
            \draw[color4] (-1.35-1.4,0)--(-1.35+0.7,0);
            \draw[color3] (1.35-0.7,0)--(1.35+1.4,0);
            \begin{scope}[scale=1.03]
                \draw (0.5,1.4) parabola (-1.65,0);
                \begin{scope}[xscale=-1]
                    \draw (0.5,1.4) parabola (-1.65,0); 
                \end{scope}
                \draw[white,fill=white] (-0.95,0.1)rectangle(0.95,1.75);
                \myparabola{0.9}{-0.75}{1.2}{0.75}{1.2}
                \myparabola{1.1}{-0.5}{0.8}{0.5}{0.8}
                \draw (-1,0.74) to[out=90, in=220](-0.75,1.2);
                \draw (-1,0.74) to[out=-30, in=220](-0.5,0.8);
                \begin{scope}[xscale=-1]
                    \draw (-1,0.74) to[out=90, in=220](-0.75,1.2);
                    \draw (-1,0.74) to[out=-30, in=220](-0.5,0.8);
                \end{scope}
            \end{scope}
        \end{tikzpicture}
        \ar@{=>}[r]_-{\epsilon_2}
        &
        \begin{tikzpicture}[scale=0.4,baseline=0]
            \draw[color4] (-1.35-1.4,0)--(-1.35+0.7,0);
            \draw[color3] (1.35-0.7,0)--(1.35+1.4,0);
            \myparabola{1.5}{-1.75}{0}{1.75}{0}
        \end{tikzpicture}
    }
}
\eqnn{\label{eq.omega_unit}
    \diagram@=3.25pc{
        \begin{tikzpicture}[scale=0.4,baseline=0]
            \mygrid{ 
                \draw[gray!30, step=0.5, opacity = 0.5] (-3,-4) grid (2,3);
                \foreach \x in {-3,-2,-1,0,1,2}
                    \draw (\x,3) node[above] {\small $\x$};
                \foreach \y in {-1,0,1,2,3}
                    \draw (3,\y) node[right] {\small $\y$};
            }
            \myparabola{1.5}{-1.75}{0}{1.75}{0}
            \draw[white, fill=white](-1.2,0)rectangle(1.2,1.5);
            \draw[fill=white] (-1.2,0.65)circle[radius=0.2];
            \draw[fill=white] (1.2,0.65)circle[radius=0.2];
            \myparabola{1.5}{-1.2}{0}{1.2}{0}
            \draw[color4] (-1.35-1,0)--(-1.35+0.7,0);
            \draw[color3] (1.35-0.7,0)--(1.35+1,0);
        \end{tikzpicture}
        \ar@{=>}[d]_{\epsilon_0}
        &
        \begin{tikzpicture}[scale=0.4,baseline=0]
            \draw[color4] (-1.35-0.7,0)--(-1.35+0.7,0);
            \draw[color3] (1.35-0.7,0)--(1.35+0.7,0);
            \myparabola{1.5}{-1.35}{0}{1.35}{0}
        \end{tikzpicture}
        \ar@{=>}[dd]^{\id}
        \ar@{=>}[r]^-{\lambda_{\wr,\cN}^{-1}\star\lambda_{\wr,\cM}^R}
        \ar@{=>}[l]_-{\lambda_{\wr,\cN}^{-1}\star\lambda_{\wr,\cM}^R}
        &
        \begin{tikzpicture}[scale=0.4,baseline=0]
            \mygrid{ 
                \draw[gray!30, step=0.5, opacity = 0.5] (-3,-4) grid (2,3);
                \foreach \x in {-3,-2,-1,0,1,2}
                    \draw (\x,3) node[above] {\small $\x$};
                \foreach \y in {-1,0,1,2,3}
                    \draw (3,\y) node[right] {\small $\y$};
            }
            \myparabola{1.5}{-1.2}{0}{1.2}{0}
            \draw[white, fill=white] (-0.6,0)rectangle(0.6,1);
            \draw[fill=white] (-0.6,0.65)circle[radius=0.2];
            \draw[fill=white] (0.6,0.65)circle[radius=0.2];
            \myparabola{1.5}{-1.75}{0}{1.75}{0}
            \draw[color4] (-1.35-1,0)--(-1.35+0.7,0);
            \draw[color3] (1.35-0.7,0)--(1.35+1,0);
        \end{tikzpicture}
        \ar@{=>}[d]^{\epsilon_0}
        \\
        \begin{tikzpicture}[scale=0.4,baseline=0]
            \draw[color4] (-1.35-1,0)--(-1.35+0.7,0);
            \draw[color3] (1.35-0.7,0)--(1.35+1,0);
            \myparabola{1.5}{-1.2}{0}{1.2}{0}
            \myparabola{1.5}{-1.75}{0}{1.75}{0}
        \end{tikzpicture}
        \ar@{=>}[d]_{\alpha_{\wr,\cN}^{-1}\star\alpha_{\wr,\cM}^R}
        &
        &
        \begin{tikzpicture}[scale=0.4,baseline=0]
            \draw[color4] (-1.35-1,0)--(-1.35+0.7,0);
            \draw[color3] (1.35-0.7,0)--(1.35+1,0);
            \myparabola{1.5}{-1.2}{0}{1.2}{0}
            \myparabola{1.5}{-1.75}{0}{1.75}{0}
        \end{tikzpicture}
        \ar@{=>}[d]^{\alpha_{\wr,\cN}^{-1}\star\alpha_{\wr,\cM}^R}
        \\
        \begin{tikzpicture}[scale=0.4,baseline=0]
            \draw[color4] (-1.35-1,0)--(-1.35+0.7,0);
            \draw[color3] (1.35-0.7,0)--(1.35+1,0);
            \begin{scope}[scale=0.909]
                \draw (0.5,1.4) parabola (-1.65,0);
                \begin{scope}[xscale=-1]
                    \draw (0.5,1.4) parabola (-1.65,0); 
                \end{scope}
                \draw[white,fill=white] (-0.95,0.1)rectangle(0.95,1.75);
                \myparabola{0.9}{-0.75}{1.2}{0.75}{1.2}
                \myparabola{1.1}{-0.5}{0.8}{0.5}{0.8}
                \draw (-1,0.74) to[out=90, in=220](-0.75,1.2);
                \draw (-1,0.74) to[out=-30, in=220](-0.5,0.8);
                \begin{scope}[xscale=-1]
                    \draw (-1,0.74) to[out=90, in=220](-0.75,1.2);
                    \draw (-1,0.74) to[out=-30, in=220](-0.5,0.8);
                \end{scope}
            \end{scope}
        \end{tikzpicture}
        \ar@{=>}[r]_-{\epsilon_2}
        &
        \begin{tikzpicture}[scale=0.4,baseline=0]
            \draw[color4] (-1.35-0.7,0)--(-1.35+0.7,0);
            \draw[color3] (1.35-0.7,0)--(1.35+0.7,0);
            \myparabola{1.5}{-1.35}{0}{1.35}{0}
        \end{tikzpicture}
        &
        \begin{tikzpicture}[scale=0.4,baseline=0]
            \draw[color4] (-1.35-1,0)--(-1.35+0.7,0);
            \draw[color3] (1.35-0.7,0)--(1.35+1,0);
            \begin{scope}[scale=0.909]
                \draw (0.5,1.4) parabola (-1.65,0);
                \begin{scope}[xscale=-1]
                    \draw (0.5,1.4) parabola (-1.65,0); 
                \end{scope}
                \draw[white,fill=white] (-0.95,0.1)rectangle(0.95,1.75);
                \myparabola{0.9}{-0.75}{1.2}{0.75}{1.2}
                \myparabola{1.1}{-0.5}{0.8}{0.5}{0.8}
                \draw (-1,0.74) to[out=90, in=220](-0.75,1.2);
                \draw (-1,0.74) to[out=-30, in=220](-0.5,0.8);
                \begin{scope}[xscale=-1]
                    \draw (-1,0.74) to[out=90, in=220](-0.75,1.2);
                    \draw (-1,0.74) to[out=-30, in=220](-0.5,0.8);
                \end{scope}
            \end{scope}
        \end{tikzpicture}
        \ar@{=>}[l]^-{\epsilon_2}   
    }
}
\caption{Commutative diagrams invovling $\MnCN$. Each commutative diagram is not a commutative diagram of pro-tensor network maps, but becomes one under the interpretation introduced in Section \ref{sub.is_probimonad}. They are proved in Figure~\ref{fig.pf_of_omega_monad}.}
\label{fig.MnCN}
\end{figure}
\begin{figure}[htbp]
\centering
\eqnn{\label{eq.omega_bimonad1}
    \diagram@R=4pc@C=2.5pc{
        \begin{tikzpicture}[scale=0.4,baseline=0]
            \draw[color3] (-1.35-0.7,0)--(-1.35+0.7,0);
            \draw[color3] (1.35-0.7,0)--(1.35+1,0);
            \draw[color3] (4.35-0.7,0)--(4.35+0.7,0);
            \myparabola{1.5}{-1.75}{0}{4.75}{0}
            \myparabola{1.5}{-1.05}{0}{4.05}{0}
        \end{tikzpicture}
        \ar@{=>}[d]_-{\eta_{\wr,\cM}}
        \ar@{=>}[r]^-{\alpha_{\wr,\cM}^{-1}\star\alpha_{\wr,\cM}^R}
        &
        \begin{tikzpicture}[scale=0.4,baseline=0]
            \draw[color3] (-1.35-0.7,0)--(-1.35+0.7,0);
            \draw[color3] (1.35-0.7,0)--(1.35+1,0);
            \draw[color3] (4.35-0.7,0)--(4.35+0.7,0);
            \begin{scope}[xshift=1.5cm,scale=1.73]
                \draw (0.5,1.4) parabola (-1.65,0);
                \begin{scope}[xscale=-1]
                    \draw (0.5,1.4) parabola (-1.65,0); 
                \end{scope}
                \draw[white,fill=white] (-0.95,0.1)rectangle(0.95,1.75);
                \myparabola{0.9}{-0.75}{1.2}{0.75}{1.2}
                \myparabola{1.1}{-0.5}{0.8}{0.5}{0.8}
                \draw (-1,0.74) to[out=90, in=220](-0.75,1.2);
                \draw (-1,0.74) to[out=-30, in=220](-0.5,0.8);
                \begin{scope}[xscale=-1]
                    \draw (-1,0.74) to[out=90, in=220](-0.75,1.2);
                    \draw (-1,0.74) to[out=-30, in=220](-0.5,0.8);
                \end{scope}
            \end{scope}
        \end{tikzpicture}
        \ar@{=>}[rr]^{\epsilon_2}
        &
        &
        \begin{tikzpicture}[scale=0.4,baseline=0]
            \draw[color3] (-1.35-0.7,0)--(-1.35+0.7,0);
            \draw[color3] (1.35-0.7,0)--(1.35+1,0);
            \draw[color3] (4.35-0.7,0)--(4.35+0.7,0);
            \myparabola{1.5}{-1.35}{0}{4.35}{0}
        \end{tikzpicture}
        \ar@{=>}[d]^{\eta_{\wr,\cM}}
        \\
        \begin{tikzpicture}[scale=0.4,baseline=0]
            \draw[color3] (-1.35-0.7,0)--(-1.35+0.7,0);
            \draw[color3] (1.35-0.7,0)--(1.35+1,0);
            \draw[color3] (4.35-0.7,0)--(4.35+0.7,0);
            \myparabola{1.5}{-1.75}{0}{4.75}{0}
            \myparabola{1.5}{-1.05}{0}{1.05}{0}
            \myparabola{1.5}{1.95}{0}{4.05}{0}
        \end{tikzpicture}
        \ar@{=>}[r]_-{\eta_{\wr,\cM}}
        &
        \begin{tikzpicture}[scale=0.4,baseline=0]
            \draw[color3] (-1.35-0.7,0)--(-1.35+0.7,0);
            \draw[color3] (1.35-0.7,0)--(1.35+1,0);
            \draw[color3] (4.35-0.7,0)--(4.35+0.7,0);
            \myparabola{1.5}{-1.4}{0}{1.4}{0}
            \myparabola{1.5}{1.6}{0}{4.4}{0}
            \myparabola{1.5}{-1.05}{0}{1.05}{0}
            \myparabola{1.5}{1.95}{0}{4.05}{0}
        \end{tikzpicture}
        \ar@{=>}[r]_-{\raisebox{-1.5em}{$\scriptstyle \alpha_{\wr,\cM}^{-1}\star\alpha_{\wr,\cM}^R\star\alpha_{\wr,\cM}^{-1}\star\alpha_{\wr,\cM}^R$}}
        &
        \begin{tikzpicture}[scale=0.4,baseline=0]
            \draw[color3] (-1.35-0.7,0)--(-1.35+0.7,0);
            \draw[color3] (1.35-0.7,0)--(1.35+1,0);
            \draw[color3] (4.35-0.7,0)--(4.35+0.7,0);
            \begin{scope}[scale=0.818]
                \draw (0.5,1.4) parabola (-1.65,0);
                \begin{scope}[xscale=-1]
                    \draw (0.5,1.4) parabola (-1.65,0); 
                \end{scope}
                \draw[white,fill=white] (-0.95,0.1)rectangle(0.95,1.75);
                \myparabola{0.9}{-0.75}{1.2}{0.75}{1.2}
                \myparabola{1.1}{-0.5}{0.8}{0.5}{0.8}
                \draw (-1,0.74) to[out=90, in=220](-0.75,1.2);
                \draw (-1,0.74) to[out=-30, in=220](-0.5,0.8);
                \begin{scope}[xscale=-1]
                    \draw (-1,0.74) to[out=90, in=220](-0.75,1.2);
                    \draw (-1,0.74) to[out=-30, in=220](-0.5,0.8);
                \end{scope}
            \end{scope}
            \begin{scope}[xshift=3cm,scale=0.818]
                \draw (0.5,1.4) parabola (-1.65,0);
                \begin{scope}[xscale=-1]
                    \draw (0.5,1.4) parabola (-1.65,0); 
                \end{scope}
                \draw[white,fill=white] (-0.95,0.1)rectangle(0.95,1.75);
                \myparabola{0.9}{-0.75}{1.2}{0.75}{1.2}
                \myparabola{1.1}{-0.5}{0.8}{0.5}{0.8}
                \draw (-1,0.74) to[out=90, in=220](-0.75,1.2);
                \draw (-1,0.74) to[out=-30, in=220](-0.5,0.8);
                \begin{scope}[xscale=-1]
                    \draw (-1,0.74) to[out=90, in=220](-0.75,1.2);
                    \draw (-1,0.74) to[out=-30, in=220](-0.5,0.8);
                \end{scope}
            \end{scope}
        \end{tikzpicture}
        \ar@{=>}[r]_-{\scriptstyle\epsilon_2\star\epsilon_2}
        &
        \begin{tikzpicture}[scale=0.4,baseline=0]
            \draw[color3] (-1.35-0.7,0)--(-1.35+0.7,0);
            \draw[color3] (1.35-0.7,0)--(1.35+1,0);
            \draw[color3] (4.35-0.7,0)--(4.35+0.7,0);
            \myparabola{1.5}{-1.35}{0}{1.35}{0}
            \myparabola{1.5}{1.65}{0}{4.35}{0}
        \end{tikzpicture}
    }
}
\eqnn{\label{eq.omega_bimonad2}
    \diagram@C=2.5pc{
    \begin{tikzpicture}[scale=0.4,baseline=0]
        \draw[color3] (-1.35-0.7,0)--(-1.35+0.7,0);
        \draw[color3] (1.35-0.7,0)--(1.35+1,0);
        \draw[color3] (4.35-0.7,0)--(4.35+0.7,0);
    \end{tikzpicture}
    \ar@{=>}[d]_{\lambda_{\wr,\cM}^{-1}\star\lambda_{\wr,\cM}^R\star\lambda_{\wr,\cM}^{-1}\star\lambda_{\wr,\cM}^R}
    \ar@{=>}[r]^-{\lambda_{\wr,\cM}^{-1}\star\lambda_{\wr,\cM}^R}
    &
    \begin{tikzpicture}[scale=0.4,baseline=0]
        \mygrid{ 
            \draw[gray!30, step=0.5, opacity = 0.5] (-3,-4) grid (6,3);
            \foreach \x in {-3,-2,-1,0,1,2,3,4,5,6}
                \draw (\x,3) node[above] {\small $\x$};
            \foreach \y in {-1,0,1,2,3}
                \draw (3,\y) node[right] {\small $\y$};
        }
        \myparabola{1.5}{-1.35}{0}{4.35}{0}
        \draw[white,fill=white] (-0.75,0)rectangle(3.75,2.5);
        \draw[fill=white] (-0.75,0.8)circle[radius=0.2];
        \draw[fill=white] (3.75,0.8)circle[radius=0.2];
        \draw[color3] (-1.35-0.7,0)--(-1.35+0.7,0);
        \draw[color3] (1.35-0.7,0)--(1.35+1,0);
        \draw[color3] (4.35-0.7,0)--(4.35+0.7,0);
    \end{tikzpicture}
    \ar@{=>}[r]^-{\epsilon_0}
    &
    \begin{tikzpicture}[scale=0.4,baseline=0]
        \draw[color3] (-1.35-0.7,0)--(-1.35+0.7,0);
        \draw[color3] (1.35-0.7,0)--(1.35+1,0);
        \draw[color3] (4.35-0.7,0)--(4.35+0.7,0);
        \myparabola{1.5}{-1.35}{0}{4.35}{0}
    \end{tikzpicture}
    \ar@{=>}[d]^{\eta_{\wr,\cM}}
    \\
    \begin{tikzpicture}[scale=0.4,baseline=0]
        \myparabola{1.5}{-1.35}{0}{1.35}{0}
        \myparabola{1.5}{1.65}{0}{4.35}{0}
        \draw[fill=white,white] (-1.35+0.7,0)rectangle(1.35-0.7,2.5);
        \draw[fill=white,white](1.35+1,0)rectangle(4.35-0.7,2.5);
        \draw[fill=white] (-1.35+0.7,0.8)circle[radius=0.2];
        \draw[fill=white] (1.35-0.7,0.8)circle[radius=0.2];
        \draw[fill=white] (1.35+1,0.8)circle[radius=0.2];
        \draw[fill=white] (4.35-0.7,0.8)circle[radius=0.2];
        \draw[color3] (-1.35-0.7,0)--(-1.35+0.7,0);
        \draw[color3] (1.35-0.7,0)--(1.35+1,0);
        \draw[color3] (4.35-0.7,0)--(4.35+0.7,0);
        \mygrid{ 
            \draw[gray!30, step=0.5, opacity = 0.5] (-3,-4) grid (6,3);
            \foreach \x in {-3,-2,-1,0,1,2,3,4,5,6}
                \draw (\x,3) node[above] {\small $\x$};
            \foreach \y in {-1,0,1,2,3}
                \draw (3,\y) node[right] {\small $\y$};
        }
    \end{tikzpicture}
    \ar@{=>}[rr]_-{\epsilon_0\star\epsilon_0}
    & &
    \begin{tikzpicture}[scale=0.4,baseline=0]
        \draw[color3] (-1.35-0.7,0)--(-1.35+0.7,0);
        \draw[color3] (1.35-0.7,0)--(1.35+1,0);
        \draw[color3] (4.35-0.7,0)--(4.35+0.7,0);
        \myparabola{1.5}{-1.35}{0}{1.35}{0}
        \myparabola{1.5}{1.65}{0}{4.35}{0}
    \end{tikzpicture}
    }
}
\eqnn{\label{eq.omega_bimonad_ass}
    \diagram{
        \begin{tikzpicture}[scale=0.4,baseline=0]
            \draw[color3] (-1.35-0.7,0)--(-1.35+0.7,0);
            \draw[color3] (1.35-0.7,0)--(1.35+1,0);
            \draw[color3] (4.35-0.7,0)--(4.35+1,0);
            \draw[color3] (7.35-0.7,0)--(7.35+0.7,0);
            \myparabola{1.35}{-1.35}{0}{7.35}{0}
        \end{tikzpicture}
        \ar@{=>}[r]^-{\eta_{\wr,\cM}}
        \ar@{=>}[d]_-{\eta_{\wr,\cM}}
        &
        \begin{tikzpicture}[scale=0.4,baseline=0]
            \draw[color3] (-1.35-0.7,0)--(-1.35+0.7,0);
            \draw[color3] (1.35-0.7,0)--(1.35+1,0);
            \draw[color3] (4.35-0.7,0)--(4.35+1,0);
            \draw[color3] (7.35-0.7,0)--(7.35+0.7,0);
            \myparabola{1.5}{-1.35}{0}{1.35}{0}
            \myparabola{1.5}{1.65}{0}{7.35}{0}
        \end{tikzpicture}
        \ar@{=>}[d]^-{\eta_{\wr,\cM}}
        \\
        \begin{tikzpicture}[scale=0.4,baseline=0]
            \draw[color3] (-1.35-0.7,0)--(-1.35+0.7,0);
            \draw[color3] (1.35-0.7,0)--(1.35+1,0);
            \draw[color3] (4.35-0.7,0)--(4.35+1,0);
            \draw[color3] (7.35-0.7,0)--(7.35+0.7,0);
            \myparabola{1.5}{-1.35}{0}{4.35}{0}
            \myparabola{1.5}{4.65}{0}{7.35}{0}
        \end{tikzpicture}
        \ar@{=>}[r]_-{\eta_{\wr,\cM}}
        &
        \begin{tikzpicture}[scale=0.4,baseline=0]
            \draw[color3] (-1.35-0.7,0)--(-1.35+0.7,0);
            \draw[color3] (1.35-0.7,0)--(1.35+1,0);
            \draw[color3] (4.35-0.7,0)--(4.35+1,0);
            \draw[color3] (7.35-0.7,0)--(7.35+0.7,0);
            \myparabola{1.5}{-1.35}{0}{1.35}{0}
            \myparabola{1.5}{1.65}{0}{4.35}{0}
            \myparabola{1.5}{4.65}{0}{7.35}{0}
        \end{tikzpicture}
    }
}
\eqnn{\label{eq.omega_bimonad_unit}
    \ctikz{[scale=1.1,xscale=1.7]
        \node (1_1) at (0,2){$
            \ctikz{[scale=0.4]
                \draw[color3] (-1.35-0.7,0)--(-1.35+0.7,0);
                \draw[color3] (1.35-0.7,0)--(4.35+0.7,0);
                \myparabola{1.5}{-1.35}{0}{4.35}{0}
            }$};
        \node (2_1) at (0,0){$
            \ctikz{[scale=0.4]
                \draw[color3] (-1.35-0.7,0)--(-1.35+0.7,0);
                \draw[color3] (1.35-0.7,0)--(4.35+0.7,0);
                \myparabola{1.5}{-1.35}{0}{1.35}{0}
                \myparabola{1.5}{1.65}{0}{4.35}{0}
            }$};
        \node (2_2) at (2,0){$
            \ctikz{[scale=0.4]
                \draw[color3] (-1.35-0.7,0)--(-1.35+0.7,0);
                \draw[color3] (1.35-0.7,0)--(4.35+0.7,0);
                \myparabola{1.5}{-1.35}{0}{1.35}{0}
            }$};
            \draw[nat] (1_1)--(2_1)node[midway, left]{\eta_{\wr,\cM}};
            \draw[nat] (2_1)--(2_2)node[midway, below]{\epsilon_{\wr,\cM}};
            \draw[nat](1_1)--(2_2)node[midway, above right]{\id};
    }
    \quad
    \ctikz{[scale=1.1,xscale=1.7]
        \node (1_1) at (0,2){$
            \ctikz{[scale=0.4,xscale=-1]
                \draw[color3] (-1.35-0.7,0)--(-1.35+0.7,0);
                \draw[color3] (1.35-0.7,0)--(4.35+0.7,0);
                \myparabola{1.5}{-1.35}{0}{4.35}{0}
            }$};
        \node (2_1) at (0,0){$
            \ctikz{[scale=0.4,xscale=-1]
                \draw[color3] (-1.35-0.7,0)--(-1.35+0.7,0);
                \draw[color3] (1.35-0.7,0)--(4.35+0.7,0);
                \myparabola{1.5}{-1.35}{0}{1.35}{0}
                \myparabola{1.5}{1.65}{0}{4.35}{0}
            }$};
        \node (2_2) at (2,0){$
            \ctikz{[scale=0.4,xscale=-1]
                \draw[color3] (-1.35-0.7,0)--(-1.35+0.7,0);
                \draw[color3] (1.35-0.7,0)--(4.35+0.7,0);
                \myparabola{1.5}{-1.35}{0}{1.35}{0}
            }$};
            \draw[nat] (1_1)--(2_1)node[midway, left]{\eta_{\wr,\cM}};
            \draw[nat] (2_1)--(2_2)node[midway, below]{\epsilon_{\wr,\cM}};
            \draw[nat] (1_1)--(2_2)node[midway, above right]{\id};
    }
}
\caption{Commutative diagrams involving $\MnCM$. Each commutative diagram is not a commutative diagram of pro-tensor network maps, but becomes one under the interpretation introduced in Section \ref{sub.is_probimonad}. For example, \eqref{eq.omega_bimonad_ass} should be interpretated as the commutative diagram \eqref{eq.omega_bimonad_ass_explain} for arbitrary pro-tensors $F,G,H\:\cM\arprof\cM$.}
\label{fig.MnCM}
\end{figure}

Strictly speaking, Figures~\ref{fig.MnCN} and \ref{fig.MnCM} do not fit into our discussion in Section~\ref{sec.network} as an overall ordering of external legs is missing. There are multiple ways to interpret these pro-tensor networks, which are canonically equivalent to each other. One way is to bend and/or braid the legs (while maintaining underlying graph, the pattern of connections) to rearrange the pro-tensor network so that it can be put on an open disk with an overall orientation. For example, the pro-tensor network $\cM\Omega_\cC\cN$ becomes $\MCN$ after bending the legs. Alternatively, a pro-tensor network with holes inside can be thought of as a cocontinuous $\cV$-functor mapping the pro-tensors filled in the holes to the contracted pro-tensor. Such cocontinuous $\cV$-functor precomposed with the Yoneda embedding can then be identified with the contraction of the pro-tensor network with holes inside. Still taking $\cM\Omega_\cC\cN$ as an example, the alternative interpretation views $\cM\Omega_\cC\cN$ as a $\cV$-functor from $[\cM\ot\cN^\op,\bcV]$ to $[\cM\ot\cN^\op,\bcV]$; precomposing the Yoneda embedding one gets a $\cV$-functor $\cM^\op\ot \cN\to  [\cM\ot\cN^\op,\bcV]$, and then by internal hom adjunction a $\cV$-functor  $\cM^\op\ot \cN\ot \cM\ot\cN^\op\to \bcV $, which is by definition a profunctor $\cM^\op\ot \cN\arprof \cM^\op\ot \cN$. See also Remark~\ref{rmk.mcn}.

Similarly, Figures~\ref{fig.MnCN} and \ref{fig.MnCM} can be understood as the corresponding commutative diagram of pro-tensor network maps with holes filled with arbitrary (but fixed) pro-tensors $\cM\arprof\cN$ ($\cM\arprof\cM$). For example, \eqref{eq.omega_bimonad_ass} in Figure \ref{fig.MnCM} should be understood as the commutative diagram
\eqnn{\label{eq.omega_bimonad_ass_explain}
    \diagram{
        \begin{tikzpicture}[scale=0.4,baseline=0]
            \draw[color3] (-1.35-0.7,0)--(7.35+0.7,0);
            \myparabola{1.35}{-1.35}{0}{7.35}{0}
            \mynode{0}{0}{0.5}{0.5}{H}
            \mynode{3}{0}{0.5}{0.5}{G}
            \mynode{6}{0}{0.5}{0.5}{F}
        \end{tikzpicture}
        \ar@{=>}[r]^-{\eta_{\wr,\cM}}
        \ar@{=>}[d]_-{\eta_{\wr,\cM}}
        &
        \begin{tikzpicture}[scale=0.4,baseline=0]
            \draw[color3] (-1.35-0.7,0)--(7.35+0.7,0);
            \myparabola{1.5}{-1.35}{0}{1.35}{0}
            \myparabola{1.5}{1.65}{0}{7.35}{0}
            \mynode{0}{0}{0.5}{0.5}{H}
            \mynode{3}{0}{0.5}{0.5}{G}
            \mynode{6}{0}{0.5}{0.5}{F}
        \end{tikzpicture}
        \ar@{=>}[d]^-{\eta_{\wr,\cM}}
        \\
        \begin{tikzpicture}[scale=0.4,baseline=0]
            \draw[color3] (-1.35-0.7,0)--(7.35+0.7,0);
            \myparabola{1.5}{-1.35}{0}{4.35}{0}
            \myparabola{1.5}{4.65}{0}{7.35}{0}
            \mynode{0}{0}{0.5}{0.5}{H}
            \mynode{3}{0}{0.5}{0.5}{G}
            \mynode{6}{0}{0.5}{0.5}{F}
        \end{tikzpicture}
        \ar@{=>}[r]_-{\eta_{\wr,\cM}}
        &
        \begin{tikzpicture}[scale=0.4,baseline=0]
            \draw[color3] (-1.35-0.7,0)--(7.35+0.7,0);
            \myparabola{1.5}{-1.35}{0}{1.35}{0}
            \myparabola{1.5}{1.65}{0}{4.35}{0}
            \myparabola{1.5}{4.65}{0}{7.35}{0}
            \mynode{0}{0}{0.5}{0.5}{H}
            \mynode{3}{0}{0.5}{0.5}{G}
            \mynode{6}{0}{0.5}{0.5}{F}
        \end{tikzpicture}
    }
}
of pro-tensor network maps for arbitrary pro-tensors $F,G,H\:\cM\arprof\cM$.

With this interpretation, \eqref{eq.omega_ass} and \eqref{eq.omega_unit} are proved by the two commutative diagrams in Figure \ref{fig.pf_of_omega_monad}, respectively, while \eqref{eq.omega_bimonad1} and \eqref{eq.omega_bimonad2} are proved by the two commutative diagrams in Figure \ref{fig.pf_of_omega_bimonad}, respectively (with the diagrams in Figure \ref{fig.pf_of_omega_monad} and Figure \ref{fig.pf_of_omega_bimonad} also understood in the manner described above). \eqref{eq.omega_bimonad_ass} follows from Theorem \ref{thm.local}, and \eqref{eq.omega_bimonad_unit} follows from \eqref{eq.zig_zag_m_module}.

\begin{figure}[htbp]
\[
    \ctikz{[scale=1.5,xscale=1.9]
        \node (1_1) at (0,6){\begin{tikzpicture}[scale = 0.7 ,baseline=(current bounding box.center)]
    \draw[color4,thick] (0.3,-1.5)--(2,-1.5);
         \draw[color3] (2.8,-1.5) --(4.5,-1.5);
        \draw  (1,-1.5)
  .. controls (2.4,-0.2) ..
  (3.8,-1.5);

\draw  (1.35,-1.5)
  .. controls (2.4,-0.5) ..
  (3.45,-1.5);

\draw  (0.65,-1.5)
  .. controls (2.4,0.1) ..
  (4.15,-1.5);
    \end{tikzpicture}};
        \node (1_2) at (2,6){\begin{tikzpicture}[scale = 0.7 ,baseline=(current bounding box.center)]
    \draw[color4,thick] (0.3,-1.5)--(2,-1.5);
         \draw[color3] (2.8,-1.5) --(4.5,-1.5);
         \draw  (1.35,-1.5)
  .. controls (2.4,-0.6) ..
  (3.45,-1.5);

\draw  (1.82-0.15,-0.6)--(2.983+0.15,-0.6);

\draw  (0.65,-1.5)
  .. controls (2.4,0.1) ..
  (4.15,-1.5);
    \end{tikzpicture}};
        \node (1_3) at (4,6){\begin{tikzpicture}[scale = 0.7 ,baseline=(current bounding box.center)]
    \draw[color4,thick] (0.3,-1.5)--(2,-1.5);
    \draw[color3] (2.8,-1.5) --(4.5,-1.5);
    \draw  (1.35,-1.5)
      .. controls (2.4,-0.6) ..
      (3.45,-1.5);
    \draw  (0.65,-1.5)
      .. controls (2.4,0) ..
      (4.15,-1.5);
    \end{tikzpicture}};
        \node (2_2) at (2,4){\begin{tikzpicture}[scale = 0.7 ,baseline=(current bounding box.center)]
    \draw[color4,thick] (0.3,-1.5)--(2,-1.5);
    \draw[color3] (2.8,-1.5) --(4.5,-1.5);
    \draw  (1.62+0.05,-0.6)--(2.983+0.15,-0.6);
    \draw  (0.65,-1.5)
      .. controls (2.4,0.1) ..
      (4.15,-1.5);
    \draw  (1.233,-1)--(3.567,-1);
    \end{tikzpicture}};
        \node (2_3) at (4,4){\begin{tikzpicture}[scale = 0.7 ,baseline=(current bounding box.center)]
    \draw[color4,thick] (0.4,-1.5)--(2,-1.5);
    \draw[color3] (2.8,-1.5) --(4.4,-1.5);
    \draw  (1,-1.5)
      .. controls (2.4,-0.3) ..
      (3.8,-1.5);
    \draw  (1.7,-0.9)--(3.1,-0.9);
    \end{tikzpicture}};
        \node (3_1) at (0,2){\begin{tikzpicture}[scale = 0.7 ,baseline=(current bounding box.center)]
    \draw[color4,thick] (0.3,-1.5)--(2,-1.5);
    \draw[color3] (2.8,-1.5) --(4.5,-1.5);
    \draw  (1.35,-1.5)
      .. controls (2.4,-0.6) ..
      (3.45,-1.5);
    \draw  (0.65,-1.5)
      .. controls (2.4,0) ..
      (4.15,-1.5);
    \draw  (1.875,-1.05)--(2.925,-1.05);
    \end{tikzpicture}};
        \node (3_2) at (2,2){\begin{tikzpicture}[scale = 0.7 ,baseline=(current bounding box.center)]
    \draw[color4,thick] (0.3,-1.5)--(2,-1.5);
    \draw[color3] (2.8,-1.5) --(4.5,-1.5);
    \draw  (0.65,-1.5)
      .. controls (2.4,0) ..
      (4.15,-1.5);
    \draw  (1.233,-1)--(3.567,-1);
    \draw  (1.82,-1)
      .. controls (2.4,-0.5) ..
      (2.983,-1);
    \end{tikzpicture}};
        \node (4_1) at (0,0){\begin{tikzpicture}[scale = 0.7 ,baseline=(current bounding box.center)]
    \draw[color4,thick] (0.3,-1.5)--(2,-1.5);
    \draw[color3] (2.8,-1.5) --(4.5,-1.5);
    \draw  (1.35,-1.5)
      .. controls (2.4,-0.6) ..
      (3.45,-1.5);
    \draw  (0.65,-1.5)
      .. controls (2.4,0) ..
      (4.15,-1.5);
    \end{tikzpicture}};
        \node (4_2) at (2,0){\begin{tikzpicture}[scale = 0.7 ,baseline=(current bounding box.center)]
    \draw[color4,thick] (0.4,-1.5)--(2,-1.5);
    \draw[color3] (2.8,-1.5) --(4.4,-1.5);
    \draw  (1,-1.5)
      .. controls (2.4,-0.3) ..
      (3.8,-1.5);
    \draw  (1.7,-0.9)--(3.1,-0.9);
    \end{tikzpicture}};
        \node (4_3) at (4,0){\begin{tikzpicture}[scale = 0.7 ,baseline=(current bounding box.center)]
    \draw[color4,thick] (0.4,-1.5)--(2,-1.5);
    \draw[color3] (2.8,-1.5) --(4.4,-1.5);
    \draw  (1,-1.5)
      .. controls (2.4,-0.3) ..
      (3.8,-1.5);
    \end{tikzpicture}};
        \draw [nat] (1_1)--(1_2)node[midway, above]{\alpha_{\wr,\cN}^{-1}\star\alpha_{\wr,\cM}^R};
        \draw [nat] (1_2)--(1_3)node[midway, above]{\epsilon_2};
        \draw[nat] (1_3)--(2_3)node[midway, right]{\alpha_{\wr,\cN}^{-1}\star\alpha_{\wr,\cM}^R};
        \draw[nat] (1_2)--(2_2)node[midway, right]{\alpha_{\wr,\cN}^{-1}\star\alpha_{\wr,\cM}^R};
        \draw[nat] (2_2)--(2_3)node[midway, above]{\epsilon_2};
        \draw[nat] (2_2)--(3_2)node[midway, right]{\alpha\star(\alpha^R)^{-1}};
        \draw[nat] (3_2)--(4_2)node[midway, right]{\epsilon_2};
        \draw[nat] (4_2)--(4_3)node[midway, below]{\epsilon_2};
        \draw[nat] (4_1)--(4_2)node[midway, below]{\alpha_{\wr,\cN}^{-1}\star\alpha_{\wr,\cM}^R};
        \draw[nat] (3_1)--(4_1)node[midway, left]{\epsilon_2};
        \draw[nat] (1_1)--(3_1)node[midway, left]{\epsilon_2};
        \draw[nat] (2_3)--(4_3)node[midway, right]{\epsilon_2};
        \draw[nat] (3_1)--(3_2)node[midway, above]{\alpha_{\wr,\cN}^{-1}\star\alpha_{\wr,\cM}^R};
        \draw (1,4)node{\tiny (\eqref{eq.pentagon_graph}+Theorem \ref{thm.local})};
        \draw (3.2,2)node{\tiny (\eqref{eq.right_dual_of_alpha}+Theorem \ref{thm.local})};
        \draw (1,1)node{\tiny (Theorem \ref{thm.local})};
        \draw (3,5)node{\tiny (Theorem \ref{thm.local})};
    }
\]
\[
    \ctikz{[scale=1.25,xscale=2.25]
        \node (1_1) at (0,4){\begin{tikzpicture}[scale = 0.7 ,baseline=(current bounding box.center)]
        \draw[color4,thick] (0.4,-1.5)--(2,-1.5);
        \draw[color3] (2.8,-1.5) --(4.4,-1.5);
        \draw  (1,-1.5)
          .. controls (2.4,-0.3) ..
          (3.8,-1.5);
    \end{tikzpicture}};
        \node (3_1) at (0,0){\begin{tikzpicture}[scale = 0.7 ,baseline=(current bounding box.center)]
        \draw[color4,thick] (0.4,-1.5)--(2,-1.5);
        \draw[color3] (2.8,-1.5) --(4.4,-1.5);
        \draw  (1,-1.5)
          .. controls (2.4,-0.3) ..
          (3.8,-1.5);
    \end{tikzpicture}};
        \node (1_3) at (4,4){\begin{tikzpicture}[scale = 0.7 ,baseline=(current bounding box.center)]
        \draw[color4,thick] (0.4,-1.5)--(2,-1.5);
        \draw[color3] (2.8,-1.5) --(4.4,-1.5);
        \draw  (1,-1.5)
          .. controls (2.4,-0.3) ..
          (3.8,-1.5);
        \draw  (1.55,-1.5)--(2.2,-0.95);
        \draw[fill = white] (2.2,-0.95) circle [radius = 0.1];
        \draw  (2.6,-0.95)--(3.25,-1.5);
        \draw[fill = white] (2.6,-0.95) circle [radius = 0.1];
    \end{tikzpicture}};
        \node (2_2) at (2,2){\begin{tikzpicture}[scale = 0.7 ,baseline=(current bounding box.center)]
        \draw[color4,thick] (0.4,-1.5)--(2,-1.5);
        \draw[color3] (2.8,-1.5) --(4.4,-1.5);
        \draw  (1,-1.5)
          .. controls (2.4,-0.2) ..
          (3.8,-1.5);
        \draw  (1.7,-0.9)--(2.2,-0.9);
        \draw[fill = white] (2.2,-0.9) circle [radius = 0.1];
        \draw  (3.1,-0.9)--(2.6,-0.9);
        \draw[fill = white] (2.6,-0.9) circle [radius = 0.1];
    \end{tikzpicture}};
        \node (2_3) at (4,2){\begin{tikzpicture}[scale = 0.7 ,baseline=(current bounding box.center)]
        \draw[color4,thick] (0.3,-1.5)--(2,-1.5);
        \draw[color3] (2.8,-1.5) --(4.5,-1.5);
        \draw  (1.35,-1.5)
          .. controls (2.4,-0.6) ..
          (3.45,-1.5);
        \draw  (0.65,-1.5)
          .. controls (2.4,0) ..
          (4.15,-1.5);
    \end{tikzpicture}};
        \node (3_3) at (4,0){\begin{tikzpicture}[scale = 0.7 ,baseline=(current bounding box.center)]
        \draw[color4,thick] (0.4,-1.5)--(2,-1.5);
        \draw[color3] (2.8,-1.5) --(4.4,-1.5);
        \draw  (1,-1.5)
          .. controls (2.4,-0.2) ..
          (3.8,-1.5);
        \draw (1.7,-0.9)--(3.1,-0.9);
    \end{tikzpicture}};
        \draw[nat] (1_1)--(3_1)node[midway, left]{\id};
        \draw[nat] (1_1)--(1_3)node[midway, above]{\lambda_{\wr,\cN}^R\star\lambda_{\wr,\cM}^{-1}};
        \draw[nat] (1_3)--(2_3)node[midway, right]{\epsilon_0};
        \draw[nat] (2_3)--(3_3)node[midway, right]{\alpha_{\wr,\cN}^{-1}\star\alpha_{\wr,\cM}^R};
        \draw[nat] (3_3)--(3_1)node[midway, below]{\epsilon_2};
        \draw[nat] (1_1)--(2_2)node[midway, below left]{\rho^{-1}\star\rho^R};
        \draw[nat] (2_2)--(3_1)node[midway, above left]{\rho\star(\rho^R)^{-1}};
        \draw[nat] (2_2)--(3_3)node[midway, above right]{\epsilon_0};
        \draw[nat] (1_3)--(2_2)node[midway, below right]{\alpha_{\wr,\cN}^{-1}\star\alpha_{\wr,\cM}^R};
        \draw (2,0.8)node{\tiny(\eqref{eq.right_dual_of_lambda})};
        \draw (2,3)node{\tiny(\eqref{eq.triangle_graph_module} +Theorem \ref{thm.local})};
        \draw (3.1,2)node{\tiny(Theorem \ref{thm.local})};
    }
\]
\caption{The proof of~\eqref{eq.omega_ass} and the right square of \eqref{eq.omega_unit}. The left square of \eqref{eq.omega_unit} can be similarly proved using \eqref{eq.triangle_other_module}.}
\label{fig.pf_of_omega_monad}
\end{figure}
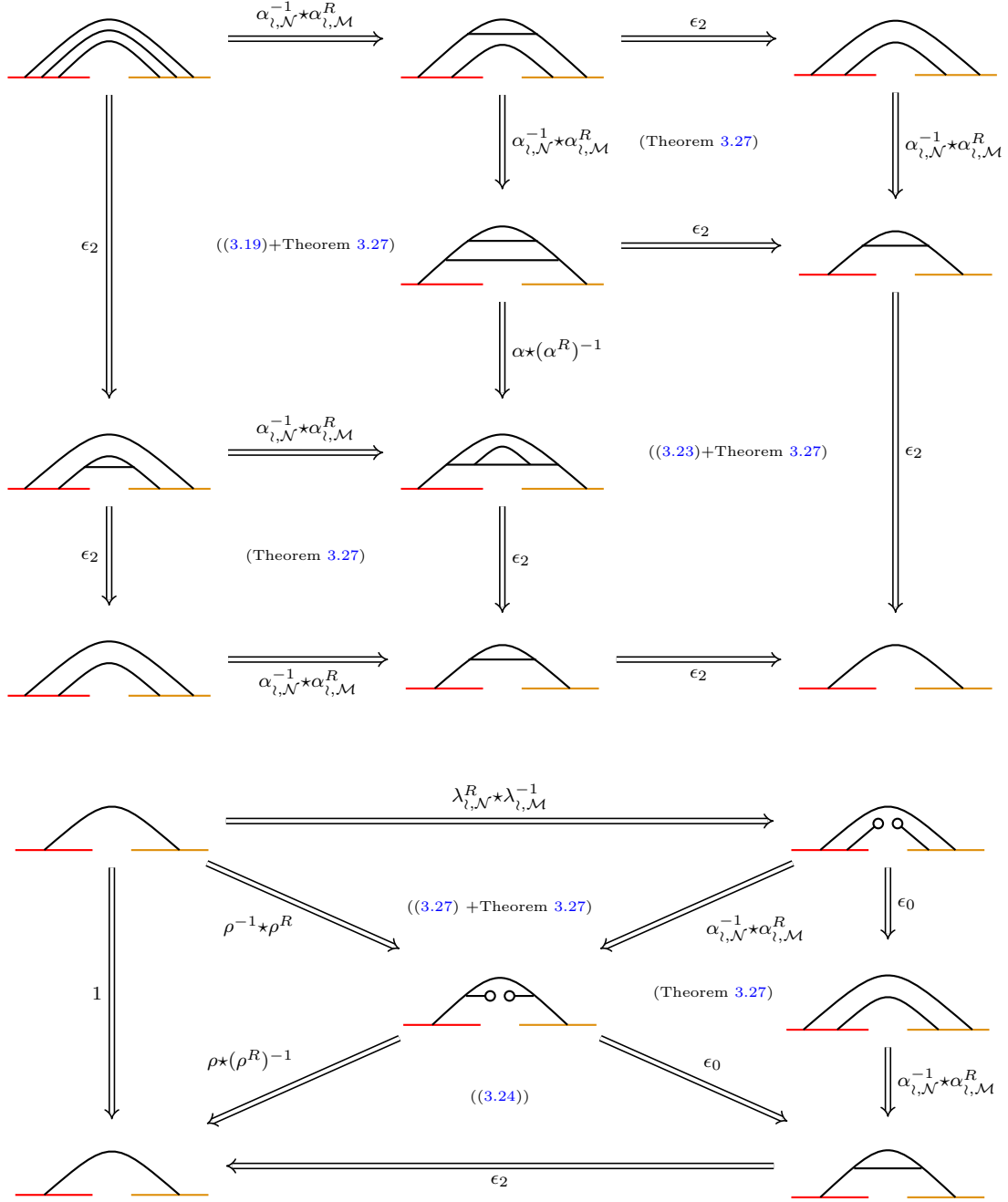

\begin{figure}[htbp]
\[
    \hspace{-4.5pc}
    \ctikz{[scale=1.5,xscale=1.8]
        \mygrid{ 
            \draw[gray!30, step=1, opacity = 0.5] (0,0) grid (6,6);
        }
        \node (1_1) at (0,6){$
        \ctikz{[scale=0.4]
            \draw[color3] (-1.35-0.7,0)--(-1.35+0.7,0);
            \draw[color3] (1.35-0.7,0)--(1.35+1,0);
            \draw[color3] (4.35-0.7,0)--(4.35+0.7,0);
            \myparabola{1.5}{-1.75}{0}{4.75}{0}
            \myparabola{1.5}{-1.05}{0}{4.05}{0}
        }$};
        \node (1_4) at (6,6){$
        \ctikz{[scale=0.4]
            \draw[color3] (-1.35-0.7,0)--(-1.35+0.7,0);
            \draw[color3] (1.35-0.7,0)--(1.35+1,0);
            \draw[color3] (4.35-0.7,0)--(4.35+0.7,0);
            \begin{scope}[xshift=1.5cm,scale=1.73]
                \draw (0.5,1.4) parabola (-1.65,0);
                \begin{scope}[xscale=-1]
                    \draw (0.5,1.4) parabola (-1.65,0); 
                \end{scope}
                \draw[white,fill=white] (-0.95,0.1)rectangle(0.95,1.75);
                \myparabola{0.9}{-0.75}{1.2}{0.75}{1.2}
                \myparabola{1.1}{-0.5}{0.8}{0.5}{0.8}
                \draw (-1,0.74) to[out=90, in=220](-0.75,1.2);
                \draw (-1,0.74) to[out=-30, in=220](-0.5,0.8);
                \begin{scope}[xscale=-1]
                    \draw (-1,0.74) to[out=90, in=220](-0.75,1.2);
                    \draw (-1,0.74) to[out=-30, in=220](-0.5,0.8);
                \end{scope}
            \end{scope}
        }$};
        \node (2_1) at (0,4){$
        \ctikz{[scale=0.4]
            \draw[color3] (-1.35-0.7,0)--(-1.35+0.7,0);
            \draw[color3] (1.35-0.7,0)--(1.35+1,0);
            \draw[color3] (4.35-0.7,0)--(4.35+0.7,0);
            \myparabola{1.5}{-1.75}{0}{4.75}{0}
            \myparabola{1.5}{-1.05}{0}{1.05}{0}
            \myparabola{1.5}{1.95}{0}{4.05}{0}
        }$};
        \node (2_2) at (2,4){$
        \ctikz{[scale=0.4]
            \begin{scope}[xshift=-1.5cm]
            \draw[color3] (-1.35-0.7,0)--(-1.35+0.7,0);
            \draw[color3] (1.35-0.7,0)--(1.35+1,0);
            \draw[color3] (4.35-0.7,0)--(4.35+0.7,0);
            \end{scope}
            \myparabola{1.5}{-3.25}{0}{3.25}{0}
            \myparabola{1.5}{-2.55}{0}{2.55}{0}
            \draw[white,fill=white] (-1,1)rectangle(1,2.5);
            \draw (-1.05,2.175)to[out=30, in=140](-0.5,2.1);
            \draw (-1.05,1.585)to[out=40, in =240](-0.5,2.1);
            \begin{scope}[xscale=-1]
                \draw (-1.05,2.175)to[out=30, in=140](-0.5,2.1);
                \draw (-1.05,1.585)to[out=40, in =240](-0.5,2.1);
            \end{scope}
            \myparabola{0.5}{-0.5}{2.1}{0.5}{2.1}
            \mygrid{ 
                \draw[gray!30, step=0.5, opacity = 0.5] (-3,-4) grid (3,3);
                \foreach \x in {-3,-2,-1,0,1,2,3,4,5,6}
                    \draw (\x,3) node[above] {\small $\x$};
                \foreach \y in {-1,0,1,2,3}
                    \draw (3,\y) node[right] {\small $\y$};
            }
        }$};
        \node (2_3) at (4,4){$
        \ctikz{[scale=0.4]
            \begin{scope}[xshift=-1.5cm]
                \draw[color3] (-1.35-0.7,0)--(-1.35+0.7,0);
                \draw[color3] (1.35-0.7,0)--(1.35+1,0);
                \draw[color3] (4.35-0.7,0)--(4.35+0.7,0);
            \end{scope}
            \myparabola{1.5}{-2.9}{0}{2.9}{0}
            \draw[white, fill=white] (-2,1)rectangle(-1,2);
            \draw (-2.05,1.1) to[out=-10,in=260](-0.95,1.95);
            \draw (-2.05,1.1) to[out=90,in=160](-0.95,1.95);
            \begin{scope}[xscale=-1]
                \draw[white, fill=white] (-2,1)rectangle(-1,2);
                \draw (-2.05,1.1) to[out=-10,in=260](-0.95,1.95);
                \draw (-2.05,1.1) to[out=90,in=160](-0.95,1.95);
            \end{scope}
            \mygrid{ 
                \draw[gray!30, step=0.5, opacity = 0.5] (-3,-4) grid (3,3);
                \foreach \x in {-3,-2,-1,0,1,2,3,4,5,6}
                    \draw (\x,3) node[above] {\small $\x$};
                \foreach \y in {-1,0,1,2,3}
                    \draw (3,\y) node[right] {\small $\y$};
            }
        }$};
        \node (3_2) at (2,2){$
        \ctikz{[scale=0.4]
            \draw[color3] (-1.35-0.7,0)--(-1.35+0.7,0);
            \draw[color3] (1.35-0.7,0)--(1.35+1,0);
            \draw[color3] (4.35-0.7,0)--(4.35+0.7,0);
            \begin{scope}[xshift=-0.2cm,xscale=-1,scale=0.85]
                \draw (0.5,1.4) parabola (-1.65,0);
                \draw[white,fill=white] (-0.95,0.1)rectangle(0.55,1.75);
                \myparabola{1.2}{-0.75}{1.2}{1.85}{0}
                \myparabola{1.2}{-0.5}{0.8}{1.35}{0}
                \draw (-1,0.74) to[out=90, in=220](-0.75,1.2);
                \draw (-1,0.74) to[out=-30, in=220](-0.5,0.8);
            \end{scope}
            \begin{scope}[xshift=3.2cm,scale=0.85]
                \draw (0.5,1.4) parabola (-1.65,0);
                \draw[white,fill=white] (-0.95,0.1)rectangle(0.55,1.75);
                \myparabola{1.2}{-0.75}{1.2}{1.85}{0}
                \myparabola{1.2}{-0.5}{0.8}{1.35}{0}
                \draw (-1,0.74) to[out=90, in=220](-0.75,1.2);
                \draw (-1,0.74) to[out=-30, in=220](-0.5,0.8);
            \end{scope}
        }$};
        \node (3_3) at (4,2){$
        \ctikz{[scale=0.4]
            \draw[color3] (-1.35-0.7,0)--(-1.35+0.7,0);
            \draw[color3] (1.35-0.7,0)--(1.35+1,0);
            \draw[color3] (4.35-0.7,0)--(4.35+0.7,0);
            \begin{scope}[xshift=1.5cm,scale=1.73]
                \draw (0.5,1.4) parabola (-1.65,0);
                \begin{scope}[xscale=-1]
                    \draw (0.5,1.4) parabola (-1.65,0); 
                \end{scope}
                \draw[white,fill=white] (-0.95,0.1)rectangle(0.95,1.75);
                \myparabola{0.9}{-0.75}{1.2}{0.75}{1.2}
                \myparabola{1.1}{-0.5}{0.8}{0.5}{0.8}
                \draw (-1,0.74) to[out=90, in=220](-0.75,1.2);
                \draw (-1,0.74) to[out=-30, in=220](-0.5,0.8);
                \begin{scope}[xscale=-1]
                    \draw (-1,0.74) to[out=90, in=220](-0.75,1.2);
                    \draw (-1,0.74) to[out=-30, in=220](-0.5,0.8);
                \end{scope}
            \end{scope}
        }$};
        \node (3_4) at (6,2){$
        \ctikz{[scale=0.4]
            \draw[color3] (-1.35-0.7,0)--(-1.35+0.7,0);
            \draw[color3] (1.35-0.7,0)--(1.35+1,0);
            \draw[color3] (4.35-0.7,0)--(4.35+0.7,0);
            \myparabola{1.5}{-1.35}{0}{4.35}{0}
        }$};
        \node (4_1) at (0,0){$
        \ctikz{[scale=0.4]
            \draw[color3] (-1.35-0.7,0)--(-1.35+0.7,0);
            \draw[color3] (1.35-0.7,0)--(1.35+1,0);
            \draw[color3] (4.35-0.7,0)--(4.35+0.7,0);
            \myparabola{1.5}{-1.4}{0}{1.4}{0}
            \myparabola{1.5}{1.6}{0}{4.4}{0}
            \myparabola{1.5}{-1.05}{0}{1.05}{0}
            \myparabola{1.5}{1.95}{0}{4.05}{0}
        }$};
        \node (4_2) at (2,0){$
        \ctikz{[scale=0.4]
            \draw[color3] (-1.35-0.7,0)--(-1.35+0.7,0);
            \draw[color3] (1.35-0.7,0)--(1.35+1,0);
            \draw[color3] (4.35-0.7,0)--(4.35+0.7,0);
            \begin{scope}[scale=0.818]
                \draw (0.5,1.4) parabola (-1.65,0);
                \begin{scope}[xscale=-1]
                    \draw (0.5,1.4) parabola (-1.65,0); 
                \end{scope}
                \draw[white,fill=white] (-0.95,0.1)rectangle(0.95,1.75);
                \myparabola{0.9}{-0.75}{1.2}{0.75}{1.2}
                \myparabola{1.1}{-0.5}{0.8}{0.5}{0.8}
                \draw (-1,0.74) to[out=90, in=220](-0.75,1.2);
                \draw (-1,0.74) to[out=-30, in=220](-0.5,0.8);
                \begin{scope}[xscale=-1]
                    \draw (-1,0.74) to[out=90, in=220](-0.75,1.2);
                    \draw (-1,0.74) to[out=-30, in=220](-0.5,0.8);
                \end{scope}
            \end{scope}
            \begin{scope}[xshift=3cm,scale=0.818]
                \draw (0.5,1.4) parabola (-1.65,0);
                \begin{scope}[xscale=-1]
                    \draw (0.5,1.4) parabola (-1.65,0); 
                \end{scope}
                \draw[white,fill=white] (-0.95,0.1)rectangle(0.95,1.75);
                \myparabola{0.9}{-0.75}{1.2}{0.75}{1.2}
                \myparabola{1.1}{-0.5}{0.8}{0.5}{0.8}
                \draw (-1,0.74) to[out=90, in=220](-0.75,1.2);
                \draw (-1,0.74) to[out=-30, in=220](-0.5,0.8);
                \begin{scope}[xscale=-1]
                    \draw (-1,0.74) to[out=90, in=220](-0.75,1.2);
                    \draw (-1,0.74) to[out=-30, in=220](-0.5,0.8);
                \end{scope}
            \end{scope}
        }$};
        \node (4_4) at (6,0){$
        \ctikz{[scale=0.4]
            \draw[color3] (-1.35-0.7,0)--(-1.35+0.7,0);
            \draw[color3] (1.35-0.7,0)--(1.35+1,0);
            \draw[color3] (4.35-0.7,0)--(4.35+0.7,0);
            \myparabola{1.5}{-1.35}{0}{1.35}{0}
            \myparabola{1.5}{1.65}{0}{4.35}{0}
        }$};
        \draw[nat] (1_1)--(1_4)node[midway, above]{\alpha_{\wr,\cM}^{-1}\star\alpha_{\wr,\cM}^R};
        \draw[nat] (1_4)--(3_4)node[midway, right]{\epsilon_2};
        \draw[nat] (1_1)--(2_1)node[midway, left]{\eta_{\wr,\cM}};
        \draw[nat] (2_1)--(4_1)node[midway, left]{\eta_{\wr,\cM}};
        \draw[nat] (4_1)--(4_2)node[midway, below=0.25em]{\alpha_{\wr,\cM}^{-1}\star\alpha_{\wr,\cM}^R\star\alpha_{\wr,\cM}^{-1}\star\alpha_{\wr,\cM}^R};
        \draw[nat] (4_2)--(4_4) node[midway, below=0.25em]{\epsilon_2\star\epsilon_2};
        \draw[nat] (3_4)--(4_4)node[midway, right]{\eta_{\wr,\cM}};
        \draw[nat] (1_1)--(2_2)node[midway, below left]{\eta_2};
        \draw[nat] (1_4)--(2_3)node[midway, above left]{\eta_2};
        \draw[nat] (2_2)--(2_3)node[midway, above]{\alpha_{\wr,\cM}^{-1}\star\alpha_{\wr,\cM}^R};
        \draw[nat] (2_2)--(3_2)node[midway, left]{\eta_{\wr,\cM}};
        \draw[nat] (3_2)--(4_1)node[midway, above left]{\alpha_{\wr,\cM}^R\star\alpha_{\wr,\cM}^{-1}};
        \draw[nat] (2_3)--(4_2)node[midway, below right]{\eta_{\wr,\cM}};
        \draw[nat] (2_3)--(3_3)node[midway, left]{\epsilon_2};
        \draw[nat] (3_3)--(3_4)node[midway, below]{\epsilon_2};
        \draw[nat] (1_4)--(3_3)node[midway, below right]{\id};
        \draw[nat] (3_2)--(4_2)node[midway, left]{\alpha_{\wr,\cM}^{-1}\star\star\alpha_{\wr,\cM}^R};
        \draw (1,3)node{\tiny (\eqref{eq.right_dual_of_alpha_module})};
        \draw (3,5)node{\tiny (Theorem \ref{thm.local})};
        \draw (2.75,2.75)node{\tiny (Theorem \ref{thm.local})};
        \draw (4,1.1)node{\tiny (Theorem \ref{thm.local})};
        \draw (4.75,4.1)node{\tiny (\eqref{eq.zig_zag_m}};
    }
\]
\[
    \hspace{-4.25pc}
    \ctikz{[scale=1.5,xscale=1.8]
        \mygrid{ 
            \draw[gray!30, step=1, opacity = 0.5] (0,0) grid (6,6);
        }
        \node (1_1) at (0,4){$
        \ctikz{[scale=0.4]
            \draw[color3] (-1.35-0.7,0)--(-1.35+0.7,0);
            \draw[color3] (1.35-0.7,0)--(1.35+1,0);
            \draw[color3] (4.35-0.7,0)--(4.35+0.7,0);
        }$};
        \node (1_2) at (2,4){$
        \ctikz{[scale=0.4]
            \mygrid{ 
                \draw[gray!30, step=0.5, opacity = 0.5] (-3,-4) grid (6,3);
                \foreach \x in {-3,-2,-1,0,1,2,3,4,5,6}
                    \draw (\x,3) node[above] {\small $\x$};
                \foreach \y in {-1,0,1,2,3}
                    \draw (3,\y) node[right] {\small $\y$};
            }
            \myparabola{1.5}{-1.35}{0}{4.35}{0}
            \draw[white,fill=white] (-0.75,0)rectangle(3.75,2.5);
            \draw[fill=white] (-0.75,0.8)circle[radius=0.2];
            \draw[fill=white] (3.75,0.8)circle[radius=0.2];
            \draw[color3] (-1.35-0.7,0)--(-1.35+0.7,0);
            \draw[color3] (1.35-0.7,0)--(1.35+1,0);
            \draw[color3] (4.35-0.7,0)--(4.35+0.7,0);
        }$};
        \node (1_4) at (6,4){$
        \ctikz{[scale=0.4]
            \draw[color3] (-1.35-0.7,0)--(-1.35+0.7,0);
            \draw[color3] (1.35-0.7,0)--(1.35+1,0);
            \draw[color3] (4.35-0.7,0)--(4.35+0.7,0);
            \myparabola{1.5}{-1.35}{0}{4.35}{0}
        }$};
        \node (2_2) at (2,2){$
        \ctikz{[scale=0.4]
            \mygrid{ 
                \draw[gray!30, step=0.5, opacity = 0.5] (-3,-4) grid (6,3);
                \foreach \x in {-3,-2,-1,0,1,2,3,4,5,6}
                    \draw (\x,3) node[above] {\small $\x$};
                \foreach \y in {-1,0,1,2,3}
                    \draw (3,\y) node[right] {\small $\y$};
            }
            \myparabola{1.5}{-1.35}{0}{4.35}{0}
            \draw[white,fill=white] (-0.75,0)rectangle(3.75,3);
            \draw[fill=white] (-0.75,0.8)circle[radius=0.2];
            \draw[fill=white] (3.75,0.8)circle[radius=0.2];
            \draw (1.35-0.7,0.8)--(1.35+1,0.8);
            \draw[fill=white](1.35-0.7,0.8)circle[radius=0.2];
            \draw[fill=white](1.35+1,0.8)circle[radius=0.2];
            \draw[color3] (-1.35-0.7,0)--(-1.35+0.7,0);
            \draw[color3] (1.35-0.7,0)--(1.35+1,0);
            \draw[color3] (4.35-0.7,0)--(4.35+0.7,0);
        }$};
        \node (2_3) at (4,2){$
        \ctikz{[scale=0.4]
            \mygrid{ 
                \draw[gray!30, step=0.5, opacity = 0.5] (-3,-4) grid (6,3);
                \foreach \x in {-3,-2,-1,0,1,2,3,4,5,6}
                    \draw (\x,3) node[above] {\small $\x$};
                \foreach \y in {-1,0,1,2,3}
                    \draw (3,\y) node[right] {\small $\y$};
            }
            \myparabola{1.5}{-1.35}{0}{4.35}{0}
            \draw[white,fill=white] (-0.75,0)rectangle(3.75,3);
            \draw[fill=white] (-0.75,0.8)circle[radius=0.2];
            \draw[fill=white] (3.75,0.8)circle[radius=0.2];
            \draw[color3] (-1.35-0.7,0)--(-1.35+0.7,0);
            \draw[color3] (1.35-0.7,0)--(1.35+1,0);
            \draw[color3] (4.35-0.7,0)--(4.35+0.7,0);
        }$};
        \node (3_1) at (0,0){$
        \ctikz{[scale=0.4]
            \myparabola{1.5}{-1.35}{0}{1.35}{0}
            \myparabola{1.5}{1.65}{0}{4.35}{0}
            \draw[fill=white,white] (-1.35+0.7,0)rectangle(1.35-0.7,2.5);
            \draw[fill=white,white](1.35+1,0)rectangle(4.35-0.7,2.5);
            \draw[fill=white] (-1.35+0.7,0.8)circle[radius=0.2];
            \draw[fill=white] (1.35-0.7,0.8)circle[radius=0.2];
            \draw[fill=white] (1.35+1,0.8)circle[radius=0.2];
            \draw[fill=white] (4.35-0.7,0.8)circle[radius=0.2];
            \draw[color3] (-1.35-0.7,0)--(-1.35+0.7,0);
            \draw[color3] (1.35-0.7,0)--(1.35+1,0);
            \draw[color3] (4.35-0.7,0)--(4.35+0.7,0);
            \mygrid{ 
                \draw[gray!30, step=0.5, opacity = 0.5] (-3,-4) grid (6,3);
                \foreach \x in {-3,-2,-1,0,1,2,3,4,5,6}
                    \draw (\x,3) node[above] {\small $\x$};
                \foreach \y in {-1,0,1,2,3}
                    \draw (3,\y) node[right] {\small $\y$};
            }
        }$};
        \node (3_4) at (6,0){$
        \ctikz{[scale=0.4]
            \draw[color3] (-1.35-0.7,0)--(-1.35+0.7,0);
            \draw[color3] (1.35-0.7,0)--(1.35+1,0);
            \draw[color3] (4.35-0.7,0)--(4.35+0.7,0);
            \myparabola{1.5}{-1.35}{0}{1.35}{0}
            \myparabola{1.5}{1.65}{0}{4.35}{0}
        }$};
        \draw[nat] (1_1)--(1_2)node[midway, above]{\lambda_{\wr,\cM}^{-1}\star\lambda_{\wr,\cM}^R};
        \draw[nat] (1_2)--(1_4)node[midway, above]{\epsilon_0};
        \draw[nat] (1_4)--(3_4)node[midway, right]{\eta_{\wr,\cM}};
        \draw[nat] (1_1)--(3_1)node[pos=0.2, right]{\lambda_{\wr,\cM}^{-1}\star\lambda_{\wr,\cM}^R\star\lambda_{\wr,\cM}^{-1}\star\lambda_{\wr,\cM}^R};
        \draw[nat] (3_1)--(3_4)node[midway, below]{\epsilon_0\star\epsilon_0};
        \draw[nat] (1_2)--(2_2)node[midway, left]{\eta_0};
        \draw[nat] (1_2)--(3_1)node[midway, above left]{\lambda_{\wr,\cM}^R\star\lambda_{\wr,\cM}^{-1}};
        \draw[nat] (2_2)--(3_1)node[midway, below right]{\eta_{\wr,\cM}};
        \draw[nat] (2_2)--(2_3)node[midway, below]{\epsilon_0};
        \draw[nat] (1_2)--(2_3)node[midway, above right]{\id};
        \draw[nat] (2_3)--(1_4)node[midway, below right]{\epsilon_0};
        \draw (1.35,2)node{\tiny (\eqref{eq.right_dual_of_lambda_module}};
        \draw (2.65,2.65)node{\tiny (\eqref{eq.zig_zag_u})};
        \draw (3.5,1)node{\tiny (Theorem \ref{thm.local})};
    }
\]
\caption{Proof of \eqref{eq.omega_bimonad1} and \eqref{eq.omega_bimonad2}}
\label{fig.pf_of_omega_bimonad}
\end{figure}

The method to prove the commutative diagrams in Figure \ref{fig.pf_of_omega_bimonad} also lead us to the fact that $\MCM$ is a probimonad.
\begin{proposition}\label{prp.is_bimonad}
    The structure $(M,\tau,U,\nu)$ defined in \eqref{eq.probimonad_of_MCM_1} and \eqref{eq.probimonad_of_MCM_2} fits into a probimonad structure on the promonad $\MCM$.
    \begin{proof}
        Define $\alpha^\bullet$ as the invertible pro-tensor network map
        \[
            \ctikz{[scale=0.5]
                \draw[color3] (0,0)--(-1,0);
                \draw[color3] (-1,0)arc(270:90:0.5);
                \draw[color3] (-1,1)--(0,1);
                \draw[color3] (0,2)--(-3,2);
                \draw[color3] (-3,2)arc(270:90:0.5);
                \draw[color3] (-3,3)--(0,3);
                \draw[color3] (0,4)--(-5,4);
                \draw[color3] (0,-1)--(-5,-1);
            }
            \Rightarrow
            \ctikz{[scale=0.5,yscale=-1]
                \draw[color3] (0,0)--(-1,0);
                \draw[color3] (-1,0)arc(270:90:0.5);
                \draw[color3] (-1,1)--(0,1);
                \draw[color3] (0,2)--(-3,2);
                \draw[color3] (-3,2)arc(270:90:0.5);
                \draw[color3] (-3,3)--(0,3);
                \draw[color3] (0,4)--(-5,4);
                \draw[color3] (0,-1)--(-5,-1);
            }
        \]
        formed by compositions of $\sigma$ and $\sigma'$'s defined in Example \ref{ex.swapper}. Also set
        \[
        \lambda^\bullet\defdtobe(
        \ctikz{[scale=0.5]
            \basetwoturn
            \draw[color3] (0,-1)--(-4,-1);
        }
        \stackrel{\zeta_\cM}{\Rightarrow}
        \ctikz{[scale=0.5,yscale=2]
            \draw[color3] (0,-1)--(-2,-1);
            \draw[color3] (0,0)--(-2,0);
        }
        ),\quad
        \rho^\bullet\defdtobe(
        \ctikz{[scale=0.5,yscale=-1]
            \basetwoturn
            \draw[color3] (0,-1)--(-4,-1);
        }
        \stackrel{\chi_\cM}{\Rightarrow}
        \ctikz{[scale=0.5,yscale=-2]
            \draw[color3] (0,-1)--(-2,-1);
            \draw[color3] (0,0)--(-2,0);
        }
        ).
        \]
        It is enough to show that $(\MCM,M,\tau,U,\nu,\alpha^\bullet,\lambda^\bullet,\rho^\bullet)$ is a probimonad, i.e., render the diagrams in \eqref{eq.tau_intertwine}-\eqref{eq.pentagon_and_triangle_promonoidal} in Figure \ref{fig.bimonad_axiom} commutative. The left diagram in \eqref{eq.tau_intertwine} can be proved commutative in essentially the same manner as \eqref{eq.omega_bimonad1}, modulo applications of Theorem \ref{thm.local} and \eqref{eq.zigzagger}. Similarly, it can be verified that the right diagram in \eqref{eq.tau_intertwine} essentially follows from \eqref{eq.omega_bimonad2}. The two diagrams in \eqref{eq.nu_intertwine}, the diagram in \eqref{eq.alpha_is_monad_2mor}, the two diagrams in \eqref{eq.lambda_rho_are_monad_2mor} essentially follow from the left diagram in \eqref{eq.right_dual_of_alpha_module}, the left diagram in \eqref{eq.right_dual_of_lambda_module}, the diagram in \eqref{eq.omega_bimonad_ass}, the two diagrams in \eqref{eq.omega_bimonad_unit}, respectively. Finally, that $\alpha^\bullet,\lambda^\bullet,\rho^\bullet$ satisfy \eqref{eq.pentagon_and_triangle_promonoidal} is well-known \cite[pp. 36-37]{Day_1970_b}; it can also be verified directly using \eqref{eq.zigzagger}.
    \end{proof}
\end{proposition}
\subsection{Finishing the proof of Theorem \ref{thm.unenriched_Kitaev_Kong} and Theorem \ref{thm.unenriched_mon_Kitaev_Kong}}
\begin{proof}[Finishing the proof of Theorem \ref{thm.unenriched_Kitaev_Kong}]
    It remains to establish an equivalence between $\Lmd_{\MnCN}$ and $\Lmd_{\MCN}(\ast)$. To this end, we construct a pair of mutually inverse functors with the focus mainly on the objects' assignments. These functors are essentially variants of the ones in Theorem \ref{thm.higher_transpose}. To be concrete, let
\[
(\ctikz{[scale=0.333]
    \draw[color4] (-2,-0.299)--(0,-0.299);
    \draw[color3] (-2,0.299)--(0,0.299);
    \mynode{0}{0}{0.6}{0.6}{}
}
,
\ctikz{[scale=0.333]
    \mygrid{ 
        \draw[gray!30, step=0.5, opacity = 0.5] (-5,-4) grid (1,3);
        \foreach \x in {-3,-2,-1,0,1,2,3}
            \draw (\x,3) node[above] {\small $\x$};
        \foreach \y in {-1,0,1,2,3}
            \draw (3,\y) node[right] {\small $\y$};
    }
    \begin{scope}[yshift=0.299cm, scale=1]
        \basetwoturn
    \end{scope}
    \draw[color4] (-4,-0.299)--(0,-0.299);
    \mynode{0}{0}{0.6}{0.6}{}
    \myparabola{2.25}{-3.7}{-0.299}{-2.25}{1.299}
}
\stackrel{\kappa}{\Rightarrow}
\ctikz{[scale=0.333]
    \draw[color4] (-2,-0.299)--(0,-0.299);
    \draw[color3] (-2,0.299)--(0,0.299);
    \mynode{0}{0}{0.6}{0.6}{}
}
)
\]
be an $\MCN$-module from $\ast$, satisfying the module axioms \eqref{eq.ModuleoOverMonad1} and \eqref{eq.ModuleoOverMonad2}. Then we define an $\MnCN$-module
\[
(
\ctikz{[scale=0.333]
    \begin{scope}[yshift=0.299cm,scale=1]
        \baseoneturn
    \end{scope}
    \draw[color4] (-3,-0.299)--(0,-0.299);
    \mynode{0}{0}{0.6}{0.6}{}
},
\ctikz{[scale=0.333]
            \myparabola{2}{-2.5}{-0.299}{-0.5}{1.299}
            \begin{scope}[yshift=0.299cm,scale=1]
                \baseoneturn
            \end{scope}
            \draw[color4] (-3,-0.299)--(0,-0.299);
            \mynode{0}{0}{0.6}{0.6}{}
        }
\stackrel{\chi_\cM^{-1}}{\Rightarrow}
\ctikz{[scale=0.333]
    \myparabola{2}{-2.5}{-0.299}{-0.5}{1.099}
    \begin{scope}[yshift=0.299cm,scale=0.8]
        \basethreeturn
    \end{scope}
    \draw[color4] (-3,-0.299)--(0,-0.299);
    \mynode{0}{0}{0.6}{0.6}{}
    }
\stackrel{\kappa}{\Rightarrow}
\ctikz{[scale=0.333]
    \begin{scope}[yshift=0.299cm,scale=1]
        \baseoneturn
    \end{scope}
    \draw[color4] (-3,-0.299)--(0,-0.299);
    \mynode{0}{0}{0.6}{0.6}{}
}.
)
\]
That this $\MnCN$-module is indeed well-defined is verified in Figure \ref{fig.monad_to_omega}. This construction naturally extends to a functor $\Phi'\:\Lmd_{\MCN}(\ast)\to\Lmd_{\MnCN}$.

\begin{figure}[htbp]
    \centering
\[
    \ctikz{[scale=1.5]
        \mygrid{
            \draw[gray!30, step=0.5, opacity = 0.5] (-3,-3) grid (3,3);
            \foreach \x in {-3,-2,-1,0,1,2,3}
                \draw (\x,3) node[above] {\small $\x$};
            \foreach \y in {-3,-2,-1,0,1,2,3}
                \draw (3,\y) node[right] {\small $\y$};
        }
        \node (1_1) at (-2.5,2){$
        \ctikz{[scale=0.333]
            \myparabola{2}{-2.5}{-0.299}{-0.5}{1.299}
            \draw[white,fill=white] (-2.25,0)rectangle(-1,2);
            \draw[fill=white] (-2.25,0.3)circle[radius=0.1];
            \draw[fill=white] (-1.05,1.67)circle[radius=0.1];
            \begin{scope}[yshift=0.299cm,scale=1]
                \baseoneturn
            \end{scope}
            \draw[color4] (-3,-0.299)--(0,-0.299);
            \mynode{0}{0}{0.6}{0.6}{}
            \mygrid{ 
            \draw[gray!30, step=0.5, opacity = 0.5] (-3,-1) grid (3,3);
            \foreach \x in {-3,-2,-1,0,1,2,3}
                \draw (\x,3) node[above] {\small $\x$};
            \foreach \y in {-1,0,1,2,3}
                \draw (3,\y) node[right] {\small $\y$};
            }
        }$};
        \node (1_2) at (0,2){$
        \ctikz{[scale=0.333]
            \myparabola{2}{-2.5}{-0.299}{-0.5}{1.299}
            \begin{scope}[yshift=0.299cm,scale=1]
                \baseoneturn
            \end{scope}
            \draw[color4] (-3,-0.299)--(0,-0.299);
            \mynode{0}{0}{0.6}{0.6}{}
            \mygrid{ 
            \draw[gray!30, step=0.5, opacity=0.3] (-3,-1) grid (3,3);
            \foreach \x in {-3,-2,-1,0,1,2,3}
                \draw (\x,3) node[above] {\small $\x$};
            \foreach \y in {-1,0,1,2,3}
                \draw (3,\y) node[right] {\small $\y$};
            }
        }$};
        \node (1_3) at (2.5,2){$
        \ctikz{[scale=0.333]
            \myparabola{2}{-2.5}{-0.299}{-0.5}{1.099}
            \begin{scope}[yshift=0.299cm,scale=0.8]
                \basethreeturn
            \end{scope}
            \draw[color4] (-3,-0.299)--(0,-0.299);
            \mynode{0}{0}{0.6}{0.6}{}
        }$};
        \node (2_2) at (0,0){$
        \ctikz{[scale=0.333]
            \myparabola{2}{-2.5}{-0.299}{-0.5}{1.099}
            \draw[white,fill=white] (-2.25,0)rectangle(-1,2);
            \draw[fill=white] (-2.25,0.3)circle[radius=0.1];
            \draw[fill=white] (-1,1.5)circle[radius=0.1];
            \begin{scope}[yshift=0.299cm,scale=0.8]
                \basethreeturn
            \end{scope}
            \draw[color4] (-3,-0.299)--(0,-0.299);
            \mynode{0}{0}{0.6}{0.6}{}
            \mygrid{ 
            \draw[gray!30, step=0.5, opacity = 0.5] (-3,-1) grid (3,3);
            \foreach \x in {-3,-2,-1,0,1,2,3}
                \draw (\x,3) node[above] {\small $\x$};
            \foreach \y in {-1,0,1,2,3}
                \draw (3,\y) node[right] {\small $\y$};
            }
        }$};
        \node (3_1) at (-2.5,-2.5){$
        \ctikz{[scale=0.333]
            \begin{scope}[yshift=0.299cm,scale=1]
                \baseoneturn
            \end{scope}
            \draw[color4] (-3,-0.299)--(0,-0.299);
            \mynode{0}{0}{0.6}{0.6}{}
            \mygrid{ 
            \draw[gray!30, step=0.5, opacity=0.5] (-3,-1) grid (3,3);
            \foreach \x in {-3,-2,-1,0,1,2,3}
                \draw (\x,3) node[above] {\small $\x$};
            \foreach \y in {-1,0,1,2,3}
                \draw (3,\y) node[right] {\small $\y$};
            }
        }$};
        \node (3_2) at (0,-2.5){
        $\ctikz{[scale=0.333]
            \begin{scope}[yshift=0.299cm,scale=0.8]
                \basethreeturn
            \end{scope}
            \draw[color4] (-3,-0.299)--(0,-0.299);
            \mynode{0}{0}{0.6}{0.6}{}
            }$
        };
        \node (3_3) at (2.5,-2.5){$
        \ctikz{[scale=0.333]
            \begin{scope}[yshift=0.299cm,scale=1]
                \baseoneturn
            \end{scope}
            \draw[color4] (-3,-0.299)--(0,-0.299);
            \mynode{0}{0}{0.6}{0.6}{}
            \mygrid{ 
            \draw[gray!30, step=0.5, opacity=0.5] (-3,-1) grid (3,3);
            \foreach \x in {-3,-2,-1,0,1,2,3}
                \draw (\x,3) node[above] {\small $\x$};
            \foreach \y in {-1,0,1,2,3}
                \draw (3,\y) node[right] {\small $\y$};
            }
        }$};
        \draw[nat] (1_1)--(1_2) node[midway, above]{\epsilon_0};
        \draw[nat] (1_2)--(1_3) node[midway, above] {\chi_\cM^{-1}};
        \draw[nat] (1_3)--(3_3) node[midway, right] {\kappa};
        \draw[nat] (1_1)--(2_2) node[midway, below left] {\chi_\cM^{-1}};
        \draw[nat] (2_2)--(1_3) node[midway, below right] {\epsilon_0};
        \draw[nat] (1_1)--(3_1) node[midway, left] {\lambda_{\wr,\cN}\star(\lambda_{\wr,\cN}^R)^{-1}};
        \draw[nat] (2_2)--(3_2) node[midway, right]{\lambda_{\wr,\cN}\star(\lambda_{\wr,\cN}^R)^{-1}};
        \draw[nat] (3_1)--(3_2) node[midway, above]{\chi_\cM^{-1}};
        \draw[nat] (3_2)--(3_3) node[midway, above]{\zeta_\cM}; 
        \draw[nat] (3_1)..controls (0,-4)..(3_3) node[midway, below] {\id};
        \draw (0,1)node{\tiny (Theorem \ref{thm.local})};
        \draw (-1.5,-0.75)node{\tiny (Theorem \ref{thm.local})};
        \draw (1.5,-0.75)node{\tiny( \eqref{eq.ModuleoOverMonad2})};
        \draw (0,-3.3)node{\tiny(\eqref{eq.zigzagger})};
    }
\]
\[
    \ctikz{[scale=1.35,xscale=1.5,yscale=1.1]
        \mygrid{
            \draw[gray!30, step=0.5, opacity = 0.5] (0,0) grid (6,6);
            \foreach \x in {0,1,2,3,4,5,6}
                \draw (\x,6) node[above] {\small $\x$};
            \foreach \y in {0,1,2,3,4,5,6}
                \draw (6,\y) node[right] {\small $\y$};
        }
        \node (1_1) at (0,6){$
        \ctikz{[scale=0.333]
            \myparabola{2}{-2.4}{-0.299}{-0.6}{1.299}
            \myparabola{2}{-2.8}{-0.299}{-0.2}{1.299}
            \begin{scope}[yshift=0.299cm,scale=1]
                \baseoneturn
            \end{scope}
            \draw[color4] (-3,-0.299)--(0,-0.299);
            \mynode{0}{0}{0.6}{0.6}{}
            \mygrid{ 
            \draw[gray!30, step=0.5, opacity=0.5] (-3,-1) grid (3,3);
            \foreach \x in {-3,-2,-1,0,1,2,3}
                \draw (\x,3) node[above] {\small $\x$};
            \foreach \y in {-1,0,1,2,3}
                \draw (3,\y) node[right] {\small $\y$};
            }
        }$};
        \node (1_3) at (4,6){$
        \ctikz{[scale=0.333]
            \myparabola{2.5}{-1.9}{-0.299}{-0.4}{1.099}
            \myparabola{2.05}{-2.6}{-0.299}{0.75}{2.699}
            \begin{scope}[yshift=0.299cm,scale=0.8]
                \basethreeturn
            \end{scope}
            \draw[color4] (-3,-0.299)--(0,-0.299);
            \mynode{0}{0}{0.6}{0.6}{}
            \mygrid{ 
            \draw[gray!30, step=0.5, opacity=0.5] (-3,-1) grid (3,3);
            \foreach \x in {-3,-2,-1,0,1,2,3}
                \draw (\x,3) node[above] {\small $\x$};
            \foreach \y in {-1,0,1,2,3}
                \draw (3,\y) node[right] {\small $\y$};
            }
        }$};
        \node (1_4) at (6,6){$
        \ctikz{[scale=0.333]
            \myparabola{2}{-2.5}{-0.299}{-0.5}{1.299}
            \begin{scope}[yshift=0.299cm,scale=1]
                \baseoneturn
            \end{scope}
            \draw[color4] (-3,-0.299)--(0,-0.299);
            \mynode{0}{0}{0.6}{0.6}{}
            \mygrid{ 
            \draw[gray!30, step=0.5, opacity=0.5] (-3,-1) grid (3,3);
            \foreach \x in {-3,-2,-1,0,1,2,3}
                \draw (\x,3) node[above] {\small $\x$};
            \foreach \y in {-1,0,1,2,3}
                \draw (3,\y) node[right] {\small $\y$};
            }
        }$};
        \node (2_2) at (2,4){$
        \ctikz{[scale=0.333]
            \myparabola{2}{-2.4}{-0.299}{-0.6}{1.099}
            \myparabola{2}{-2.8}{-0.299}{-0.2}{1.099}
            \begin{scope}[yshift=0.299cm,scale=0.8]
                \basethreeturn
            \end{scope}
            \draw[color4] (-3,-0.299)--(0,-0.299);
            \mynode{0}{0}{0.6}{0.6}{}
            \mygrid{ 
                \draw[gray!30, step=0.5, opacity=0.5] (-3,-1) grid (3,3);
                \foreach \x in {-3,-2,-1,0,1,2,3}
                    \draw (\x,3) node[above] {\small $\x$};
                \foreach \y in {-1,0,1,2,3}
                    \draw (3,\y) node[right] {\small $\y$};
            }
        }$};
        \node (2_3) at (4,4){$
        \ctikz{[scale=0.333]
            \myparabola{2.5}{-1.9}{-0.299}{-0.4}{0.999}
            \myparabola{2}{-2.6}{-0.299}{0.75}{2.399}
            \begin{scope}[yshift=0.299cm,scale=0.7]
                \basefiveturn
            \end{scope}
            \draw[color4] (-3,-0.299)--(0,-0.299);
            \mynode{0}{0}{0.6}{0.6}{}
            \mygrid{ 
            \draw[gray!30, step=0.5, opacity=0.5] (-3,-1) grid (3,3);
            \foreach \x in {-3,-2,-1,0,1,2,3}
                \draw (\x,3) node[above] {\small $\x$};
            \foreach \y in {-1,0,1,2,3}
                \draw (3,\y) node[right] {\small $\y$};
            }
        }$};
        \node (2_4) at (6,4){$
        \ctikz{[scale=0.333]
            \myparabola{2}{-2.5}{-0.299}{-0.5}{1.099}
            \begin{scope}[yshift=0.299cm,scale=0.8]
                \basethreeturn
            \end{scope}
            \draw[color4] (-3,-0.299)--(0,-0.299);
            \mynode{0}{0}{0.6}{0.6}{}
        }$};
        \node (3_2) at (2,2){$
        \ctikz{[scale=0.333]
            \begin{scope}[yshift=-0.299cm,xshift=0.2cm,scale=0.524,xscale=1.1]
                \myparabola{2}{-4.1}{0}{-1.65}{2.666}
                \myparabola{2.1}{-4.6}{0}{-1.15}{2.666}
                \draw[white,fill=white] (-2,0.1)rectangle(0,4);
                \draw[white, fill=white](-5,0)rectangle(-4.2,2);
                \draw[white, fill=white] (-5,0)rectangle(-3.84,1.25);
                \draw (-4.4,0)to[out=84, in=258](-4.2,0.6);
                \draw (-4.2,0.6)to[out=90,in=245](-4,1.52);
                \draw (-4.2,0.6)to[out=0, in =255] (-3.7,1.13);
                \draw (-2.1,2.92)to[out=-30,in=235](-1.7,3);
                \draw (-2.1,3.375)to[out=-20,in=120](-1.7,3);
                \draw (-1.7,3)to[out=-45,in=120](-1.4,2.666);
            \end{scope}
            \begin{scope}[yshift=0.299cm,scale=0.8]
                \basethreeturn
            \end{scope}
            \draw[color4] (-3,-0.299)--(0,-0.299);
            \mynode{0}{0}{0.6}{0.6}{}
            \mygrid{ 
                \draw[gray!30, step=0.5, opacity=0.5] (-3,-1) grid (3,3);
                \foreach \x in {-3,-2,-1,0,1,2,3}
                    \draw (\x,3) node[above] {\small $\x$};
                \foreach \y in {-1,0,1,2,3}
                    \draw (3,\y) node[right] {\small $\y$};
            }
        }$};
        \node (4_1) at (0,0){$
        \ctikz{[scale=0.333]
            \begin{scope}[yshift=-0.299cm,xshift=0.2cm,scale=0.599,xscale=1.1]
                \myparabola{2}{-4.1}{0}{-1.65}{2.666}
                \myparabola{2.1}{-4.6}{0}{-1.15}{2.666}
                \draw[white,fill=white] (-2,0.1)rectangle(0,4);
                \draw[white, fill=white](-5,0)rectangle(-4.2,2);
                \draw[white, fill=white] (-5,0)rectangle(-3.84,1.25);
                \draw (-4.4,0)to[out=84, in=258](-4.2,0.6);
                \draw (-4.2,0.6)to[out=90,in=245](-4,1.52);
                \draw (-4.2,0.6)to[out=0, in =255] (-3.7,1.13);
                \draw (-2.1,2.92)to[out=-30,in=235](-1.7,3);
                \draw (-2.1,3.375)to[out=-20,in=120](-1.7,3);
                \draw (-1.7,3)to[out=-45,in=120](-1.4,2.666);
            \end{scope}
            \begin{scope}[yshift=0.299cm,scale=1]
                \baseoneturn
            \end{scope}
            \draw[color4] (-3,-0.299)--(0,-0.299);
            \mynode{0}{0}{0.6}{0.6}{}
            \mygrid{ 
            \draw[gray!30, step=0.5, opacity=0.5] (-3,-1) grid (3,3);
            \foreach \x in {-3,-2,-1,0,1,2,3}
                \draw (\x,3) node[above] {\small $\x$};
            \foreach \y in {-1,0,1,2,3}
                \draw (3,\y) node[right] {\small $\y$};
            }
        }$};
        \node (4_2) at (2,0){$
        \ctikz{[scale=0.333]
            \myparabola{2}{-2.5}{-0.299}{-0.5}{1.299}
            \begin{scope}[yshift=0.299cm,scale=1]
                \baseoneturn
            \end{scope}
            \draw[color4] (-3,-0.299)--(0,-0.299);
            \mynode{0}{0}{0.6}{0.6}{}
            \mygrid{ 
            \draw[gray!30, step=0.5, opacity=0.5] (-3,-1) grid (3,3);
            \foreach \x in {-3,-2,-1,0,1,2,3}
                \draw (\x,3) node[above] {\small $\x$};
            \foreach \y in {-1,0,1,2,3}
                \draw (3,\y) node[right] {\small $\y$};
            }
        }$};
        \node (4_3) at (4,0){$
        \ctikz{[scale=0.333]
            \myparabola{2}{-2.5}{-0.299}{-0.5}{1.099}
            \begin{scope}[yshift=0.299cm,scale=0.8]
                \basethreeturn
            \end{scope}
            \draw[color4] (-3,-0.299)--(0,-0.299);
            \mynode{0}{0}{0.6}{0.6}{}
        }$};
        \node (4_4) at (6,0){$
        \ctikz{[scale=0.333]
            \begin{scope}[yshift=0.299cm,scale=1]
                \baseoneturn
            \end{scope}
            \draw[color4] (-3,-0.299)--(0,-0.299);
            \mynode{0}{0}{0.6}{0.6}{}
            \mygrid{ 
            \draw[gray!30, step=0.5, opacity=0.5] (-3,-1) grid (3,3);
            \foreach \x in {-3,-2,-1,0,1,2,3}
                \draw (\x,3) node[above] {\small $\x$};
            \foreach \y in {-1,0,1,2,3}
                \draw (3,\y) node[right] {\small $\y$};
            }
        }$};
        \draw[nat] (1_1)--(1_3) node[midway, above]{\chi_\cM^{-1}};
        \draw[nat] (1_3)--(1_4) node[midway, above]{\kappa};
        \draw[nat] (2_3)--(2_4) node[midway, below]{\kappa};
        \draw[nat] (2_3)--(2_2) node[midway, below]{\chi_\cM^{-1}};
        \draw[nat] (4_1)--(4_2) node[midway, below]{\epsilon_2};
        \draw[nat] (4_2)--(4_3) node[midway, below]{\chi_\cM^{-1}};
        \draw[nat] (4_3)--(4_4) node[midway, below]{\kappa};
        \draw[nat] (1_1)--(4_1) node[midway, left]{\alpha_{\wr,\cN}^{-1}\star\alpha_{\wr,\cM}^R};
        \draw[nat] (2_2)--(3_2) node[midway, right]{\alpha_{\wr,\cN}^{-1}\star\alpha_{\wr,\cM}^R};
        \draw[nat] (1_3)--(2_3) node[midway, right]{\chi_\cM^{-1}};
        \draw[nat] (1_4)--(2_4) node[midway, right]{\chi_\cM^{-1}};
        \draw[nat] (2_4)--(4_4) node[midway, right] {\kappa};
        \draw[nat] (1_1)--(2_2) node[midway, above right]{\chi_\cM^{-1}};
        \draw[nat] (4_1)--(3_2) node[midway, above left]{\chi_\cM^{-1}};
        \draw[nat] (3_2)--(4_3) node[midway, above right]{\epsilon_2};
        \draw (4.25,2) node {\tiny(\eqref{eq.ModuleoOverMonad1})};
        \draw (5,5) node {\tiny (Theorem \ref{thm.local})};
        \draw (2.5,5) node {\tiny (Theorem \ref{thm.local})};
        \draw (1,3) node {\tiny (Theorem \ref{thm.local})};
        \draw (1.75,0.75) node {\tiny (Theorem \ref{thm.local})};
    }
\]
    \caption{}
    \label{fig.monad_to_omega}
\end{figure}

Conversely, given an $\MnCN$-module
\[
(
\begin{tikzpicture}[scale=0.5,baseline=0]
    \draw[color3] (2,0)--(0,0);
    \draw[color4] (0,0)--(-2,0);
    \mynode{0}{0}{0.4}{0.4}{}
\end{tikzpicture},
\begin{tikzpicture}[scale=0.5,baseline=0]
    \draw[color3] (2,0)--(0,0);
    \draw[color4] (0,0)--(-2,0);
    \mynode{0}{0}{0.4}{0.4}{}
    \myparabola{1.5}{-1.35}{0}{1.35}{0}
\end{tikzpicture}
\stackrel{\omega}{\Rightarrow}
\begin{tikzpicture}[scale=0.5,baseline=0]
    \draw[color3] (2,0)--(0,0);
    \draw[color4] (0,0)--(-2,0);
    \mynode{0}{0}{0.4}{0.4}{}
\end{tikzpicture}
),
\]
we define an $\MCN$-module from $\ast$
\[
(
\ctikz{[scale=0.5]
    \draw[color3] (2,0)--(0,0);
    \draw[color4] (-2,0)--(0,0);
    \draw[color3] (2,0) arc (-90:90:1);
    \draw[color3] (2,2)--(-2,2);
    \mynode{0}{0}{0.4}{0.4}{}
}
,
    \ctikz{[scale=0.333]
        \begin{scope}[scale=1.333,xscale=-1]
            \basethreeturn
        \end{scope}
        \draw[color3] (0,0)--(-2,0);
        \draw[color4] (-2,0)--(-5,0);
        \mynode{-2}{0}{0.6}{0.6}{}
        \myparabola{2}{-4.5}{0}{-1.5}{2.666}
    }
    \stackrel{\chi_\cM}{\Rightarrow}
    \ctikz{[scale=0.5]
        \draw[color3] (2,0)--(0,0);
        \draw[color4] (-2,0)--(0,0);
        \draw[color3] (2,0) arc (-90:90:1);
        \draw[color3] (2,2)--(-2,2);
        \myparabola{1.5}{-1.35}{0}{1.35}{0}
        \mynode{0}{0}{0.4}{0.4}{}
    }
    \stackrel{\omega}{\Rightarrow}
    \ctikz{[scale=0.5]
        \draw[color3] (2,0)--(0,0);
        \draw[color4] (-2,0)--(0,0);
        \draw[color3] (2,0) arc (-90:90:1);
        \draw[color3] (2,2)--(-2,2);
        \mynode{0}{0}{0.4}{0.4}{}
    }
).
\]
That this indeed defines an $\MCN$-module is verified  in Figure \ref{fig.omega_to_monad}. The construction easily extends to a functor $\Psi'\:\Lmd_{\MnCN}\to\Lmd_{\MCN}(\ast)$. 

\begin{figure}[htbp]
\centering
\[
    \ctikz{[scale=1.5,xscale=1.5]
        \mygrid{
            \draw[gray!30, step=0.5, opacity = 0.5] (0,0) grid (6,6);
            \foreach \x in {0,1,2,3,4,5,6}
                \draw (\x,6) node[above] {\small $\x$};
            \foreach \y in {0,1,2,3,4,5,6}
                \draw (6,\y) node[right] {\small $\y$};
        }
        \node (1_1) at (0,4){$
        \ctikz{[scale=0.5]
            \mygrid{
                \draw[gray!30, step=0.5, opacity=0.5] (-3,-1) grid (3,3);
                \foreach \x in {-3,-2,-1,0,1,2,3}
                    \draw (\x,3) node[above] {\small $\x$};
                \foreach \y in {-1,0,1,2,3}
                    \draw (3,\y) node[right] {\small $\y$};
            }
            \draw[color3] (0,0)--(2,0);
            \draw[color4] (-2,0)--(0,0);
            \draw[color3] (2,0) arc (-90:90:1);
            \draw[color3] (2,2)--(-2,2);
            \mynode{0}{0}{0.4}{0.4}{}
        }$};
        \node (1_2) at (2,4){$
        \ctikz{[scale=0.333]
            \mygrid{ 
                \draw[gray!30, step=0.5, opacity=0.5] (-5,-1) grid (3,3);
                \foreach \x in {-3,-2,-1,0,1,2,3}
                    \draw (\x,3) node[above] {\small $\x$};
                \foreach \y in {-1,0,1,2,3}
                    \draw (3,\y) node[right] {\small $\y$};
            }
            \begin{scope}[scale=1.333,xscale=-1]
                \basethreeturn
            \end{scope}
            \draw[color3] (0,0)--(-1,0);
            \draw[color4] (-1,0)--(-5,0);
            \mynode{-1}{0}{0.6}{0.6}{}
        }$};
        \node (1_3) at (4,4){$
        \ctikz{[scale=0.333]
            \mygrid{ 
                \draw[gray!30, step=0.5, opacity=0.5] (-5,-1) grid (3,3);
                \foreach \x in {-3,-2,-1,0,1,2,3}
                    \draw (\x,3) node[above] {\small $\x$};
                \foreach \y in {-1,0,1,2,3}
                    \draw (3,\y) node[right] {\small $\y$};
            }
            \myparabola{2}{-4.5}{0}{-1.5}{2.666}
            \draw[white, fill=white] (-4.1,0)rectangle(-2.3,4);
            \draw[fill=white] (-4.1,0.9)circle[radius=0.15];
            \draw[fill=white] (-2.3,3.15)circle[radius=0.15];
            \begin{scope}[scale=1.333,xscale=-1]
                \basethreeturn
            \end{scope}
            \draw[color3] (0,0)--(-2,0);
            \draw[color4] (-2,0)--(-5,0);
            \mynode{-2}{0}{0.6}{0.6}{}
        }$};
        \node (2_3) at (4,2){$
        \ctikz{[scale=0.333]
            \mygrid{ 
                \draw[gray!30, step=0.5, opacity=0.5] (-5,-1) grid (3,3);
                \foreach \x in {-3,-2,-1,0,1,2,3}
                    \draw (\x,3) node[above] {\small $\x$};
                \foreach \y in {-1,0,1,2,3}
                    \draw (3,\y) node[right] {\small $\y$};
            }
            \begin{scope}[scale=1.333,xscale=-1]
                \basethreeturn
            \end{scope}
            \draw[color3] (0,0)--(-2,0);
            \draw[color4] (-2,0)--(-5,0);
            \mynode{-2}{0}{0.6}{0.6}{}
            \myparabola{2}{-4.5}{0}{-1.5}{2.666}
        }$};
        \node (2_2) at (2,2){$
        \ctikz{[scale=0.5]
            \mygrid{ 
                \draw[gray!30, step=0.5, opacity=0.5] (-3,-1) grid (3,3);
                \foreach \x in {-3,-2,-1,0,1,2,3}
                    \draw (\x,3) node[above] {\small $\x$};
                \foreach \y in {-1,0,1,2,3}
                    \draw (3,\y) node[right] {\small $\y$};
            }
            \myparabola{1.5}{-1.35}{0}{1.35}{0}
            \draw[white,fill=white](-0.9,0)rectangle(0.9,2);
            \draw[fill=white] (0.9,0.55)circle[radius=0.1];
            \draw[fill=white] (-0.9,0.55)circle[radius=0.1];
            \draw[color3] (0,0)--(2,0);
            \draw[color4] (0,0)--(-2,0);
            \draw[color3] (2,0) arc (-90:90:1);
            \draw[color3] (2,2)--(-2,2);
            \mynode{0}{0}{0.4}{0.4}{}
        }$};
        \node (3_1) at (0,0){$
        \ctikz{[scale=0.5]
            \mygrid{ 
                \draw[gray!30, step=0.5, opacity=0.5] (-3,-1) grid (3,3);
                \foreach \x in {-3,-2,-1,0,1,2,3}
                    \draw (\x,3) node[above] {\small $\x$};
                \foreach \y in {-1,0,1,2,3}
                    \draw (3,\y) node[right] {\small $\y$};
            }
            \draw[color3] (0,0)--(2,0);
            \draw[color4] (-2,0)--(0,0);
            \draw[color3] (2,0) arc (-90:90:1);
            \draw[color3] (2,2)--(-2,2);
            \mynode{0}{0}{0.4}{0.4}{}
        }$};
        \node (3_3) at (4,0){$
        \ctikz{[scale=0.5]
            \mygrid{ 
                \draw[gray!30, step=0.5, opacity=0.5] (-3,-1) grid (3,3);
                \foreach \x in {-3,-2,-1,0,1,2,3}
                    \draw (\x,3) node[above] {\small $\x$};
                \foreach \y in {-1,0,1,2,3}
                    \draw (3,\y) node[right] {\small $\y$};
            }
            \draw[color3] (2,0)--(0,0);
            \draw[color4] (-2,0)--(0,0);
            \draw[color3] (2,0) arc (-90:90:1);
            \draw[color3] (2,2)--(-2,2);
            \myparabola{1.5}{-1.35}{0}{1.35}{0}
            \mynode{0}{0}{0.4}{0.4}{}
        }$};
        \draw[nat] (1_1)--(1_2)node[midway, above]{\zeta_\cM^{-1}};
        \draw[nat] (1_2)--(1_3)node[midway, above]{\lambda_{\wr,\cN}^{-1}\star\lambda_{\wr,\cM}^R};
        \draw[nat] (1_3)--(2_3)node[midway, right]
        {\epsilon_0};
        \draw[nat] (2_2)--(3_3)node[midway, above]{\epsilon_0};
        \draw[nat] (1_3)--(2_2)node[midway, below right]{\chi_\cM};
        \draw[nat] (2_3)--(3_3)node[midway, right]{\chi_\cM};
        \draw[nat] (3_3)--(3_1)node[midway, below]{\omega};
        \draw[nat] (1_1)--(2_2)node[midway, below left]{\lambda_{\wr,\cN}^{-1}\star\lambda_{\wr,\cM}^R};
        \draw[nat] (2_2)--(3_1)node[midway, above left, xshift = 8pt]{\lambda_{\wr,\cN}\star(\lambda_{\wr,\cM}^R)^{-1}};
        \draw[nat] (1_1)--(3_1)node[midway, below left]{\id};
        \draw (3.2,2) node{\tiny(Theorem \ref{thm.local})};
        \draw (2,3) node{\tiny(\eqref{eq.zigzagger})};
        \draw (2,0.8) node{\tiny(\eqref{eq.intuition_uni})};
    }
\]
\[
    \ctikz{[scale=1.5,xscale=1.5,yscale=1.2]
        \mygrid{
            \draw[gray!30, step=0.5, opacity = 0.5] (0,0) grid (6,6);
            \foreach \x in {0,1,2,3,4,5,6}
                \draw (\x,6) node[above] {\small $\x$};
            \foreach \y in {0,1,2,3,4,5,6}
                \draw (6,\y) node[right] {\small $\y$};
        }
        \node (1_1) at (0,4){$
        \ctikz{[scale=0.333]
            \mygrid{ 
                \draw[gray!30, step=0.5, opacity=0.5] (-6,-1) grid (3,6);
                \foreach \x in {-3,-2,-1,0,1,2,3}
                    \draw (\x,3) node[above] {\small $\x$};
                \foreach \y in {-1,0,1,2,3}
                    \draw (3,\y) node[right] {\small $\y$};
            }
            \begin{scope}[scale=1,xscale=-1]
                \basefiveturn
            \end{scope}
            \draw[color3] (0,0)--(-1,0);
            \draw[color4] (-1,0)--(-6,0);
            \mynode{-1}{0}{0.6}{0.6}{}
            \myparabola{3}{-3}{0}{-1.25}{2}
            \myparabola{3}{-5}{0}{-2.5}{4}
        }$};
        \node (1_2) at (2,4){$
        \ctikz{[scale=0.333]
            \mygrid{ 
                \draw[gray!30, step=0.5, opacity=0.5] (-5,-1) grid (3,3);
                \foreach \x in {-3,-2,-1,0,1,2,3}
                    \draw (\x,3) node[above] {\small $\x$};
                \foreach \y in {-1,0,1,2,3}
                    \draw (3,\y) node[right] {\small $\y$};
            }
            \begin{scope}[scale=1.333,xscale=-1]
                \basethreeturn
            \end{scope}
            \draw[color3] (0,0)--(-1,0);
            \draw[color4] (-1,0)--(-5,0);
            \mynode{-1}{0}{0.6}{0.6}{}
            \myparabola{2}{-4.1}{0}{-1.65}{2.666}
            \myparabola{2.1}{-4.6}{0}{-1.15}{2.666}
        }$};
        \node (1_3) at (4,4){$
        \ctikz{[scale=0.333]
            \mygrid{ 
                \draw[gray!30, step=0.5, opacity=0.5] (-5,-1) grid (3,5);
                \foreach \x in {-5,-4,-3,-2,-1,0,1,2,3}
                    \draw (\x,5) node[above] {\small $\x$};
                \foreach \y in {-1,0,1,2,3}
                    \draw (3,\y) node[right] {\small $\y$};
            }
            \myparabola{2}{-4.1}{0}{-1.65}{2.666}
            \myparabola{2.1}{-4.6}{0}{-1.15}{2.666}
            \draw[white,fill=white] (-2,0.1)rectangle(0,4);
            \draw[white, fill=white](-5,0)rectangle(-4.1,2);
            \draw[white, fill=white] (-5,0)rectangle(-3.8,1.25);
            \draw (-4.4,0)to[out=84, in=258](-4.2,0.6);
            \draw (-4.2,0.6)to[out=90,in=245](-4,1.52);
            \draw (-4.2,0.6)to[out=0, in =255] (-3.7,1.13);
            \draw (-2.05,2.91)to[out=-30,in=235](-1.7,3);
            \draw (-2.05,3.375)to[out=-35,in=120](-1.7,3);
            \draw (-1.7,3)to[out=-45,in=120](-1.4,2.666);
            \begin{scope}[scale=1.333,xscale=-1]
                \basethreeturn
            \end{scope}
            \draw[color3] (0,0)--(-1,0);
            \draw[color4] (-1,0)--(-5,0);
            \mynode{-1}{0}{0.6}{0.6}{}
        }$};
        \node (1_4) at (6,4){$
        \ctikz{[scale=0.333]
            \mygrid{ 
                \draw[gray!30, step=0.5, opacity=0.5] (-5,-1) grid (3,3);
                \foreach \x in {-3,-2,-1,0,1,2,3}
                    \draw (\x,3) node[above] {\small $\x$};
                \foreach \y in {-1,0,1,2,3}
                    \draw (3,\y) node[right] {\small $\y$};
            }
            \begin{scope}[scale=1.333,xscale=-1]
                \basethreeturn
            \end{scope}
            \draw[color3] (0,0)--(-2,0);
            \draw[color4] (-2,0)--(-5,0);
            \mynode{-2}{0}{0.6}{0.6}{}
            \myparabola{2}{-4.5}{0}{-1.5}{2.666}
        }$};
        \node (2_1) at (0,2){$
        \ctikz{[scale=0.333]
            \mygrid{ 
                \draw[gray!30, step=0.5, opacity=0.5] (-5,-1) grid (3,3);
                \foreach \x in {-3,-2,-1,0,1,2,3}
                    \draw (\x,3) node[above] {\small $\x$};
                \foreach \y in {-1,0,1,2,3}
                    \draw (3,\y) node[right] {\small $\y$};
            }
            \begin{scope}[scale=1.333,xscale=-1]
                \basethreeturn
            \end{scope}
            \draw[color3] (0,0)--(-2,0);
            \draw[color4] (-2,0)--(-5,0);
            \mynode{-2}{0}{0.6}{0.6}{}
            \myparabola{1.5}{-3.4}{0}{-0.6}{0}
            \myparabola{2}{-4.5}{0}{-1.5}{2.666}
        }$};
        \node (2_2) at (2,2){$
        \ctikz{[scale=0.5]
            \mygrid{ 
                \draw[gray!30, step=0.5, opacity=0.5] (-3,-1) grid (3,3);
                \foreach \x in {-3,-2,-1,0,1,2,3}
                    \draw (\x,3) node[above] {\small $\x$};
                \foreach \y in {-1,0,1,2,3}
                    \draw (3,\y) node[right] {\small $\y$};
            }
            \draw[color3] (2,0)--(0,0);
            \draw[color4] (-2,0)--(0,0);
            \draw[color3] (2,0) arc (-90:90:1);
            \draw[color3] (2,2)--(-2,2);
            \myparabola{1.5}{-1}{0}{1}{0}
            \myparabola{1.5}{-1.6}{0}{1.6}{0}
            \mynode{0}{0}{0.4}{0.4}{}
        }$};
        \node (2_3) at (4,2){$
        \ctikz{[scale=0.5]
            \mygrid{ 
                \draw[gray!30, step=0.5, opacity=0.5] (-3,-1) grid (3,3);
                \foreach \x in {-3,-2,-1,0,1,2,3}
                    \draw (\x,3) node[above] {\small $\x$};
                \foreach \y in {-1,0,1,2,3}
                    \draw (3,\y) node[right] {\small $\y$};
            }
            \draw[color3] (2,0)--(0,0);
            \draw[color4] (-2,0)--(0,0);
            \draw[color3] (2,0) arc (-90:90:1);
            \draw[color3] (2,2)--(-2,2);
            \begin{scope}[scale=0.909]
                \draw (0.5,1.4) parabola (-1.65,0);
                \begin{scope}[xscale=-1]
                    \draw (0.5,1.4) parabola (-1.65,0); 
                \end{scope}
                \draw[white,fill=white] (-0.95,0.1)rectangle(0.95,1.75);
                \myparabola{0.9}{-0.75}{1.2}{0.75}{1.2}
                \myparabola{1.1}{-0.5}{0.8}{0.5}{0.8}
                \draw (-1,0.74) to[out=90, in=220](-0.75,1.2);
                \draw (-1,0.74) to[out=-30, in=220](-0.5,0.8);
                \begin{scope}[xscale=-1]
                    \draw (-1,0.74) to[out=90, in=220](-0.75,1.2);
                    \draw (-1,0.74) to[out=-30, in=220](-0.5,0.8);
                \end{scope}
            \end{scope}
            \mynode{0}{0}{0.4}{0.4}{}
        }$};
        \node (2_4) at (6,2){$
        \ctikz{[scale=0.5]
            \mygrid{ 
                \draw[gray!30, step=0.5, opacity=0.5] (-3,-1) grid (3,3);
                \foreach \x in {-3,-2,-1,0,1,2,3}
                    \draw (\x,3) node[above] {\small $\x$};
                \foreach \y in {-1,0,1,2,3}
                    \draw (3,\y) node[right] {\small $\y$};
            }
            \draw[color3] (2,0)--(0,0);
            \draw[color4] (-2,0)--(0,0);
            \draw[color3] (2,0) arc (-90:90:1);
            \draw[color3] (2,2)--(-2,2);
            \myparabola{1.5}{-1.35}{0}{1.35}{0}
            \mynode{0}{0}{0.4}{0.4}{}
        }$};
        \node (3_1) at (0,0){$
        \ctikz{[scale=0.333]
            \mygrid{ 
                \draw[gray!30, step=0.5, opacity=0.5] (-5,-1) grid (3,3);
                \foreach \x in {-3,-2,-1,0,1,2,3}
                    \draw (\x,3) node[above] {\small $\x$};
                \foreach \y in {-1,0,1,2,3}
                    \draw (3,\y) node[right] {\small $\y$};
            }
            \begin{scope}[scale=1.333,xscale=-1]
                \basethreeturn
            \end{scope}
            \draw[color3] (0,0)--(-2,0);
            \draw[color4] (-2,0)--(-5,0);
            \mynode{-2}{0}{0.6}{0.6}{}
            \myparabola{2}{-4.5}{0}{-1.5}{2.666}
        }$};
        \node (3_2) at (2,0){$
        \ctikz{[scale=0.5]
            \mygrid{ 
                \draw[gray!30, step=0.5, opacity=0.5] (-3,-1) grid (3,3);
                \foreach \x in {-3,-2,-1,0,1,2,3}
                    \draw (\x,3) node[above] {\small $\x$};
                \foreach \y in {-1,0,1,2,3}
                    \draw (3,\y) node[right] {\small $\y$};
            }
            \draw[color3] (2,0)--(0,0);
            \draw[color4] (-2,0)--(0,0);
            \draw[color3] (2,0) arc (-90:90:1);
            \draw[color3] (2,2)--(-2,2);
            \myparabola{1.5}{-1.35}{0}{1.35}{0}
            \mynode{0}{0}{0.4}{0.4}{}
        }$};
        \node (3_4) at (6,0){$
        \ctikz{[scale=0.5]
            \mygrid{ 
                \draw[gray!30, step=0.5, opacity=0.5] (-3,-1) grid (3,3);
                \foreach \x in {-3,-2,-1,0,1,2,3}
                    \draw (\x,3) node[above] {\small $\x$};
                \foreach \y in {-1,0,1,2,3}
                    \draw (3,\y) node[right] {\small $\y$};
            }
            \draw[color3] (2,0)--(0,0);
            \draw[color4] (-2,0)--(0,0);
            \draw[color3] (2,0) arc (-90:90:1);
            \draw[color3] (2,2)--(-2,2);
            \mynode{0}{0}{0.4}{0.4}{}
        }$};
        \draw[nat] (1_1)--(1_2) node[midway, above]{\chi_\cM};
        \draw[nat] (1_2)--(1_3) node[midway, above]{\alpha_{\wr,\cN}^{-1}\star\alpha_{\wr,\cM}^R};
        \draw[nat] (1_3)--(1_4) node[midway, above]{\epsilon_2};
        \draw[nat] (2_1)--(2_2) node[midway,above]{\chi_\cM};
        \draw[nat] (2_2)--(2_3) node[midway,above]{\alpha_{\wr,\cN}^{-1}\star\alpha_{\wr,\cM}^R};
        \draw[nat] (2_3)--(2_4) node[midway, above]{\epsilon_2};
        \draw[nat] (3_1)--(3_2) node[midway, below]{\chi_\cM};
        \draw[nat] (3_2)--(3_4) node[midway, below]{\omega};
        \draw[nat] (1_1)--(2_1) node[midway, left]{\chi_\cM};
        \draw[nat] (1_2)--(2_2) node[midway, left]{\chi_\cM};
        \draw[nat] (1_3)--(2_3) node[midway, left]{\chi_\cM};
        \draw[nat] (1_4)--(2_4) node[midway, right]{\chi_\cM};
        \draw[nat] (2_1)--(3_1) node[midway, left]{\omega};
        \draw[nat] (2_2)--(3_2) node[midway, left]{\omega};
        \draw[nat] (2_4)--(3_4) node[midway, right]{\omega};
        \draw (1,1) node {\tiny(Theorem \ref{thm.local})};
        \draw (1,3) node {\tiny(Theorem \ref{thm.local})};
        \draw (3,3) node {\tiny(Theorem \ref{thm.local})};
        \draw (5,3) node {\tiny(Theorem \ref{thm.local})};
        \draw (4,1) node {\tiny(\eqref{eq.intuition_asso})};
    }
\]
    \caption{}
    \label{fig.omega_to_monad}
\end{figure}

Finally, it is not hard to see that $\Phi'$ and $\Psi'$ are mutually inverses to each other, mimicking the proof of Theorem \ref{thm.higher_transpose}.
\end{proof}

\begin{proof}[Finishing the proof of Theorem \ref{thm.unenriched_mon_Kitaev_Kong}]
It suffices to show that $\Phi'\:\Lmd_{\MCM}(\ast)\to\Lmd_{\MnCM}$ constructed in the proof of Theorem \ref{thm.unenriched_Kitaev_Kong} is monoidal. To this end, let
    \[
(\ctikz{[scale=0.667]
    \draw[color3] (-1.5,-0.2)--(0,-0.2);
    \draw[color3] (-1.5,0.2)--(0,0.2);
    \mynode{0}{0}{0.3}{0.3}{P}
}
,
\ctikz{[scale=0.667]
    \mygrid{ 
        \draw[gray!30, step=0.5, opacity = 0.5] (-5,-4) grid (1,3);
        \foreach \x in {-3,-2,-1,0,1,2,3}
            \draw (\x,3) node[above] {\small $\x$};
        \foreach \y in {-1,0,1,2,3}
            \draw (3,\y) node[right] {\small $\y$};
    }
    \begin{scope}[yshift=0.2cm, scale=0.75]
        \basetwoturn
    \end{scope}
    \draw[color3] (-3,-0.2)--(0,-0.2);
    \mynode{0}{0}{0.3}{0.3}{P}
    \myparabola{2.25}{-2.8}{-0.2}{-1.75}{0.95}
}
\stackrel{\kappa_P}{\Rightarrow}
\ctikz{[scale=0.667]
    \draw[color3] (-1.5,-0.2)--(0,-0.2);
    \draw[color3] (-1.5,0.2)--(0,0.2);
    \mynode{0}{0}{0.3}{0.3}{P}
}
)
\quad\text{and}\quad
(\ctikz{[scale=0.667]
    \draw[color3] (-1.5,-0.2)--(0,-0.2);
    \draw[color3] (-1.5,0.2)--(0,0.2);
    \mynode{0}{0}{0.3}{0.3}{Q}
}
,
\ctikz{[scale=0.667]
    \mygrid{ 
        \draw[gray!30, step=0.5, opacity = 0.5] (-5,-4) grid (1,3);
        \foreach \x in {-3,-2,-1,0,1,2,3}
            \draw (\x,3) node[above] {\small $\x$};
        \foreach \y in {-1,0,1,2,3}
            \draw (3,\y) node[right] {\small $\y$};
    }
    \begin{scope}[yshift=0.2cm, scale=0.75]
        \basetwoturn
    \end{scope}
    \draw[color3] (-3,-0.2)--(0,-0.2);
    \mynode{0}{0}{0.3}{0.3}{Q}
    \myparabola{2.25}{-2.8}{-0.2}{-1.75}{0.95}
}
\stackrel{\kappa_Q}{\Rightarrow}
\ctikz{[scale=0.667]
    \draw[color3] (-1.5,-0.2)--(0,-0.2);
    \draw[color3] (-1.5,0.2)--(0,0.2);
    \mynode{0}{0}{0.3}{0.3}{Q}
}
)
\]
be objects in $\Lmd_{\MCM}(\ast)$. One can see that $\Phi'(P)\conv \Phi'(Q)$ and $\Phi'(P\conv Q)$ have underlying pro-tensors
\[
    \ctikz{[scale=0.667]
        \mygrid{ 
            \draw[gray!30, step=0.5, opacity = 0.5] (-5,-4) grid (1,3);
            \foreach \x in {-3,-2,-1,0,1,2,3}
                \draw (\x,3) node[above] {\small $\x$};
            \foreach \y in {-1,0,1,2,3}
                \draw (3,\y) node[right] {\small $\y$};
        }   
        \begin{scope}[yshift=0.2cm, scale=0.5]
            \baseoneturn
        \end{scope}
        \draw[color3] (-1.5,-0.2)--(0,-0.2);
        \mynode{0}{0}{0.3}{0.3}{Q}
        \begin{scope}[yshift=1.1cm, xshift=1cm, scale=0.5]
            \baseoneturn
        \end{scope}
        \mynode{1}{0.9}{0.3}{0.3}{P}
    }
    \quad\text{and}\quad
    \ctikz{[scale=0.667]
        \mygrid{ 
            \draw[gray!30, step=0.5, opacity = 0.5] (-5,-4) grid (1,3);
            \foreach \x in {-3,-2,-1,0,1,2,3}
                \draw (\x,3) node[above] {\small $\x$};
            \foreach \y in {-1,0,1,2,3}
                \draw (3,\y) node[right] {\small $\y$};
        }   
        \begin{scope}[yshift=0.2cm, xshift=-0.1cm,scale=0.5]
            \draw[color3] (-1,0)--(0,0);
            \draw[color3] (-1,0) arc (270:90:0.5);
            \draw[color3] (-1,1)--(0,1);
        \end{scope}
        \draw[color3] (-1.5,-0.2)--(0,-0.2);
        \mynode{0}{0}{0.3}{0.3}{Q}
        \begin{scope}[yshift=1.1cm, xshift=-0.4cm, scale=0.5]
            \baseoneturn
            \draw[color3] (1.5,1)--(2.5,1);
            \draw[color3] (0,0)--(0.5,0);
        \end{scope}
        \mynode{0}{0.9}{0.3}{0.3}{P}
    }
\]
respectively. Then one can verify that the invertible pro-tensor network map
\eqnn{\label{eq.proof.thm.unenriched_mon_Kitaev_Kong}
{\Phi'_2}_{P,Q}\:\ctikz{[scale=0.667]
    \mygrid{ 
        \draw[gray!30, step=0.5, opacity = 0.5] (-5,-4) grid (1,3);
        \foreach \x in {-3,-2,-1,0,1,2,3}
            \draw (\x,3) node[above] {\small $\x$};
        \foreach \y in {-1,0,1,2,3}
            \draw (3,\y) node[right] {\small $\y$};
    }   
    \begin{scope}[yshift=0.2cm, scale=0.5]
        \baseoneturn
    \end{scope}
    \draw[color3] (-1.5,-0.2)--(0,-0.2);
    \mynode{0}{0}{0.3}{0.3}{Q}
    \begin{scope}[yshift=1.1cm, xshift=1cm, scale=0.5]
        \baseoneturn
    \end{scope}
    \mynode{1}{0.9}{0.3}{0.3}{P}
}
\stackrel{\sim}{\Rightarrow}
\ctikz{[scale=0.667]
    \mygrid{ 
        \draw[gray!30, step=0.5, opacity = 0.5] (-5,-4) grid (1,3);
        \foreach \x in {-3,-2,-1,0,1,2,3}
            \draw (\x,3) node[above] {\small $\x$};
        \foreach \y in {-1,0,1,2,3}
            \draw (3,\y) node[right] {\small $\y$};
    }   
    \begin{scope}[yshift=0.2cm, xshift=-0.1cm,scale=0.5]
        \draw[color3] (-1,0)--(0,0);
        \draw[color3] (-1,0) arc (270:90:0.5);
        \draw[color3] (-1,1)--(0,1);
    \end{scope}
    \draw[color3] (-1.5,-0.2)--(0,-0.2);
    \mynode{0}{0}{0.3}{0.3}{Q}
    \begin{scope}[yshift=1.1cm, xshift=-0.4cm, scale=0.5]
        \baseoneturn
        \draw[color3] (1.5,1)--(2.5,1);
        \draw[color3] (0,0)--(0.5,0);
    \end{scope}
    \mynode{0}{0.9}{0.3}{0.3}{P}
}
    }
formed by the composition of $\sigma$ and $\sigma'$'s defined in Example \ref{ex.swapper} is an invertible $\MCM$-module homomorphism. Indeed, the $\MnCM$-action on $\Phi'(P)\conv\Phi'(Q)$ is given by 
\eqnn{\label{eq2.proof.thm.unenriched_mon_Kitaev_Kong}
\diagram@C=3pc{
\ctikz{[scale=0.667]
    \mygrid{ 
        \draw[gray!30, step=0.5, opacity = 0.5] (-5,-4) grid (1,3);
        \foreach \x in {-3,-2,-1,0,1,2,3}
            \draw (\x,3) node[above] {\small $\x$};
        \foreach \y in {-1,0,1,2,3}
            \draw (3,\y) node[right] {\small $\y$};
    }   
    \begin{scope}[yshift=0.2cm, scale=0.5]
        \baseoneturn
    \end{scope}
    \draw[color3] (-1.5,-0.2)--(0,-0.2);
    \mynode{0}{0}{0.3}{0.3}{Q}
    \begin{scope}[yshift=1.1cm, xshift=1cm, scale=0.5]
        \baseoneturn
    \end{scope}
    \mynode{1}{0.9}{0.3}{0.3}{P}
    \myparabola{2}{-1.25}{-0.2}{0.75}{1.6}
}
\ar@{=>}[r]^-{\eta_\wr}
&
\ctikz{[scale=0.667]
    \mygrid{ 
        \draw[gray!30, step=0.5, opacity = 0.5] (-5,-4) grid (1,3);
        \foreach \x in {-3,-2,-1,0,1,2,3}
            \draw (\x,3) node[above] {\small $\x$};
        \foreach \y in {-1,0,1,2,3}
            \draw (3,\y) node[right] {\small $\y$};
    }   
    \begin{scope}[yshift=0.2cm, scale=0.5]
        \baseoneturn
    \end{scope}
    \draw[color3] (-1.5,-0.2)--(0,-0.2);
    \mynode{0}{0}{0.3}{0.3}{Q}
    \begin{scope}[yshift=1.1cm, xshift=1cm, scale=0.5]
        \baseoneturn
    \end{scope}
    \mynode{1}{0.9}{0.3}{0.3}{P}
    \myparabola{2.25}{-1.25}{-0.2}{-0.375}{0.7}
    \myparabola{2.25}{-0.2}{0.7}{0.75}{1.6}
}
\ar@{=>}[r]^-{\chi_\cM^{-1}\star\chi_\cM^{-1}}
&
\ctikz{[scale=0.667]
    \mygrid{ 
        \draw[gray!30, step=0.5, opacity = 0.5] (-2,-1) grid (9,6);
        \foreach \x in {-2,-1,0,1,2,3,4,5,6,7,8,9}
            \draw (\x,3) node[above] {\small $\x$};
        \foreach \y in {-1,0,1,2,3,4,5,6}
            \draw (-2,\y) node {\small $\y$};
    }   
    \begin{scope}[yshift=0.2cm, scale=0.5]
        \basethreeturn
    \end{scope}
    \draw[color3] (-1.5,-0.2)--(0,-0.2);
    \mynode{0}{0}{0.3}{0.3}{Q}
    \begin{scope}[yshift=2.1cm, xshift=1.7cm, scale=0.5]
        \basethreeturn
    \end{scope}
    \mynode{1.7}{1.9}{0.3}{0.3}{P}
    \myparabola{2.25}{-1.25}{-0.2}{-0.375}{0.7}
    \myparabola{2.25}{0.5}{1.7}{1.45}{2.6}
}
\ar@{=>}[r]^-{\kappa_Q\star\kappa_P}
&
\ctikz{[scale=0.667]
    \mygrid{ 
        \draw[gray!30, step=0.5, opacity = 0.5] (-5,-4) grid (1,3);
        \foreach \x in {-3,-2,-1,0,1,2,3}
            \draw (\x,3) node[above] {\small $\x$};
        \foreach \y in {-1,0,1,2,3}
            \draw (3,\y) node[right] {\small $\y$};
    }   
    \begin{scope}[yshift=0.2cm, scale=0.5]
        \baseoneturn
    \end{scope}
    \draw[color3] (-1.5,-0.2)--(0,-0.2);
    \mynode{0}{0}{0.3}{0.3}{Q}
    \begin{scope}[yshift=1.1cm, xshift=1cm, scale=0.5]
        \baseoneturn
    \end{scope}
    \mynode{1}{0.9}{0.3}{0.3}{P}
    \myparabola{2}{-1.25}{-0.2}{0.75}{1.6}
}
}
}
while the $\MnCM$-action on $\Phi'(P\conv Q)$ is also given by \eqref{eq2.proof.thm.unenriched_mon_Kitaev_Kong} under the identification ${\Phi'_2}_{P,Q}$. Moreover, ${\Phi'_2}_{P,Q}$ is natural in $P$ and $Q$. Furthermore, it can be verified that 
\[
    \Phi'_0\defdtobe\chi_\cM\:
    \ctikz{[scale=0.667]
        \draw[color3] (-1.5,0)--(1.5,0);
    }
    \Rightarrow
    \ctikz{[scale=0.5,xscale=-1]
        \basetwoturn
    }
\]
is an invertible $\MnCM$-module morphism $J\stackrel{\sim}{\Rightarrow}\Phi'(J')$, where $J$ and $J'$ are respectively the tensor unit of $\Lmd_{\MnCM}$ and $\Lmd_{\MCM}(\ast)$. Finally, one can check that $(\{{\Phi'_2}_{P,Q}\}
,\Phi'_0)$ satisfies the hexagon and quadrandgle equalities of a monoidal funtor, thus serves as a monoidal structure of $\Phi$. This implies that $\Phi'$ is a monoidal equivalence.
\end{proof}

\bibliographystyle{JHEP}
\bibliography{biblio}
\end{document}